\documentclass{lmcs}
\pdfoutput=1

\usepackage{lastpage}
\lmcsdoi{22}{2}{29}
\lmcsheading{}{\pageref{LastPage}}{}{}%
{May~12,~2025}{Jun.~19,~2026}{}

\input{macros}


\begin{document}

\title{The calculus of neo-Peircean relations}%

\author{Filippo Bonchi\lmcsorcid{0000-0002-3433-723X}}[a]
\author{Alessandro Di Giorgio\lmcsorcid{0000-0002-6428-6461}}[b]
\author{Nathan Haydon\lmcsorcid{0000-0002-1604-9832}}[c]
\author{Pawel Sobocinski\lmcsorcid{0000-0002-7992-9685}}[b]

\address{University of Pisa, Italy}
\email{filippo.bonchi@unipi.it}

\address{Tallinn University of Technology, Estonia}
\email{alessandro.digiorgio@taltech.ee, pawel.sobocinski@taltech.ee}

\address{University of Waterloo, Canada}
\email{nhaydon@uwaterloo.ca}

\begin{abstract}
The calculus of relations was introduced by De Morgan and Peirce during the second half of the 19th century, as an extension of Boole's algebra of classes. Later developments on quantification theory by Frege and Peirce himself, paved the way to what is known today as first-order logic, causing the calculus of relations to be long forgotten.
This was until 1941, when Tarski raised the question on the existence of a complete axiomatisation for it. This question found only negative answers: there is no finite axiomatisation for the calculus of relations and many of its fragments, as shown later by several no-go theorems.
In this paper we show that -- by moving from traditional syntax (cartesian) to a diagrammatic one (monoidal) -- it is possible to have complete axiomatisations for the full calculus. The no-go theorems are circumvented by the fact that our calculus, named the calculus of neo-Peircean relations, is more expressive than the calculus of relations and, actually, as expressive as first-order logic. The axioms are obtained by combining two well known categorical structures: cartesian and linear bicategories. %

\end{abstract}

\maketitle

\tikzset{baseline=-0.5ex,>=stealth'}
\tikzset{every picture/.append style={scale=0.4}}

\section{Introduction}

The modern understanding of first-order logic ($\FOL$) %
is the result of an evolution with contributions from many philosophers and mathematicians. %
Amongst these, particularly relevant for our exposition is the calculus of relations ($\CR$) by Charles S. Peirce~\cite{peirce1897_the-logic-of-relatives}.
Peirce, inspired by De Morgan~\cite{de1860syllogism}, proposed a relational analogue of Boole's algebra~\cite{boole1847mathematical}: a rigorous mathematical language for combining relations with  operations governed by algebraic laws.

With the rise of first-order logic, Peirce's calculus was forgotten until Tarski, who in~\cite{tarski1941calculus} recognised its algebraic flavour.
In the introduction to~\cite{tarski1988formalization}, written shortly before his death, Tarski put much emphasis on two key features of $\CR$: (a) its lack of quantifiers and (b) its sole deduction rule of substituting equals by equals. The calculus, however, comes with two great shortcomings: (c) it is strictly less expressive than $\FOL$~\cite{lowenheim1915moglichkeiten} and (d) it is \emph{not} axiomatisable~\cite{monk}.

Despite these limitations, $\CR$ had ---and continues to have--- a great impact in computer science, e.g., in the theory of databases~\cite{codd1983relational} and in the semantics of programming languages~\cite{pratt1976semantical,hoare1986weakest,lassen1998relational,bird1996algebra,DBLP:journals/pacmpl/LagoG22a}.
Indeed, the lack of quantifiers avoids the usual burden of bindings, scopes and capture-avoid substitutions (see~\cite{DBLP:journals/fac/GabbayP02,pitts2013nominal,pfenning1988higher,hofmann1999semantical,fiore1999abstract,goncharov2023towards} for some theories developed to address specifically the issue of bindings). This feature, together with purely equational proofs, makes $\CR$ particularly suitable for proof assistants~\cite{pous2013kleene, pous2016automata, krauss2012regular}.

Less influential in computer science, there are two others quantifier-free alternatives to $\FOL$ that are worth mentioning: first, \emph{predicate functor logic} ($\PF$)~\cite{QUINE1971309} that was thought by Quine as the first-order logic analogue of combinatory logic~\cite{curry1958combinatory} for the $\lambda$-calculus; %
second, Peirce's \emph{existential graphs} ($\EG$s)~\cite{roberts1973_the-existential-graphs-of-charles-s.-peirce} and, in particular, its fragment named \emph{system $\beta$}. In this system $\FOL$ formulas are \emph{diagrams} and the deduction system consists of rules for their manipulation.  Peirce's work on $\EG$s remained unpublished during his lifetime.

Diagrams have been used as formal entities since the dawn of computer science, e.g.\
in the %
B\"ohm-Jacopini theorem~\cite{bohm1966flow}. More recently,
the spatial nature of mobile computations led Milner to move from the traditional term-based syntax of process calculi to bigraphs~\cite{milner2009space}. Similarly, the impossibility of copying quantum information and, more generally, the paradigm-shift of treating data as a physical resource (see e.g.~\cite{orchard2019quantitative,gaboardi2016combining}), has led to the use~\cite{BaezErbele-CategoriesInControl,DBLP:journals/pacmpl/BonchiHPSZ19,Bonchi2015,CoeckeDuncanZX2011,Fong2015,DBLP:journals/corr/abs-2009-06836,Ghica2016,DBLP:conf/lics/MuroyaCG18,Piedeleu2021,DBLP:journals/jacm/BonchiGKSZ22} of \emph{string diagrams}~\cite{joyal1991geometry,Selinger2009} as syntax.
 String diagrams, formally arrows of a freely generated  symmetric (strict) monoidal category, combine the rigour of traditional terms with a visual and intuitive graphical representation.
 Like traditional terms, they can be equipped with a compositional semantics.

\smallskip

In this paper, we introduce the calculus of \emph{neo-Peircean relations}, a string diagrammatic account of $\FOL$ that has several key features:

\begin{enumerate}
\item Its diagrammatic syntax is closely related to Peirce's $\EG$s, but it can also be given through a context free grammar  equipped with an elementary type system;
\item It is quantifier-free and, differently than $\FOL$, its compositional semantics can be given by few simple rules: see \eqref{fig:semantics};
\item Terms and predicates are not treated as separate syntactic and semantic entities;
\item Its sole deduction rule is substituting equals by equals, like $\CR$, but differently, it features a complete axiomatisation;
\item The axioms are those of well-known algebraic structures, also occurring in different fields such as linear algebra \cite{interactinghopf} or quantum foundations \cite{CoeckeDuncanZX2011};
\item It allows for compositional encodings of $\FOL$, $\CR$ and $\PF$;
\item String diagrams disambiguate interesting corner cases where traditional $\FOL$ encounters difficulties. One perk is that we allow empty models ---forbidden in classical treatments--- leading to (slightly) more general G\"odel completeness;
\item The corner case of empty models coincides with \emph{propositional} models and in that case our axiomatisation simplifies to the deep inference Calculus of Structures~\cite{DBLP:phd/de/Brunnler2003,guglielmi2007system}. %
\end{enumerate}
By returning to the algebraic roots of logic we preserve $\CR$'s benefits (a) and (b) while overcoming its limitations (c) and (d). %

\begin{figure*}
		$\begin{array}{ll}
		  \Circ{c}  \, \Coloneqq\!\!\!\!  & \copierCirc[+]       \mid
										 \discardCirc[+]      \mid
														 \boxCirc[+]{R}       \mid
														 \codiscardCirc[+]    \mid
														 \cocopierCirc[+]     \mid
														 \emptyCirc[+]        \mid
														 \idCirc[+]           \mid
														 \symmCirc[+]         \mid
														 \seqCirc[+]{c}{c}    \mid
														 \tensorCirc[+]{c}{c} \mid  \notag \\[14pt]
						& \copierCirc[-]        \mid
										  \discardCirc[-]       \mid
											\boxOpCirc[-]{R}        \mid
											\codiscardCirc[-]     \mid
											\cocopierCirc[-]      \mid
											\emptyCirc[-]         \mid
											\idCirc[-]            \mid
											\symmCirc[-]          \mid
											\seqCirc[-]{c}{c}     \mid
											\tensorCirc[-]{c}{c}
		\end{array}$
	\caption{Diagrammatic syntax of $\NPR$ %
	}\label{fig:diagsyntax}
\end{figure*}

\paragraph{Cartesian syntax} To ease the reader into this work, we show how traditional terms appear as string diagrams. Consider a (single-sorted) signature $\sign$ consisting of a unary symbol $f$ and two binary symbols $g$ and $h$. The term $h(\,g(f(x_3),f(x_3)), x_1\,)$ corresponds to the string diagram on the left below.
\[ 
    \InputIfFileExists{cartesianTerm.tikz}{}{\input{tikz/cartesianTerm.tikz}}
 \qquad \qquad 
    \InputIfFileExists{cartesianTerm2.tikz}{}{\input{tikz/cartesianTerm2.tikz}}
 \]
A difference w.r.t. traditional syntax tree is the explicit treatment of copying and discarding. The discharger $\scalebox{0.8}{\discardCirc[+]}$ informs us that the variable $x_2$ does not appear in the term; the copier $\scalebox{0.8}{\copierCirc[+]}$ makes clear that the variable $x_3$ is shared by two sub-terms. The string diagram on the right represents the same term if one admits the equations
\begin{equation}\label{eq:natintro}\tag{Nat}

    \InputIfFileExists{axioms/cb/plus/copierLaxNat1.tikz}{}{\input{tikz/axioms/cb/plus/copierLaxNat1.tikz}}
 = 
    \InputIfFileExists{axioms/cb/plus/copierLaxNat2.tikz}{}{\input{tikz/axioms/cb/plus/copierLaxNat2.tikz}}
\; \text{ and }\; 
    \InputIfFileExists{axioms/cb/plus/discardLaxNat.tikz}{}{\input{tikz/axioms/cb/plus/discardLaxNat.tikz}}
 = \discardCirc[+]\text{ .}
\end{equation}
Fox~\cite{fox1976coalgebras} showed that \eqref{eq:natintro} together with axioms asserting that copier and discard form a \emph{comonoid} (\eqref{ax:comPlusAssoc}, \eqref{ax:comPlusUnit}, \eqref{ax:comPlusComm} in Fig.~\ref{fig:cb axioms}) force the monoidal category of string diagrams to be  \emph{cartesian} ($\otimes$ is the categorical product): actually, it is the \emph{free} cartesian category on $\sign$, as a monoidal category. %

\paragraph{Functorial semantics} The work of Lawvere~\cite{LawvereOriginalPaper} illustrates the deep connection of syntax with semantics, explaining why cartesian syntax is so well-suited to functional structures, but also hinting at its limitations %
when denoting other structures, e.g. relations. Given an algebraic theory $\T{T}$ in the universal algebraic sense, i.e., a signature $\sign$ with a set of equations $E$, one can freely generate a cartesian category $\Cat{L}_{\T{T}}$. %
\emph{Models} --in the standard algebraic
sense-- are in one-to-one correspondence with cartesian functors $\mathcal{M}$ from  $\Cat{L}_{\T{T}}$ to $\Cat{Set}$, the category of sets and functions. More generally, models of the theory in any cartesian category $\Cat{C}$ are cartesian functors $\mathcal{M}\colon \Cat{L}_{\T{T}} \to \Cat{C}$.
By taking $\Cat{C}$ to be $\Relp$, the category of sets and relations, one could wish to use the same approach for relational theories but any such attempt would raise a fundamental issue: the categorical product in $\Relp$ is not given by cartesian product of sets ($\times$), but rather by their disjoint union ($+$). Intuitively,  any cartesian functor $\mathcal{M}\colon \Cat{L}_{\T{T}} \to \Relp$ maps the binary symbol $g$ from the example above into a relation of type $X+X \to X$ for some set $X$, rather than, as one would desire, a relation of type  $X \times X \to X$.

\paragraph{Cartesian bicategories} An evolution of Lawvere's approach for relational structures is developed in~\cite{DBLP:journals/corr/abs-1711-08699,GCQ,seeber2020logical}. Departing from cartesian syntax, it uses string diagrams generated by the \emph{first} row of the grammar in Fig.~\ref{fig:diagsyntax},
where $R$ is taken from a monoidal signature $\sign$ -- a set of symbols equipped with both an arity and also a \emph{coarity} -- and can be thought of as akin to relation symbols of $\FOL$. The diagrams are subject to the laws of cartesian bicategories~\cite{carboni1987cartesian} in Fig.~\ref{fig:cb axioms}: $\scalebox{0.8}{\copierCirc[+]}$ and $\scalebox{0.8}{\discardCirc[+]}$ form a comonoid, but the category of diagrams is not cartesian since the equations in \eqref{eq:natintro} hold laxly (\eqref{ax:comPlusLaxNat}, \eqref{ax:discPlusLaxNat}). The diagrams $\scalebox{0.8}{\cocopierCirc[+]}$ and $\scalebox{0.8}{\codiscardCirc[+]}$ form a \emph{monoid} (\eqref{ax:monPlusAssoc}, \eqref{ax:monPlusUnit}, \eqref{ax:monPlusComm}) and are right \emph{adjoint} to copier and discard. Furthermore, monoids and comonoids together satisfy \emph{special Frobenius} equations (\eqref{ax:plusSpecFrob},\eqref{ax:plusFrob}). The category of diagrams $\CB$ is the free cartesian bicategory generated by $\sign$ and, like in Lawvere's functorial semantics, models are morphisms of cartesian bicategories $\mathcal{M}\colon \CB \to \Relp$. Importantly, the laws of cartesian bicategories provide a complete axiomatisation for $\Relp$, meaning that $c,d$ in $\CB$ are provably equal with the laws of cartesian bicategories iff $\mathcal{M}(c)=\mathcal{M}(d)$ for all models $\mathcal{M}$.

The (co)monoid structures allow one to express existential quantification: for instance, the $\FOL$ formula $\exists x_2. P(x_1,x_2) \wedge Q(x_2)$ is depicted as the diagram below. 
\[
    \InputIfFileExists{regularFormula.tikz}{}{\input{tikz/regularFormula.tikz}}
\]
The expressive power of $\CB$ is, however, limited to the existential-conjunctive fragment of $\FOL$. %

\paragraph{Cocartesian bicategories} To express the universal-disjunctive fragment, we consider the category $\CCB$  of string diagrams generated by the \emph{second} row of the grammar in Fig.~\ref{fig:diagsyntax} %
and subject to the laws of cocartesian bicategories in Fig.~\ref{fig:cocb axioms}:  those of cartesian bicategories but with the reversed order $\geq$. The diagrams of $\CCB$ are photographic negative of those in $\CB$.
 To explain this change of colour, note that sets and relations form \emph{another} category: $\Relm$.  Composition $\seq[-]$ in $\Relm$ is the De Morgan dual of the usual relational composition: 
 \[ R\seq[+]S \defeq \{(x,z) \mid \exists y . (x,y)\in R \wedge (y,z)\in S\}\]
  but
\[ R\seq[-]S \defeq \{(x,z) \mid \forall y . (x,y)\in R \vee (y,z)\in S\}.\]
 While $\Relp$ is a cartesian bicategory, $\Relm$ is \emph{cocartesian}. Interestingly, the ``black'' composition $\seq[-]$ was used in Peirce's approach~\cite{peirce1883_studies-in-logic.-by-members-of-the-johns-hopkins-university} to relational algebra.%

\medskip

Just as $\CB$ is complete with respect to $\Relp$, dually, $\CCB$ is complete w.r.t. $\Relm$. The former accounts for the existential-conjunctive fragment of $\FOL$; the latter for its universal-disjunctive fragment. This raises a natural question:

\smallskip
\begin{center}
	How do white and black structures combine into a complete account of first-order logic?
\end{center}

\paragraph{Linear bicategories} %
Although $\Relp$ and $\Relm$ have the same objects and arrows, there
are two different compositions ($\seq[+]$ and $\seq[-]$). The appropriate categorical structures to deal with these situations are \emph{linear bicategories} introduced in \cite{cockett2000introduction} as a horizontal categorification of linearly distributive categories \cite{de1991dialectica,cockett1997weakly}. The laws of linear bicategories are in Fig.~\ref{fig:closed lin axioms}: the key law is \emph{linearly distributivity} of $\seq[+]$ over $\seq[-]$ (\eqref{ax:leftLinDistr}, \eqref{ax:rightLinDistr}), that was already known to hold for relations since the work of Peirce \cite{peirce1883_studies-in-logic.-by-members-of-the-johns-hopkins-university}.
Another crucial property observed by Peirce is that for any $R\subseteq X \times Y$, the relation $\rla{R}\subseteq Y \times X\defeq \{(y,x) \mid (x,y) \notin R\}$ is its \emph{linear adjoint}. This operation has an intuitive graphical depiction: given $\Circ{c}$, take its mirror image $\CircOp[+]{c}$ and then its photographic negative $\CircOp[-]{c}$. For instance, the linear adjoint of $\boxCirc[+]{R}$ is $\boxOpCirc[-]{R}$.

\paragraph{First-order bicategories} The final step is to characterise how cartesian, cocartesian and linear bicategories combine: %
\textit{(i)} white and black (co)monoids are linear adjoints that \textit{(ii)} satisfy a ``linear'' version of the Frobenius law. We dub the result \emph{first-order bicategories}.
	We shall see that this is a complete axiomatisation for first-order logic, yet all of the algebraic machinery is compactly summarised %
in the following picture, named ``the Tao of Logic''.
\[
		\xymatrix@C=12pt@R=12pt{
			\copierCirc[+]\;\; \discardCirc[+]  \ar@{<->}[dd]^{\text{lin. adj.}} \ar[rr]_{\text{left adj.}} & &
			\cocopierCirc[+] \; \;  \codiscardCirc[+]    \ar@{<->}[dd]_{\text{lin. adj.}}    \ar@{--}@(ul,ur)[ll]_{\text{spec. Frob.}}
			\\ \\
				\cocopierCirc[-] \; \;  \codiscardCirc[-] \ar@{--}@/^1pc/[uu]^{\text{lin. Frob.}} &&
				\copierCirc[-]\;\; \discardCirc[-] \ar[ll]_{\text{right adj.}}  \ar@{--}@(dl,dr)[ll]^{\text{spec. Frob.}} \ar@{--}@/_1pc/[uu]_{\text{lin. Frob.}}
		}
\]

\paragraph{Functorial semantics for first-order theories} In the spirit of functorial semantics, we take the free first-order bicategory $\LCB[\T{T}]$ generated by a theory $\T{T}$ and observe that models of $\T{T}$ in a first-order bicategory $\Cat{C}$ are
morphisms %
$\mathcal{M}\colon \LCB[\T{T}] \to \Cat{C}$. Taking $\Cat{C}=\Rel$, the first-order bicategory of sets and relations, these are models in the sense of $\FOL$ with one notable exception: in $\FOL$ models with the empty domain are forbidden. %
As we shall see, theories with empty models are exactly the propositional theories. %

\paragraph{Completeness}  We prove that the laws of first-order bicategories provide a complete axiomatisation for first-order logic. The proof is a diagrammatic adaptation of Henkin's proof \cite{henkin_1949} of G\"odel's completeness theorem. However, in order to properly consider models with an empty domain, we make a slight additional step to go beyond G\"odel completeness.

\paragraph{A taste of diagrammatic logic} Before we introduce the calculus of neo-Peircean relations, we start with a short worked example to give the reader a taste of using the calculus to prove a non-trivial result of first-order logic. Doing so lets us illustrate the methodology of proof within the calculus, which is sometimes referred to as diagrammatic reasoning or string diagram surgery.

Let $R$ be a symbol with arity $2$ and coarity $0$. %
The two diagrams below correspond to $\FOL$ formulas
$\exists x.\,\forall y.\,R(x,y)$ and
$\forall y.\,\exists x.\,R(x,y)$: see \S~\ref{sec:fol} for a dictionary of translating between $\FOL$ and diagrams. %
\[
	
    \InputIfFileExists{existsForallR.tikz}{}{\input{tikz/existsForallR.tikz}}
 \seminclusion 
    \InputIfFileExists{forallExistsR.tikz}{}{\input{tikz/forallExistsR.tikz}}

\]
It is well-known that $\exists x.\,\forall y.\,R(x,y) \models \forall y.\,\exists x.\,R(x,y)$, i.e.\ in any model, if the first formula evaluates to true then so does the second. Within our calculus, this statement is expressed as the above inequality.
This can be proved by mean of the axiomatisation we introduce in this work: %
\input{tikz/proofs/forallExists.tex}
The two central steps (Prop.~\ref{prop:maps}) rely on the particularly good behaviour of \emph{maps}, intuitively those relations that are functional. In particular $\scalebox{0.8}{\discardCirc[+]}$ is an  example. The details are not important at this stage.

\paragraph{Synopsis} We begin by recalling $\CR$ in \S~\ref{sec:calculusrelations}. The calculus of neo-Peircean relations is introduced in \S~\ref{sec:NPR}, together with the statement of our main result (Theorem~\ref{thm:completeness}). We recall (co)cartesian and linear bicategories in \S~\ref{sec:cartesianbi} and \S~\ref{sec:linbic}. The categorical structures most important for our work are first-order bicategories, introduced in \S~\ref{sec:fobic}. In \S~\ref{sec:theories} we consider first-order theories, the diagrammatic version of the deduction theorem (Theorem~\ref{th:deduction}) and some subtle differences with $\FOL$ that play an important role in the proof of completeness in \S~\ref{sec:completeness}.
In \S~\ref{sec:fol} we discuss the relationship of $\NPR$ with $\FOL$, while in \S~\ref{sec:encodings} with $\CR$, $\PF$ and $\EG$s. In \S~\ref{app:peirce}, we give a closer look to the work of Peirce, illustrating how fo-bicategories provide solid categorical foundations to most of his intuitions.

The reader can find in Appendix~\ref{app:additional} two figures reporting the axioms of the calculus of neo-Peircean relations: one in forms of terms and the other in form of diagrams. Appendix~\ref{app:dictionary} summarises the correspondence between term-based and diagrammatic notation, as well as, all the various diagrammatic conventions adopted through the paper. The remaining appendices are devoted to technical proofs.  
This paper is an extended version of~\cite{DBLP:conf/lics/Bonchi0H024}: it additionally contains proofs and detailed comparisons with the work of Quine and Peirce.

\section{The Calculus of Binary Relations}\label{sec:calculusrelations}
The calculus %
of binary relations, in the original presentation given by Peirce~\cite{peirce1883_studies-in-logic.-by-members-of-the-johns-hopkins-university}, features two forms of relational compositions $\seq[+][][]$ and  $\seq[-][][]$, defined for all relations $R\subseteq X \times Y$ and $S \subseteq Y \times Z$ as
\begin{equation}\label{eq:seqRel}
   \begin{aligned}
      \seq[+][R][S] &\defeq \{(x,z) \mid \exists y\!\in\! Y \,.\, (x,y)\in R  \wedge (y,z)\in S\}\subseteq X\times Z \; \text{and} \\
      \seq[-][R][S] &\defeq \{(x,z) \mid \forall y\!\in\! Y \,.\, (x,y)\in R \vee (y,z)\in S\}\subseteq X\times Z
   \end{aligned}
   \end{equation}
with units the equality and the difference relations respectively, defined for all sets $X$ as
\begin{equation}\label{eq:idRel}
      \id[+][X]\! \defeq \! \{(x,y) \!\mid\! x=y\}\!\subseteq\! X \times X \text{ and }
      \id[-][X] \!\defeq \! \{(x,y) \!\mid \! x \neq y\}\! \subseteq\! X \times X \text{.}
\end{equation}
Beyond the usual %
union $\cup$, intersection $\cap$, and their units $\bot$ and $\top$, the calculus also features two unary operations $\op{(\cdot)}$ and $\nega{(\cdot)}$ denoting the opposite and the complement:
\begin{equation}\label{def:opcom}\op{R} \defeq \{ (y,x) \mid (x,y)\in R \}\subseteq Y\times X \text{ and }\nega{R} \defeq \{ (x,y) \mid (x,y) \notin R \}\subseteq X \times Y\text{.}\end{equation}
In summary, its syntax is given by the following context free grammar
\begin{equation}\tag{$\CRS$} \label{eq:calculusofrelation}
   \begin{array}{rc c@{\;\;\; \mid \;\;}c@{\;\; \mid \;\;\;}c@{\;\;\; \mid \;\;}c@{\;\; \mid \;\;\;}c@{\;\;\; \mid \;\;\;}c}
   E & \Coloneqq & R      & \id[+] & \seq[+][E][E] & \id[-] &  \seq[-][E][E] & \\
     &     & \op{E} & \top   & E \cap E      & \bot   &  E \cup E      & \nega{E}
   \end{array}
   \end{equation}
where $R$ is taken from a given set $\Sigma$ of generating symbols.
The semantics is defined w.r.t. a  \emph{relational interpretation} $\interpretation$, that is, a set $X$ together with a binary relation $\rho(R)\subseteq X \times X$ for each $R\in \Sigma$.
The following inductive definition assigns a binary relation to each expression. %
\begin{equation}\label{eq:sematicsExpr}
   \hspace*{-1.6em}
   {\small
   \begin{array}{r@{\,}c@{\,}l r@{\,}c@{\,}l r@{\,}c@{\,}l r@{\,}c@{\,}l}
      \dsemRel{\id[+]} & \defeq & \id[+][X] &
      \dsemRel{E_1 \seq[+] E_2} & \defeq & \dsemRel{E_1} \seq[+] \dsemRel{E_2} &
      \dsemRel{\top} & \defeq & X \times X &
      \dsemRel{E_1 \cap E_2} & \defeq & \dsemRel{E_1} \cap \dsemRel{E_2} \\
      \dsemRel{\id[-]} & \defeq & \id[-][X] &
      \dsemRel{E_1 \seq[-] E_2} & \defeq & \dsemRel{E_1} \seq[-] \dsemRel{E_2} &
      \dsemRel{\bot} & \defeq & \varnothing &
      \dsemRel{E_1 \cup E_2} & \defeq & \dsemRel{E_1} \cup \dsemRel{E_2} \\
      \multicolumn{12}{c}{
         \begin{array}{c@{\qquad}c@{\qquad}c}
            \dsemRel{R} \defeq \rho(R) &
            \dsemRel{\op{E}} \defeq \op{\dsemRel{E}} &
            \dsemRel{\nega{E}} \defeq \nega{\dsemRel{E}} \\
         \end{array}
      }
   \end{array}
   }
\end{equation}
Two expressions $E_1$, $E_2$ are said to be \emph{equivalent}, written $E_1 \equiv_{\CR}E_2$, if and only if $\dsemRel{E_1} = \dsemRel{E_2}$, for all interpretations $\interpretation$. Inclusion, denoted by $\minorExpression$, is defined analogously by replacing $=$ with $\subseteq$.
For instance, the following inclusions hold, witnessing the fact that $\seq[+][][]$ \emph{linearly distributes} over $\seq[-][][]$.
\begin{equation}\label{eq:distributivityExpres}
\seq[+][R][(\seq[-][S][T])] \minorExpression  (R\seq[+][][] S) \seq[-][][] T  \qquad \seq[+][(\seq[-][R][S])][T] \minorExpression  \seq[-][R][(\seq[+][S][T])]
\end{equation}
Along with the boolean laws, in `Note B' \cite{peirce1883_studies-in-logic.-by-members-of-the-johns-hopkins-university} Peirce states \eqref{eq:distributivityExpres} and stresses its importance. However, since $\seq[-][R][S] \equiv_{\CR} \nega{\seq[+][\nega{R}][\nega{S}]}$ and $\id[-] \equiv_{\CR} \nega{\id[+]}$, both $\seq[-]$ and $\id[-]$ are often considered redundant, for instance ommited by Tarski~\cite{tarski1941calculus} and in much of the later work. %

Tarski asked whether $\equiv_{\CR}$ can be axiomatised, i.e., is there a basic set of laws from which one can prove all the valid equivalences? Unfortunately, %
there is no finite complete axiomatisations for the whole calculus~\cite{monk} nor for several fragments, e.g.,~\cite{hodkinson2000axiomatizability,redko1964defining,freyd1990categories,doumane2020non,DBLP:conf/stacs/Pous18}. %

Our work returns to the same problem, but from a radically different perspective: we see the calculus of relations as a sub-calculus of a more general system for arbitrary (i.e.\ not merely binary) relations.
The latter is strictly more expressive than $\CRS$ -- actually it is as expressive as first-order logic ($\FOL$)-- but allows for an elementary complete axiomatisation based on the interaction of two influential algebraic structures%
: that of linear bicategories and cartesian bicategories. %

\begin{table*}[ht]{
	\resizebox{\textwidth}{!}{
	\tiny
	\begin{tabular}{@{}c@{}}
		\toprule
		$
		\begin{array}{@{}c}
			\begin{array}{c@{\qquad\quad}c@{\qquad\quad}c@{\qquad\quad}c@{\qquad\quad}c@{\qquad\quad}c@{\qquad\quad}c}
				{\id[][0]\colon 0 \to 0}
				&
				{\id[][1] \colon 1 \to 1}
				&
				{\symm[][1][1] \colon 2 \to 2}
				&
				\copier[][1]\colon 1 \to 2
				&
				\discard[][1]\colon 1 \to 0
				&
				\cocopier[][1]\colon 2 \to 1
				&
				\codiscard[][1]\colon 0 \to 1
			\end{array}
			\\
			\begin{array}{ccc}
				\inferrule{\ari(R) =n \and \coar(R)=n}{R^\circ \colon n \to m} \qquad \inferrule{\ari(R) =n \and \coar(R)=m}{R^\bullet \colon m \to n}
				&
				\inferrule{c \colon n_1 \to m_1 \and d \colon n_2 \to m_2}{\tensor[][c][d] \colon n_1 + n_2 \to m_1 + m_2}
				&
				\inferrule{c \colon n \to m \and d \colon m \to o}{\seq[][c][d] \colon n \to o}
			\end{array}
		\end{array}
		$
		\\
		\midrule
		$\arraycolsep=3pt
		\begin{array}{@{}llll}
			\begin{array}{l}
				\id[][0]=\id[][0] \\
				\id[][1+n]=\id[][1] \tensor[] \id[][n] \\
			\end{array}
			&
			\begin{array}{l}
				\symm[][0][0]=\id[][0]
				\\
				\symm[][1][1+1]=(\symm[][1][n] \tensor[] \id[][1]) \seq[](\id[][n] \tensor[] \symm[][1][1])
			\end{array}
			&
			\multicolumn{2}{l}{
				\begin{array}{l}
					\symm[][1][0]=\symm[][0][1]=\id[][1] \\
					\symm[][m+1][n]=(\id[][1] \tensor[] \symm[][m][n]) \seq[] (\symm[][1][n] \tensor[] \id[][m])
				\end{array}
			}
			\\
			\begin{array}{l}
				\copier[][0]=\id[][0]
				\\
				\copier[][1+n]=(\copier[][1]\tensor[]\copier[][n] )\seq[] (\id[][1]\tensor[] \symm[][1][n] \tensor[] \id[][n])
			\end{array}
			&
			\begin{array}{l}
				\cocopier[][0]=\id[][0]
				\\
				\cocopier[][1+n]= (\id[][1]\tensor[] \symm[][1][n] \tensor[] \id[][n]) \seq[] (\cocopier[][1]\tensor[]\cocopier[][n] )
			\end{array}
			&
			\begin{array}{l}
				\discard[][0]=\id[][0]
				\\
				\discard[][1+n]=\discard[][1]\tensor[] \discard[][n]
			\end{array}
			&
			\begin{array}{l}
				\codiscard[][0]=\id[][0]
				\\
				\codiscard[][n+1]= \codiscard[][1]\tensor[] \codiscard[][n]
			\end{array}
		\end{array}$
		\\
		\midrule
		\begin{tabular}{cccc}
			\multicolumn{4}{c}{
				\begin{tabular}{c@{\quad\qquad}c@{\quad\qquad}c}
					$\seq[][a][(\seq[][b][c])] = \seq[][(\seq[][a][b])][c]$
					&
					$\seq[][\id[][n]][c]=c=\seq[][c][\id[][m]]$
					&
					$\seq[][(\tensor[][a][b])][ (\tensor[][c][d])] = \tensor[][(\seq[][a][c] )][(\seq[][b][d])]$
				\end{tabular}
			}
		\\
		$\tensor[][a][(\tensor[][b][c])] = \tensor[][(\tensor[][a][b])][c]$
		&
		$\tensor[][\id[][0]][c] = c = \tensor[][c][\id[][0]] $
		&
		$(c \tensor[][][] \id[][o]) \seq[][][] \symm[][m][o]  = \symm[][n][o] \seq[][][] (\id[][o] \tensor[][][] c)$
		&
		$\symm[][1][1] \seq[][][] \symm[][1][1] = \id[][2]$
		\end{tabular}
		\\
		\bottomrule
	\end{tabular}
	}
	}
	\caption{Typing rules (top); inductive definitions of syntactic sugar (middle); structural congruence (bottom)}
	\label{fig:typingrules}
	\label{fig:sugar}
	\label{fig:freestricmmoncatax}
\end{table*}

\section{Neo-Peircean Relations}\label{sec:NPR} %
Here we introduce the calculus of \emph{neo-Peircean relations} ($\NPR$). %

The first step is to move from binary relations $R\subseteq X \times X$ to relations $R\subseteq X^n \times X^m$ where, for any $n \in \nat$, $X^n$ denotes the set of row vectors $(x_1, \dots, x_n)$ with all $x_i\in X$. In particular, $X^0$ is the one element set $\singleton\defeq\{\star\}$. Considering these kinds of relations allows us to identify two novel fundamental constants: the \emph{copier} $\copier[+][X] \subseteq X \times X^2$, which is the diagonal function $\langle \id[+][X], \id[+][X]\rangle \colon X \to X \times X$ considered as a relation, and the \emph{discharger} $\discard[+][X] \subseteq X \times \singleton$ which is, similarly, the unique function from $X$ to $\singleton$. By combining them with opposite and complement we obtain, in total, 8 basic relations.
\begin{equation}\label{eq:comonoidsREL}
	\hspace*{-1em}
		\begin{tabular}{rcl rcl}
			$\copier[+][X]$   & $\!\!\!\!\!\defeq\!\!\!\!\!$ & $\{(x, \; (y,z)) \mid x=y \wedge x=z\} \subseteq X \times (X \times X)$  	    &  $\discard[+][X]$   & $\!\!\!\!\!\defeq\!\!\!\!\!$ & $\{(x, \star) \mid x\in X\} \subseteq X \times \singleton$ \\
			$\cocopier[+][X]$ & $\!\!\!\!\!\defeq\!\!\!\!\!$ & $\{((y,z),\; x) \mid x=y \wedge x=z\} \subseteq (X \times X) \times X$         &  $\codiscard[+][X]$ & $\!\!\!\!\!\defeq\!\!\!\!\!$ &  $\{(\star,x) \mid x\in X\} \subseteq \singleton \times X$ \\
			\midrule
			$\copier[-][X]$   & $\!\!\!\!\!\defeq\!\!\!\!\!$ & $\{(x, \; (y,z)) \mid x\neq y \vee x \neq z\} \subseteq X \times (X \times X)$   &  $\discard[-][X]$   & $\!\!\!\!\!\defeq\!\!\!\!\!$ & $\varnothing  \subseteq X \times \singleton$ \\
			$\cocopier[-][X]$ & $\!\!\!\!\!\defeq\!\!\!\!\!$ & $\{((y,z), \; x  ) \mid x\neq y \vee x \neq z\} \subseteq (X \times X) \times X$ &  $\codiscard[-][X]$ & $\!\!\!\!\!\defeq\!\!\!\!\!$ & $\varnothing \subseteq \singleton \times X$
		\end{tabular}
\end{equation}
Together with the compositions $\seq[+]$ and $\seq[-]$ and the identities $\id[+][X]$ and $\id[-][X]$ from~\eqref{eq:seqRel} and~\eqref{eq:idRel}, there are black and white \emph{symmetries}:
\begin{equation}\label{eq:symmetries}
\begin{tabular}{rcl}
	$\symm[+][X][Y]$ & $\defeq$ & $\{(\;(x_1,y_1), (y_2,x_2)\;) \mid x_1=x_2 \wedge y_1=y_2\} \subseteq (X \times Y) \times (Y\times X)$\\
	$\symm[-][X][Y]$ & $\defeq$ & $ \{(\;(x_1,y_1), (y_2,x_2)\;) \mid x_1\neq x_2 \vee y_1 \neq y_2\} \subseteq (X \times Y) \times (Y\times X)$ \text{.}
	\end{tabular}
\end{equation}
The calculus does \emph{not} feature the boolean operators nor the opposite and the complement: these can be derived using the above structure and two \emph{monoidal products} $\tensor[+]$ and $\tensor[-]$, defined for $R\subseteq X \times Y$ and $S \subseteq V \times W$  as
\begin{equation}\label{eq:tensorREL}
	\begin{tabular}{rcl}
	$R \tensor[+]S$ & $\defeq$ & $\{ ( \,(x,v), (y,w) \,) \mid (x,y)\in R \wedge (v,w) \in S \} \subseteq (X \times V) \times (Y\times W)$\\
	$R \tensor[-]S$ & $\defeq$ & $\{ ( \,(x,v), (y,w) \,) \mid (x,y)\in R \vee (v,w) \in S \}\subseteq (X \times V) \times (Y\times W)$ \text{.}
	\end{tabular}
\end{equation}

\noindent\paragraph{Syntax}
Terms are defined by the following context free grammar
\begin{equation}\tag{$\NPR$}\label{eq:syntax} %
	\begin{array}{rc c@{\;\; \mid \;\;}c@{\;\; \mid \;\;}c@{\;\; \mid \;\;}c@{\;\; \mid \;\;}c@{\;\; \mid \;\;}c@{\;\; \mid \;\;}c@{\;\; \mid \;\;}c@{\;\; \mid \;\;}c@{\;\; \mid \;\;}c}
		c  & \Coloneqq  & \copier[+][1] & \discard[+][1] & R^\circ & \codiscard[+][1] & \cocopier[+][1] & \id[+][0] & \id[+][1] & \symm[+][1][1] & c \seq[+] c & c \tensor[+] c \mid \\
		   &            & \copier[-][1] & \discard[-][1] & R^\bullet & \codiscard[-][1] & \cocopier[-][1] & \id[-][0] & \id[-][1] & \symm[-][1][1] & c \seq[-] c & \!\!\! c \tensor[-] c \\
	  \end{array}
  \end{equation}
where $R$, like in $\CRS$, belongs to a fixed set $\sign$ of \emph{generators}. Differently than in $\CRS$, each $R\in \sign$ comes with two natural numbers: arity $\ari(R)$ and coarity $\coar(R)$. The tuple $(\sign, \ari, \coar)$, hereafter referred simply as $\sign$, is a \emph{monoidal signature}. Intuitively, every $R\in \sign$ represents some relation $R\subseteq X^{\ari(R)} \times X^{\coar(R)}$. %

In the first row of~\eqref{eq:syntax} there are eight constants and two operations: white composition ($\seq[+]$) and white monoidal product ($\tensor[+]$). These, together with identities ($\id[+][0]$ and $\id[+][1]$) and symmetry ($\symm[+][1][1]$) are typical of symmetric monoidal categories. Apart from $R^{\switchLabelS{+}}$, the constants are the copier ($\copier[+][1]$), discharger ($\discard[+][1]$) and their opposite cocopier ($\cocopier[+][1]$) and codischarger ($\codiscard[+][1]$). The second row contains the ``black'' versions of the same constants and operations. %
Note that our syntax does not have variables, no quantifiers, nor the usual associated meta-operations like capture-avoiding substitution.

We shall refer to the terms generated by the first row as the \emph{white fragment}, while to those of second row as the \emph{black fragment}. Sometimes, we use the gray colour to be agnostic w.r.t. white or black. The rules in top of Table \ref{fig:typingrules} assigns to each term at most one type $n \to m$. We consider only those terms that can be typed. For all $n,m \in \nat$, $\id[][n]\colon n \to n$, $\symm[][n][m] \colon n+m \to m+n$, $\copier[][n]\colon n \to n+n$, $\cocopier[][n]\colon n+n \to n$, $\discard[][n]\colon n \to 0$ and $\codiscard[][n]\colon 0 \to n$ are inductively defined as in middle of Table~\ref{fig:sugar}. %

\paragraph{Semantics}
As for $\CRS$, the semantics of $\NPR$ needs an interpretation $\interpretation=(X,\rho)$: a set $X$, the \emph{semantic domain}, and $\rho(R) \subseteq X^{\ari(R)} \times X^{\coar(R)}$ for each $R\in\Sigma$. The semantics of terms is defined inductively as follows.%
\begin{equation}\label{fig:semantics}\arraycolsep=1.4pt%
	{%
	\begin{array}{l@{\qquad}l@{\qquad}l@{\qquad}l}
		\interpretationFunctor ( \copier[][1] ) \defeq \copier[][X]
		&
		\interpretationFunctor ( \discard[][1] ) \defeq \discard[][X]
		&
		\interpretationFunctor ( \cocopier[][1] ) \defeq \cocopier[][X]
		&
		\interpretationFunctor ( \codiscard[][1] ) \defeq \codiscard[][X]
		\\
		\interpretationFunctor ( \id[][0] ) \defeq \id[][\singleton]
		&
		\interpretationFunctor ( \id[][1] ) \defeq \id[][X]
		&
		\interpretationFunctor ( \symm[][1][1] ) \defeq \symm[][X][X]
		&
		\interpretationFunctor (R^{\switchLabelS{+}} ) \defeq \rho(R)
		\\
		\multicolumn{3}{l}{
			\begin{array}{l@{\qquad}r}
				\!\interpretationFunctor ( c \!\seq[]\! d ) \defeq \interpretationFunctor ( c) \! \seq[] \! \interpretationFunctor ( d )
				&
				\interpretationFunctor ( c \!\tensor[]\! d ) \defeq \interpretationFunctor ( c) \! \tensor[] \! \interpretationFunctor ( d )
			\end{array}
		}
		&
		\interpretationFunctor (R^{\switchLabelS{-}} ) \defeq \op{\nega{\rho(R)}} %
	\end{array}
	}
\end{equation}

\noindent 
 The constants and operations appearing on the right-hand-side of the above equations are amongst those defined in \eqref{eq:seqRel}, \eqref{eq:idRel}, \eqref{def:opcom}, \eqref{eq:comonoidsREL},  \eqref{eq:symmetries} and \eqref{eq:tensorREL}. %
A simple inductive argument confirms that $\interpretationFunctor$ maps terms $c$ of type $n \to m$ to relations $R \subseteq X^n \times X^m$.
\begin{rem}
In particular, $\id[][0] \colon 0 \to 0$ is sent to $\id[][\singleton]\subseteq \singleton \times \singleton$, since $X^0=\singleton$ independently of $X$. Note that there are only two relations on the singleton set $\singleton=\{\star\}$: the relation $\{(\star,\star)\} \subseteq \singleton \times \singleton$ and the empty relation $\varnothing\subseteq \singleton \times \singleton$. These are $\id[+][\singleton]$ and $\id[-][\singleton]$ since
\[
\begin{array}{rcccl}
\id[+][\singleton] & \stackrel{\eqref{eq:idRel}}{=} & \{(x,y) \in \singleton \times \singleton  \mid x=y\}   &=& \{(\star,\star)\}\quad  \text{ and}\\
\id[-][\singleton] & \stackrel{\eqref{eq:idRel}}{=} & \{(x,y) \in \singleton \times \singleton\mid x \neq y\}  &= &\varnothing \text{.}
\end{array}
\]
 It is worth emphasising that $\id[+][\singleton]$ and $\id[-][\singleton]$ provide \emph{truth} and \emph{falsity} independently of the chosen domain of interpretation. %
\end{rem}
\begin{exa}\label{eq:intersectionandtop}
Take $\sign$ with two symbols $R$ and $S$ with arity and coarity $1$. From Table~\ref{fig:typingrules}, the two terms below have type $1 \to 1$.
\begin{equation*}\label{eq:exampletopint}
\discard[+][1] \seq[+] \codiscard[+][1] \qquad   \copier[+][1] \seq[+] ( (R^{\switchLabelS{+}} \tensor[+] S^{\switchLabelS{+}})\seq[+] \cocopier[+][1])
\end{equation*}
For any interpretation $\interpretation = (X, \rho)$, the semantics of the leftmost term, $\interpretationFunctor(\discard[+][1] \seq[+] \codiscard[+][1])$, is the top relation $X\times X$, denoted in $\CRS$ by $\top$:
\begin{align*}
	\interpretationFunctor(\discard[+][1] \seq[+] \codiscard[+][1]) &= \discard[+][X] \seq[+] \codiscard[+][X] \tag{\ref{fig:semantics}}\\
	& = \{(x, \star) \!\mid\! x\in X\} \seq[+] \{(\star, x) \!\mid\! x\in X\} \tag{\ref{eq:comonoidsREL}}\\
	& = \{(x,y) \!\mid\! x,y\in X\} \tag{\ref{eq:seqRel}}\\
	& = X \!\times\! X \\
	& = \dsemRel{\top}\text{.} \tag{\ref{eq:sematicsExpr}}
\end{align*}
Similarly, %
$\interpretationFunctor( \copier[+][1] \seq[+] ( (R^{\switchLabelS{+}} \tensor[+] S^{\switchLabelS{+}})\seq[+] \cocopier[+][1]) = \rho(R) \cap \rho(S)$ which is denoted in $\CRS$ by $R\cap S$.

We leave to the reader to check that the following two terms
\[\discard[-][1] \seq[-] \codiscard[-][1] \qquad   \copier[-][1] \seq[-] ( (R^{\switchLabelS{+}} \tensor[-] S^{\switchLabelS{+}})\seq[-] \cocopier[-][1])\]
denote, instead, the bottom relation $\varnothing$ and $\rho(R) \cup \rho(S)$ corresponding to the $\CRS$ expressions $\bot$ and $R\cup S$. In \S~\ref{ssec:CRrevisited}, we will illustrate in detail an encoding of the whole $\CRS$ into $\NPR$.
\end{exa}

\begin{exa}\label{ex:alternation} %
We now illustrate how $\NPR$ can express alternation of quantifiers. Let $\sign$ be a signature with one symbol $R$ with arity $2$ and coarity $0$.
Consider the term $(\id[-][1] \tensor[-] \codiscard[-][1]) \; \seq[-]  R^{\switchLabelS{+}}$ which, by Table~\ref{fig:typingrules}, has type $1 \to 0$.
For all interpretations $\interpretation = (X, \rho)$, 
\[ \interpretationFunctor((\id[-][1] \tensor[-] \codiscard[-][1])  \seq[-]  R^{\switchLabelS{+}} )= \{(x,\star) \!\mid\! \forall y \in X .\, (x,y)\in \rho(R) \} \subseteq X \times \singleton \] 
Indeed:
\begin{align*}
	\interpretationFunctor((\id[-][1] \tensor[-] \codiscard[-][1])  \seq[-]  R^{\switchLabelS{+}} ) &= (\id[-][X] \tensor[-] \codiscard[-][X])  \seq[-] \rho(R) \tag{\ref{fig:semantics}}\\
	& =  ( \; \{(x, x') \!\mid\! x\neq x'\} \tensor[-]  \varnothing \;) \seq[-] \rho(R)  \tag{\ref{eq:idRel}, \, \ref{eq:comonoidsREL}}\\
	& = \{(x,(x',y)) \!\mid\! x \neq x'\} \seq[-] \rho(R) \tag{\ref{eq:tensorREL}}\\
	& = \{(x,\star) \!\mid\! \forall (x',y)\in X^2 .\, x\neq x' 	\vee (x',y)\in \rho(R) \} \tag{\ref{eq:seqRel}} \\
	& = \{(x,\star) \!\mid\! \forall y \in X .\, (x,y)\in \rho(R) \} 
\end{align*}
Observe the universal quantification of the variable $y$. To existentially quantify over $x$, one can now take the term $\codiscard[+][1] \seq[+]  ((\id[-][1] \tensor[-] \codiscard[-][1])  \seq[-]  R^{\switchLabelS{+}})$ of  type $0\to 0$.
Depending on the interpretation $\interpretation = (X, \rho)$, it may denote either $\id[+][\singleton]$  or $\id[-][\singleton]$: 
\[ \interpretationFunctor( \, \codiscard[+][1] \seq[+]  ((\id[-][1] \tensor[-] \codiscard[-][1])  \seq[-]  R^{\switchLabelS{+}}) \, ) = \begin{cases} \{(\star,\star)\} & \text{if } \exists x\in X. \, \forall y \in X .\, (x,y)\in \rho(R) \\ \varnothing & \text{otherwise} \end{cases}\]
Indeed:
\begin{align*}
	\interpretationFunctor( \, \codiscard[+][1] \seq[+]  ((\id[-][1] \tensor[-] \codiscard[-][1])  \seq[-]  R^{\switchLabelS{+}}) \, ) &= \codiscard[+][X] \seq[+]  ( (\id[-][X] \tensor[-] \codiscard[-][X])  \seq[-] \rho(R) ) \tag{\ref{fig:semantics}}\\
	& =  \codiscard[+][X] \seq[+]  \,  \{(x,\star) \!\mid\! \forall y \in X .\, (x,y)\in \rho(R) \} 
\,  \tag{derivation above}\\
	& = \{(\star,x) \!\mid\! x \in X\} \seq[+]  \,  \{(x,\star) \!\mid\! \forall y \in X .\, (x,y)\in \rho(R) \}  \tag{\ref{eq:comonoidsREL}}\\
	& = \{(\star,\star) \!\mid\! \exists x\in X. \, \forall y \in X .\, (x,y)\in \rho(R) \} \tag{\ref{eq:seqRel}} 
\end{align*}
One can switch the order of quantifiers by considering the term $ \codiscard[-][1] \seq[-]  (( \codiscard[+][1] \tensor[+] \id[+][1] )  \seq[+]  R^{\switchLabelS{+}})$. Computations analogous to those above confirm the following equality.
\[ \interpretationFunctor( \, \codiscard[-][1] \seq[-]  (( \codiscard[+][1] \tensor[+] \id[+][1] )  \seq[+]  R^{\switchLabelS{+}}) \, ) = \begin{cases} \{(\star,\star)\} & \text{if } \forall y \in X .\, \exists x\in X. \,  (x,y)\in \rho(R) \\ \varnothing & \text{otherwise} \end{cases}\]

\end{exa}

\begin{rem} %
We will see in \S~\ref{sec:fol} that $\NPR$ is as expressive as $\FOL$. We draw the reader's attention to the simplicity of the inductive definition of semantics compared to the traditional $\FOL$ approach where variables and quantifiers make the definition more involved.
Moreover, in traditional accounts, the domain of an interpretation is required to be a \emph{non-empty} set. In our calculus this is unnecessary and it is \emph{not} a mere technicality: in \S~\ref{sec:theories} we shall see that empty models capture the propositional calculus. %
\end{rem}

\paragraph{Semantic equivalence} Like in $\CRS$, \emph{semantic equivalence} plays a key role. For all terms $c,d\colon n \to m$, we write
\begin{equation}\tag{$\semequivalence$}
c \semequivalence d \text{ iff, for all interpretations }\interpretation\text{,} \; \interpretationFunctor (c) = \interpretationFunctor(d)\text{.}
\end{equation}
\emph{Semantic inclusion}, written $\seminclusion$, is defined analogously replacing $=$ with $\subseteq$.

By definition $\semequivalence$ and $\seminclusion$ only relate terms of the same type. Throughout the paper, we will encounter several relations amongst terms of the same type. To avoid any confusion with the relations denoted by the terms, we call them \emph{well-typed relations} and use symbols $\wtrel$ rather than the usual $R, S, T$. In the following, we write $c \basicR d$ for $(c,d)\in \basicR$ and $\pcong{\wtrel}$ for the smallest precongruence (w.r.t.  $\tensor[+]$, $\tensor[-]$, $\seq[+]$ and $\seq[-]$,) generated by $\basicR$, i.e., the relation inductively generated as
\begin{equation}\label{eq:pc}\arraycolsep=2pt
{%
\begin{array}{c@{}c@{}c}
\inferrule*[right=($id$)]{c \wtrel d}{c \,\pcong{\wtrel} \,d}
\quad\quad\quad
&
\inferrule*[right=($r$)]{-}{c\,  \pcong{\wtrel} \,c}
\quad\quad\quad
&
\inferrule*[right=($t$)]{a \, \pcong{\wtrel} \,b \quad b \, \pcong{\wtrel}\, c}{a  \,\pcong{\wtrel} \,c}
\\[5px]
\multicolumn{3}{c}{
\inferrule*[right=($   {\seq}   $)]{c_1  \,\pcong{\wtrel} \,c_2 \quad d_1 \, \pcong{\wtrel} \, d_2}{c_1\seq[] d_1\,  \pcong{\wtrel} \, c_2 \seq[] d_2}
\qquad
\inferrule*[right=($   {\tensor}  $)]{c_1 \, \pcong{\wtrel} \,c_2 \quad d_1 \, \pcong{\wtrel} \, d_2}{c_1\tensor[] d_1 \, \pcong{\wtrel} \, c_2 \tensor[] d_2}
}
\end{array}
}
\end{equation}
Similarly, we will use $\congr{\wtrel}$ for the smallest congruence generated by $\basicR$, namely the well typed relation inductively generated by adding the symmetric closure to the rules above:
\[\inferrule*[right=($s$)]{a \, \congr{\wtrel} \,b }{b  \,\congr{\wtrel} \,a}
\]
A well-typed congruence that we will often encounter through the paper is \emph{structural congruence}, in symbols $\structuralcong$. This is defined as
\begin{equation}\tag{$\structuralcong$}
\structuralcong \defeq \congr{\mathbb{SMC}}\text{,}
\end{equation}
where $\mathbb{SMC}$  is the well-typed relation obtained by substituting $a,b,c,d$ in the bottom of Table \ref{fig:freestricmmoncatax} with terms of the appropriate type.
The name $\mathbb{SMC}$ reflects the fact that these are exactly the laws of (strict) symmetric monoidal categories. The reader can easily check that each of the equality in Table \ref{fig:freestricmmoncatax} is sound with respect to $\semequivalence$ and thus
\[\structuralcong \subseteq \semequivalence\text{.}\]
Intuitively, structural congruence identifies terms that are semantically equivalent for trivial --structural-- reasons, pretty much like $\alpha$-renaming in most of the languages with variables and binders.

\paragraph{Diagrams}
Terms of $\NPR$ enjoy an elegant diagrammatic representation inspired by
string diagrams \cite{joyal1991geometry,Selinger2009}. %
A term $c\colon n \to m$ is drawn as a diagram with $n$ ports on the left and $m$ ports on the right; $\seq[]$ is depicted as horizontal composition while $\tensor[]$ by vertically ``stacking'' diagrams. The two compositions  $\seq[+]$ and $\seq[-]$ and two monoidal products $\tensor[+]$ and $\tensor[-]$ are distinguished with different colours. %
All constants in the white fragment have white background, mutatis mutandis for the black fragment: for instance $\id[+][1]$ and $\id[-][1]$ are drawn
$\idCirc[+]$ and $\idCirc[-]$. Indeed, the diagrammatic version of \eqref{eq:syntax} is the grammar in Fig.\ref{fig:diagsyntax}, reported below.
\[
		\begin{array}{ll}
		  \Circ{c}  \, \Coloneqq\!\!\!\!  & \copierCirc[+]       \mid
										 \discardCirc[+]      \mid
														 \boxCirc[+]{R}       \mid
														 \codiscardCirc[+]    \mid
														 \cocopierCirc[+]     \mid
														 \emptyCirc[+]        \mid
														 \idCirc[+]           \mid
														 \symmCirc[+]         \mid
														 \seqCirc[+]{c}{c}    \mid
														 \tensorCirc[+]{c}{c} \mid  \notag \\[14pt]
						& \copierCirc[-]        \mid
										  \discardCirc[-]       \mid
											\boxOpCirc[-]{R}        \mid
											\codiscardCirc[-]     \mid
											\cocopierCirc[-]      \mid
											\emptyCirc[-]         \mid
											\idCirc[-]            \mid
											\symmCirc[-]          \mid
											\seqCirc[-]{c}{c}     \mid
											\tensorCirc[-]{c}{c}
		\end{array}
\]
The diagrams $\boxCirc[+]{R}$ and $\boxOpCirc[-]{R}$ above correspond to the terms $R^{\switchLabelS{+}}$ and $R^{\switchLabelS{-}}$ for a symbol $R\in \sign$ with arity and coarity $1$. For symbols with arity $n$ and coarity $m$, we draw the corresponding diagrams with $n$ ports on the left and $m$ ports on the right. For instance, for $R$ of arity $2$ and coarity $0$, $R^{\switchLabelS{+}}$ is drawn as $\begin{tikzpicture}
	\begin{pgfonlayer}{nodelayer}
		\node [{boxStyle/+}] (129) at (0, -0.025) {$R$};
		\node [style=none] (134) at (-1, -0.275) {};
		\node [style=none] (135) at (-1, 0.225) {};
		\node [style=none] (136) at (-1, -0.875) {};
		\node [style=none] (137) at (-1, 0.85) {};
		\node [style=none] (138) at (1, 0.85) {};
		\node [style=none] (139) at (1, -0.875) {};
		\node [style=none] (140) at (0, -0.275) {};
		\node [style=none] (141) at (0, 0.225) {};
	\end{pgfonlayer}
	\begin{pgfonlayer}{edgelayer}
		\draw [{bgStyle/+}] (138.center)
			 to (137.center)
			 to (136.center)
			 to (139.center)
			 to cycle;
		\draw [style={wStyle/+}] (134.center) to (140.center);
		\draw [style={wStyle/+}] (141.center) to (135.center);
	\end{pgfonlayer}
\end{tikzpicture}
$.

Diagrams built only from the first row of the grammar above are string diagrams \cite{joyal1991geometry,Selinger2009} with a white background. Diagrams built only from the second row are string diagrams with a black background. For instance, the terms  $\discard[+][1] \seq[+] \codiscard[+][1]$ and $\copier[+][1] \seq[+] ( (R^{\switchLabelS{+}} \tensor[+] S^{\switchLabelS{+}})\seq[+] \cocopier[+][1])$ from Example \ref{eq:intersectionandtop} are drawn  as
\[  \topCirc \qquad \text{ and } \qquad \intersectionCirc{R}{S}\text{.}\]
Note that one diagram may correspond to more than one term:  for instance the diagram on the right above
does not only represent the term $\copier[+][1] \seq[+] ( (R^{\switchLabelS{+}} \tensor[+] S^{\switchLabelS{+}})\seq[+] \cocopier[+][1])$,  but also  $(\copier[+][1] \seq[+]  (R^{\switchLabelS{+}} \tensor[+] S^{\switchLabelS{+}}))\seq[+] \cocopier[+][1]$. Indeed, it is clear that traditional term-based syntax carries more information than the diagrammatic one (e.g. associativity). From the point of view of the semantics, however, this bureaucracy is irrelevant and is conveniently discarded by the diagrammatic notation.

In general, diagrams of $\NPR$ can have nested white and black backgrounds. For instance, 
\[

    \InputIfFileExists{existsForallR.tikz}{}{\input{tikz/existsForallR.tikz}}
 \qquad\text{ and }\qquad 
    \InputIfFileExists{forallExistsR.tikz}{}{\input{tikz/forallExistsR.tikz}}

\]
correspond to the terms $\codiscard[+][1] \seq[+]  ((\id[-][1] \tensor[-] \codiscard[-][1])  \seq[-]  R^{\switchLabelS{+}})$ and $ \codiscard[-][1] \seq[-]  (( \codiscard[+][1] \tensor[+] \id[+][1] )  \seq[+]  R^{\switchLabelS{+}}) $ from Example \ref{ex:alternation}.
To obtain a term from an arbitrary diagram, we first decompose it into layers of nested white and black regions. Within each layer, we then slice either through disconnected components—yielding a tensor—or through wires—yielding a sequential composition. For example, consider the diagram in Figure~\ref{fig:diagrams-terms} (left) and its tiled decomposition (center). When deriving its term representation (right), we start from the innermost layer and progressively plug it into the next one, identifying at each stage the other subterms and their compositions by slicing. Note that we do not give a topological~\cite{joyal1991geometry} nor a combinatorial~\cite{DBLP:journals/jacm/BonchiGKSZ22} interpretation for our diagrams; the latter is left as future work, see \S\ref{sec:future} --- we therefore cannot yet claim that the diagrams in this paper ought to be treated as first-class mathematical objects.
However, by following the procedure outlined above, one can obtain a term from any diagram. Moreover, this procedure is sound: the usual ``only connectivity matters'' topological transformations within each layer are sound since, individually, both the black and white fragments are symmetric monoidal.

\begin{figure}
	\[
	\begin{array}{c@{\qquad\qquad}c@{\qquad\qquad}c}
		\begin{tikzpicture}[scale=0.85]
               \begin{pgfonlayer}{nodelayer}
                   \node [{boxStyle/+}, scale=0.85] (129) at (-0.25, 0) {$R$};
                   \node [style=none] (134) at (-2, 0.25) {};
                   \node [style={dotStyle/-}] (135) at (-1.5, -0.25) {};
                   \node [style=none] (136) at (-2, 0.85) {};
                   \node [style=none] (137) at (-2, -0.875) {};
                   \node [style=none] (138) at (0.75, -0.875) {};
                   \node [style=none] (139) at (0.75, 0.85) {};
                   \node [style=none] (140) at (-1, 0.25) {};
                   \node [style=none] (141) at (-1, -0.25) {};
                   \node [style={dotStyle/+}] (142) at (-2.5, 0.25) {};
                   \node [style=none] (143) at (-3, 1.1) {};
                   \node [style=none] (144) at (-3, -1.125) {};
                   \node [style=none] (145) at (1, -1.125) {};
                   \node [style=none] (146) at (1, 1.1) {};
                   \node [style=none] (147) at (-4, 1.35) {};
                   \node [style=none] (148) at (-4, -1.375) {};
                   \node [style=none] (149) at (1.25, -1.375) {};
                   \node [style=none] (150) at (1.25, 1.35) {};
                   \node [style=none] (151) at (-1, 0.6) {};
                   \node [style=none] (152) at (-1, -0.625) {};
                   \node [style=none] (153) at (0.5, -0.625) {};
                   \node [style=none] (154) at (0.5, 0.6) {};
                   \node [style=none] (155) at (-0.25, 0.25) {};
                   \node [style=none] (156) at (-0.25, -0.25) {};
               \end{pgfonlayer}
               \begin{pgfonlayer}{edgelayer}
                   \draw [{bgStyle/-}] (149.center)
                        to (148.center)
                        to (147.center)
                        to (150.center)
                        to cycle;
                   \draw [{bgStyle/+}] (145.center)
                        to (144.center)
                        to (143.center)
                        to (146.center)
                        to cycle;
                   \draw [{bgStyle/-}] (138.center)
                        to (137.center)
                        to (136.center)
                        to (139.center)
                        to cycle;
                   \draw [style={wStyle/+}] (134.center) to (142);
                   \draw [style={wStyle/-}] (134.center) to (140.center);
                   \draw [style={wStyle/-}] (141.center) to (135);
                   \draw [{bgStyle/+}] (153.center)
                        to (152.center)
                        to (151.center)
                        to (154.center)
                        to cycle;
                   \draw [style={wStyle/+}] (140.center) to (155.center);
                   \draw [style={wStyle/+}] (141.center) to (156.center);
               \end{pgfonlayer}
     \end{tikzpicture}
	 &
\begin{tikzpicture}[scale=0.85]
	\begin{pgfonlayer}{nodelayer}
		\node [style=none] (14) at (-2.5, -0.9) {};
		\node [style=none] (15) at (-5.25, -3.625) {};
		\node [style=none] (16) at (1.95, -3.625) {};
		\node [style=none] (17) at (4.55, -0.9) {};
		\node [style=none] (31) at (0.35, 4.675) {};
		\node [style=none] (32) at (0.075, 4.4) {};
		\node [style=none] (42) at (0.775, 5.1) {};
		\node [style=none] (43) at (-0.475, 3.875) {};
		\node [style=none] (44) at (1.525, 3.875) {};
		\node [style=none] (45) at (2.75, 5.1) {};
		\node [style=none] (46) at (0.85, 4.675) {};
		\node [style=none] (47) at (0.575, 4.4) {};
		\node [style=none] (49) at (-0.7, 3) {};
		\node [style={dotStyle/-}] (50) at (-0.775, 2.5) {};
		\node [style=none] (51) at (-0.1, 3.6) {};
		\node [style=none] (52) at (-1.85, 1.875) {};
		\node [style=none] (53) at (1.65, 1.875) {};
		\node [style=none] (54) at (3.4, 3.6) {};
		\node [style=none] (55) at (0.4, 3) {};
		\node [style=none] (56) at (-0.1, 2.5) {};
		\node [style=none] (71) at (-0.725, 0.75) {};
		\node [style={dotStyle/+}] (79) at (-1.4, 0.75) {};
		\node [style=none] (80) at (-1, 1.6) {};
		\node [style=none] (81) at (-3.25, -0.625) {};
		\node [style=none] (82) at (1.8, -0.625) {};
		\node [style=none] (83) at (4, 1.6) {};
		\node [style=none] (87) at (0.775, 3.35) {};
		\node [style=none] (88) at (-0.475, 2.125) {};
		\node [style=none] (89) at (1.525, 2.125) {};
		\node [style=none] (90) at (2.775, 3.35) {};
		\node [style=none] (93) at (-0.1, 1.35) {};
		\node [style=none] (94) at (-1.85, -0.375) {};
		\node [style=none] (95) at (1.65, -0.375) {};
		\node [style=none] (96) at (3.4, 1.35) {};
		\node [style=none] (97) at (-1, -1.15) {};
		\node [style=none] (98) at (-3.25, -3.375) {};
		\node [style=none] (99) at (1.8, -3.375) {};
		\node [style=none] (100) at (4, -1.15) {};
		\node [style=none, xslant=0.7, scale=0.6] (101) at (1.15, 4.5) {$R$};
		\node [style=none] (102) at (1.05, 4.85) {};
		\node [style=none] (103) at (0.3, 4.125) {};
		\node [style=none] (104) at (1.425, 4.125) {};
		\node [style=none] (107) at (1.9, 4.6) {};
		\node [style=none] (108) at (1.65, 4.85) {};
	\end{pgfonlayer}
	\begin{pgfonlayer}{edgelayer}
		\draw [{bgStyle/-}] (16.center)
			 to (15.center)
			 to (14.center)
			 to (17.center)
			 to cycle;
		\draw [densely dashed, lightgray] (80.center) to (97.center);
		\draw [{bgStyle/+}] (82.center)
			 to (81.center)
			 to (80.center)
			 to (83.center)
			 to cycle;
		\draw [densely dashed, lightgray] (51.center) to (93.center);
		\draw [{bgStyle/-}] (53.center)
			 to (52.center)
			 to (51.center)
			 to (54.center)
			 to cycle;
		\draw [densely dashed, lightgray] (42.center) to (87.center);
		\draw [{bgStyle/+}] (44.center)
			 to (43.center)
			 to (42.center)
			 to (45.center)
			 to cycle;
		\draw [style={wStyle/+}] (31.center) to (46.center);
		\draw [style={wStyle/+}] (32.center) to (47.center);
		\draw [style={wStyle/-}] (49.center) to (55.center);
		\draw [style={wStyle/-}] (56.center) to (50);
		\draw [style={wStyle/+}] (71.center) to (79);
		\draw [{bgStyle/g}] (89.center)
			 to (88.center)
			 to (87.center)
			 to (90.center)
			 to cycle;
		\draw [{bgStyle/g}] (95.center)
			 to (94.center)
			 to (93.center)
			 to (96.center)
			 to cycle;
		\draw [{bgStyle/g}] (97.center)
			 to (100.center)
			 to (99.center)
			 to (98.center)
			 to cycle;
		\draw [densely dashed, lightgray] (43.center) to (88.center);
		\draw [densely dashed, lightgray] (45.center) to (90.center);
		\draw [densely dashed, lightgray] (44.center) to (89.center);
		\draw [densely dashed, lightgray] (54.center) to (96.center);
		\draw [densely dashed, lightgray] (53.center) to (95.center);
		\draw [densely dashed, lightgray] (52.center) to (94.center);
		\draw [densely dashed, lightgray] (83.center) to (100.center);
		\draw [densely dashed, lightgray] (82.center) to (99.center);
		\draw [densely dashed, lightgray] (81.center) to (98.center);
		\draw [{wStyle/+}] (103.center)
			 to (102.center)
			 to (108.center)
			 to (107.center)
			 to (104.center)
			 to cycle;
	\end{pgfonlayer}
\end{tikzpicture}
	 &
	 \begin{array}{c@{}c@{}c@{}c@{}c}
		& & & & R^{\switchLabelS{+}}
		\\
		& & & &
			\begin{tikzpicture}[scale=0.85]
			\begin{pgfonlayer}{nodelayer}
				\node [style=none] (110) at (0.05, 1) {};
				\node [style=none] (111) at (0.05, 0) {};
			\end{pgfonlayer}
			\begin{pgfonlayer}{edgelayer}
				\draw [densely dashed, lightgray] (110.center) to (111.center);
			\end{pgfonlayer}
		\end{tikzpicture}
		\\[-7pt]
		\multicolumn{4}{r}{(\id[-][1] \tensor[-] \codiscard[-][1]) \; \seq[-] \!\!} & \gbox
		\\
		& &  &
			\begin{tikzpicture}[scale=0.85]
			\begin{pgfonlayer}{nodelayer}
				\node [style=none] (110) at (0.05, 1) {};
				\node [style=none] (111) at (0.05, 0) {};
			\end{pgfonlayer}
			\begin{pgfonlayer}{edgelayer}
				\draw [densely dashed, lightgray] (110.center) to (111.center);
			\end{pgfonlayer}
		\end{tikzpicture}
		&
		\\[-7pt]
		\multicolumn{3}{r}{\codiscard[+][1] \; \seq[+] \!\!\!\!\!} & \gbox &
		\\
		& &
			\begin{tikzpicture}[scale=0.85]
			\begin{pgfonlayer}{nodelayer}
				\node [style=none] (110) at (0.05, 1) {};
				\node [style=none] (111) at (0.05, 0) {};
			\end{pgfonlayer}
			\begin{pgfonlayer}{edgelayer}
				\draw [densely dashed, lightgray] (110.center) to (111.center);
			\end{pgfonlayer}
		\end{tikzpicture}
		\\[-7pt]
		\multicolumn{2}{r}{\id[-][0] \; \seq[-] \!} & \gbox &
	 \end{array}
	\end{array}
	\]
	\caption{A diagram, its tiled decomposition and its term derivation.}
	\label{fig:diagrams-terms}
\end{figure}

It is possible to perform calculations with either terms or diagrams, but we usually prefer working with diagrams because of the lesser bureaucratic overhead and the compelling visual intuitions afforded by the notation. Moreover, as we will illustrate in the following sections,  diagrams allow for a visual representation of opposite ($\op{\cdot}$) and complement ($\nega{\cdot}$) operations: the \emph{mirror image} of a diagram --obtained by swapping left and right-- provides the opposite, while the \emph{photographic negative} --obtained by swapping white and black-- provides the complement.

Nevertheless, for typographical reasons, it is sometimes convenient to use terms, and thus, the reader should become familiar with both notations. To ease this transition, in the early stages of the paper, we will occasionally present key notions in both notations. Additionally, Table \ref{tab:dictionary} in Appendix \ref{app:dictionary} provides a summary of the diagrammatic conventions, which the reader may find useful.

\paragraph{Axioms} %
Figure~\ref{fig:textual axioms} in Appendix~\ref{app:additional} illustrates a complete system of axioms for the semantic inclusion $\seminclusion$.
Let $\mathbb{FOB}$ be the well-typed relation obtained by substituting $a,b,c,d$ in Fig. \ref{fig:textual axioms} with terms of the appropriate type
and and call  \emph{syntactic inclusion},
written $\syninclusion$, the precongruence generated by $\mathbb{FOB}$ and $\structuralcong$. In symbols,
\begin{equation}\tag{$\syninclusion$}
\syninclusion \defeq \pcong{\mathbb{FOB} \;\cup \structuralcong}\text{.}
\end{equation}
We will also write $\synequivalence \defeq \syninclusion \cap \syninclusionop$ for the \emph{syntactic equivalence}. Our main result is:
\begin{thm}\label{thm:completeness}
For all terms $c,d\colon n \to m$, $c\syninclusion d$ iff $c\seminclusion d$.
\end{thm}
\noindent 
The axiomatisation is far from minimal and is redundant in several respects. We chose the more verbose presentation in order to emphasise both the underlying categorical structures and the various dualities that we will highlight in the next sections. We confined the complete axiomatisation to the appendix because the axioms in Figure \ref{fig:textual axioms} appear also in Figures~\ref{fig:cb axioms}, \ref{fig:cocb axioms}, \ref{fig:closed lin axioms}, \ref{fig:fo bicat axioms}  in form of diagrams. This allows for a more modular presentation that places each axiom in its proper context, highlighting its provenance from one of the categorical structures involved.

\paragraph{Proofs as diagrams rewrites}
Proofs in $\NPR$ are rather different from those of traditional proof systems: since the only inference rules are those in~\eqref{eq:pc}, any proof of $c\syninclusion d$ consists of a sequence of applications of axioms. %
As an example consider the proof in Fig. \ref{fig:forallExists full} which is the same of \eqref{eq:forallexists} from the Introduction, but only using axioms.
Note that, when applying axioms, we are in fact performing diagram rewriting: an instance of the left hand side of an axiom is found within a larger diagram and replaced with the right hand side. Since such rewrites can happen anywhere, there is a close  connection between proofs in $\NPR$ and the work on \emph{deep inference}~\cite{hughes2021combinatorial,DBLP:phd/de/Brunnler2003,guglielmi2007system} -- see Example \ref{subsub:propo calc}.%

\input{tikz/proofs/forallExistsFull.tex}

\medskip

The theory of neo-Peircean relations is both rich and elegant, built upon well-established concepts from category theory. For this reason, we now move away from a traditional, term-based exposition and adopt the more abstract framework of category theory. In the next two sections, we provide the necessary categorical background --covering cartesian and linear bicategories --before introducing first-order bicategories in \S~\ref{sec:fobic}. We will then return to $\NPR$ in \S~\ref{sec:freely}, where we explain how it gives rise to a first-order bicategory, specifically the freely generated one: any equivalence that holds in $\NPR$ also holds in arbitrary fo-bicategories.

We hope that this section, which deliberately avoids categorical language, has been enjoyable even for readers who are not experts in category theory and that it has sparked their curiosity to continue reading.

\section{(Co)Cartesian Bicategories}\label{sec:cartesianbi}

Although the term bicategory might seem ominous, the beasts considered in this paper are actually quite simple. %
We consider \emph{poset-enriched symmetric monoidal categories}:  every homset carries a partial order $\leq$, and composition $\seq[]$ and monoidal product $\tensor$ are monotone. That is, if $a\leq b$ and $c\leq d$ then $a\seq[] c \leq b\seq[] d$ and $a\tensor[] c \leq b\tensor[] d$. A \emph{poset-enriched symmetric monoidal functor} is a (strong, and usually strict) symmetric monoidal functor that preserves the order $\leq$.
The notion of \emph{adjoint arrows}, which will play a key role, amounts to the following: for $c \colon X \to Y$ and $d \colon Y \to X$,  $c$ is  \emph{left adjoint} to $d$, or $d$ is \emph{right adjoint} to $c$, written $d \vdash c$, if $\id[][X] \leq c \seq[] d$  and $d \seq[] c \leq \id[][Y]$.

For a poset-enriched symmetric monoidal category $(\Cat{C}, \tensor[][][], \unittensor)$, we will write $\opposite{\Cat{C}}$ for the category having the same objects as $\Cat{C}$ but homsets $\opposite{\Cat{C}}[X,Y] \defeq \Cat{C}[Y,X]$: ordering, identities and monoidal product are defined as in $\Cat{C}$, while composition $c \seq[] d$ in $\opposite{\Cat{C}}$ is $d \seq[] c$ in $\Cat{C}$.
Similarly, we will write  $\co{\Cat{C}}$ to denote the category having the same objects and arrows of $\Cat{C}$ but equipped with the reversed ordering $\geq$. Composition, identities and monoidal product are defined as in $\Cat{C}$.
In this paper, we will often tacitly use the facts that, by definition, both $\opposite{(\opposite{\Cat{C}})}$ and $\co{(\co{\Cat{C}})}$ are $\Cat{C}$ and that $\opposite{(\co{\Cat{C}})}$ is $\co{(\opposite{\Cat{C}})}$. %

All (poset-enriched) monoidal categories considered throughout this paper are tacitly assumed to be strict \cite{mac_lane_categories_1978}, i.e.\ $(X\tensor[] Y)\tensor[] Z = X \tensor[] (Y \tensor[] Z)$ and $\unittensor \tensor[] X = X =X \tensor[] \unittensor$ for all objects $X,Y,Z$. This is harmless:  strictification~\cite{mac_lane_categories_1978} (see~\cite[Thm 2.6.4]{johnson2024bimonoidal} for the enriched case) allows to transform any (enriched) monoidal category into a strict one, enabling the sound use of string diagrams that  will be exploited in this and the next sections. These are like the diagrams of $\NPR$ in \S~\ref{sec:NPR} but are interpreted as arrows of the categorical structures of interest and wires are labeled by objects of such categories.
For instance,  the diagrams
\[{\copierCirc[+][X]}, \qquad {\discardCirc[+][X]}, \qquad  {\cocopierCirc[+][X]} \qquad \text{and} \qquad {\codiscardCirc[+][X]}\]
represent arrows
\[\copier[+][X]\colon X \to X\tensor[+]X, \quad \discard[+][X] \colon X \to \unittensor, \quad \cocopier[+][X]\colon X\tensor[+]X \to X\quad \text{and}\quad \codiscard[+][X] \colon \unittensor \to X \]
of cartesian bicategories, introduced below. %

\begin{defi}\label{def:cartesian bicategory}
A \emph{cartesian bicategory} $(\Cat{C}, \tensor[+], \unittensor, \copier[+], \discard[+], \cocopier[+], \codiscard[+])$, shorthand $(\Cat{C}, \copier[+], \cocopier[+])$,
is a poset-enriched symmetric monoidal category $(\Cat{C}, \tensor[+], \unittensor)$ and, for every object $X$ in $\Cat{C}$,
arrows $\copier[+][X]\colon X \to X\tensor[+]X$, $\discard[+][X]\colon X \to \unittensor$, $\cocopier[+][X] \colon X \tensor[+]X \to X$,  $\codiscard[+][X]\colon \unittensor \to X$ such that
\begin{enumerate}
\item $(\copier[+][X], \discard[+][X])$ is a comonoid and $(\cocopier[+][X], \codiscard[+][X])$ a monoid, i.e., the following hold:
\[
\begin{array}{rclrcl}
          \copier[+][X] \seq[+] (\id[+][X] \tensor[+] \copier[+][X]) &\stackrel{(\copier[+]\text{-as})}{=}& \copier[+][X] \seq[+] (\copier[+][X] \tensor[+] \id[+][X]) \; & \; (\id[+][X] \tensor[+] \cocopier[+][X]) \seq[+] \cocopier[+][X] &\stackrel{(\cocopier[+]\text{-as})}{=}& (\cocopier[+][X] \tensor[+] \id[+][X]) \seq[+] \cocopier[+][X] \\
          \copier[+][X] \seq[+] (\id[+][X] \tensor[+] \discard[+][X]) &\stackrel{(\copier[+]\text{-un})}{=}& \id[+][X]  \; &\; (\id[+][X] \tensor[+] \codiscard[+][X]) \seq[+] \cocopier[+][X] &\stackrel{(\cocopier[+]\text{-un})}{=}& \id[+][X] \\
          \copier[+][X] \seq[+] \symm[+][X][X] &\stackrel{(\copier[+]\text{-co})}{=}& \copier[+][X] \;&\; \symm[+][X][X] \seq[+] \cocopier[+][X] &\stackrel{(\cocopier[+]\text{-co})}{=}& \cocopier[+][X] \\
\end{array}
\]
\item the comonoid $(\copier[+][X], \discard[+][X])$ is lax natural, i.e., for all arrows $c \colon X \to Y$ the following hold: %
\[
\begin{array}{rclrcl}
                  c \seq[+] \copier[+][X] &\stackrel{\eqref*{ax:comPlusLaxNat}}{\leq}& \copier[+][Y] \seq[+] (c \tensor[+] c)
                  \quad&\quad
                  c \seq[+] \discard[+][X] &\stackrel{\eqref*{ax:discPlusLaxNat}}{\leq}& \discard[+][Y]
\end{array}
\]
\item $(\copier[+][X], \discard[+][X])$ are left adjoints to $(\cocopier[+][X], \codiscard[+][X])$, i.e.:
\[
                  \begin{array}{r@{\,\,}c@{\,\,}l @{\qquad} r@{\,\,}c@{\,\,}l @{\qquad} r@{\,\,}c@{\,\,}l @{\qquad} r@{\,\,}c@{\,\,}l}
                      \cocopier[+][X] \seq[+] \copier[+][X]
                      &\stackrel{\eqref*{ax:plusCocopyCopy}}{\leq}&
                      (\id[+][X] \tensor[+] \id[+][X])
                     &
                      \id[+][X]
                      &\stackrel{\eqref*{ax:plusCopyCocopy}}{\leq}&
                      \copier[+][X] \seq[+] \cocopier[+][X]
                     &
                      \id[+][X]
                      &\stackrel{\eqref*{ax:plusDiscCodisc}}{\leq}&
                      \discard[+][X] \seq[+] \codiscard[+][X]
                      &
                      \codiscard[+][X] \seq[+] \discard[+][X]
                      &\stackrel{\eqref*{ax:plusCodiscDisc}}{\leq}&
                      \id[+][0]
                  \end{array}
\]
\item $(\copier[+][X], \discard[+][X])$ and  $(\cocopier[+][X], \codiscard[+][X])$ form special Frobenius algebras, i.e.:
\[
\begin{array}{rclrcl}(\copier[+][X] \tensor[+] \id[+][X]) \seq[+] (\id[+][X] \tensor[+] \cocopier[+][X]) &\stackrel{(\text{F}^{\switchLabelS{+}})}{=}& (\id[+][X] \tensor[+] \copier[+][X]) \seq[+] (\cocopier[+][X] \tensor[+] \id[+][X]) \quad &\quad  \copier[+][X] \seq[+] \cocopier[+][X] &\stackrel{(\text{S}^{\switchLabelS{+}})}{=}& \id[+][X] \\
\end{array}
\]
\item $(\copier[+][X], \discard[+][X])$ and $(\cocopier[+][X], \codiscard[+][X])$ satisfy the usual coherence conditions: %
\[%
\begin{array}{r@{\;}c@{\;}l @{\qquad} r@{\;}c@{\;}l @{\qquad} r@{\;}c@{\;}l @{\qquad} r@{\;}c@{\;}l}
\copier[+][\unittensor]    &=& \id[+][\unittensor] & \copier[+][X\tensor[+]Y]      &=& (\copier[+][X] \tensor[+] \copier[+][Y]) \seq[+]( \id[+][X] \tensor[+]\symm[+][X][Y] \tensor[+] \id[+][Y]) & \discard[+][\unittensor]  &=& \id[+][\unittensor] & \discard[+][X \tensor[+] Y]  &=& \discard[+][X] \tensor[+]\discard[+][Y]\\
\cocopier[+][\unittensor]  &=& \id[+][\unittensor] & \cocopier[+][X\tensor[+]Y]    &=& ( \id[+][X] \tensor[+]\symm[+][X][Y] \tensor[+] \id[+][Y]) \seq[+] (\cocopier[+][X] \tensor[+] \cocopier[+][Y]) & \codiscard[+][\unittensor] &=& \id[+][\unittensor] &  \codiscard[+][X \tensor[+] Y] &=& \codiscard[+][X] \tensor[+]\codiscard[+][Y]
\end{array}
\]
\end{enumerate}
$\Cat{C}$ is a \emph{cocartesian bicategory} if  $\co{\Cat{C}}$ is a cartesian bicategory. %
A \emph{morphism of (co)cartesian bicategories} is a poset-enriched strong symmetric monoidal functor preserving the specified monoid and comonoid structures. \end{defi}
\begin{rem}
In the original presentation of~\cite{carboni1987cartesian}, the structures in Definition \ref{def:cartesian bicategory} are named cartesian bicategories \emph{of relations}. Here we have chosen to go for brevity and just call them cartesian bicategories.

Fig. \ref{fig:cb axioms} illustrates the axioms of cartesian bicategories by means of diagrams: the axioms on the top-left corner are those of comonoids and monoids; on the bottom-left corner, there are the axioms of adjointness; bottom-right illustrates lax naturality and top-right the special Frobenius. The axioms of coherence --point (5) in Definition \ref{def:cartesian bicategory}-- that we purposely do not display in diagrams, corresponds to the inductive definitions of $\copier[+][n]$, $\cocopier[+][n]$, $\discard[+][n]$ and $\codiscard[+][n]$ in Tab.~\ref{fig:sugar}: $\tensor[+] \mapsto +$,$\unittensor \mapsto 0$, $X\mapsto 1$ and $Y\mapsto n$.
\end{rem}

\begin{figure*}[t]
\mylabel{ax:comPlusAssoc}{$\copier[+]$-as}
\mylabel{ax:comPlusUnit}{$\copier[+]$-un}
\mylabel{ax:comPlusComm}{$\copier[+]$-co}
\mylabel{ax:monPlusAssoc}{$\cocopier[+]$-as}
\mylabel{ax:monPlusUnit}{$\cocopier[+]$-un}
\mylabel{ax:monPlusComm}{$\cocopier[+]$-co}
\mylabel{ax:plusSpecFrob}{S$^\circ$}
\mylabel{ax:plusFrob}{F$^\circ$}
\mylabel{ax:comPlusLaxNat}{$\copier[+]$-nat}
\mylabel{ax:discPlusLaxNat}{$\discard[+]$-nat}
\mylabel{ax:plusCodiscDisc}{$\epsilon\discard[+]$}
\mylabel{ax:plusDiscCodisc}{$\eta\discard[+]$}
\mylabel{ax:plusCocopyCopy}{$\epsilon\!\copier[+]$}
\mylabel{ax:plusCopyCocopy}{$\eta\!\copier[+]$}
\centering
\resizebox{\textwidth}{!}{
$\begin{array}{c|c}
        \begin{array}{c@{\!}c@{\!}c c@{\!}c@{\!}c}
                {
    \InputIfFileExists{axiomsNEW/cb/plus/comAssoc1.tikz}{}{\input{tikz/axiomsNEW/cb/plus/comAssoc1.tikz}}
}   & \Leq{\eqref*{ax:comPlusAssoc}} &  {
    \InputIfFileExists{axiomsNEW/cb/plus/comAssoc2.tikz}{}{\input{tikz/axiomsNEW/cb/plus/comAssoc2.tikz}}
}
                &
                {
    \InputIfFileExists{axiomsNEW/cb/plus/monAssoc1.tikz}{}{\input{tikz/axiomsNEW/cb/plus/monAssoc1.tikz}}
}   & \Leq{\eqref*{ax:monPlusAssoc}} &  {
    \InputIfFileExists{axiomsNEW/cb/plus/monAssoc2.tikz}{}{\input{tikz/axiomsNEW/cb/plus/monAssoc2.tikz}}
}
                \\
                {
    \InputIfFileExists{axiomsNEW/cb/plus/comUnit.tikz}{}{\input{tikz/axiomsNEW/cb/plus/comUnit.tikz}}
}     & \Leq{\eqref*{ax:comPlusUnit}}  &  {\idCirc[+][X]}
                &
                {
    \InputIfFileExists{axiomsNEW/cb/plus/monUnit.tikz}{}{\input{tikz/axiomsNEW/cb/plus/monUnit.tikz}}
}     & \Leq{\eqref*{ax:monPlusUnit}}  &  {\idCirc[+][X]}
                \\
                {
    \InputIfFileExists{axiomsNEW/cb/plus/comComm.tikz}{}{\input{tikz/axiomsNEW/cb/plus/comComm.tikz}}
}     & \Leq{\eqref*{ax:comPlusComm}}  &  {\copierCirc[+][X]}
                &
                {
    \InputIfFileExists{axiomsNEW/cb/plus/monComm.tikz}{}{\input{tikz/axiomsNEW/cb/plus/monComm.tikz}}
}     & \Leq{\eqref*{ax:monPlusComm}}  &  {\cocopierCirc[+][X]}
        \end{array}
        &
        \begin{array}{c@{}c@{}c}
                {
    \InputIfFileExists{axiomsNEW/cb/plus/frob1.tikz}{}{\input{tikz/axiomsNEW/cb/plus/frob1.tikz}}
}       & \Leq{\eqref*{ax:plusFrob}}     &  {
    \InputIfFileExists{axiomsNEW/cb/plus/frob2.tikz}{}{\input{tikz/axiomsNEW/cb/plus/frob2.tikz}}
}
                \\
                {
    \InputIfFileExists{axiomsNEW/cb/plus/specFrob.tikz}{}{\input{tikz/axiomsNEW/cb/plus/specFrob.tikz}}
}    & \Leq{\eqref*{ax:plusSpecFrob}} &  {\idCirc[+][X]}
        \end{array}
        \\
        \midrule
        \begin{array}{c@{\!}c@{\!}c @{\qquad} c@{\!}c@{\!}c}
                {
    \InputIfFileExists{axiomsNEW/cb/plus/cocopierCopier.tikz}{}{\input{tikz/axiomsNEW/cb/plus/cocopierCopier.tikz}}
} & \Lleq{\eqref*{ax:plusCocopyCopy}}& {
    \InputIfFileExists{axiomsNEW/id2P.tikz}{}{\input{tikz/axiomsNEW/id2P.tikz}}
}
                &
                {
    \InputIfFileExists{axiomsNEW/cb/plus/codiscDisc.tikz}{}{\input{tikz/axiomsNEW/cb/plus/codiscDisc.tikz}}
} & \Lleq{\eqref*{ax:plusCodiscDisc}}& {\emptyCirc[+]}
                \\
                {\idCirc[+][X]} &\Lleq{\eqref*{ax:plusCopyCocopy}}& {
    \InputIfFileExists{axiomsNEW/cb/plus/specFrob.tikz}{}{\input{tikz/axiomsNEW/cb/plus/specFrob.tikz}}
}
                &
                {\idCirc[+][X]} &\Lleq{\eqref*{ax:plusDiscCodisc}}& {
    \InputIfFileExists{axiomsNEW/top.tikz}{}{\input{tikz/axiomsNEW/top.tikz}}
}
        \end{array}
        &
        \begin{array}{c@{}c@{}c}
                {
    \InputIfFileExists{axiomsNEW/cb/plus/copierLaxNat1.tikz}{}{\input{tikz/axiomsNEW/cb/plus/copierLaxNat1.tikz}}
} &\Lleq{\eqref*{ax:comPlusLaxNat}}& 
    \InputIfFileExists{axiomsNEW/cb/plus/copierLaxNat2.tikz}{}{\input{tikz/axiomsNEW/cb/plus/copierLaxNat2.tikz}}

                \\
                {
    \InputIfFileExists{axiomsNEW/cb/plus/discardLaxNat.tikz}{}{\input{tikz/axiomsNEW/cb/plus/discardLaxNat.tikz}}
} &\Lleq{\eqref*{ax:discPlusLaxNat}}& {\discardCirc[+][X][X]}
        \end{array}
\end{array}$
}
\caption{Axioms of cartesian bicategories}\label{fig:cb axioms}
\end{figure*}

The archetypal example of a cartesian bicategory is  $(\Relp, \copier[+], \cocopier[+])$. $\Relp$ the bicategory of sets and relations ordered by inclusion $\subseteq$ with white composition $\seq[+]$ and identities $\id[+]$ defined as in \eqref{eq:seqRel} and \eqref{eq:idRel}. The monoidal product on objects is the cartesian product of sets with unit $\unittensor$ the singleton set $\singleton$. On arrows, $\tensor[+]$  is defined as in \eqref{eq:tensorREL}. It is immediate to check that, for every set $X$, the arrows $\copier[+][X]$, $\discard[+][X]$ defined in \eqref{eq:comonoidsREL} form a comonoid in $\Relp$, while $\cocopier[+][X]$, $\codiscard[+][X]$ a monoid. Simple computations also proves all the (in)equalities in Definition.~\ref{def:cartesian bicategory}. %

Lax naturality of the comonoid $(\copier[+][X],\discard[+][X])$ is the most interesting to show: %
since for any relation $R\subseteq X \times Y$
\[R \seq[+] \copier[+][Y] = \{(x,(y,y)) \mid (x,y)\in R\}  \; \subseteq  \; \{(x,(y,z)) \mid (x,y)\in R \wedge (x,z)\in R\} = \copier[+][X] \seq[+] (R \tensor[+] R)\]
and
\[R\seq[+] \discard[+][Y] = \{(x,\star) \mid \exists y\in X \;. \; (x,y)\in R\}  \; \subseteq  \; \{(x,\star) \mid x\in X\} = \discard[+][X]\text{,}\]
the axioms \eqref{ax:comPlusLaxNat} and  \eqref{ax:discPlusLaxNat} hold in $\Relp$. The reversed inclusions are also interesting to consider: $R \seq[+] \copier[+][Y] \supseteq \copier[+][X] \seq[+] (R \tensor[+] R)$ holds iff the relation $R$ is single valued namely, for all $x\in X$, there is \emph{at most} one $y\in Y$ such that $(x,y)\in R$,
while $R\seq[+] \discard[+][Y] \supseteq \discard[+][X]$ iff $R$ is total  i.e., for all $x\in X$, there is \emph{at least} one $y\in Y$ such that $(x,y)\in R$. That is, the two inequalities in Definition \ref{def:cartesian bicategory}.(2) are equalities iff the relation $R$ is a \emph{function}. This justifies the following:

\begin{defi}\label{def:maps} %
Let $(\Cat{C}, \copier[+], \cocopier[+])$ be a cartesian bicategory.
An arrow $c\colon X \to Y$ is a \emph{map} if  \[c \seq[+] \copier[+][Y] \; \geq \; \copier[+][X] \seq[+] (c \tensor[+] c) \qquad \text{ and } \quad c\seq[+] \discard[+][Y] \; \geq \; \discard[+][X]\text{.}\] In diagrams,
\[ 
    \InputIfFileExists{axiomsNEW/cb/plus/copierLaxNat1.tikz}{}{\input{tikz/axiomsNEW/cb/plus/copierLaxNat1.tikz}}
 \geq 
    \InputIfFileExists{axiomsNEW/cb/plus/copierLaxNat2.tikz}{}{\input{tikz/axiomsNEW/cb/plus/copierLaxNat2.tikz}}
 \qquad \text{ and }\quad 
    \InputIfFileExists{axiomsNEW/cb/plus/discardLaxNat.tikz}{}{\input{tikz/axiomsNEW/cb/plus/discardLaxNat.tikz}}
 \!\!\!\geq \discardCirc[+][X] \]
The category of maps of $\Cat{C}$, denoted by $\texttt{Map}(\Cat{C})$, is the subcategory of $\Cat{C}$ having as objects those of $\Cat{C}$ and as arrows only the maps.
\end{defi}

\begin{prop}
$\texttt{Map}(\Cat{C})$ is a cartesian category: the final object is $\unittensor$ and the product is $\tensor[+]$.
\end{prop}
\begin{proof}
        See Theorem 1.6 in~\cite{carboni1987cartesian}.
\end{proof}
The cartesianity of $\texttt{Map}(\Cat{C})$ provides several properties of maps that will be useful later.
\begin{lem}\label{lemma:cb maps}\label{prop:nary maps} In a cartesian bicategory $(\Cat{C}, \copier[+], \cocopier[+])$ the following holds:
        \begin{enumerate}
        \item For all objects $X$, $\id[+][X]\colon X \to X$, $\copier[+][X]\colon X \to X \tensor[+] X$ and $\discard[+][X]\colon X \to \unittensor$ are maps;
        \item For maps $a$ and $b$ properly typed, $a \seq[+] b$ and $a \tensor[+] b$ are maps;
        \item If $a \colon \unittensor \to \unittensor $ is a map, then $a=\id[+][\unittensor]$;
        \item If $a \colon \unittensor \to X\tensor[+] Y $ is a map, then there are maps $c\colon \unittensor \to X, d\colon \unittensor \to Y$ such that $a=c\tensor[+]d$.
        \end{enumerate}
\end{lem}
\begin{proof}
        See Theorem 1.6 in~\cite{carboni1987cartesian}.
\end{proof}
\noindent 
In a cartesian bicategory $(\Cat{C}, \copier[+], \cocopier[+])$, for any object X, there are arrows \[\cupCircL{X} \colon \unittensor \to X\tensor[+]X \text{ and } \capCircL{X} \colon X\tensor[+]X \to \unittensor\] 
such that the yanking conditions hold, i.e. \[\opIdCirc[+]{X} = \idCirc[+][X] = \oppIdCirc[+]{X}\text{.}\]
These arrows make $\Cat{C}$ a self-dual compact closed category.
\begin{lem}\label{lem:self dual coco}
Any cartesian bicategory $(\Cat{C}, \copier[+], \cocopier[+])$ is \emph{self-dual compact closed}.
\end{lem}
\begin{proof}
See Theorem 2.4 in~\cite{carboni1987cartesian}.
\end{proof}
\noindent 
Moreover, one can take $\opposite{\Cat{C}}$, swap monoids and comonoids and thus, obtain a cartesian bicategory $(\opposite{\Cat{C}}, \cocopier[+],\copier[+])$. Most importantly, the self-dual compact closed structure yields an  identity-on-objects isomorphism %
$\op{(\cdot)}\colon \Cat{C} \to \opposite{\Cat{C}}$ defined for all arrows $c \colon X \to Y$ as %
\begin{equation}\label{eqdagger}
\op{c} \defeq \daggerCirc[+]{c}[Y][X]
\end{equation}%
Note that in \S~\ref{sec:calculusrelations}, we used the same symbol $\op{(\cdot)}$ to denote the converse relation. This is no accident:  in $\Relp$, $\op{R}$ as in \eqref{eqdagger} is exactly $\{(y,x) \mid (x,y)\in R\}$. %

\begin{prop}\label{prop:opcartesianfunctor}
$\op{(\cdot)}\colon \Cat{C} \to \opposite{\Cat{C}}$ is an isomorphism of cartesian bicategories, namely the laws in the first three rows of Table \ref{table:daggerproperties}.(a) hold.
\end{prop}
\begin{proof}%
        See Theorem 2.4 in~\cite{carboni1987cartesian}.
\end{proof}

\begin{lem}\label{lm:opfunctor}
        Let $\mathcal{F}\colon \Cat{C_1} \to \Cat{C_2}$ be a morphism of cartesian bicategories. Then, for all $c\colon X \to Y$, $\op{\mathcal{F}(c)} = \mathcal{F}(\op{c}).$
        \end{lem}
        \begin{proof}
        See Remark 2.9 in~\cite{carboni1987cartesian}.
\end{proof}

The following result generalises the well-known fact that a relation $R$ is a function if and only if it is left adjoint to $\op{R}$.

\begin{prop}\label{prop:map adj}
In a cartesian bicategory,  an arrow $c\colon X \to Y$ is a map iff $\op{c} \vdash c$, namely
   \[     \id[+][X] \leq c \seq[+] \op{c} \qquad\text{and}\qquad \op{c} \seq[+] c \leq \id[+][Y]\text{.}\]
\end{prop}
\begin{proof}%
See Lemma 2.5 in~\cite{carboni1987cartesian}.
\end{proof}
\noindent 
Hereafter, we write $\CircOp[+]{c}$ for $\op{\Circ[+]{c}}$ and we call it the \emph{mirror image} of $\Circ[+]{c}$. With this notation, we can nicely express the following result that we will often use in our diagrammatic proofs. %
\begin{lem}\label{lm:wrong way}
	In a cartesian bicategory, the following holds for every arrow $c \colon X \to Y$.
	\[
\begin{tikzpicture}
	\begin{pgfonlayer}{nodelayer}
		\node [{boxStyle/+}] (134) at (0.5, 0.5) {$c$};
		\node [style=none] (135) at (-1.25, 0) {};
		\node [style=none] (136) at (-1.25, 1.225) {};
		\node [style=none] (137) at (-1.25, -1) {};
		\node [style=none] (138) at (1.5, -1) {};
		\node [style=none] (139) at (1.5, 1.225) {};
		\node [style={dotStyle/+}] (144) at (-0.5, 0) {};
		\node [style=none] (145) at (0.25, -0.5) {};
		\node [style=none] (146) at (1.5, 0.5) {};
		\node [style=none] (147) at (1.5, -0.5) {};
		\node [style=none] (150) at (0.25, 0.5) {};
		\node [style=label] (151) at (-1.675, 0) {$X$};
		\node [style=label] (152) at (1.925, 0.5) {$Y$};
		\node [style=label] (153) at (1.925, -0.5) {$X$};
	\end{pgfonlayer}
	\begin{pgfonlayer}{edgelayer}
		\draw [{bgStyle/+}] (138.center)
			 to (137.center)
			 to (136.center)
			 to (139.center)
			 to cycle;
		\draw [style={wStyle/+}, bend left] (145.center) to (144);
		\draw [style={wStyle/+}] (144) to (135.center);
		\draw [style={wStyle/+}] (134) to (146.center);
		\draw [style={wStyle/+}] (147.center) to (145.center);
		\draw [style={wStyle/+}] (134) to (150.center);
		\draw [style={wStyle/+}, bend right] (150.center) to (144);
	\end{pgfonlayer}
\end{tikzpicture}
    \leq
    \begin{tikzpicture}
	\begin{pgfonlayer}{nodelayer}
		\node [{boxOpStyle/+}] (134) at (-0.75, -0.25) {$c$};
		\node [style={boxStyle/+}] (135) at (-3, 0.25) {$c$};
		\node [style=none] (136) at (-4, 1.225) {};
		\node [style=none] (137) at (-4, -1) {};
		\node [style=none] (138) at (0.25, -1) {};
		\node [style=none] (139) at (0.25, 1.225) {};
		\node [style={dotStyle/+}] (144) at (-2, 0.25) {};
		\node [style=none] (145) at (-1.25, 0.75) {};
		\node [style=none] (146) at (0.25, 0.75) {};
		\node [style=none] (147) at (0.25, -0.25) {};
		\node [style=none] (150) at (-1.25, -0.25) {};
		\node [style=none] (152) at (-4, 0.25) {};
		\node [style=label] (153) at (-4.425, 0.25) {$X$};
		\node [style=label] (154) at (0.675, -0.25) {$X$};
		\node [style=label] (155) at (0.675, 0.75) {$Y$};
	\end{pgfonlayer}
	\begin{pgfonlayer}{edgelayer}
		\draw [{bgStyle/+}] (138.center)
			 to (137.center)
			 to (136.center)
			 to (139.center)
			 to cycle;
		\draw [style={wStyle/+}, bend right] (145.center) to (144);
		\draw [style={wStyle/+}] (144) to (135);
		\draw [style={wStyle/+}] (134) to (150.center);
		\draw [style={wStyle/+}, bend left] (150.center) to (144);
		\draw [style={wStyle/+}] (134) to (147.center);
		\draw [style={wStyle/+}] (146.center) to (145.center);
		\draw [style={wStyle/+}] (135) to (152.center);
	\end{pgfonlayer}
\end{tikzpicture}      
\]
\end{lem}
\begin{proof}
	See e.g. Lemma 4.3 in~\cite{DBLP:journals/corr/abs-1711-08699}.
\end{proof}

For all objects $X,Y$ and arrows $c,d\colon X \to Y$, one can define $c \sqcap d$ and $\top$ as follows. %
\begin{equation}\label{eq:def:cap}c \sqcap d \defeq \intersectionCirc{c}{d}[X][Y] \qquad  \top \defeq \topCirc[X][Y]
\end{equation}
We have already seen in Example~\ref{eq:intersectionandtop} that these diagrams, when interpreted in $\Relp$, denote respectively intersection and top. It is easy to show that in any cartesian bicategory $\Cat{C}$, $\sqcap$ and $\top$ form a \emph{meet-semilattice with top}, namely
$\sqcap$ is associative, commutative, idempotent and has $\top$ as unit.
\begin{lem}\label{lemma meet}%
For all arrows $c,d,e\colon X \to Y$ of a cartesian bicategory, the following hold.
\[(c \sqcap d) \sqcap e = c \sqcap (d \sqcap e) \qquad c \sqcap \top = c \qquad c \sqcap d = d \sqcap c \qquad c \sqcap c =c \]
\end{lem}
\begin{proof} The first three equalities --- associativity, unitality and commutativity --- follow directly from \eqref{ax:comPlusAssoc}, \eqref{ax:comPlusUnit}, \eqref{ax:comPlusComm}, as well as  \eqref{ax:monPlusAssoc}, \eqref{ax:monPlusUnit}, \eqref{ax:monPlusComm}. The last equality, idempotency, is proved diagrammatically below, where we show the two inclusions separately:
        \[
        \intersectionCirc{c}{c}[X][Y]
    \stackrel{\eqref{ax:plusDiscCodisc}}{\leq}
    \begin{tikzpicture}
	\begin{pgfonlayer}{nodelayer}
		\node [{dotStyle/+}] (107) at (2.25, 0) {};
		\node [style=none] (108) at (1.65, 0.575) {};
		\node [style=none] (109) at (1.65, -0.575) {};
		\node [style=none] (122) at (2.775, -1.2) {};
		\node [style=none] (123) at (2.775, 1.2) {};
		\node [style=none] (124) at (-1.5, 1.2) {};
		\node [style=none] (125) at (-1.5, -1.2) {};
		\node [style=none] (127) at (2.775, 0) {};
		\node [{dotStyle/+}] (128) at (-1, 0) {};
		\node [style=none] (129) at (-0.4, 0.575) {};
		\node [style=none] (130) at (-0.4, -0.575) {};
		\node [style=none] (131) at (-1.5, 0) {};
		\node [style=none] (133) at (2.775, -0.825) {};
		\node [style={boxStyle/+}] (134) at (0, 0.575) {$c$};
		\node [style={boxStyle/+}] (135) at (0, -0.575) {$c$};
		\node [style=label] (136) at (-1.975, 0) {$X$};
		\node [style=label] (137) at (3.25, 0) {$Y$};
		\node [{dotStyle/+}] (139) at (1, -0.575) {};
		\node [{dotStyle/+}] (140) at (1.5, -0.575) {};
	\end{pgfonlayer}
	\begin{pgfonlayer}{edgelayer}
		\draw [{bgStyle/+}] (124.center)
			 to (123.center)
			 to (122.center)
			 to (125.center)
			 to cycle;
		\draw [{wStyle/+}, bend right=45] (109.center) to (107);
		\draw [{wStyle/+}, bend right=45] (107) to (108.center);
		\draw [{wStyle/+}] (107) to (127.center);
		\draw [{wStyle/+}, bend left=45] (130.center) to (128);
		\draw [{wStyle/+}, bend left=45] (128) to (129.center);
		\draw [{wStyle/+}] (128) to (131.center);
		\draw [style={wStyle/+}] (129.center) to (134);
		\draw [style={wStyle/+}] (130.center) to (135);
		\draw [style={wStyle/+}] (134) to (108.center);
		\draw (139) to (135);
		\draw (140) to (109.center);
	\end{pgfonlayer}
\end{tikzpicture}
    \stackrel{\eqref{ax:discPlusLaxNat}}{\leq}
    \begin{tikzpicture}
	\begin{pgfonlayer}{nodelayer}
		\node [{dotStyle/+}] (107) at (1, 0) {};
		\node [style=none] (108) at (0.4, 0.575) {};
		\node [style=none] (109) at (0.4, -0.575) {};
		\node [style=none] (122) at (1.525, -1.2) {};
		\node [style=none] (123) at (1.525, 1.2) {};
		\node [style=none] (124) at (-1.5, 1.2) {};
		\node [style=none] (125) at (-1.5, -1.2) {};
		\node [style=none] (127) at (1.525, 0) {};
		\node [{dotStyle/+}] (128) at (-1, 0) {};
		\node [style=none] (129) at (-0.4, 0.575) {};
		\node [style=none] (130) at (-0.4, -0.575) {};
		\node [style=none] (131) at (-1.5, 0) {};
		\node [style=none] (133) at (1.525, -0.825) {};
		\node [style={boxStyle/+}] (134) at (0, 0.575) {$c$};
		\node [style=label] (136) at (-1.975, 0) {$X$};
		\node [style=label] (137) at (2, 0) {$Y$};
		\node [{dotStyle/+}] (156) at (-0.275, -0.575) {};
		\node [{dotStyle/+}] (157) at (0.225, -0.575) {};
	\end{pgfonlayer}
	\begin{pgfonlayer}{edgelayer}
		\draw [{bgStyle/+}] (124.center)
			 to (123.center)
			 to (122.center)
			 to (125.center)
			 to cycle;
		\draw [{wStyle/+}, bend right=45] (109.center) to (107);
		\draw [{wStyle/+}, bend right=45] (107) to (108.center);
		\draw [{wStyle/+}] (107) to (127.center);
		\draw [{wStyle/+}, bend left=45] (130.center) to (128);
		\draw [{wStyle/+}, bend left=45] (128) to (129.center);
		\draw [{wStyle/+}] (128) to (131.center);
		\draw [style={wStyle/+}] (129.center) to (134);
		\draw [style={wStyle/+}] (134) to (108.center);
		\draw (109.center) to (157);
		\draw (156) to (130.center);
	\end{pgfonlayer}
\end{tikzpicture}
    \stackrel{\eqref{ax:comPlusUnit}, \eqref{ax:monPlusUnit}}{=}
    \boxCirc[+]{c}[X][Y]
\]
and
\[
    \boxCirc[+]{c}[X][Y]
    \stackrel{\eqref{ax:plusSpecFrob}}{=}
    \begin{tikzpicture}
	\begin{pgfonlayer}{nodelayer}
		\node [{dotStyle/+}] (107) at (1.75, 0) {};
		\node [style=none] (108) at (1.15, 0.575) {};
		\node [style=none] (109) at (1.15, -0.575) {};
		\node [style=none] (122) at (2.275, -1.2) {};
		\node [style=none] (123) at (2.275, 1.2) {};
		\node [style=none] (124) at (-1.75, 1.2) {};
		\node [style=none] (125) at (-1.75, -1.2) {};
		\node [style=none] (127) at (2.275, 0) {};
		\node [{dotStyle/+}] (128) at (0.25, 0) {};
		\node [style=none] (129) at (0.85, 0.575) {};
		\node [style=none] (130) at (0.85, -0.575) {};
		\node [style=none] (131) at (-1.75, 0) {};
		\node [style=none] (133) at (2.275, -0.825) {};
		\node [style=label] (136) at (-2.225, 0) {$X$};
		\node [style=label] (137) at (2.75, 0) {$Y$};
		\node [style={boxStyle/+}] (138) at (-0.75, 0) {$c$};
	\end{pgfonlayer}
	\begin{pgfonlayer}{edgelayer}
		\draw [{bgStyle/+}] (124.center)
			 to (123.center)
			 to (122.center)
			 to (125.center)
			 to cycle;
		\draw [{wStyle/+}, bend right=45] (109.center) to (107);
		\draw [{wStyle/+}, bend right=45] (107) to (108.center);
		\draw [{wStyle/+}] (107) to (127.center);
		\draw [{wStyle/+}, bend left=45] (130.center) to (128);
		\draw [{wStyle/+}, bend left=45] (128) to (129.center);
		\draw [{wStyle/+}] (128) to (131.center);
		\draw (130.center) to (109.center);
		\draw (129.center) to (108.center);
	\end{pgfonlayer}
\end{tikzpicture}
    \stackrel{\eqref{ax:comPlusLaxNat}}{\leq}
    \intersectionCirc{c}{c}[X][Y]. \qedhere
\]
\end{proof}
\begin{rem}\label{rem:enrichment}Note that, however, $\Cat{C}$ is usually \emph{not} enriched over meet-semilattices since $\seq[+]$ distributes only laxly over $\sqcap$. Indeed, in $\Relp$,  \[R\seq[+](S \cap T) \subseteq (R \seq[+] S)\cap(R\seq[+]T)\] holds for all (properly typed) relations  $R,S,T$, but the reverse does not.
\end{rem}

\medskip

\begin{figure*}[t]
        \mylabel{ax:comMinusAssoc}{$\copier[-]$-as}
        \mylabel{ax:comMinusUnit}{$\copier[-]$-un}
        \mylabel{ax:comMinusComm}{$\copier[-]$-co}
        \mylabel{ax:monMinusAssoc}{$\cocopier[-]$-as}
        \mylabel{ax:monMinusUnit}{$\cocopier[-]$-un}
        \mylabel{ax:monMinusComm}{$\cocopier[-]$-co}
        \mylabel{ax:minusSpecFrob}{S$^\bullet$}
        \mylabel{ax:minusFrob}{F$^\bullet$}
        \mylabel{ax:comMinusLaxNat}{$\copier[-]$-nat}
        \mylabel{ax:discMinusLaxNat}{$\discard[-]$-nat}
        \mylabel{ax:minusCodiscDisc}{$\eta\codiscard[-]$}
        \mylabel{ax:minusDiscCodisc}{$\epsilon\codiscard[-]$}
        \mylabel{ax:minusCocopyCopy}{$\eta\!\cocopier[-]$}
        \mylabel{ax:minusCopyCocopy}{$\epsilon\!\cocopier[-]$}
        \centering
        \resizebox{\textwidth}{!}{
        $
        $
        $\begin{array}{c|c}
                \begin{array}{c@{\!}c@{\!}c c@{\!}c@{\!}c}
                        {
    \InputIfFileExists{axiomsNEW/cb/minus/comAssoc1.tikz}{}{\input{tikz/axiomsNEW/cb/minus/comAssoc1.tikz}}
}   & \Leq{\eqref*{ax:comMinusAssoc}} &  {
    \InputIfFileExists{axiomsNEW/cb/minus/comAssoc2.tikz}{}{\input{tikz/axiomsNEW/cb/minus/comAssoc2.tikz}}
}
                        &
                        {
    \InputIfFileExists{axiomsNEW/cb/minus/monAssoc1.tikz}{}{\input{tikz/axiomsNEW/cb/minus/monAssoc1.tikz}}
}   & \Leq{\eqref*{ax:monMinusAssoc}} &  {
    \InputIfFileExists{axiomsNEW/cb/minus/monAssoc2.tikz}{}{\input{tikz/axiomsNEW/cb/minus/monAssoc2.tikz}}
}
                        \\
                        {
    \InputIfFileExists{axiomsNEW/cb/minus/comUnit.tikz}{}{\input{tikz/axiomsNEW/cb/minus/comUnit.tikz}}
}     & \Leq{\eqref*{ax:comMinusUnit}}  &  {\idCirc[-][X]}
                        &
                        {
    \InputIfFileExists{axiomsNEW/cb/minus/monUnit.tikz}{}{\input{tikz/axiomsNEW/cb/minus/monUnit.tikz}}
}     & \Leq{\eqref*{ax:monMinusUnit}}  &  {\idCirc[-][X]}
                        \\
                        {
    \InputIfFileExists{axiomsNEW/cb/minus/comComm.tikz}{}{\input{tikz/axiomsNEW/cb/minus/comComm.tikz}}
}     & \Leq{\eqref*{ax:comMinusComm}}  &  {\copierCirc[-][X]}
                        &
                        {
    \InputIfFileExists{axiomsNEW/cb/minus/monComm.tikz}{}{\input{tikz/axiomsNEW/cb/minus/monComm.tikz}}
}     & \Leq{\eqref*{ax:monMinusComm}}  &  {\cocopierCirc[-][X]}
                \end{array}
                &
                \begin{array}{c@{}c@{}c}
                        {
    \InputIfFileExists{axiomsNEW/cb/minus/frob1.tikz}{}{\input{tikz/axiomsNEW/cb/minus/frob1.tikz}}
}       & \Leq{\eqref*{ax:minusFrob}}     &  {
    \InputIfFileExists{axiomsNEW/cb/minus/frob2.tikz}{}{\input{tikz/axiomsNEW/cb/minus/frob2.tikz}}
}
                        \\
                        {
    \InputIfFileExists{axiomsNEW/cb/minus/specFrob.tikz}{}{\input{tikz/axiomsNEW/cb/minus/specFrob.tikz}}
}    & \Leq{\eqref*{ax:minusSpecFrob}} &  {\idCirc[-][X]}
                \end{array}
                \\
                \midrule
                \begin{array}{@{\!\!\!\!}c@{\!}c@{\!}c @{\qquad} c@{\!}c@{\!}c}
                        {
    \InputIfFileExists{axiomsNEW/cb/minus/specFrob.tikz}{}{\input{tikz/axiomsNEW/cb/minus/specFrob.tikz}}
}     &\Lleq{\eqref*{ax:minusCopyCocopy}}& {\idCirc[-][X]}
                        &
                        {
    \InputIfFileExists{axiomsNEW/bottom.tikz}{}{\input{tikz/axiomsNEW/bottom.tikz}}
}     &\Lleq{\eqref*{ax:minusDiscCodisc}}& {\idCirc[-][X]}
                        \\
                        {
    \InputIfFileExists{axiomsNEW/id2M.tikz}{}{\input{tikz/axiomsNEW/id2M.tikz}}
} & \Lleq{\eqref*{ax:minusCocopyCopy}}& {
    \InputIfFileExists{axiomsNEW/cb/minus/cocopierCopier.tikz}{}{\input{tikz/axiomsNEW/cb/minus/cocopierCopier.tikz}}
}
                        &
                        {\emptyCirc[-]} & \Lleq{\eqref*{ax:minusCodiscDisc}}& {
    \InputIfFileExists{axiomsNEW/cb/minus/codiscDisc.tikz}{}{\input{tikz/axiomsNEW/cb/minus/codiscDisc.tikz}}
}
                \end{array}
                &
                \begin{array}{c@{}c@{}c}
                        
    \InputIfFileExists{axiomsNEW/cb/minus/copierLaxNat2.tikz}{}{\input{tikz/axiomsNEW/cb/minus/copierLaxNat2.tikz}}
 &\Lleq{\eqref*{ax:comMinusLaxNat}}& {
    \InputIfFileExists{axiomsNEW/cb/minus/copierLaxNat1.tikz}{}{\input{tikz/axiomsNEW/cb/minus/copierLaxNat1.tikz}}
}
                        \\
                        {\discardCirc[-][X][X]} &\Lleq{\eqref*{ax:discMinusLaxNat}}& {
    \InputIfFileExists{axiomsNEW/cb/minus/discardLaxNat.tikz}{}{\input{tikz/axiomsNEW/cb/minus/discardLaxNat.tikz}}
}
                \end{array}
        \end{array}$}
        \caption{Axioms of cocartesian bicategories}\label{fig:cocb axioms}
        \end{figure*}

Let us now turn to \emph{co}cartesian bicategories. Our main example is $(\Relm, \copier[-], \cocopier[-])$. $\Relm$ is the bicategory of sets and relations ordered by $\subseteq$ with composition $\seq[-]$, identities $\id[-]$ and $\tensor[-]$ defined as in \eqref{eq:seqRel}, \eqref{eq:idRel} and \eqref{eq:tensorREL}. Comonoids $(\copier[-][X],\discard[-][X])$ and  monoids $(\cocopier[-][X] , \codiscard[-][X])$ are those of~\eqref{eq:comonoidsREL}. To see that $\Relm$ is a cocartesian bicategory, observe that the complement $\nega{(\cdot)}$ is a poset-enriched symmetric monoidal isomorphism $\nega{(\cdot)} \colon \co{(\Relp)} \to \Relm$ preserving (co)monoids.

We draw arrows of cocartesian bicategories in black: $\copier[-][X]$,$\discard[-][X]$, $\cocopier[-][X]$ and $\codiscard[-][X]$ are drawn  ${\copierCirc[-][X]}$, ${\discardCirc[-][X]}$, ${\cocopierCirc[-][X]}$ and ${\codiscardCirc[-][X]}$. Following this convention, the axioms of cocartesian bicategories are in Fig.~\ref{fig:cocb axioms}; they can also be obtained from Fig.~\ref{fig:cb axioms} by inverting both the colours and the order.

It is not surprising that in a cocartesian bicategory $\Cat{C}$, every homset $\Cat{C}[X,Y]$ carries a join semi-lattice with bottom, where $c\sqcup d$ and $\bot$ are defined for all arrows $c,d\colon X \to Y$ as follows.
\begin{equation}\label{eq:def:cup}
c \sqcup d \defeq \unionCirc{c}{d}[X][Y]
\qquad
\bot \defeq \bottomCirc[X][Y]
\end{equation}
\begin{lem}\label{lemma join}
For all arrows $c,d,e\colon X \to Y$ of a cocartesian bicategory, the following hold.
\[(c \sqcup d) \sqcup e = c \sqcup (d \sqcup e) \qquad c \sqcup \bot = c \qquad c \sqcup d = d \sqcup c \qquad c \sqcup c =c \]
\end{lem}
\begin{proof}
        The proof is analogous to that of Lemma~\ref{lemma meet}, but relies on the axioms in Figure~\ref{fig:cocb axioms}.
\end{proof}

\section{Linear Bicategories}\label{sec:linbic}
We have seen that $\Relp$ forms a cartesian bicategory, and $\Relm$ a cocartesian bicategory. Categorically, they are remarkably similar --- as evidenced by the isomorphism $\nega{(\cdot)}$ --- but from a logical viewpoint they represent two complementary parts of $\FOL$: $\Relp$ the existential conjunctive fragment, and $\Relm$ the universal disjunctive fragment. 

To discover the full story, we must merge them into one entity and study the algebraic interactions between them. However, the coexistence of two different compositions $\seq[+]$ and $\seq[-]$ brings us out of the realm of ordinary categories. The solution is linear bicategories~\cite{cockett2000introduction}. %
Here $\seq[+]$ linearly distributes over $\seq[-]$, as in Peirce's calculus.
To keep our development easier, we stick to the poset-enriched case and rely on diagrams, using white and black to distinguish $\seq[+]$ and $\seq[-]$.

\begin{defi}\label{def:linear bicategory} %
A \emph{linear bicategory} $(\Cat{C}, \seq[+], \id[+], \seq[-], \id[-])$ consists of two poset-enriched categories $(\Cat{C}, \seq[+], \id[+])$ and $(\Cat{C}, \seq[-], \id[-])$ with the same objects, arrows and orderings but possibly different identities and compositions such that  $\seq[+]$ linearly distributes over $\seq[-]$, i.e., the following hold.
\[ a \seq[+] (b \seq[-] c) \stackrel{\eqref*{ax:leftLinDistr}}{\leq} (a \seq[+] b) \seq[-] c \qquad (a \seq[-] b) \seq[+] c \stackrel{\eqref*{ax:rightLinDistr}}{\leq} a \seq[-] (b \seq[+] c) 
\]
\end{defi}
\begin{defi}\label{def:monlinear bicategory}
A \emph{symmetric monoidal linear bicategory} $(\Cat{C}, \seq[+], \id[+], \seq[-], \id[-], \tensor[+],\symm[+], \tensor[-], \symm[-], \unittensor)$, shortly  $(\Cat{C},\tensor[+], \tensor[-], \unittensor)$, consists of
a linear bicategory $(\Cat{C}, \seq[+], \id[+], \seq[-], \id[-])$ and  two poset-enriched symmetric monoidal categories $(\Cat{C}, \tensor[+],  \unittensor)$ and $(\Cat{C}, \tensor[-], \unittensor)$ such that \  $\tensor[+]$ and $\tensor[-]$ agree on objects, i.e., $X \tensor[+]Y= X\tensor[-]Y$, share the same unit $\unittensor$ and
\begin{enumerate}
\item there are linear strengths for $(\tensor[+],\tensor[-])$, i.e.,: %
\[\begin{array}{rclrcl}
	 (a \seq[-] b) \tensor[+] (c \seq[-] d) &\stackrel{\eqref*{ax:linStrn1}}{\leq}& (a \tensor[+] c) \seq[-] (b \tensor[-] d)
          &
          (a \tensor[-] c) \seq[+] (b \tensor[+] d) &\stackrel{\eqref*{ax:linStrn3}}{\leq}& (a \seq[+] b) \tensor[-] (c \seq[+] d)          
          \\
	 (a \seq[-] b) \tensor[+] (c \seq[-] d) &\stackrel{\eqref*{ax:linStrn2}}{\leq}& (a \tensor[-] c) \seq[-] (b \tensor[+] d)
          &
          (a \tensor[+] c) \seq[+] (b \tensor[-] d) &\stackrel{\eqref*{ax:linStrn4}}{\leq}& (a \seq[+] b) \tensor[-] (c \seq[+] d)
\end{array}\]

\item the black tensor $\tensor[-]$  preserves $\id[+]$ colaxly and $\tensor[+]$  preserves $\id[-]$ laxly, i.e.,;
\[\begin{array}{rclrcl}
          \id[+][X\tensor[+]Y] &\stackrel{\eqref*{ax:tensorMinusIdPlus}}{\leq}& \id[+][X] \tensor[-] \id[+][Y] 
          & 
          \id[-][X] \tensor[+] \id[-][Y] &\stackrel{\eqref*{ax:tensorPlusIdMinus}}{\leq}& \id[-][X\tensor[-]Y] 
\end{array}\]
\end{enumerate}
A \emph{morphism of symmetric monoidal linear bicategories} $\mathcal{F}\colon (\Cat{C_1}, \tensor[+], \tensor[-],  \unittensor) \to (\Cat{C_2},\tensor[+], \tensor[-], \unittensor)$ consists of two poset-enriched symmetric monoidal  functors $\mathcal{F}^\circ \colon(\Cat{C_1},  \tensor[+], \unittensor) \to (\Cat{C_2}, \tensor[+],  \unittensor)$ and $\mathcal{F}^\bullet \colon(\Cat{C_1}, \ \tensor[-], \unittensor) \to (\Cat{C_2},  \tensor[-],  \unittensor)$ that agree on objects and arrows, namely $\mathcal{F}^{\circ} (X) = \mathcal{F}^{\bullet}(X)$ and $\mathcal{F}^{\circ} (c) = \mathcal{F}^{\bullet}(c)$ for all objects $X$ and arrows $c$. %
\end{defi}
\begin{rem}
In the literature %
 $\seq[+]$, $\id[+]$, $\seq[-]$ and $\id[-]$ are written with the linear logic notation $\otimes$, $\top$, $\oplus$ and $\bot$. %
Modulo this, %
the traditional notion of linear bicategory (Definition 2.1 in \cite{cockett2000introduction}) coincides with the one in Definition \ref{def:linear bicategory} whenever the 2-structure is collapsed to a poset. %
Monoidal products on linear bicategories are not well studied although %
the axioms in Definition~\ref{def:monlinear bicategory}.(1) already appeared in \cite{naeimabadiconstructing}. They are the linear strengths of the pair $(\tensor[+], \tensor[-])$ seen as a linear functor (Definition 2.4 in \cite{cockett2000introduction}), a notion of morphism that crucially differs from ours on the fact that the $\mathcal{F}^\circ$ and $\mathcal{F}^\bullet$ may not coincide on arrows. Instead the inequalities  \eqref{ax:tensorPlusIdMinus} and \eqref{ax:tensorMinusIdPlus}  are, to the best of our knowledge, novel. Beyond being natural, they are crucial for Lemma \ref{lm:linearly distr cat} below. 

Fig. \ref{fig:closed lin axioms} illustrates the diagrams corresponding to the axioms of Definition \ref{def:linear bicategory} in the top-left corner and Definition \ref{def:monlinear bicategory} in the bottom.
\end{rem}

\begin{figure}[t]
    \mylabel{ax:leftLinDistr}{$\delta_l$}
    \mylabel{ax:rightLinDistr}{$\delta_r$}
    \mylabel{ax:linStrn1}{$\nu^\circ_l$}
    \mylabel{ax:linStrn2}{$\nu^\circ_r$}
    \mylabel{ax:linStrn3}{$\nu^\bullet_l$}
    \mylabel{ax:linStrn4}{$\nu^\bullet_r$}
    \mylabel{ax:tensorPlusIdMinus}{$\tensor[+]^\bullet$}
    \mylabel{ax:tensorMinusIdPlus}{$\tensor[-]^\circ$}
    \mylabel{ax:tauSymmPlus}{$\tau\symm[+]$}
    \mylabel{ax:tauRPlus}{$\tau R^\circ$}
    \mylabel{ax:gammaSymmPlus}{$\gamma\symm[+]$}
    \mylabel{ax:gammaRPlus}{$\gamma R^\circ$}
    \mylabel{ax:tauSymmMinus}{$\tau\symm[-]$}
    \mylabel{ax:tauRMinus}{$\tau R^\bullet$}
    \mylabel{ax:gammaSymmMinus}{$\gamma\symm[-]$}
    \mylabel{ax:gammaRMinus}{$\gamma R^\bullet$}
    \[
        \resizebox{\textwidth}{!}{$
        \begin{array}{@{}c@{}c@{}c@{}}
            \multicolumn{3}{c}{
                \begin{array}{@{}c@{}|c@{}}
                    \begin{array}{@{}c@{}c@{}c@{}}
                        
    \InputIfFileExists{axiomsNEW/leftLinDistr1.tikz}{}{\input{tikz/axiomsNEW/leftLinDistr1.tikz}}
  & \!\!\Lleq{\eqref*{ax:leftLinDistr}}\!\!  & 
    \InputIfFileExists{axiomsNEW/leftLinDistr2.tikz}{}{\input{tikz/axiomsNEW/leftLinDistr2.tikz}}
 \\[15pt]
                        
    \InputIfFileExists{axiomsNEW/rightLinDistr1.tikz}{}{\input{tikz/axiomsNEW/rightLinDistr1.tikz}}
 & \!\!\Lleq{\eqref*{ax:rightLinDistr}}\!\! & 
    \InputIfFileExists{axiomsNEW/rightLinDistr2.tikz}{}{\input{tikz/axiomsNEW/rightLinDistr2.tikz}}

                    \end{array}
                    &
                    \begin{array}{@{}c@{}c@{}c@{}c@{}c@{}c@{}}
                        
    \InputIfFileExists{axiomsNEW/idXYP.tikz}{}{\input{tikz/axiomsNEW/idXYP.tikz}}
 & \!\!\Lleq{\eqref*{ax:tauSymmPlus}}\!\!   & 
    \InputIfFileExists{axiomsNEW/linadj/symMsym.tikz}{}{\input{tikz/axiomsNEW/linadj/symMsym.tikz}}
 & 
    \InputIfFileExists{axiomsNEW/linadj2/symPsym.tikz}{}{\input{tikz/axiomsNEW/linadj2/symPsym.tikz}}
     & \!\!\Lleq{\eqref*{ax:gammaSymmPlus}}\!\!   & 
    \InputIfFileExists{axiomsNEW/idXYM.tikz}{}{\input{tikz/axiomsNEW/idXYM.tikz}}
 \\
                        
    \InputIfFileExists{axiomsNEW/idXYP.tikz}{}{\input{tikz/axiomsNEW/idXYP.tikz}}
 & \!\!\Lleq{\eqref*{ax:tauSymmMinus}}\!\!   & 
    \InputIfFileExists{axiomsNEW/linadj2/symMsym.tikz}{}{\input{tikz/axiomsNEW/linadj2/symMsym.tikz}}
 & 
    \InputIfFileExists{axiomsNEW/linadj/symPsym.tikz}{}{\input{tikz/axiomsNEW/linadj/symPsym.tikz}}
     & \!\!\Lleq{\eqref*{ax:gammaSymmMinus}}\!\!   & 
    \InputIfFileExists{axiomsNEW/idXYM.tikz}{}{\input{tikz/axiomsNEW/idXYM.tikz}}
 \\
                        \idCirc[+][X] & \!\!\Lleq{\eqref*{ax:tauRPlus}}\!\!  & 
    \InputIfFileExists{axiomsNEW/linadj/rMrop.tikz}{}{\input{tikz/axiomsNEW/linadj/rMrop.tikz}}
 & 
    \InputIfFileExists{axiomsNEW/linadj/ropPr.tikz}{}{\input{tikz/axiomsNEW/linadj/ropPr.tikz}}
       & \!\!\Lleq{\eqref*{ax:gammaRPlus}}\!\!  & \idCirc[-][Y] \\
                        \idCirc[+][Y] & \!\!\Lleq{\eqref*{ax:tauRMinus}}\!\!  & 
    \InputIfFileExists{axiomsNEW/linadj2/rMrop.tikz}{}{\input{tikz/axiomsNEW/linadj2/rMrop.tikz}}
 & 
    \InputIfFileExists{axiomsNEW/linadj2/ropPr.tikz}{}{\input{tikz/axiomsNEW/linadj2/ropPr.tikz}}
       &\!\!\Lleq{\eqref*{ax:gammaRMinus}}\!\!  & \idCirc[-][X]
                    \end{array}
                \end{array}
            } \\
            \midrule
                \begin{array}{@{} c @{} c @{} c @{}}
                    
    \InputIfFileExists{axiomsNEW/linStr1_1.tikz}{}{\input{tikz/axiomsNEW/linStr1_1.tikz}}
 & \!\!\Lleq{\eqref*{ax:linStrn1}}\!\! & 
    \InputIfFileExists{axiomsNEW/linStr1_2.tikz}{}{\input{tikz/axiomsNEW/linStr1_2.tikz}}
 \\[20pt]
                    
    \InputIfFileExists{axiomsNEW/linStr3_1.tikz}{}{\input{tikz/axiomsNEW/linStr3_1.tikz}}
 & \!\!\Lleq{\eqref*{ax:linStrn3}}\!\! & 
    \InputIfFileExists{axiomsNEW/linStr3_2.tikz}{}{\input{tikz/axiomsNEW/linStr3_2.tikz}}

                \end{array}
            &
                \begin{array}{@{} c @{} c @{} c @{}}
                    
    \InputIfFileExists{axiomsNEW/linStr1_1.tikz}{}{\input{tikz/axiomsNEW/linStr1_1.tikz}}
 & \!\!\Lleq{\eqref*{ax:linStrn2}}\!\! & 
    \InputIfFileExists{axiomsNEW/linStr2_2.tikz}{}{\input{tikz/axiomsNEW/linStr2_2.tikz}}
 \\[20pt]
                    
    \InputIfFileExists{axiomsNEW/linStr4_1.tikz}{}{\input{tikz/axiomsNEW/linStr4_1.tikz}}
 & \!\!\Lleq{\eqref*{ax:linStrn4}}\!\! & 
    \InputIfFileExists{axiomsNEW/linStr3_2.tikz}{}{\input{tikz/axiomsNEW/linStr3_2.tikz}}

                \end{array}
            &
            \begin{array}{@{} c @{} c @{} c @{}}
                
    \InputIfFileExists{axiomsNEW/id2Pbig.tikz}{}{\input{tikz/axiomsNEW/id2Pbig.tikz}}
   & \!\!\Lleq{\eqref*{ax:tensorMinusIdPlus}}\!\!  & 
    \InputIfFileExists{axiomsNEW/idPMinusidP.tikz}{}{\input{tikz/axiomsNEW/idPMinusidP.tikz}}
 \\[20pt]
                
    \InputIfFileExists{axiomsNEW/idMPlusidM.tikz}{}{\input{tikz/axiomsNEW/idMPlusidM.tikz}}
 & \!\!\Lleq{\eqref*{ax:tensorPlusIdMinus}}\!\! & 
    \InputIfFileExists{axiomsNEW/id2Mbig.tikz}{}{\input{tikz/axiomsNEW/id2Mbig.tikz}}

            \end{array}
        \end{array}
        $}
    \]
    \caption{Axioms of closed symmetric monoidal linear bicategories}\label{fig:closed lin axioms}
\end{figure}

All linear bicategories in this paper are symmetric monoidal. We therefore omit the adjective \emph{symmetric monoidal} and refer to them simply as linear bicategories.
For a linear bicategory $(\Cat{C}, \tensor[+], \tensor[-], \unittensor)$, we will often refer to $(\Cat{C}, \tensor[+], \unittensor)$ as the \emph{white structure}, shorthand $\Cat{C}^\circ$, and to $(\Cat{C}, \tensor[-], \unittensor)$ as the \emph{black structure}, $\Cat{C}^\bullet$. Note that a morphism $\mathcal{F}$ is a mapping of objects and arrows that preserves the ordering, the white and black structures; thus we write $\mathcal{F}$ for both $\mathcal{F}^\circ$ and $\mathcal{F}^\bullet$.

If $(\Cat{C}, \tensor[+], \tensor[-], \unittensor)$ is  linear bicategory then  $(\opposite{\Cat{C}}, \tensor[+], \tensor[-], \unittensor)$ is  a linear bicategory. Similarly $(\co{\Cat{C}}, \tensor[-], \tensor[+], \unittensor)$, the bicategory obtained from $\Cat{C}$ by reversing the ordering and swapping the white and the black structure, is a linear bicategory. %

Our main example is the linear bicategory $(\Rel,\tensor[+],\tensor[-],\singleton)$ of sets and relations ordered by $\subseteq$. The white structure is the symmetric monoidal category $(\Relp, \tensor[+], \singleton)$, introduced in the previous section and the black structure is $(\Relm, \tensor[-], \singleton)$. Observe that the two have the same objects, arrows and ordering. The white and black monoidal products $\tensor[+]$ and $\tensor[-]$ agree on objects and are the cartesian product of sets. As common unit object, they have the singleton set $\singleton$.
We already observed in \eqref{eq:distributivityExpres} that the white composition $\seq[+]$ distributes over $\seq[-]$ and thus \eqref{ax:leftLinDistr} and \eqref{ax:rightLinDistr} hold. By using the definitions in \eqref{eq:seqRel}, \eqref{eq:idRel} and \eqref{eq:tensorREL}, the reader can easily check also the inequalities in Definition \ref{def:monlinear bicategory}.(1) and (2).

\begin{lem}\label{lm:mix cat}\label{lm:linearly distr cat}%
	Let $(\Cat{C}, \tensor[+], \tensor[-], \unittensor)$ be a linear bicategory. For all arrows $a,b,c$ the following hold: %
	\[ (1) \; \id[-][\unittensor] \leq \id[+][\unittensor] \qquad (2) \; a \tensor[+] b \leq a \tensor[-] b  \qquad (3) \; (a \tensor[-] b) \tensor[+] c \leq a \tensor[-] (b \tensor[+] c). \]
\end{lem}
\begin{proof}%
	The proof of $(1)$ is on the left and $(2)$ on the right:
	
	\noindent\begin{minipage}{0.48\linewidth}
		\begin{align*}
			\id[-][\unittensor] &= \id[-][\unittensor] \seq[+] \id[+][\unittensor] \\
					  &= \id[-][\unittensor] \seq[+] (\id[-][\unittensor] \seq[-] \id[+][\unittensor]) \\ 
					  &\leq (\id[-][\unittensor] \seq[+] \id[-][\unittensor]) \seq[-] \id[+][\unittensor] \tag{\ref{ax:leftLinDistr}} \\ 
					  &= (\id[-][\unittensor] \tensor[+] \id[-][\unittensor]) \seq[-] \id[+][\unittensor] \tag{SMC} \\ 
					  &\leq (\id[-][\unittensor] \tensor[-] \id[-][\unittensor]) \seq[-] \id[+][\unittensor] \tag{\ref{ax:tensorPlusIdMinus}}\\ 
					  &= \id[+][\unittensor]
		\end{align*}
		\end{minipage}\quad \vline
		\begin{minipage}{0.48\linewidth}
		\begin{align*}
			a \tensor[+] b =\; & (a \seq[-] \id[-]) \tensor[+] (b \seq[-] \id[-]) \\
			\leq\; & (a \tensor[-] b) \seq[-] (\id[-] \tensor[+] \id[-]) \tag{\ref{ax:linStrn2}} \\
			\leq\; & (a \tensor[-] b) \seq[-] (\id[-] \tensor[-] \id[-]) \tag{\ref{ax:tensorPlusIdMinus}} \\
			=\; & a \tensor[-] b
		\end{align*}
	\end{minipage}

	The proof of $(3)$ is given diagrammatically as follows:
	\input{tikz/proofs/linearlyDistrCat.tex}
	\qedhere
\end{proof}

\begin{rem}
As $\tensor[+]$ linearly distributes over $\tensor[-]$, it may seem that symmetric monoidal linear bicategories of Definition~\ref{def:monlinear bicategory} are linearly distributive~\cite{de1991dialectica,cockett1997weakly}. Moreover (1), (2) of Lemma \ref{lm:mix cat} may suggest that they are mix categories \cite{cockett1997proof}. This is not the case: functoriality of $\tensor[+]$ over  $\seq[-]$ and of $\tensor[-]$ over $\seq[+]$ fails in general.
\end{rem}

\subsection{Closed linear bicategories}
In \S~\ref{sec:cartesianbi}, we recalled adjoints of arrows in bicategories; in linear bicategories one can define \emph{linear} adjoints.
For $a \colon X \to Y$ and $b \colon Y \to X$, $a$ is  \emph{left linear adjoint} to $b$, or $b$ is \emph{right linear adjoint} to $a$, written $b \Vdash a$, if 
$\id[+][X] \leq a \seq[-] b$ and $b \seq[+] a \leq \id[-][Y]$.

Next we discuss some properties of right linear adjoints. Those of left adjoints are analogous but they do not feature in our exposition since in the categories of interest --- in next section --- left and right linear adjoint coincide. As expected, linear adjoints are unique.
\begin{lem}\label{lemma:uniquenessla}
If $b \Vdash a$ and $c \Vdash a$, then $b=c$.
\end{lem}
\begin{proof}%
    By the following two derivations.
    
    \noindent\noindent\begin{minipage}{0.48\linewidth}
    \begin{align*}
            b & =  b \seq[+] \id[+][X] \\
             &\leq b \seq[+] (a \seq[-] c) \tag{$c \Vdash a$}\\
             &\leq (b \seq[+] a )\seq[-] c \tag{\ref{ax:leftLinDistr}}\\
             & \leq  \id[-][Y] \seq[-] c \tag{$b \Vdash a$}\\
             &= c
    \end{align*}
    \end{minipage}\;\;\vline
    \begin{minipage}{0.5\linewidth}
    \begin{align*}
            c & =  c \seq[+] \id[+][X] \\
             &\leq c \seq[+] (a \seq[-] b) \tag{$b \Vdash a$}\\
             &\leq (c \seq[+] a )\seq[-] b \tag{\ref{ax:leftLinDistr}}\\
             & \leq  \id[-][Y] \seq[-] b \tag{$c \Vdash a$}\\
             &= b \tag*{\qedhere}
    \end{align*}\qedhere
    \end{minipage}
    \end{proof}

By virtue of the above result we can write $\rla{a}\colon Y \to X$ for \emph{the} right linear adjoint of $a\colon X \to Y$. %
With this notation one can write the \emph{left residual} of $b\colon Z \to Y$ by $a\colon X \to Y$ as $b\seq[-]\rla{a} \colon Z \to X$. The left residual is the greatest arrow $Z \to X$ making the diagram below commute laxly in $\Cat{C}^\circ$, namely if $c\seq[+] a \leq b$ then $c\leq b\seq[-]\rla{a}$.

\[\xymatrix{
    X \ar[r]^a & Y \\
    Z \ar[ru]_b \ar[u]|{b \seq[-]\rla{a}} \ar@(ul,dl)[u]^{c}
    } \]

When $a$ and $b$ have the same type and $c$ is the identity, the principle of residuation provides an alternative characterisation of the order on the arrows. 

\begin{lem}[Residuation]\label{lm:residuation}
$a \leq b$ iff $\id[+][X] \leq b \seq[-] \rla{a} $. %
\end{lem}
\begin{proof}%
    In the leftmost derivation we prove $a \leq b \Rightarrow \id[+][X] \leq b \seq[-] \rla{a} $ and in the rightmost $a \leq b \Leftarrow \id[+][X] \leq b \seq[-] \rla{a} $.
    
    \noindent\noindent\begin{minipage}{0.48\linewidth}
    \begin{align*}
            \id[+][X] & \leq   a \seq[-] \rla{a} \tag{$\rla{a} \Vdash a$} \\
             &\leq b \seq[-] \rla{a}  \tag{$a \leq b$}
    \end{align*}
    \end{minipage}\;\;\vline
    \begin{minipage}{0.5\linewidth}
    \begin{align*}
            a & =  \id[+][X] \seq[+]  a  \\
             &\leq (b \seq[-] \rla{a} ) \seq[+] a \tag{$\id[+][X] \leq b \seq[-] \rla{a} $}\\
             &\leq b \seq[-] ( \rla{a}  \seq[+] a )  \tag{\ref{ax:rightLinDistr}}\\
             & \leq  b \seq[-] \id[-][Y]   \tag{$\rla{a} \Vdash a$}\\
             &= b\tag*{\qedhere}
    \end{align*}
    \end{minipage}

\end{proof}

\begin{defi}
A linear bicategory $(\Cat{C},\tensor[+],\tensor[-],\unittensor)$ is said to be \emph{closed} if every $a\colon X \to Y$ has both a left and a right linear adjoint and the white symmetry is both left and right linear adjoint to the black symmetry, i.e., the following inequalities hold.
\[               
  \id[+][n+m] \stackrel{\eqref*{ax:tauSymmPlus}}{\leq} \symm[+][n][m] \seq[-] \symm[-][m][n]
                      \;        
                      \symm[-][n][m] \seq[+] \symm[+][m][n] \stackrel{\eqref*{ax:gammaSymmPlus}}{\leq} \id[-][n+m]
                      \;
                      \id[+][n+m] \stackrel{\eqref*{ax:tauSymmMinus}}{\leq} \symm[-][n][m] \seq[-] \symm[+][m][n]
                      \;
                      \symm[+][n][m] \seq[+] \symm[-][m][n] \stackrel{\eqref*{ax:gammaSymmMinus}}{\leq} \id[-][n+m]
\]
\end{defi}
\begin{rem}
The top-right corner of Fig.~\ref{fig:closed lin axioms} contains the diagrammatic representation of the four axioms for the symmetries. The remaining four axioms are intended for $\NPR$: for all generators $R\in \sign$, $R^\circ$ is both left and right linear adjoint to $R^\bullet$. As we will formally show in \S~\ref{sec:freely}, these conditions guarantee that \emph{all} diagrams of $\NPR$ have left and right linear adjoints and thus they give rise to a \emph{closed} linear bicategory.
\end{rem}

As expected, $(\Rel,\tensor[+],\tensor[-],\singleton)$ is a closed linear bicategory: both left and right linear adjoints of a relation $R \subseteq X \times Y $ are given by $\op{\nega{R}}=\{(y,x) \mid (x,y) \notin R \} \subseteq Y \times X$. With this, it is easy to see that $\symm[-] \Vdash \symm[+] \Vdash \symm[-]$ in $\Rel$.

Observe that if a linear bicategory $(\Cat{C},\tensor[+],\tensor[-],\unittensor)$ is closed, then also $(\opposite{\Cat{C}},\tensor[+],\tensor[-],\unittensor)$ and $(\co{\Cat{C}}, \tensor[-], \tensor[+], \unittensor)$ are closed.
The assignment $a \mapsto \rla{a}$ gives rise to an identity-on-objects functor $\rla{(\cdot)} \colon \Cat{C} \to \opposite{(\co{\Cat{C}})}$.

\begin{prop}\label{prop:rlamorphism}\label{lm:linadjmonoidal}\label{lm:linadjfunct}
$\rla{(\cdot)} \colon \Cat{C} \to \opposite{(\co{\Cat{C}})}$ is a morphism of linear bicategories, i.e., the laws in the first two columns of Table \ref{table:rlaproperties}.(b) hold.
\end{prop}
\begin{proof}
The laws for $\seq[+]$, $\id[+]$, $\seq[-]$ and $\id[-]$ are well-known, e.g. from~\cite{cockett2000introduction}. For convenience of the reader, we report anyway their proofs in Appendix \ref{app:linbic}. The remaining cases are illustrated below where $a$ and $b$ range over arbitrary morpshims $a\colon X_1 \to Y_1$ and $b\colon X_2 \to Y_2$.

    \begin{itemize}
    \item $\rla{(a \tensor[+] b )} = \rla{a} \tensor[-] \rla{b}$ . The following two derivations
    
    \noindent\begin{minipage}{0.44\linewidth}
    \begin{align*}
                & \id[+][X_1 \tensor[+] X_2] \\
              = & \id[+][X_1]\tensor[+] \id[+][X_2] \\
           \leq & (a \seq[-] \rla{a}) \tensor[+] (b \seq[-] \rla{b}) \tag{$\rla{a}\Vdash a \; ,\; \rla{b}\Vdash b$} \\
           \leq & (a \tensor[+] b) \seq[-] (\rla{a} \tensor[-] \rla{b}) \tag{\ref{ax:linStrn4}}
    \end{align*}
    \end{minipage}\; \vline
    \begin{minipage}{0.54\linewidth}
    \begin{align*}
             &(\rla{a} \tensor[-] \rla{b}) \seq[+] (a \tensor[+] b)  \\
        \leq &(\rla{a} \seq[+] a) \tensor[-] (\rla{b} \seq[+] b) \tag{\ref{ax:linStrn3}} \\
        \leq &\id[-][Y_1]\tensor[-] \id[-][Y_2] \tag{$\rla{a}\Vdash a \; ,\; \rla{b}\Vdash b$} \\
         =   &\id[-][Y_1 \tensor[-] Y_2]
     \end{align*}
    \end{minipage}
    show that $(\rla{a} \tensor[-] \rla{b}) \Vdash (a \tensor[+] b)$. Thus, by Lemma \ref{lemma:uniquenessla}, $\rla{(a \tensor[+] b)} = \rla{b} \tensor[-] \rla{a}$.

    \item $\rla{(a \tensor[-] b )} = \rla{a} \tensor[+] \rla{b}$. The following two derivations
    
    \noindent\begin{minipage}{0.44\linewidth}
    \begin{align*}
             &\id[+][X_1 \tensor[+] X_2] \\
            =&\id[+][X_1]\tensor[+] \id[+][X_2] \\
        \leq &(a \seq[-] \rla{a}) \tensor[+] (b \seq[-] \rla{b}) \tag{$\rla{a}\Vdash a \; ,\; \rla{b}\Vdash b$} \\
        \leq &(a \tensor[-] b) \seq[-] (\rla{a} \tensor[+] \rla{b}) \tag{\ref{ax:linStrn2}}
    \end{align*}
    \end{minipage}\; \vline
    \begin{minipage}{0.54\linewidth}
    \begin{align*}
             &(\rla{a} \tensor[+] \rla{b}) \seq[+] (a \tensor[-] b)   \\
        \leq &(\rla{a} \seq[+] a) \tensor[-] (\rla{b} \seq[+] b) \tag{\ref{ax:linStrn3}} \\
        \leq &\id[-][Y_1]\tensor[-] \id[-][Y_2] \tag{$\rla{a}\Vdash a \; ,\; \rla{b}\Vdash b$} \\
        =    &\id[-][Y_1 \tensor[-] Y_2]
     \end{align*}
    \end{minipage}
    show that $(\rla{a} \tensor[+] \rla{b}) \Vdash (a \tensor[-] b)$. Thus, by Lemma \ref{lemma:uniquenessla}, $\rla{(a \tensor[-] b)} = \rla{b} \tensor[+] \rla{a}$.    
    \item $\rla{(\symm[+])} = \symm[-]$. By axioms \eqref{ax:tauSymmPlus} and \eqref{ax:gammaSymmPlus}.
    \item $\rla{(\symm[-])} = \symm[+]$. By axioms \eqref{ax:tauSymmMinus} and \eqref{ax:gammaSymmMinus}. \qedhere
    \end{itemize}
\end{proof}

\noindent 
We conclude our exposition of closed linear bicategories with the following result, stating that $\rla{(\cdot)}$ commutes with any other morphism of closed linear bicategories.

\begin{lem}\label{lm:adjfunctor}
    Let $\mathcal{F}\colon \Cat{C_1} \to \Cat{C_2}$ be a morphism of closed linear bicategories. Then, for all $a\colon X \to Y$ in $\Cat{C_1}$,
    $\rla{\mathcal{F}(a)} = \mathcal{F}(\rla{a})$.
\end{lem}	
\begin{proof}
    The following two derivations %

    \noindent\begin{minipage}{0.48\linewidth}
    \begin{align*}
            \id[+][X] & =  \mathcal{F}(\id[+][X]) \\
             &\leq \mathcal{F}(a \seq[-] \rla{a}) \tag{$\rla{a} \Vdash a$}\\
             &= \mathcal{F}(a) \seq[-] \mathcal{F}(\rla{a}) 
    \end{align*}
    \end{minipage}\quad \vline
    \begin{minipage}{0.48\linewidth}
    \begin{align*}
             \mathcal{F}(\rla{a}) \seq[+] \mathcal{F}(a)   = \;\; &\mathcal{F}( \rla{a} \seq[+] a )  \\
             \leq\;\; &\mathcal{F}( \id[-][Y] ) \tag{$\rla{a} \Vdash a$}\\
             = \;\;&\id[-][Y]
    \end{align*}
    \end{minipage}

    show that $\mathcal{F}(\rla{a}) \Vdash \mathcal{F}(a)$. Thus, by Lemma \ref{lemma:uniquenessla}, we conclude that $\rla{\mathcal{F}(a)} = \mathcal{F}(\rla{a})$. %
\end{proof}
\noindent 
Hereafter, the diagram obtained from $\Circ{c}$, by taking its mirror image $\CircOp[+]{c}$ and then its photographic negative $\CircOp[-]{c}$ will denote $\rla{\Circ{c}}$.

\section{First-Order Bicategories}\label{sec:fobic}

Here we focus on the most important and novel part of the axiomatisation.
Indeed, having introduced the two main ingredients, cartesian and linear bicategories, it is time to fire up the Bunsen burner.
The remit of this section is to
 understand how the cartesian and the linear bicategory structures interact in the context of relations.
We introduce \emph{first-order bicategories} that make these interactions precise. The resulting axioms echo those of cartesian bicategories but in the linear bicategory setting: recall that in a cartesian bicategory the monoid and comonoids are adjoint and satisfy the Frobenius law. Here, the white and black (co)monoids are again related, but by \emph{linear} adjunctions; moreover, they also satisfy appropriate ``linear'' counterparts of the Frobenius equations.

\begin{figure}[t]
    \mylabel{ax:tauCopierPlus}{$\tau\!\copier[+]$}
    \mylabel{ax:tauDiscardPlus}{$\tau\discard[+]$}
    \mylabel{ax:tauCocopierPlus}{$\tau\!\cocopier[+]$}
    \mylabel{ax:tauCodiscardPlus}{$\tau\codiscard[+]$}

    \mylabel{ax:gammaCopierPlus}{$\gamma\!\copier[+]$}
    \mylabel{ax:gammaDiscardPlus}{$\gamma\discard[+]$}
    \mylabel{ax:gammaCocopierPlus}{$\gamma\!\cocopier[+]$}
    \mylabel{ax:gammaCodiscardPlus}{$\gamma\codiscard[+]$}

    \mylabel{ax:tauCopierMinus}{$\tau\!\copier[-]$}
    \mylabel{ax:tauDiscardMinus}{$\tau\discard[-]$}
    \mylabel{ax:tauCocopierMinus}{$\tau\!\cocopier[-]$}
    \mylabel{ax:tauCodiscardMinus}{$\tau\codiscard[-]$}

    \mylabel{ax:gammaCopierMinus}{$\gamma\!\copier[-]$}
    \mylabel{ax:gammaDiscardMinus}{$\gamma\discard[-]$}
    \mylabel{ax:gammaCocopierMinus}{$\gamma\!\cocopier[-]$}
    \mylabel{ax:gammaCodiscardMinus}{$\gamma\codiscard[-]$}

    \mylabel{ax:bwFrob}{F$\tiny{\begin{array}{@{}c@{}c@{}} \bullet & \\[-3pt] & \circ \end{array}}$}
    \mylabel{ax:bwFrob2}{F$\tiny{\begin{array}{@{}c@{}c@{}} \circ & \\[-3pt] & \bullet \end{array}}$}
    \mylabel{ax:wbFrob}{F$\tiny{\begin{array}{@{}c@{}c@{}}  & \circ  \\[-3pt] \bullet & \end{array}}$}
    \mylabel{ax:wbFrob2}{F$\tiny{\begin{array}{@{}c@{}c@{}}  & \bullet  \\[-3pt] \circ & \end{array}}$}
    \[
        \resizebox{\textwidth}{!}{$
        \begin{array}{@{}c@{}c@{}c@{}c@{}}
            \begin{array}{@{}c@{}c@{}c@{}}
                \idCirc[+][X] & \!\!\Lleq{\eqref*{ax:tauCopierPlus}}\!\! & 
    \InputIfFileExists{axiomsNEW/linadj/comMmon.tikz}{}{\input{tikz/axiomsNEW/linadj/comMmon.tikz}}
 \\
                \idCirc[+][X] & \!\!\Lleq{\eqref*{ax:tauDiscardPlus}}\!\! & 
    \InputIfFileExists{axiomsNEW/linadj/bangMcobang.tikz}{}{\input{tikz/axiomsNEW/linadj/bangMcobang.tikz}}
 \\
                
    \InputIfFileExists{axiomsNEW/id2P.tikz}{}{\input{tikz/axiomsNEW/id2P.tikz}}
 & \!\!\Lleq{\eqref*{ax:tauCocopierPlus}}\!\! &  
    \InputIfFileExists{axiomsNEW/linadj/monMcom.tikz}{}{\input{tikz/axiomsNEW/linadj/monMcom.tikz}}
 \\
                \emptyCirc[+]         & \!\!\Lleq{\eqref*{ax:tauCodiscardPlus}}\!\! & 
    \InputIfFileExists{axiomsNEW/linadj/cobangMbang.tikz}{}{\input{tikz/axiomsNEW/linadj/cobangMbang.tikz}}

            \end{array} &
            \begin{array}{@{}c@{}c@{}c@{}}
                
    \InputIfFileExists{axiomsNEW/linadj2/monPcom.tikz}{}{\input{tikz/axiomsNEW/linadj2/monPcom.tikz}}
     & \!\!\Lleq{\eqref*{ax:gammaCopierPlus}}\!\! & 
    \InputIfFileExists{axiomsNEW/id2M.tikz}{}{\input{tikz/axiomsNEW/id2M.tikz}}
 \\
                
    \InputIfFileExists{axiomsNEW/linadj2/cobangPbang.tikz}{}{\input{tikz/axiomsNEW/linadj2/cobangPbang.tikz}}
 & \!\!\Lleq{\eqref*{ax:gammaDiscardPlus}}\!\! & \emptyCirc[-]  \\
                
    \InputIfFileExists{axiomsNEW/linadj/comPmon.tikz}{}{\input{tikz/axiomsNEW/linadj/comPmon.tikz}}
     & \!\!\Lleq{\eqref*{ax:gammaCocopierPlus}}\!\! & \idCirc[-][X] \\
                
    \InputIfFileExists{axiomsNEW/linadj/bangPcobang.tikz}{}{\input{tikz/axiomsNEW/linadj/bangPcobang.tikz}}
 & \!\!\Lleq{\eqref*{ax:gammaCodiscardPlus}}\!\! & \idCirc[-][X]
            \end{array} &
            \begin{array}{@{}c@{}c@{}c@{}}
                \idCirc[+][X] & \!\!\Lleq{\eqref*{ax:tauCopierMinus}}\!\! & 
    \InputIfFileExists{axiomsNEW/linadj2/comMmon.tikz}{}{\input{tikz/axiomsNEW/linadj2/comMmon.tikz}}
 \\
                \idCirc[+][X] & \!\!\Lleq{\eqref*{ax:tauDiscardMinus}}\!\! & 
    \InputIfFileExists{axiomsNEW/linadj2/bangMcobang.tikz}{}{\input{tikz/axiomsNEW/linadj2/bangMcobang.tikz}}
 \\
                
    \InputIfFileExists{axiomsNEW/id2P.tikz}{}{\input{tikz/axiomsNEW/id2P.tikz}}
 & \!\!\Lleq{\eqref*{ax:tauCocopierMinus}}\!\! &  
    \InputIfFileExists{axiomsNEW/linadj2/monMcom.tikz}{}{\input{tikz/axiomsNEW/linadj2/monMcom.tikz}}
 \\
                \emptyCirc[+]         & \!\!\Lleq{\eqref*{ax:tauCodiscardMinus}}\!\! & 
    \InputIfFileExists{axiomsNEW/linadj2/cobangMbang.tikz}{}{\input{tikz/axiomsNEW/linadj2/cobangMbang.tikz}}

            \end{array} &\!\!\!\!\!
            \begin{array}{@{}c@{}c@{}c@{}}
                
    \InputIfFileExists{axiomsNEW/linadj/monPcom.tikz}{}{\input{tikz/axiomsNEW/linadj/monPcom.tikz}}
     & \!\!\Lleq{\eqref*{ax:gammaCopierMinus}}\!\! & 
    \InputIfFileExists{axiomsNEW/id2M.tikz}{}{\input{tikz/axiomsNEW/id2M.tikz}}
 \\
                
    \InputIfFileExists{axiomsNEW/linadj/cobangPbang.tikz}{}{\input{tikz/axiomsNEW/linadj/cobangPbang.tikz}}
 & \!\!\Lleq{\eqref*{ax:gammaDiscardMinus}}\!\! & \emptyCirc[-]  \\
                
    \InputIfFileExists{axiomsNEW/linadj2/comPmon.tikz}{}{\input{tikz/axiomsNEW/linadj2/comPmon.tikz}}
     & \!\!\Lleq{\eqref*{ax:gammaCocopierMinus}}\!\! & \idCirc[-][X] \\
                
    \InputIfFileExists{axiomsNEW/linadj2/bangPcobang.tikz}{}{\input{tikz/axiomsNEW/linadj2/bangPcobang.tikz}}
 & \!\!\Lleq{\eqref*{ax:gammaCodiscardMinus}}\!\! & \idCirc[-][X]
            \end{array} \\
            \midrule
            \multicolumn{4}{c}{
            \begin{array}{@{}c@{}c@{}c@{}c@{}c@{}c@{}c@{}c@{}c@{}c@{}c@{}c}
                
    \InputIfFileExists{axiomsNEW/bwS2.tikz}{}{\input{tikz/axiomsNEW/bwS2.tikz}}
 & \!\!\Leq{\eqref*{ax:bwFrob}}\!\! & 
    \InputIfFileExists{axiomsNEW/bwZ2.tikz}{}{\input{tikz/axiomsNEW/bwZ2.tikz}}
  & 
    \InputIfFileExists{axiomsNEW/bwS.tikz}{}{\input{tikz/axiomsNEW/bwS.tikz}}
 & \!\!\Leq{\eqref*{ax:bwFrob2}}\!\! & 
    \InputIfFileExists{axiomsNEW/bwZ.tikz}{}{\input{tikz/axiomsNEW/bwZ.tikz}}
 & 
    \InputIfFileExists{axiomsNEW/wbS2.tikz}{}{\input{tikz/axiomsNEW/wbS2.tikz}}
 & \!\!\Leq{\eqref*{ax:wbFrob}}\!\! & 
    \InputIfFileExists{axiomsNEW/wbZ2.tikz}{}{\input{tikz/axiomsNEW/wbZ2.tikz}}
  & 
    \InputIfFileExists{axiomsNEW/wbS.tikz}{}{\input{tikz/axiomsNEW/wbS.tikz}}
 & \!\!\Leq{\eqref*{ax:wbFrob2}}\!\! & 
    \InputIfFileExists{axiomsNEW/wbZ.tikz}{}{\input{tikz/axiomsNEW/wbZ.tikz}}

            \end{array}
            }
        \end{array}
        $}
    \]
    \caption{Additional axioms for fo-bicategories}\label{fig:fo bicat axioms}
\end{figure}

\begin{defi}\label{def:fobicategory}
    A \emph{first-order bicategory} $(\Cat{C},\tensor[+],\tensor[-], \unittensor, \copier[+], \discard[+], \cocopier[+], \codiscard[+], \copier[-], \discard[-], \cocopier[-], \codiscard[-])$, shorthand \emph{fo-bicategory} $(\Cat{C},\copier[+], \cocopier[+],  \copier[-], \cocopier[-])$, consists of
    \begin{itemize}
\item a closed linear bicategory $(\Cat{C}, \tensor[+], \tensor[-],\unittensor)$,
\item a cartesian bicategory $(\Cat{C},\copier[+], \cocopier[+])$ and
\item a cocartesian bicategory $(\Cat{C}, \copier[-], \cocopier[-])$, such that
\end{itemize}
\begin{enumerate}
\item  the white comonoid  $(\copier[+], \discard[+])$ is left and right linear adjoint to black monoid $(\cocopier[-], \codiscard[-])$ and the white monoid $(\cocopier[+], \codiscard[+])$ is left and right linear adjoint to black comonoid $(\copier[-], \discard[-])$, i.e.,%
  \[
  \begin{array}{cc}
                  \begin{array}{r@{\,\,}c@{\,\,}l r@{\,\,}c@{\,\,}l}
                      \id[+][X] &\stackrel{\eqref*{ax:tauCopierPlus}}{\leq}& \copier[+][X] \seq[-] \cocopier[-][X]
                      &
                      \cocopier[-][X] \seq[+] \copier[+][X] &\stackrel{\eqref*{ax:gammaCopierPlus}}{\leq}& \id[-][X\tensor[-]X]
                      \\
                      \id[+][X] &\stackrel{\eqref*{ax:tauDiscardPlus}}{\leq}& \discard[+][X] \seq[-] \codiscard[-][X]
                      &
                      \codiscard[-][X] \seq[+] \discard[+][X] &\stackrel{\eqref*{ax:gammaDiscardPlus}}{\leq}& \id[-][0]
                      \\
                      \id[+][X] &\stackrel{\eqref*{ax:tauCopierMinus}}{\leq}& \copier[-][X] \seq[-] \cocopier[+][X]
                      &
                      \cocopier[+][X] \seq[+] \copier[-][X] &\stackrel{\eqref*{ax:gammaCopierMinus}}{\leq}& \id[-][X\tensor[-]X]
                      \\
                      \id[+][X] &\stackrel{\eqref*{ax:tauDiscardMinus}}{\leq}& \discard[-][X] \seq[-] \codiscard[+][X]
                      &
                      \codiscard[+][X] \seq[+] \discard[-][X] &\stackrel{\eqref*{ax:gammaDiscardMinus}}{\leq}& \id[-][0]
                  \end{array}
                  &
                  \begin{array}{r@{\,\,}c@{\,\,}l r@{\,\,}c@{\,\,}l}
                      \id[+][X\tensor[+]X] &\stackrel{\eqref*{ax:tauCocopierPlus}}{\leq}& \cocopier[+][X] \seq[-] \copier[-][X]
                      &
                      \copier[-][X] \seq[+] \cocopier[+][X] &\stackrel{\eqref*{ax:gammaCocopierPlus}}{\leq}& \id[-][X]
                      \\
                      \id[+][0] &\stackrel{\eqref*{ax:tauCodiscardPlus}}{\leq}& \codiscard[+][X] \seq[-] \discard[-][X]
                      &
                      \discard[-][X] \seq[+] \codiscard[+][X] &\stackrel{\eqref*{ax:gammaCodiscardPlus}}{\leq}& \id[-][X]
                      \\
                      \id[+][X\tensor[+]X] &\stackrel{\eqref*{ax:tauCocopierMinus}}{\leq}& \cocopier[-][X] \seq[-] \copier[+][X]
                      &
                      \copier[+][X] \seq[+] \cocopier[-][X] &\stackrel{\eqref*{ax:gammaCocopierMinus}}{\leq}& \id[-][X]
                      \\
                      \id[+][0] &\stackrel{\eqref*{ax:tauCodiscardMinus}}{\leq}& \codiscard[-][X] \seq[-] \discard[+][X]
                      &
                      \discard[+][X] \seq[+] \codiscard[-][X] &\stackrel{\eqref*{ax:gammaCodiscardMinus}}{\leq}& \id[-][0]
                  \end{array}
              \end{array}
  \]
    \item white and black (co)monoids satisfy the linear Frobenius laws, i.e.
    \[\begin{array}{r@{\,\,}c@{\,\,}l}
                  (\copier[-][X] \tensor[+] \id[+][X]) \seq[+] (\id[+][X] \tensor[+] \cocopier[+][X])
                  &\stackrel{\eqref*{ax:bwFrob}}{=}&
                  (\id[+][n] \tensor[+] \copier[+][X]) \seq[+] (\cocopier[-][X] \tensor[+] \id[+][X])
                  \\
                  (\copier[+][X] \tensor[+] \id[+][X]) \seq[+] (\id[+][X] \tensor[+] \cocopier[-][X])
                  &\stackrel{\eqref*{ax:bwFrob2}}{=}&
                  (\id[+][X] \tensor[+] \copier[-][X]) \seq[+] (\cocopier[+][X] \tensor[+] \id[+][X])
                  \\
                  (\copier[+][X] \tensor[-] \id[-][X]) \seq[-] (\id[-][X] \tensor[-] \cocopier[-][X])
                  &\stackrel{\eqref*{ax:wbFrob}}{=}&
                  (\id[-][X] \tensor[-] \copier[-][X]) \seq[-] (\cocopier[+][X] \tensor[-] \id[-][X])
                  \\
                  (\copier[-][X] \tensor[-] \id[-][X]) \seq[-] (\id[-][X] \tensor[-] \cocopier[+][X])
                  &\stackrel{\eqref*{ax:wbFrob2}}{=}&
                  (\id[-][X] \tensor[-] \copier[+][X]) \seq[-] (\cocopier[-][X] \tensor[-] \id[-][X])
              \end{array} \]

\end{enumerate}
A \emph{morphism of fo-bicategories} is a morphism of linear, cartesian and cocartesian bicategories. %
\end{defi}
\begin{rem}%
Fig. \ref{fig:fo bicat axioms} illustrates the diagrams for the axioms of linear adjointness (top) and linear Frobenius (bottom). The latter, in particular, highlights the intuitive clarity that diagrams offer compared to term-based representations, which in particular clearly exhibit the symmetries of the axiomatisation. Regarding linear adjoints, note that --- because of the symmetries involved --- expressing the two statements in point (1) of the definition above requires 16 axioms. However, while the axioms governing fo-bicategories are numerous, the information conveyed by them is elegantly and concisely captured, as illustrated by ``the Tao of Logic'' introduced earlier.
\end{rem}

We have seen that $(\Rel, \tensor[+], \tensor[-], \singleton)$ is a closed linear bicategory, $(\Relp, \copier[+], \cocopier[+])$ a cartesian bicategory and $(\Relm, \copier[-], \cocopier[-])$ a cocartesian bicategory. Given~\eqref{eq:comonoidsREL}, it is easy to confirm linear adjointness  and linear Frobenius. %

\noindent 
If $(\Cat{C},\copier[+], \cocopier[+],  \copier[-], \cocopier[-])$ is a fo-bicategory then $(\opposite{\Cat{C}},\cocopier[+], \copier[+],  \cocopier[-], \copier[-])$  and $(\co{\Cat{C}},\copier[-], \cocopier[-],  \copier[+], \cocopier[+])$ are fo-bicategories: the laws of Fig.~\ref{fig:fo bicat axioms} are closed under mirror-reflection and photographic negative.
The condition (1) in Definition \ref{def:fobicategory} entails that the morphism of linear bicategories $\rla{(\cdot)}\colon \Cat{C} \to \opposite{(\co{\Cat{C}})}$ (see Proposition~\ref{prop:rlamorphism}) is a morphism of fo-bicategories and, similarly, the condition (2) that the morphism of cartesian bicategories
$\op{(\cdot)} \colon \Cat{C} \to \opposite{\Cat{C}}$ (see Proposition~\ref{prop:opcartesianfunctor}) is a morphism of fo-bicategories.

\begin{table*}[t]
    \scalebox{0.85}{$
    \renewcommand{\arraystretch}{1.2}
    \begin{array}{@{}cccc|cc@{}}
    \toprule
    \multicolumn{4}{c|}{
        \text{(a) Properties of $\op{(\cdot)} \colon (\Cat{C},\copier[+]\!, \cocopier[+]\!,  \copier[-]\!, \cocopier[-]) \to (\opposite{\Cat{C}},\cocopier[+]\!, \copier[+]\!,  \cocopier[-]\!, \copier[-])$}
    }
    & \multicolumn{2}{c}{
        \text{(c) Interaction of $\op{\cdot},\rla{\cdot}$ with $\sqcap,\sqcup$}
    } \\ \midrule
    \multicolumn{2}{c}{\text{if }c\leq d\text{ then }\op{c} \leq \op{d}}        & \multicolumn{2}{c|}{\op{(\op{c})}= c}   &                                            \op{(c \sqcap d)} = \op{c} \sqcap \op{d}  & \op{\top} = \top            \\
    \op{(c \seq[+] d)} = \op{d} \seq[+] \op{c} & \op{(\id[+][X])} = \id[+][X]   & \op{(\cocopier[+][X])} = \copier[+][X]  & \op{(\codiscard[+][X])} = \discard[+][X] & \op{(c \sqcup d)} = \op{c} \sqcup \op{d}  & \op{\bot} = \bot      \\
    \op{(c \tensor[+] d)} = \op{c} \tensor[+] \op{d} &  \op{(\symm[+][X][Y])} = \symm[+][Y][X]   & \op{(\copier[+][X])} = \cocopier[+][X]   & \op{(\discard[+][X])} = \codiscard[+][X]  &   \rla{(c \sqcap d)} = \rla{c} \sqcup \rla{d}  & \rla{\top} = \bot      \\ \cline{1-4}
    \op{(c \seq[-] d)} = \op{d} \seq[-] \op{c} & \op{(\id[-][X])} = \id[-][X]   & \op{(\cocopier[-][X])} = \copier[-][X]  & \op{(\codiscard[-][X])} = \discard[-][X] &  \rla{(c \sqcup d)} = \rla{c} \sqcap \rla{d}  & \rla{\bot} = \top           \\
    \op{(c \tensor[-] d)} = \op{c} \tensor[-] \op{d} &  \op{(\symm[-][X][Y])} = \symm[-][Y][X]   & \op{(\copier[-][X])} = \cocopier[-][X]   & \op{(\discard[-][X])} = \codiscard[-][X]  & \multicolumn{2}{c}{\rla{(\op{c})} = \op{(\rla{c})} } \\ \midrule
    \multicolumn{4}{c|}{
        \text{(b)  Properties of $\rla{(\cdot)} \colon (\Cat{C},\copier[+]\!, \cocopier[+]\!,  \copier[-]\!, \cocopier[-]) \to (\opposite{(\co{\Cat{C}})},\cocopier[-]\!, \copier[-]\!,  \cocopier[+]\!, \copier[+])$}
    }
    & \multicolumn{2}{c}{
        \text{(d) Laws of Boolean algebras}
    } \\ \midrule
    \multicolumn{2}{c}{ \text{if }c\leq d\text{ then }\rla{c} \geq \rla{d}}     & \multicolumn{2}{c|}{ \rla{(\rla{c})}= c }    & \multicolumn{2}{c}{ c \sqcap (d \sqcup e) = (c \sqcap d) \sqcup (c \sqcap e) } \\
    \rla{(c \seq[+] d)} = \rla{d} \seq[-] \rla{c} & \multicolumn{1}{c|}{ \rla{(\id[+][X])} = \id[-][X] } & \rla{(\cocopier[+][X])} = \copier[-][X]  & \rla{(\codiscard[+][X])} = \discard[-][X] & \multicolumn{2}{c}{ c \sqcup (d \sqcap e) = (c \sqcup d) \sqcap (c \sqcup e) } \\
    \rla{(c \tensor[+] d)} = \rla{c} \tensor[-] \rla{d} & \multicolumn{1}{c|}{ \rla{(\symm[+][X][Y])} = \symm[-][Y][X] } & \rla{(\copier[+][X])} = \cocopier[-][X]  &  \rla{(\discard[+][X])} = \codiscard[-][X] &  \nega{(c \sqcap d)} = \nega{c} \sqcup \nega{d} & \nega{\top} = \bot           \\
    \rla{(c \seq[-] d)} = \rla{d} \seq[+] \rla{c} & \multicolumn{1}{c|}{ \rla{(\id[-][X])} = \id[+][X] } &  \rla{(\cocopier[-][X])} = \copier[+][X]  & \rla{(\codiscard[-][X])} = \discard[+][X]  & \nega{(c \sqcup d)} = \nega{c} \sqcap \nega{d}      & \nega{\bot} = \top          \\
    \rla{(c \tensor[-] d)} = \rla{c} \tensor[+] \rla{d} & \multicolumn{1}{c|}{ \rla{(\symm[-][X][Y])} = \symm[+][Y][X] } &  \rla{(\copier[-][X])} = \cocopier[+][X]  &  \rla{(\discard[-][X])} = \codiscard[+][X]  &  c \sqcap \nega{c} = \bot \;\;\;         &     c \sqcup \nega{c} = \top \quad\qquad     \\ \midrule
    \multicolumn{6}{c}{
        \text{(e) Enrichment over join-meet semilattices}
    }                                                       \\ \midrule
    \multicolumn{2}{c}{c \seq[+] (d \sqcup e) = (c \seq[+] d) \sqcup (c \seq[+] e)}     & \multicolumn{2}{c}{(d \sqcup e) \seq[+] c = (d \seq[+] c) \sqcup (e \seq[+] c)}     & \multicolumn{2}{c}{c \seq[+] \bot = \bot = \bot \seq[+] c } \\
    \multicolumn{2}{c}{\,\;c \tensor[+] (d \sqcup e) = (c \tensor[+] d) \sqcup (c \tensor[+] e)}     & \multicolumn{2}{c}{\;\,(d \sqcup e) \tensor[+] c = (d \tensor[+] c) \sqcup (e \tensor[+] c)}     & \multicolumn{2}{c}{c \tensor[+] \bot = \bot = \bot \tensor[+] c} \\
    \multicolumn{2}{c}{c \seq[-] (d \sqcap e) = (c \seq[-] d) \sqcap (c \seq[-] e)}     & \multicolumn{2}{c}{(d \sqcap e) \seq[-] c = (d \seq[-] c) \sqcap (e \seq[-] c)}     & \multicolumn{2}{c}{c \seq[-] \top = \top = \top \seq[-] c} \\
    \multicolumn{2}{c}{\,\;c \tensor[-] (d \sqcap e) = (c \tensor[-] d) \sqcap (c \tensor[-] e)}     & \multicolumn{2}{c}{\;\,(d \sqcap e) \tensor[-] c = (d \tensor[-] c) \sqcap (e \tensor[-] c)}     & \multicolumn{2}{c}{c \tensor[-] \top = \top = \top \tensor[-] c} \\ \bottomrule
    \end{array}
    $}
    \caption{Properties of first-order bicategories.}\label{table:daggerrlalattice}\label{table:daggerproperties}\label{table:rlaproperties}\label{table:enrichment}
\end{table*}

\begin{prop}\label{prop:opfunctor}\label{prop:rlafunctor} %
    $\op{(\cdot)} \colon\! \Cat{C} \to \opposite{\Cat{C}}$ and $\rla{(\cdot)}\colon\! \Cat{C} \to \opposite{(\co{\Cat{C}})}$ are isomorphisms of fo-bicategories, namely the laws in Table~\ref{table:daggerproperties}.(a) and (b) hold.
    \end{prop}
The proof is illustrated in Appendix \ref{app:functors}. Several useful facts easily follow.

\begin{cor}\label{cor:oplinearadjoint}\label{cor:interactiondaggerrlalattice}
    The laws in Table~\ref{table:daggerrlalattice}.(c) hold.
\end{cor}
\begin{proof}
    $\rla{(\op{c})} = \op{(\rla{c})}$ is immediate from Proposition \ref{prop:opfunctor} and Lemma \ref{lm:adjfunctor}. The other laws are derived by the definitions of $\sqcap$,  $\top$, $\sqcup$, $\bot$ in \eqref{eq:def:cap} and \eqref{eq:def:cup}, and the laws in Tables \ref{table:daggerproperties}.(a) and  \ref{table:rlaproperties}.(b). For instance,
    \input{tikz/decomposedCapNeg.tex}
\end{proof}

The next result about maps (Definition \ref{def:maps}) plays a crucial role. %
\begin{prop}\label{prop:maps}%
For all maps $f\colon X \to Y$ and arrows $c\colon Y\to Z$,
\begin{center}
(1) $f \seq[+] c = \rla{(\op{f})} \seq[-] c$ and  (2) $c \seq[+] \op{f} = c \seq[-] \rla{f}$
\end{center}
and thus in particular
\begin{equation}\label{eq:cd-maps}\tag{maps}
    \begin{array}{c@{\qquad}c@{\qquad}c@{\qquad}c}
        \input{tikz/mapsCopier} &  \input{tikz/mapsDiscard.tex} &
        \input{tikz/mapsCocopier} & \input{tikz/mapsCodiscard.tex}.
    \end{array}
\end{equation}
\end{prop}
\begin{proof}%
    The following two derivations prove the two inclusion of (1).

    \noindent\begin{minipage}{0.4\linewidth}
    \begin{align*}
f \seq[+] c =\;    &\id[+][X] \seq[+] f \seq[+]c   \\
            \leq\; &(\rla{(\op{f})}\seq[-] \op{f}) \seq[+] f \seq[+] c  \tag{$\op{f} \Vdash \rla{(\op{f})}$}\\
            \leq\; &\rla{(\op{f})} \seq[-] (\op{f} \seq[+] f \seq[+] c) \tag{\ref{ax:rightLinDistr}} \\
            \leq\; &\rla{(\op{f})} \seq[-] (\id[+][Y] \seq[+] c)   \tag{Proposition~\ref{prop:map adj}} \\
            =\;    &\rla{(\op{f})} \seq[-]  c
    \end{align*}
    \end{minipage}\; \vline
    \begin{minipage}{0.4\linewidth}
    \begin{align*}
f \seq[+] c =\;    &f \seq[+] ( \id[-][X] \seq[-] c)   \\
            \geq\; &f \seq[+] ( (\op{f} \seq[+] \rla{(\op{f})} ) \seq[-] c)  \tag{$\op{f} \Vdash \rla{(\op{f})}$}\\
            \geq\; &f \seq[+]  \op{f} \seq[+] (\rla{(\op{f})}  \seq[-] c ) \tag{\ref{ax:leftLinDistr}} \\
            \geq\; &\id[+][X] \seq[+] (\rla{(\op{f})}  \seq[-] c)  \tag{Proposition~\ref{prop:map adj}} \\ %
            =\;    &\rla{(\op{f})} \seq[-]  c
    \end{align*}
    \end{minipage}
    Note that $\op{f} \Vdash \rla{(\op{f})}$ holds since, by Proposition \ref{prop:rlafunctor}, in any fo-bicategory left and right linear adjoint coincide (namely $\rla{(\rla{a})}=a$).

    To check (2), we use Table~\ref{table:daggerproperties}.(a) and (1): $c \seq[+] \op{f} = \op{ (f \seq[+] \op{c} )} = \op{(\rla{(\op{f})} \seq[-] \op{c})}= c \seq[-] \rla{f}$. %
    For the four equivalence, one concludes by taking as map $f$ either $\copier[+]$ or $\discard[+]$.
\end{proof}

\begin{rem}
    A useful perspective on reasoning with string diagrams is that many proofs rely on higher-level proof strategies rather than step-by-step algebraic manipulation. Examples include the spider theorem~\cite{Lack2004a} and the ``only topology matters'' principle. These strategies capture whole families of low-level rewrites in a single conceptual move. 
    
    The equalities in~\eqref{eq:cd-maps} illustrate similar higher-level principles in the setting of fo-bicategories, expressing that the (co)monoid structures can freely slide across boundaries and switch colours. As an example, compare the proof in~\eqref{eq:forallexists} with its purely axiomatic version in Figure~\ref{fig:forallExists full}.
\end{rem}

Recall from Remark~\ref{rem:enrichment}, that cartesian bicategories are \emph{not} enriched over $\sqcap$-semilattices, despite the fact that all homsets carry such structures. Interestingly, in a fo-bicategory, every homset carries a proper lattice, but the white structure \emph{is} enriched over $\sqcup$ and the black structure over $\sqcap$. In $\Rel$, this is the well-known fact that $R \seq[+](S \cup T)= (R\seq[+]S)\cup (R\seq[+]T)$. %

\begin{prop}\label{prop:enrichment}
Let $(\Cat{C},\copier[+], \cocopier[+],  \copier[-], \cocopier[-])$ be a fo-bicategory. Then
\begin{itemize}
\item $(\Cat{C}, \tensor[+], \unittensor)$ is monoidally enriched over $\sqcup$-semilattices with $\bot$, while
\item $(\Cat{C}, \tensor[-], \unittensor)$  is monoidally enriched over $\sqcap$-semilattices with $\top$,
\end{itemize} that is the laws in Table~\ref{table:enrichment}.(e) hold.
\end{prop}
\begin{proof}
Below we prove that $(\Cat{C}, \tensor[+], \unittensor)$ is monoidally enriched over $\sqcup$-semilattices with $\bot$. The proofs for showing that $(\Cat{C}, \tensor[-], \unittensor)$ is monoidally enriched over $\sqcap$-semilattices with $\top$ are analogous. %
    \begin{itemize}
        \item For $a \seq[+] (b \sqcup c) = (a \seq[+] b) \sqcup (a \seq[+] c)$ we prove the two inclusions separately:
            \input{tikz/proofs/seqPlusDistrUnion}
        The proof for $(a \sqcup b) \seq[+] c = (a \seq[+] c) \sqcup (b \seq[+] c)$ is similar. %
        \item We prove the left to right inclusion of $a \seq[+] \bot = \bot$:
            \input{tikz/proofs/seqPlusDistrBot}
            The other inclusion trivially holds. $\bot \seq[+] a = \bot$ is proved analogously.
        \item The right-to-left inclusion of $a \otimes (b \sqcup c) = (a \otimes b) \sqcup (a \otimes c)$ is proved by the universal property of $\sqcup$, namely:
        since $a \tensor[+] b = a \tensor[+] (b \sqcup \bot) \leq a \tensor[+] (b \sqcup c)$ and $a \tensor[+] c = a \tensor[+] (\bot \sqcup c) \leq a \tensor[+] (b \sqcup c)$, then $(a \tensor[+] b) \sqcup (a \tensor[+] c) \leq  a \tensor[+] (b \sqcup c)$. For the other inclusion, the following holds:
        \input{tikz/proofs/tensorPlusDistrUnion.tex}
            Again, $(a \sqcup b) \otimes c = (a \otimes c) \sqcup (b \otimes c)$ is proved analogously.
        \item We prove the left to right inclusion of $a \otimes \bot = \bot$:
            \input{tikz/proofs/tensorPlusDistrBot}
            The other inclusion trivially holds. $\bot \otimes a = \bot$ is proved analogously. \qedhere
    \end{itemize}
\end{proof}

    \noindent 
    For a fo-bicategory $\Cat{C}$, we have the four isomorphisms in the diagram below, which commutes by Corollary \ref{cor:oplinearadjoint}.
    \[\xymatrix{\Cat{C} \ar[r]^{\op{(\cdot)}} \ar[d]_{\rla{(\cdot)}} & \opposite{\Cat{C}} \ar[d]^{\rla{(\cdot)}} \\
    \opposite{(\co{\Cat{C}})} \ar[r]_{\op{(\cdot)}} & \co{\Cat{C}}}\]
    We can thus define the complement as the diagonal of the square, namely
    \begin{equation}\label{def:nega}\nega{(\cdot)} \defeq \op{(\rla{(\cdot)})}\text{.}\end{equation}  In diagrams, given $\Circ[+]{c}$, its negation is $\op{(\rla{\Circ[+]{c}})} = \op{\CircOp[-]{c}} = \Circ[-]{c}$.

    Clearly $\nega{(\cdot)} \colon \Cat{C} \to \co{\Cat{C}}$ is an isomorphism of fo-bicategories.
    Moreover, it induces a Boolean algebra on each homset of $\Cat{C}$.

\begin{prop}\label{prop:bool homsets}
    Let $(\Cat{C},\copier[+], \cocopier[+],  \copier[-], \cocopier[-])$ be a fo-bicategory. Then every homset of $\Cat{C}$ is a Boolean algebra, that is the laws in Table~\ref{table:daggerrlalattice}.(d) hold. %
\end{prop}
\begin{proof}
    \begin{itemize}
        \item The De Morgan laws follow immediately from the definition of $\nega{(\cdot)}$ and Corollary~\ref{cor:interactiondaggerrlalattice}. We show only the following two:

        \noindent\begin{minipage}{0.45\textwidth}
            \begin{align*}
                \nega{c \sqcap d} &= \op{(\rla{(c \sqcap d)})} \tag{Definition of $\nega{(\cdot)}$} \\
                                  &= \op{(\rla{c})} \sqcup \op{(\rla{d})} \tag{Corollary~\ref{cor:interactiondaggerrlalattice}} \\
                                  &= \nega{c} \sqcup \nega{d} \tag{Definition of $\nega{(\cdot)}$}
            \end{align*}
        \end{minipage}
        \; \vline
        \begin{minipage}{0.45\textwidth}
            \begin{align*}
                \nega{\top} &= \op{(\rla{\top})} \tag{Definition of $\nega{(\cdot)}$}  \\
                            &= \bot \tag{Corollary~\ref{cor:interactiondaggerrlalattice}}
            \end{align*}
        \end{minipage}

        \item The distributivity of $\sqcap$ over $\sqcup$ follows from the enrichment and the definition of $\sqcap$:
        \begin{align*}
		a\sqcap (b \sqcup c) &= \copier[+]  \seq[+]  ( a  \tensor[+]  (b \sqcup c)  )  \seq[+] \cocopier[+][] \tag{\ref{eq:def:cap}}  \\
							 &= \copier[+][]  \seq[+]   ( (a \tensor[+] b) \sqcup (a  \tensor[+]  c)  )  \seq[+]  \cocopier[+][] \tag{\text{Table }\ref{table:enrichment}.(e)} \\
							 &= ( \copier[+][]  \seq[+]  (a \tensor[+] b) \seq[+] \cocopier[+][]) \sqcup ( \copier[+][]  \seq[+]  (a \tensor[+] c)  \seq[+]  \cocopier[+][] )  \tag{\text{Table }\ref{table:enrichment}.(e)} \\
							 &= (a \sqcap b) \sqcup (a \sqcap c). \tag{\ref{eq:def:cap}}
        \end{align*}
        Similarly for the distributivity of $\sqcup$ over $\sqcap$.
        \item We are left to prove the laws of excluded middle and non-contradiction, namely $\top \leq a \sqcup \nega{a}$ and $a \sqcap \nega{a} \leq \bot$. To do that, we first prove them on the identities, namely:
        \begin{equation}\label{eq:exclmiddle id}
            \topCirc \leq 
    \InputIfFileExists{axioms/exclmiddle.tikz}{}{\input{tikz/axioms/exclmiddle.tikz}}
 \quad \text{ and } \quad 
    \InputIfFileExists{axioms/noncontr.tikz}{}{\input{tikz/axioms/noncontr.tikz}}
 \leq \bottomCirc.
        \end{equation}
        For the first inequality observe that the following holds:
	    \input{tikz/proofs/LNC_id.tex}
        Thus, we can conclude by residuation (Lemma~\ref{lm:residuation}). The other inequality is proved analogously.

        Finally, observe that the following holds:
        \input{tikz/proofs/LNC.tex}
        The proof of the other law is analogous. \qedhere
    \end{itemize}
\end{proof}

\noindent 
We conclude this section with a result that extends Lemma \ref{lm:residuation} with five different possibilities to express the concept of logical entailment. It is worth emphasising that the following result stands at the core of our proofs. Once again, the diagrammatic approach proves to be an enhancement over the classical syntax. In this specific case we are looking at five (of many) different possibilities to express the ubiquitous concept of logical entailment. (1) expresses $a$ implies $b$ as a direct rewriting of the former into the latter. We have already seen that (2) corresponds to residuation. (3) corresponds to right residuation. (4) asserts the validity of the formula $\neg a \vee b$, thus it corresponds to the classical implication. Finally, (5) may look eccentric but it is actually a closed version of (3) that comes in handy if one has to consider closed diagrams.
\begin{lem}\label{lm:implications}%
In a fo-bicategory, the following are equivalent:
\[
    \begin{array}{l@{\;}l@{\;}l}
        (1) \boxCirc[+]{a}[X][Y] \!\!\leq\!\! \boxCirc[+]{b}[X][Y] & (2) \idCirc[+][X] \!\!\leq\!\! \seqCirc[-]{b}{a}[X][X][w][b] & (3) \idCirc[+][Y] \!\!\leq\!\! \seqCirc[-]{a}{b}[Y][Y][b] \\[10pt]
        \multicolumn{3}{c}{
            (4) \topCirc[X][Y] \!\!\leq\!\! \unionCirc{a}{b}[X][Y][b] \;\;\; (5)\; \emptyCirc[+] \leq \circleCirc{a}{b}[b]
        }
    \end{array}
\]
\end{lem}
\begin{proof}%
    We prove that $(1)$ is pairwise equivalent to $(2), (3)$ and $(4)$ and that $(4)$ is equivalent to $(5)$.

    \begin{enumerate}
        \item[] $(1)$ iff $(2)$ is Lemma~\ref{lm:residuation}.
        \item[] $(1)$ iff $(3)$: $a \leq b$ iff $\rla{b} \leq \rla{a}$ by the property of $\rla{(\cdot)}$ in Table~\ref{table:rlaproperties}.(b). By Lemma~\ref{lm:residuation}, $\rla{b} \leq \rla{a}$ iff $\id[+][Y] \leq \rla{a} \seq[-] \rla{(\rla{b})}$ where $\rla{(\rla{b})} = b$ by the property of $\rla{(\cdot)}$ in Table~\ref{table:rlaproperties}.(b).
        \item[] $(1)$ implies $(4)$: $\nega{a} \sqcup b \stackrel{(1)}{\geq} \nega{a} \sqcup a \stackrel{\text{Table}~\ref{table:daggerrlalattice}.(d)}{=} \top$.
        \item[] $(4)$ implies $(1)$: \input{tikz/proofs/implication/notAorB.tex}

        \item[] $(4)$ iff $(5)$ holds by means of residuation. In particular, recall that in a fo-bicategory
        \[  c\leq b\seq[-]\rla{a} \stackrel{(*_1)}{\iff} c\seq[+] a \leq b \stackrel{(*_2)}{\iff} a \leq \rla{c} \seq[-] b \]
        for all $a,b$ and $c$ properly typed. Thus, in particular:
        \[
            \topCirc[X][Y] \leq \unionCirc{a}{b}[X][Y][b]
            \stackrel{(*_1)}{\iff}
            \discardCirc[+][X] \leq  
    \InputIfFileExists{tikz/unionCirc1.tikz}{}{\input{tikz/tikz/unionCirc1.tikz}}

            \stackrel{(*_2)}{\iff}
            \emptyCirc[+]  \leq  
    \InputIfFileExists{tikz/unionCirc2.tikz}{}{\input{tikz/tikz/unionCirc2.tikz}}
.
        \]
        Using the self-dual compact closed structure of (co)cartesian bicategories it is immediate to show that the last diagram is equivalent to the right-hand side of $(5)$. \qedhere
    \end{enumerate}

\end{proof}

\subsection{The freely generated first-order bicategory}\label{sec:freely}
We now return to $\NPR$. %
Recall that $\syninclusion$ is the precongruence  obtained from the axioms in Figures~\ref{fig:cb axioms}, \ref{fig:cocb axioms}, \ref{fig:closed lin axioms} and \ref{fig:fo bicat axioms}. Its soundness (half of Theorem~\ref{thm:completeness}) is immediate since $\Rel$ is a fo-bicategory.
\begin{prop}\label{prop:soundness}
For all terms $c,d\colon n \to m$, if $c\syninclusion d$ then $c\seminclusion d$.
\end{prop}
\begin{proof}%
    Let $\interpretation =(X,\rho)$ be an interpretation of $\Sigma$. Recall that $\syninclusion$ is defined as $\pcong{\mathbb{FOB}\cup \structuralcong}$. We prove by induction on the rules in \eqref{eq:pc}, that
    \begin{center}
    if $c\syninclusion d$ then $\interpretationFunctor (c) \subseteq \interpretationFunctor (d)$.
    \end{center}
    By definition of $\seminclusion$, the above statement is equivalent to the proposition.

    The proof for the rules $(r)$ and $(t)$ is trivial. For the rule ($\seq[][][]$), suppose that $c=c_1 \seq[] c_2$ and $d=d_1 \seq[] d_2$ with $c_1 \syninclusion d_1$ and $c_2 \syninclusion d_2$. Then
    \begin{align*}
    \interpretationFunctor (c) &= \interpretationFunctor (c_1 \seq[] c_2) \\
    &= \interpretationFunctor (c_1) \seq[] \interpretationFunctor (c_2) \tag{\ref{fig:semantics}} \\
    &\subseteq  \interpretationFunctor (d_1) \seq[] \interpretationFunctor (d_2) \tag{Ind. hyp.} \\
    &= \interpretationFunctor (d_1 \seq[] d_2) \tag{\ref{fig:semantics}} \\
    &= \interpretationFunctor (d)  \\
    \end{align*}
The proof for ($\tensor[][][]$) is analogous to the one above. The only interesting case is the rule $(id)$: we should prove that if $(c,d)\in \mathbb{FOB}$, then $\interpretationFunctor (c) \subseteq \interpretationFunctor (d)$. However, we have already done most of the work: since all the axioms in $\mathbb{FOB}\cup \structuralcong$ -- with the only exception of the four stating $R^\bullet \Vdash R^\circ \Vdash R^\bullet$ (axioms \eqref{ax:tauRPlus}, \eqref{ax:gammaRPlus}, \eqref{ax:tauRMinus} and \eqref{ax:gammaRMinus}  in Figure \ref{fig:closed lin axioms})  -- are those of fo-bicategories and since $\Rel$ is a fo-bicategory, it only remains to show the soundness of those stating $R^\bullet \Vdash R^\circ \Vdash R^\bullet$. Note however that this is trivial by definition of $\interpretationFunctor (R^\bullet)$ as $\rla{\rho(R)}=\rla{(\interpretationFunctor (R^\circ))}$.
\end{proof}

Next, we show how $\NPR$ gives rise to a fo-bicategory $\LCB$. Objects are natural numbers and monoidal products $\tensor[][][]$ are defined as addition with unit object $0$.
Arrows from $n$ to $m$ are terms $c\colon n \to m$ modulo syntactic equivalence $\synequivalence$, namely $\LCB{}[n,m] \defeq \{[c]_{\synequivalence} \mid c\colon n \to m\}$. Observe that this is well defined since $\synequivalence$ is well-typed. Since $\synequivalence$ is a congruence, the operations  $\seq[]$ and  $\tensor[]$ on terms are well defined on equivalence classes: $[t_1]_{\synequivalence}\seq[][][] [t_2]_{\synequivalence} \defeq [t_1\seq[][][] t_2]_{\synequivalence}$ and $[t_1]_{\synequivalence}\tensor[][][] [t_2]_{\synequivalence} \defeq [t_1\tensor[][][] t_2]_{\synequivalence}$.
The partial order is given by the syntactic inclusion $\syninclusion$. For all objects $n\in \nat$, $\copier[][n]$, $\discard[][n]$, $\cocopier[][n]$ and $\codiscard[][n]$ are inductively defined as in Table \ref{fig:sugar}. With this structure, one can easily prove (see Appendix \ref{sec:appfreely}) the following.

\begin{prop}\label{prop:LCBfo}
$\LCB$ is a first-order bicategory.
\end{prop}

\noindent 
A  useful consequence of Proposition~\ref{prop:LCBfo} is that, for any interpretation $\interpretation=(X,\rho)$, the semantics $\interpretationFunctor$ gives rise to a morphism $\interpretationFunctor \colon \LCB \to \Rel$ of fo-bicategories: it is defined on objects as $n \mapsto X^n$ and on arrows by the inductive definition in \eqref{fig:semantics}. To see that it is a morphism, note that, by \eqref{fig:semantics}, all the structure of (co)cartesian bicategories and of linear bicategories is preserved (e.g. $\interpretationFunctor (\copier[+][1])=\copier[+][X]$). Moreover, the ordering is preserved by Prop.~\ref{prop:soundness}. Note that, by construction,
\begin{equation}\label{eq:free}
\interpretationFunctor(1)= X \text{ and }\interpretationFunctor(R^\circ)= \rho (R) \text{ for all }R\in \sign\text{.}
\end{equation}

Actually,  $\interpretationFunctor$ is the unique such morphism of fo-bicategories.
This is a consequence of a more general universal property: $\Rel$ can be replaced with an arbitrary fo-bicategory $\Cat{C}$.
To see this, we first need to generalise the notion of interpretation.

\begin{defi}\label{def:intCAT}
Let $\sign$ be a monoidal signature and $\Cat{C}$ a first-order bicategory. An \emph{interpretation} $\interpretation=(X,\rho)$ of $\sign$ in $\Cat{C}$ consists of an object $X$ of $\Cat{C}$ and an arrow $\rho(R)\colon X^{n} \to X^{m}$ for each $R \in \sign[n,m]$. %
\end{defi}

With this definition, we can state that $\LCB$ is \emph{the} fo-bicategory freely generated by $\sign$. %

\begin{prop}\label{prop:free}
Let $\sign$ be a monoidal signature, $\Cat{C}$ a first-order bicategory and $\interpretation=(X,\rho)$  an interpretation of $\sign$ in $\Cat{C}$. There exists a unique morphism of fo-bicategories $\interpretationFunctor\colon \LCB \to \Cat{C}$ such that  $\interpretationFunctor(1)= X$ and $\interpretationFunctor(R^\circ)= \rho (R)$ for all $R\in \sign$.
\end{prop}
\begin{proof}%
    Observe that the rules in \eqref{fig:semantics} defining  $\interpretationFunctor\colon \LCB \to \Rel$ also defines $\interpretationFunctor\colon \LCB \to \Cat{C}$ for an interpretation $\interpretation$ of $\sign$ in $\Cat{C}$ by fixing $\interpretationFunctor(R^\bullet)= \rla{(\interpretationFunctor(R^\circ))}$. To prove that $\interpretationFunctor$ preserve the ordering, one can use exactly the same proof of Proposition \ref{prop:soundness}. All the structure of (co)cartesian bicateries and linear bicategories is preserved by definition of $\interpretationFunctor$. %
    Thus, $\interpretationFunctor\colon \LCB \to \Cat{C}$ is a morphism of fo-bicategories. By definition, it also holds that $\interpretationFunctor(1)= X$ and $\interpretationFunctor(R^\circ)= \rho(R)$.

    To see that it is unique, observe that a morphism $\mathcal{F} \colon  \LCB \to \Cat{C}$ should map the object $0$ into $\unittensor$ (the unit object of $\tensor$) and any other natural number $n$ into $\mathcal{F}(1)^n$. Thus the only degree of freedom for the objects is the choice of where to map the natural number $1$. Similarly, for arrows, the only degree of freedom is where to map $R^\circ$ and $R^\bullet$. However, the axioms in $\mathbb{FOB}$ obliges $R^\bullet$ to be mapped into the right linear adjoint of $R^\circ$. Thus, by fixing $\mathcal{F}(1)=X$ and $\mathcal{F}(R^\circ)=\rho (R)$, $\mathcal{F}$ is forced to be $\interpretationFunctor$.
\end{proof}

\begin{rem}
Freeness is intended in the sense of \cite{Selinger2009}: consider the category where objects are pairs $(\Cat{C}, \interpretation)$ --where $\interpretation$ is an interpretation of the signature $\sign$ in the fo-bicategory $\Cat{C}$-- and arrows $\mathcal{F}\colon (\Cat{C}_1, \interpretation_1) \to (\Cat{C}_2, \interpretation_2)$ are morphism of fo-bicategories $\mathcal{F}\colon \Cat{C}_1 \to \Cat{C}_2$ preserving the interpretation, namely by fixing $\interpretation_i=(X_i,\rho_i)$,  $\mathcal{F}(X_1)=X_2$ and,  for all $R\in \sign$, $F(\rho_1(R))=\rho_2(R)$. Then, consider the obvious interpretation $\mathcal{O}=(X,\rho)$ of $\sign$ in $\LCB$ where the object $X$ of $\LCB$ is $1$ and  $\rho(R)=R^\circ$.
By Proposition \ref{prop:free}, it is immediate that $(\LCB,\mathcal{O})$ is an initial object in such category.
\end{rem}

We conclude this section with another useful consequence of the fact that $\interpretationFunctor \colon \LCB \to \Rel$ is a morphism of fo-bicategories.

\begin{lem}\label{lm:hen dagger adj}
Let $\interpretation$ be an interpretation of $\sign$ in $\Rel$ and let $c \colon n \to m$ be an arrow in $\LCB$. Then
    \[\interpretationFunctor(\op{c}) = \op{(\interpretationFunctor(c))}, \qquad \interpretationFunctor(\rla{c}) = \rla{(\interpretationFunctor(c))}, \qquad \interpretationFunctor(\nega{c}) = \nega{(\interpretationFunctor(c))}. \]
\end{lem}
\begin{proof}
    Since $\interpretationFunctor$ is a morphism of fo-bicategories the proof for $\op{(\cdot)}$ and $\rla{(\cdot)}$ follows from Lemma~\ref{lm:opfunctor} and Lemma~\ref{lm:adjfunctor}. Negation is preserved as well, since $\nega{(\cdot)} = \rla{(\op{\cdot})}$.
\end{proof}

\section{Diagrammatic First-Order Theories}\label{sec:theories}

Here we take the first steps towards completeness and show that for first-order theories, fo-bicategories play an analogous role to cartesian categories in Lawvere's functorial semantics of algebraic theories.

\begin{defi}\label{def:fo theory}
A \emph{first-order theory} $\T{T}$ is a pair $(\sign, \T{I})$ where $\sign$ is a signature and $\T{I}$ is a set of \emph{axioms}: pairs $(c,d)$ where  $c,d \colon n \to m$ are in $\LCB$.
 A \emph{model} of $\T{T}$ is an interpretation $\interpretation$ of $\sign$ where if $(c,d) \in \T{I}$, then $\interpretationFunctor(c) \subseteq \interpretationFunctor(d)$.
\end{defi}

Intuitively, each pair $(c,d)$ in the set $\T{I}$ is an axiom of the theory stating that the diagram $c$ should be included into $d$.
As expected, models of a first-order theory are interpretations of the relational symbols in the signature $\sign$, such that the inequalities in $\T{I}$ are preserved. Here we show a few meaningful example of first-order theories and comment on their models.

\begin{exa}[Theory of sets]\label{ex:full:empty theory}
The simplest case is $\T{T} = (\varnothing, \varnothing)$, where both the set of generating symbols and the set of inequations are empty. It is straightforward to verify that interpretations and models of this theory coincide. In particular, every possible set, also the empty set $\varnothing$, is a model of $\T{T}$.
\end{exa}

\begin{exa}[Theory of non-empty sets]\label{ex:full:non empty domains}
    To discard empty sets from the models of the theory $\T{T}$ defined above, it suffices to impose one single inequality.  Consider the theory \[\T{T'} = (\varnothing, \{ (\, \scalebox{0.8}{\emptyCirc[+]} , \scalebox{0.8}{
    \InputIfFileExists{axioms/cb/plus/codiscDisc.tikz}{}{\input{tikz/axioms/cb/plus/codiscDisc.tikz}}
} \!) \})\text{.}\] An interpretation $\interpretation$ of $\T{T'}$ is just a set $X$, since the set of generating symbols is empty. To see what is a model, it is necessary to understand the meaning of the only axiom in $\T{T'}$. By the definition of $\interpretationFunctor$ in \eqref{fig:semantics},

    \[ 
        \interpretationFunctor{( \scalebox{0.8}{\emptyCirc[+]} )} = \{(\star,\star)\} 
        \qquad\qquad
        \interpretationFunctor{( \scalebox{0.8}{
    \InputIfFileExists{axioms/cb/plus/codiscDisc.tikz}{}{\input{tikz/axioms/cb/plus/codiscDisc.tikz}}
}\!\!)} = \{(\star,x)\mid x\in X\} \seq[+] \{(x,\star) \mid x\in X\}
    \]

    \medskip

    Observe that $\interpretationFunctor{( \scalebox{0.8}{\emptyCirc[+]} )} = \{(\star,\star)\}$ regardless of the interpretation, since $X^0$ is always the singleton $\singleton$. Instead, $\interpretationFunctor{( \scalebox{0.8}{
    \InputIfFileExists{axioms/cb/plus/codiscDisc.tikz}{}{\input{tikz/axioms/cb/plus/codiscDisc.tikz}}
}\!\!)}$ depends on the chosen domain $X$. In particular, if $X \neq \varnothing$, then $\interpretationFunctor{( \scalebox{0.8}{
    \InputIfFileExists{axioms/cb/plus/codiscDisc.tikz}{}{\input{tikz/axioms/cb/plus/codiscDisc.tikz}}
}\!\!)}$ amounts to $\{ (\star, \star) \}$, otherwise if $X = \varnothing$, $\interpretationFunctor{( \scalebox{0.8}{
    \InputIfFileExists{axioms/cb/plus/codiscDisc.tikz}{}{\input{tikz/axioms/cb/plus/codiscDisc.tikz}}
}\!\!)} = \varnothing$.

    Therefore, the only inequality in $\T{T'}$ forces its models to be all and only non-empty sets, i.e. all those interpretations $\interpretation$ such that $\interpretationFunctor{(\scalebox{0.8}{\emptyCirc[+]} )} \subseteq \interpretationFunctor{( \scalebox{0.8}{
    \InputIfFileExists{axioms/cb/plus/codiscDisc.tikz}{}{\input{tikz/axioms/cb/plus/codiscDisc.tikz}}
}\!\!)}$.
\end{exa}

\begin{exa}[Linear orders]\label{ex:lin ord}%

Consider $\T{T}_R = (\sign_R, \T{I}_R)$, where $\sign_R = \{ R \colon 1 \to 1 \}$ and let $\T{I}_R$ be as follows:
      \[ \{\, ( \scalebox{0.8}{\idCirc[+]} , \scalebox{0.8}{\boxCirc[+]{R}}) ,\;
          (\scalebox{0.8}{\seqCirc[+]{R}{R}} , \scalebox{0.8}{\boxCirc[+]{R}}) ,\;
          (\scalebox{0.8}{
    \InputIfFileExists{linearOrdIntersection.tikz}{}{\input{tikz/linearOrdIntersection.tikz}}
} , \scalebox{0.8}{\idCirc[+]}) ,\;
          (\scalebox{0.8}{\topCirc} , \scalebox{0.8}{
    \InputIfFileExists{linearOrdUnion.tikz}{}{\input{tikz/linearOrdUnion.tikz}}
})\,\}.
      \]
An interpretation of $\T{T}_R$ is a set $X$ together with a relation $R \subseteq X \times X$. This is a model iff $R$ is reflexive (i.e., $\id[+]_{X}\subseteq R$), transitive ($R\seq[+]R \subseteq R$), antisymmetric ($R\cap \op{R}\subseteq \id[+]_X$) and total relation ($\top \subseteq R\cup \op{R}$), thus a linear order. %
\end{exa}

Monoidal signatures $\sign$, unlike usual $\FOL$ alphabets,
do not have function symbols. The reason is that, by adding the axioms below to $\mathbb{I}$, one forces a symbol $f\colon n \to 1\in\sign$ to be a function.
\begin{equation}\label{TMAP}\tag{$\TMAP{f}$} 
    \InputIfFileExists{copierFunc1.tikz}{}{\input{tikz/copierFunc1.tikz}}
 \leq 
    \InputIfFileExists{copierFunc2.tikz}{}{\input{tikz/copierFunc2.tikz}}
 \qquad \qquad \discardCirc[+][n] \leq 
    \InputIfFileExists{discardFunc.tikz}{}{\input{tikz/discardFunc.tikz}}
 \end{equation} %
We depict functions as \!\!$\scalebox{0.8}{\funcCirc[+]{f}[n]}$ and constants, being $0\to 1$ functions, as $\scalebox{0.8}{\constCirc[+]{k}}$. By the definitions of $\copier[+][0]$ and $\discard[+][0]$ in Table~\ref{fig:sugar}, the axioms \eqref{TMAP} for constants become the following.
\begin{equation}\label{TMAPk}\tag{$\TMAP{k}$}
    
    \InputIfFileExists{copierConst1.tikz}{}{\input{tikz/copierConst1.tikz}}
 \leq 
    \InputIfFileExists{copierConst2.tikz}{}{\input{tikz/copierConst2.tikz}}
 \qquad \qquad \emptyCirc[+] \leq 
    \InputIfFileExists{discardConst.tikz}{}{\input{tikz/discardConst.tikz}}

\end{equation}

\medskip

The axioms of a theory together with $\syninclusion$ form a deduction system. Formally, the \emph{deduction relation} induced by $\T{T}=(\Sigma, \T{I})$ is the closure (see \eqref{eq:pc}) of $\syninclusion \cup \;\wtrel$, i.e.
\begin{equation}\tag{$\syninclusionT{\T{T}}$}
\syninclusionT{\T{T}} \defeq \pcong{\syninclusion \cup \;\wtrel}\text{.}
\end{equation} 
We write $\synequivalenceT{\T{T}}$ for $\syninclusionT{\T{T}} \cap \synreverdedinclusionT{\T{T}}$. The following result proves that the deduction relation is sound, i.e. it preserves all models.
\begin{prop}\label{prop:soundnessoftheories}
Let $\T{T}=(\sign, \T{I})$ be a theory and $c,d\colon n \to m$ in $\LCB[\T{T}]$. If $c \syninclusionT{\T{T}} d$, then $\interpretationFunctor(c) \subseteq \interpretationFunctor(d)$ for all models $\interpretation$ of $\T{T}$.
\end{prop}
\begin{proof}
    By induction on \eqref{eq:pc}. For the rule $(id)$, we have two cases: either $(c,d) \in \syninclusion$ or $(c,d)\in \wtrel$. For $\syninclusion$, we conclude immediately by Proposition~\ref{prop:soundness}. For  $(c,d)\in \wtrel$, the inclusion $\interpretationFunctor(c) \subseteq \interpretationFunctor(d)$ holds by definition of model. The proofs for the other rules are trivial. 
\end{proof}

\begin{exa}[Theory with constants]
Consider the theory $\T{T}$ with $\sign=\{k\colon 0 \to 1\}$ and axioms $\TMAP{k}$. 
An interpretation $\interpretation$ of $\sign$ consists of a set $X$ and a relation $k\subseteq \singleton \times X$. An interpretation is a model if and only if $k$ is a function of type $\singleton \to X$. %

Intuitively, all models of such theories must have non-empty domains, witnessed by the presence of the constant $k$. Formally, non-emptiness can be proved by the following derivation:
\input{tikz/proofs/nonEmptiness.tex}
\end{exa}

\subsection{Trivial vs Contradictory theories}

The following classes of theories are important for the subtleties of completeness. It is also a remarkable example of where the syntax of $\NPR$ can be more informative than traditional $\FOL$ notation.
\begin{defi} %
Let $\T{T}$ be a first-order theory.
\begin{itemize}  
\item $\T{T}$ is \emph{contradictory} if ${\emptyCirc[+]} \syninclusionT{\T{T}} {\emptyCirc[-]}$;
\item $\T{T}$  is \emph{trivial} if ${\codiscardCirc[+]} \syninclusionT{\T{T}} {\codiscardCirc[-]}$.
\end{itemize}
\end{defi}
\noindent 
The distinction between contradictory and trivial theories is subtle. %
Triviality implies all models have empty domain: $\interpretationFunctor{({\codiscardCirc[+]})}=  \{(\star,x)\mid x\in X\}$ is included in $\interpretationFunctor{({\codiscardCirc[-]})} = \varnothing$ iff $X=\varnothing$. 
On the other hand, contradictory theories cannot have a model, not even when $X=\varnothing$: since $\interpretationFunctor{({\emptyCirc[+]})}=  \{(\star,\star)\}$ and $\interpretationFunctor{({\emptyCirc[-]})}=  \varnothing$, independently of $X$.

As expected, every contradictory theory is also trivial.

\begin{lem}\label{lemma:contraddictoryimpliestrivial}
    Let $\T{T}$ be a theory. If $\T{T}$ is contradictory then it is trivial.
\end{lem}
\begin{proof}
    Assume $\T{T}$ to be contradictory and consider the following derivation.

    \[
        \codiscardCirc[+] 
        \structuralcong 
        \begin{tikzpicture}
            \begin{pgfonlayer}{nodelayer}
                \node [style=none] (81) at (0.5, 0) {};
                \node [{dotStyle/+}] (107) at (-0.25, 0) {};
                \node [style=none] (112) at (0.5, 0.75) {};
                \node [style=none] (113) at (0.5, -0.75) {};
                \node [style=none] (114) at (-2, -0.75) {};
                \node [style=none] (115) at (-2, 0.75) {};
            \end{pgfonlayer}
            \begin{pgfonlayer}{edgelayer}
                \draw [{bgStyle/+}] (114.center)
                     to (113.center)
                     to (112.center)
                     to (115.center)
                     to cycle;
                \draw [{wStyle/+}] (81.center) to (107);
            \end{pgfonlayer}
        \end{tikzpicture}        
        \stackrel{\T{T}\text{ is contradictory}}{\syninclusionT{\T{T}}}
        \begin{tikzpicture}
            \begin{pgfonlayer}{nodelayer}
                \node [style=none] (81) at (0.5, 0) {};
                \node [{dotStyle/+}] (107) at (-0.25, 0) {};
                \node [style=none] (112) at (0.5, 0.75) {};
                \node [style=none] (113) at (0.5, -0.75) {};
                \node [style=none] (114) at (-2, -0.75) {};
                \node [style=none] (115) at (-2, 0.75) {};
                \node [style=none] (116) at (-0.75, 0.5) {};
                \node [style=none] (117) at (-0.75, -0.5) {};
                \node [style=none] (118) at (-1.75, -0.5) {};
                \node [style=none] (119) at (-1.75, 0.5) {};
            \end{pgfonlayer}
            \begin{pgfonlayer}{edgelayer}
                \draw [{bgStyle/+}] (114.center)
                     to (113.center)
                     to (112.center)
                     to (115.center)
                     to cycle;
                \draw [{wStyle/+}] (81.center) to (107);
                \draw [{bgStyle/-}] (118.center)
                     to (117.center)
                     to (116.center)
                     to (119.center)
                     to cycle;
            \end{pgfonlayer}
        \end{tikzpicture}            
        \stackrel{\text{Proposition}~\ref{prop:maps}}{\syninclusionT{\T{T}}} 
        \begin{tikzpicture}
            \begin{pgfonlayer}{nodelayer}
                \node [style=none] (81) at (0.5, 0) {};
                \node [{dotStyle/-}] (107) at (-0.25, 0) {};
                \node [style=none] (112) at (0.5, 0.75) {};
                \node [style=none] (113) at (0.5, -0.75) {};
                \node [style=none] (114) at (-2, -0.75) {};
                \node [style=none] (115) at (-2, 0.75) {};
            \end{pgfonlayer}
            \begin{pgfonlayer}{edgelayer}
                \draw [{bgStyle/-}] (114.center)
                     to (113.center)
                     to (112.center)
                     to (115.center)
                     to cycle;
                \draw [{wStyle/-}] (81.center) to (107);
            \end{pgfonlayer}
        \end{tikzpicture}     
        \structuralcong
        \codiscardCirc[-]  \qedhere
    \]
\end{proof}

\begin{rem}\label{rmk:trivialandcontradictory} %
The difference between contradictory and trivial theories is not usually \emph{seen} in $\FOL$. Indeed, as we will see later in Remark~\ref{rmk:ambiguity}, both ${\emptyCirc[+]}$ and
${\codiscardCirc[+]}$ translate to $\top$ (truth) in $\FOL$ syntax, while ${\emptyCirc[-]}$ and ${\codiscardCirc[-]}$ translate to $\bot$ (falsity).
\end{rem}

\begin{exa}[The Trivial Theory of Propositional Calculus]\label{subsub:propo calc}
Let $\T{P}=(\sign, \T{I})$, where $\sign$ contains only symbols $P,Q,R\dots$ of type $0 \to 0$ and $\T{I}=\{  ({\codiscardCirc[+]} , {\codiscardCirc[-]}) \}$. In any model of $\T{P}$, the domain $X$ must be $\varnothing$, because of the only axiom in $\T{I}$, that makes $\T{P}$ a trivial theory. A model is a mapping of each of the symbols in $\sign$ to either $\{(\star,\star)\}$ or $\varnothing$. In other words, $P,Q,R,\dots$ act as propositional variables and any model is just an assignment of boolean values. 

In $\T{P}$, like in any trivial theory, all diagrams are equal with the exception of those of type $0\to 0$ (see Lemma \ref{lemma:trivalallequal} in Appendix \ref{app:trivialprop}).
Diagrams of type $0\to 0$ are exactly propositional formulas, as illustrated below (see Proposition \ref{prop:propcalc} in Appendix \ref{app:trivialprop} for a formal statement).
\[ \propVar[+]{} \mapsto \top   \; \quad \; \propVar[-]{} \mapsto \bot   \; \quad \;  \propVar[+]{R} \mapsto R\; \quad \;  \propOpVar[-]{R} \mapsto \nega{R}  \; \quad \;  \propSeqCirc[+]{c}{d} \mapsto c \sqcap d  \; \quad  \; \propSeqCirc[-]{c}{d} \mapsto  c \sqcup d  \] 
Note that, by the axioms of symmetric monoidal categories,  $\seq[.]$ and $\tensor[]$ coincide on diagrams $0 \to 0$ and are associative, commutative and with unit $\id[.][0]$. 

For arrows of type $0\to 0$ our axiomatisation reduces to the one in Figure~\ref{fig:correspondencepropositionalcalculus}. Consider for instance \eqref{ax:comPlusLaxNat}: by definition of $\copier[+][0]$ in Table~\ref{fig:sugar}, the two diagrams of  \eqref{ax:comPlusLaxNat} in Figure~\ref{fig:cb axioms} reduce to those in Figure~\ref{fig:correspondencepropositionalcalculus} above. The rules \eqref{ax:linStrn1}, \eqref{ax:linStrn2}, \eqref{ax:linStrn3} and \eqref{ax:linStrn4} become redundant.
Interestingly, the collapsed axiomatisation corresponds to the rules of the \emph{deep inference calculus of structures} $\mathsf{SKSg}$ presented in \cite{DBLP:phd/de/Brunnler2003}. The correspondence is illustrated in Figure~\ref{fig:correspondencepropositionalcalculus}, where each axiom reports its associated $\mathsf{SKSg}$ rule on the right. 
\end{exa}

\begin{figure}[t]
  \centering
  \[
      \begin{array}{cc}
            \begin{array}{cc}
              {\propCirc[+]{c}} \Lleq{\eqref*{ax:comPlusLaxNat}} {\propTensorCirc[+]{c}{c}}  & \raisebox{-0.8em}{\ensuremath{\inferrule*[left=(c$\uparrow$)]{c}{c \wedge c}}}
            \end{array}
          & \begin{array}{cc}
              {\propCirc[+]{c}} \Lleq{\eqref*{ax:discPlusLaxNat}} {\emptyCirc[+]}            & \raisebox{-0.8em}{\ensuremath{\inferrule*[left=(w$\uparrow$)]{c}{\top}}}
            \end{array}
          \\[20pt] 
              \begin{array}{cc}
              {\propTensorCirc[-]{c}{c}} \Lleq{\eqref*{ax:comMinusLaxNat}} {\propCirc[-]{c}} & \raisebox{-0.8em}{\ensuremath{\inferrule*[left=(c$\downarrow$)]{c \vee c}{c}}}
             \end{array}
          & \begin{array}{cc}
              {\emptyCirc[-]} \Lleq{\eqref*{ax:discMinusLaxNat}} {\propCirc[-]{c}}             & \raisebox{-0.8em}{\ensuremath{\inferrule*[left=(w$\downarrow$)]{\bot}{c}}}
             \end{array} \\[15pt]
            \begin{array}{cc}
                    {\emptyCirc[+]} \stackrel{\footnotesize{\stackanchor{\eqref*{ax:tauRPlus}}{\eqref*{ax:tauRMinus}}}}{\leq}  {
    \InputIfFileExists{axiomsNEW/propositional/tauR.tikz}{}{\input{tikz/axiomsNEW/propositional/tauR.tikz}}
}
                & \raisebox{-0.8em}{\ensuremath{\inferrule*[left=(i$\downarrow$)]{\top}{c \vee \nega{c}}}}
            \end{array}
            & 
            \begin{array}{cc}
                    {
    \InputIfFileExists{axiomsNEW/propositional/gammaR.tikz}{}{\input{tikz/axiomsNEW/propositional/gammaR.tikz}}
}\stackrel{\footnotesize{\stackanchor{\eqref*{ax:gammaRPlus}}{\eqref*{ax:gammaRMinus}}}}{\leq} {\emptyCirc[-]}
                & \raisebox{-0.8em}{\ensuremath{\inferrule*[left=(i$\uparrow$)]{c \wedge \nega{c}}{\bot}}}
            \end{array}
        \\[20pt]
          \multicolumn{2}{c}{
              \begin{array}{cc}
                  {
    \InputIfFileExists{axiomsNEW/propositional/distributivity1.tikz}{}{\input{tikz/axiomsNEW/propositional/distributivity1.tikz}}
} \stackrel{\footnotesize{\stackanchor{\eqref*{ax:leftLinDistr}}{\eqref*{ax:rightLinDistr}}}}{\leq} {
    \InputIfFileExists{axiomsNEW/propositional/distributivity2.tikz}{}{\input{tikz/axiomsNEW/propositional/distributivity2.tikz}}
} 
              & \raisebox{-0.8em}{\ensuremath{\inferrule*[left=(s)]{a \wedge (b \vee c)}{(a \wedge b) \vee c}}}
              \end{array}
          }
      \end{array}
  \]
  \caption{The axioms of fo-bicategories reduce to those above for diagrams of type $0 \to 0$.}
  \label{fig:correspondencepropositionalcalculus}
\end{figure}

\subsection{Closed Theories and the Deduction Theorem}
Even though we did not establish yet a formal correspondence between formulas of $\FOL$ and diagrams of $\LCB$, one can already guess that dangling wires, either on the left or on the right of a diagram, play the role of free variables. Thus, diagrams $d\colon 0\to 0$ can be thought of as \emph{closed} formulas of $\FOL$, which also play an important role in our proof of completeness. 

Recall that a theory in $\FOL$ is usually defined as a set $\mathcal{T}$ of closed formulas that must hold in all models. With a slight abuse of notation, one can think of constructing a corresponding theory in $\LCB$, whose set of axioms is $\{(\id[+][0],d) \mid d\in \mathcal{T}\}$. Since the semantics $\interpretationFunctor$ assigns to every diagram $d \colon 0 \to 0$ a relation $R\subseteq \singleton \times \singleton$, either $\{(\star, \star)\}$ (i.e., $\id[+][\singleton]$) representing true or $\varnothing$ (i.e., $\id[-][\singleton]$) representing false, the fact that $d$ must hold in any model is indeed forced by requiring the axiom $(\id[+][0],d)$.  This leads us to the definition of another relevant class of diagrammatic first-order theories, that we call \emph{closed theories}.

\begin{defi}
A theory $\T{T} = (\sign, \T{I})$ is said to be \emph{closed} if all the pairs $(c,d) \in \T{I}$ are of the form $(\id[+][0], d)$. %
\end{defi}
For instance, the theory of sets and the theory of non-empty sets in Examples \ref{ex:full:empty theory} and \ref{ex:full:non empty domains} are closed, while the other theories encountered so far are not. However, by means of Lemma~\ref{lm:implications}, one can always translate an arbitrary theory $\T{T} = (\sign, \T{I})$ into a closed theory $\T{T}^c = (\sign, \T{I}^c)$ where
\[\T{I}^c \defeq \left\{ \left( \; \emptyCirc[+] \; , \scalebox{0.75}{\circleCirc{c}{d}[b]} \; \right) \mid (c,d) \in \T{I} \right\}\text{.}\]

\begin{prop}\label{prop:closed}
Let $\T{T} = (\sign, \T{I})$ be a theory and $a,b\colon n \to m$ in $\LCB$. Then $a \syninclusionT{\T{T}} b$ iff $a \syninclusionT{\T{T}^c} b$.
\end{prop}
\begin{proof} %
      By induction on the rules in $\eqref{eq:pc}$. %
      The base case $(id)$ is given by means of Lemma~\ref{lm:implications} and in particular from the fact that:
      \[ \boxCirc[+]{a} \syninclusionT{\T{T}} \boxCirc[+]{b} \;\text{ if and only if }\; \emptyCirc[+] \syninclusionT{\T{T}} \scalebox{0.75}{\circleCirc{a}{b}[b]} \; \text{ for any pair $(a,b) \in \T{I}$.} \]

   The base case $(r)$ and the inductive cases are trivial.
\end{proof}
\noindent 
This result allows us to safely restrict our attention to closed theories. However, note that this assumption is not actually needed for the proof of completeness.  More interestingly, it tells us that while diagrammatic first-order theories, in general, appear to be rather different from the usual $\FOL$ theories, they can always be translated into closed theories which are essentially those of $\FOL$.%

\medskip

The fact that a closed formula $d$ is derivable in $\mathcal{T}$, usually written as $\mathcal{T}\vdash d$, translates in $\LCB$ to $\id[+][0]  \syninclusionT{\T{T}} d$.
In particular, when $d$ is an implication $c \Rightarrow b$, we have $\id[+][0] \syninclusionT{\T{T}} b \seq[-] \rla{c}$ that, by Lemma \ref{lm:residuation}, is equivalent to $c  \syninclusionT{\T{T}} b$. 

In $\FOL$ it is trivial -- by modus ponens -- that if  $\mathcal{T}\vdash c \Rightarrow b$ then $\mathcal{T}\cup \{c\} \vdash b$. In $\LCB$, this fact follows by transitivity of $\syninclusionT{\T{T}}$: fix $\T{T}'=(\sign, \T{I} \cup \{(\id[+][0],c)\})$ and observe that $\id[+][0]  \syninclusionT{\T{T'}} c \syninclusionT{\T{T'}} b $.
The converse implication, namely  if $\mathcal{T}\cup \{c\} \vdash b$ then $\mathcal{T}\vdash c \Rightarrow b$, is known in $\FOL$ as the \emph{deduction theorem}. It can be generalised in $\NPR$ as follows.

\begin{thm}[Deduction theorem]\label{th:deduction}%
    Let $\T{T} = (\sign, \T{I})$ be a theory and $c \colon 0 \to 0$ in $\LCB$. Let $\T{I'} = \T{I} \cup \{(\id[+][0], c)\}$ and let $\T{T'}$ denote the theory $(\sign, \T{I'})$. Then, for every $a,b \colon n \to m$ arrows of $\LCB$,
    \begin{center}if $\;\;  \boxCirc[+]{a} \;\;\precongR{\T{T'}}\;\; \boxCirc[+]{b} \;\;$ then $\;\; 
    \InputIfFileExists{cPerId.tikz}{}{\input{tikz/cPerId.tikz}}
 \;\;\precongR{\T{T}}\;\!\!\!\! \seqCirc[-]{b}{a}[][][w][b]$\!\!\!\!.\end{center}

\end{thm}
\begin{proof}%

By induction on the rules of~\eqref{eq:pc}.   The base cases $(id)$ and $(r)$ are trivial. The inductive cases are listed below.

\begin{enumerate}
    \item[$(t)$] Assume $a \syninclusionT{\T{T'}} d$ and $d \syninclusionT{\T{T'}} b$ for some $d \colon n \to m$. Observe that $a \syninclusionT{\T{T'}} b$ by $(t)$ and $c \tensor[+] \id[+][n] \syninclusionT{\T{T}} d \seq[-] \rla{a}$ and $c \tensor[+] \id[+][n] \syninclusionT{\T{T}} b \seq[-] \rla{d}$  by inductive hypothesis. To conclude we need to show:
    \input{tikz/proofs/deductionTheorem/transitivity.tex}

    \item[$({\seq[+]})$] Assume $a = a_1 \seq[+] a_2$ and $b = b_1 \seq[+] b_2$ for some $a_1, b_1 \colon n \to l, a_2, b_2 \colon l \to m$ such that $a_1 \syninclusionT{\T{T'}} b_1$ and $a_2 \syninclusionT{\T{T'}} b_2$. By induction hypothesis $c \tensor[+] \id[+][n] \syninclusionT{\T{T}} b_1 \seq[-] \rla{a_1}$ and $c \tensor[+] \id[+][n] \syninclusionT{\T{T}} b_2 \seq[-] \rla{a_2}$. Note that:
    \input{tikz/proofs/deductionTheorem/seqPlus.tex}

    \item[$({\seq[-]})$] Assume $a_1 \syninclusionT{\T{T'}} b_1$ and $a_2 \syninclusionT{\T{T'}} b_2$ such that $a = a_1 \seq[-] a_2$ and $b = b_1 \seq[-] b_2$ for some $a_1, b_1 \colon n \to l, a_2, b_2 \colon l \to m$. Observe that $a_1 \seq[-] a_2 \syninclusionT{\T{T'}} b_1 \seq[-] b_2$ by $(\seq[-])$ and $c \tensor[+] \id[+][n] \syninclusionT{\T{T}} b_1 \seq[-] \rla{a_1}$ and $c \tensor[+] \id[+][n] \syninclusionT{\T{T}} b_2 \seq[-] \rla{a_2}$  by inductive hypothesis. To conclude we need to show:
    \input{tikz/proofs/deductionTheorem/seqMinus.tex}

    \item[$({\tensor[+]})$] Assume $a_1 \syninclusionT{\T{T'}} b_1$ and $a_2 \syninclusionT{\T{T'}} b_2$ such that $a = a_1 \tensor[+] a_2$ and $b = b_1 \tensor[+] b_2$ for some $a_1,b_1 \colon n' \to m', a_2,b_2 \colon n'' \to m''$. Observe that $a_1 \tensor[+] a_2 \syninclusionT{\T{T'}} b_1 \tensor[+] b_2$ by $(\tensor[+])$ and $c \tensor[+] \id[+][n] \syninclusionT{\T{T}} b_1 \seq[-] \rla{a_1}$ and $c \tensor[+] \id[+][n] \syninclusionT{\T{T}} b_2 \seq[-] \rla{a_2}$  by inductive hypothesis. To conclude we need to show:
    \input{tikz/proofs/deductionTheorem/tensorPlus.tex}

    \item[$({\tensor[-]})$] Assume $a_1 \precongR{\T{T'}} b_1$ and $a_2 \syninclusionT{\T{T'}} b_2$ such that $a = a_1 \tensor[-] a_2$ and $b = b_1 \tensor[-] b_2$ for some $a_1,b_1 \colon n' \to m', a_2,b_2 \colon n'' \to m''$. Observe that $a_1 \tensor[-] a_2 \syninclusionT{\T{T'}} b_1 \tensor[-] b_2$ by $(\tensor[-])$ and $c \tensor[+] \id[+][n] \syninclusionT{\T{T}} b_1 \seq[-] \rla{a_1}$ and $c \tensor[+] \id[+][n] \syninclusionT{\T{T}} b_2 \seq[-] \rla{a_2}$  by inductive hypothesis. To conclude we need to show:
    \input{tikz/proofs/deductionTheorem/tensorMinus.tex}
  \end{enumerate}
\end{proof}

\begin{cor}\label{cor:deduction}
Let $\T{T}=(\sign, \T{I})$ be a theory, $c\colon 0 \to 0$ in $\LCB$ and $\T{T}'=(\sign, \T{I}\cup \{(\id[+][0], \nega{c} ) \})$. Then $\id[+][0]\syninclusionT{\T{T}}c$ iff \ $\T{T'}$ is contradictory.
\end{cor}
\begin{proof}
        Suppose that $\T{T}'$ is contradictory, namely $\id[+][0] \syninclusionT{\T{T}'} \id[-][0]$. By the deduction theorem (Theorem \ref{th:deduction}), $\nega{c} \syninclusionT{\T{T}}  \id[-][0]$ and thus $\nega{\id[-][0]}  \syninclusionT{\T{T}}  \nega{\nega{c}}$, that is $\id[+][0] \syninclusionT{\T{T}} c$. 
        The other direction is trivial: since $\id[+][0] \syninclusionT{\T{T}'} c$ and $\id[+][0] \syninclusionT{\T{T}'} \nega{c}$, then $\id[+][0] \syninclusionT{\T{T}'} c \sqcap \nega{c} \syninclusionT{\T{T}'}  \bot = \id[-][0] $. 
\end{proof}

\subsection{Functorial Semantics for First-Order Theories}

Recall that the notion of interpretation of a signature $\sign$ in $\Rel$ has been generalised in Definition \ref{def:intCAT} to an arbitrary fo-bicategory. As expected, the same is possible also with the notion of model.
\begin{defi}
Let $\T{T}=(\sign, \T{I})$ be a theory and $\Cat{C}$ a first-order bicategory. An interpretation $\interpretation$ of $\sign$ in $\Cat{C}$ is a model iff, for all $(c,d)\in \T{I}$,
$\interpretationFunctor(c) \leq \interpretationFunctor(d)$.
\end{defi}

For any theory $\T{T}=(\sign, \T{I})$, one can construct a fo-bicategory $\LCB[\T{T}]$ similarly to the construction of $\LCB[\sign]$  illustrated in Section \ref{sec:freely}: we fix every homset $\LCB[\T{T}][n,m] \defeq \{ [d]_{\synequivalenceT{\T{T}}} \mid d\in \LCB[\sign][n,m]\}$ that is ordered by $\syninclusionT{\T{T}}$. Since, by definition, $\syninclusion \subseteq \syninclusionT{\T{T}}$, $\LCB[\T{T}]$ is a fo-bicategory.
Thus, one can consider an interpretation $\mathcal{Q}_{\T{T}}$ of $\Sigma$ in $\LCB[\T{T}]$: the domain $X$ is $1$ and $\rho (R) = [R^\circ]_{\synequivalenceT{\T{T}}}$ for all $R \in \Sigma$. By Proposition \ref{prop:free}, $\mathcal{Q}_{\T{T}}$ induces a morphism of fo-bicategories $\mathcal{Q}_{\T{T}}^\sharp \colon \LCB \to \LCB[\T{T}]$.

\begin{prop}\label{prop:modelfactor}
Let $\T{T}=(\sign, \T{I})$ be a theory, $\Cat{C}$ a fo-bicategory and $\interpretation$ an interpretation of $\Sigma$ in $\Cat{C}$. Then $\interpretation$ is a model of $\T{T}$ in $\Cat{C}$ iff $\interpretationFunctor \colon \LCB \to \Cat{C}$ factors through $\mathcal{Q}_{\T{T}}^\sharp \colon \LCB \to \LCB[\T{T}]$, namely there exists a morphism of fo-bicategories $\interpretationFunctor_{\T{T}} \colon \LCB[\T{T}] \to \Cat{C}$ such that the diagram below commutes. Moreover, $\interpretationFunctor_{\T{T}}$ is the unique such morphism.
\[\xymatrix{\LCB \ar[rd]|{\interpretationFunctor} \ar[r]^{\mathcal{Q}_{\T{T}}^\sharp}& \LCB[\T{T}] \ar@{.>}[d]^{\interpretationFunctor_{\T{T}}}\\
& \Cat{C}}\]
\end{prop}
\begin{proof}
    First, observe that a simple inductive argument allows to prove that, for all diagrams $c$ in $\LCB$, 
    \begin{equation}\label{eq:functor1}
    \mathcal{Q}_{\T{T}}^\sharp(c)=[c]_{\synequivalenceT{\T{T}}}\text{.}
    \end{equation}
    
    Now, suppose that there exists $\interpretationFunctor_{\T{T}} \colon \LCB[\T{T}] \to \Cat{C}$ making the above diagram commute and consider $(c,d)\in \T{I}$. 
    By definition, $c \syninclusionT{\T{T}} d$ and, by \eqref{eq:functor1}, 
    \begin{equation}\label{eq:functor2}
    \mathcal{Q}_{\T{T}}^\sharp (c) \syninclusionT{\T{T}} \mathcal{Q}_{\T{T}}^\sharp (d) \text{.}
    \end{equation}
    Then, the following derivation confirms that $\interpretation$ is a model of $\T{T}$ in $\Cat{C}$.
    \begin{align*}
    \interpretationFunctor(c) &= \interpretationFunctor_{\T{T}}( \mathcal{Q}_{\T{T}}^\sharp (c)) \tag{$\interpretationFunctor = \mathcal{Q}_{\T{T}}^\sharp ; \interpretationFunctor_{\T{T}}$} \\ 
    &\leq \interpretationFunctor_{\T{T}}( \mathcal{Q}_{\T{T}}^\sharp (d)) \tag{\eqref{eq:functor2} and $ \interpretationFunctor_{\T{T}}$ is a morphism}\\
    & = \interpretationFunctor(d) \tag{$\interpretationFunctor = \mathcal{Q}_{\T{T}}^\sharp ; \interpretationFunctor_{\T{T}}$} 
    \end{align*}

    Vice versa, suppose that $\interpretation$ is a model of $\T{T}$ in $\Cat{C}$. Then by definition of model, for all $(c,d)\in \T{I}$,  $\interpretationFunctor(c) \leq \interpretationFunctor(d)$. A simple inductive argument on the rules in \eqref{eq:pc} confirms that, for all diagrams $c,d$ in $\LCB$,
    \begin{center}
    if $c \syninclusionT{\T{T}} d$ then $\interpretationFunctor(c) \leq \interpretationFunctor(d)$.
    \end{center}
    In particular, if $c \synequivalenceT{\T{T}} d$ then $\interpretationFunctor(c) = \interpretationFunctor(d)$. Therefore, we are allowed to define $\interpretationFunctor_{\T{T}} ([c]_{\synequivalenceT{\T{T}}}) \defeq \interpretationFunctor (c)$ for all arrows $[c]_{\synequivalenceT{\T{T}}}$ of $\LCB[\T{T}]$ and $\interpretationFunctor_{\T{T}}(n)\defeq\interpretationFunctor(n)$ for all objects $n$ of $\LCB[\T{T}]$. The fact that $\interpretationFunctor_{\T{T}}$ preserves the ordering follows immediately from the above implication. The fact that   $\interpretationFunctor_{\T{T}}$ preserves the structure of fo-bicategories follows easily from the fact that $\interpretationFunctor$ is a morphism. Therefore $\interpretationFunctor_{\T{T}}$ is a morphism of fo-bicategories. The fact that the above diagram commutes is obvious by definition of $\interpretationFunctor_{\T{T}}$ and \eqref{eq:functor1}.

    Uniqueness follows immediately from the fact that $\mathcal{Q}_{\T{T}}^\sharp \colon \LCB \to \LCB[\T{T}]$ is an epi, namely all objects and arrows of $\LCB[\T{T}]$ are in the image of $\mathcal{Q}_{\T{T}}^\sharp$.
\end{proof}

The assignment $\interpretation \mapsto \interpretationFunctor_{\T{T}}$ provides a bijective correspondence between models and morphisms.

\begin{cor}\label{corollarymodelfunctor}
To give a model of \,$\T{T}$ in $\Cat{C}$ is to give a fo-bicategory morphism $\LCB[\T{T}] \to \Cat{C}$.
\end{cor}
\begin{proof}
    To go from models to morphisms we use the assignment $\interpretation \mapsto \interpretationFunctor_{\T{T}}$ provided by Proposition \ref{prop:modelfactor}.
    To transform morphisms into models, we need a slightly less straightforward assignment. Take a morphism of fo-bicategories $\mathcal{F}\colon \LCB[\T{T}] \to \Cat{C}$ and consider $\mathcal{Q}_{\T{T}}^\sharp ; \mathcal{F}\colon \LCB \to \Cat{C}$. This gives rise to the interpretation $\interpretation_\mathcal{F} = (X, \rho)$ defined as 
    
    \[ X \defeq \mathcal{Q}_{\T{T}}^\sharp ; \mathcal{F} (1) \qquad \text{ and } \qquad \rho(R) \defeq \mathcal{Q}_{\T{T}}^\sharp ; \mathcal{F}(R^\circ) \;\; \text{ for all } R \in \sign \]
    
    Then, by Proposition \ref{prop:free}, $\interpretation_\mathcal{F}^\sharp =\mathcal{Q}_{\T{T}}^\sharp ; \mathcal{F}$ and thus, by Proposition \ref{prop:modelfactor}, $\interpretation_\mathcal{F}$ is a model. Since $\interpretation_\mathcal{F}^\sharp =\mathcal{Q}_{\T{T}}^\sharp ; \mathcal{F}$, by the uniqueness provided by Proposition \ref{prop:modelfactor}, $(\interpretation_{\mathcal{F}})_{\T{T}}^\sharp = \mathcal{F}$.
    
    To conclude, we only need to prove that $\interpretation_{(\interpretationFunctor_{\T{T}}) } = \interpretation$. Since $\mathcal{Q}_{\T{T}}^\sharp  ; \interpretationFunctor_{\T{T}} = \interpretationFunctor$, then $\interpretation_{(\interpretationFunctor_{\T{T}}) } (R^\circ) = \mathcal{Q}_{\T{T}}^\sharp  ; \interpretationFunctor_{\T{T}} (R^\circ) = \interpretationFunctor(R^\circ) = \rho(R)$ for all $R\in \sign$. Similarly for the domain $X$.
\end{proof}

\section{Completeness}\label{sec:completeness}\label{ssec:henkinmodel}
In this section we illustrate a proof of Theorem \ref{thm:completeness}, asserting completeness of the axioms of first-order bicategories.
Our proof is divided into two main statements
\begin{equation}\label{thm:Godel}
\text{if $\T{T}$ is a non-trivial theory, then $\T{T}$ has a model}\tag{G\"odel}
\end{equation}
and
\begin{equation}\label{thm:trivialcom}
\text{\!\!if $\T{T}$ is a trivial and non-contradictory theory,  then $\T{T}$ has a model} \tag{Prop}
\end{equation}
that immediately entail
\begin{equation}\label{cor:gencompleteness}
\text{if $\T{T}$ is a non-contradictory theory, then  $\T{T}$ has a model.}\tag{General}
\end{equation}
Then, Theorem \ref{thm:completeness} easily follows by means of standard first-order logic arguments relying on the deduction theorem (Theorem \ref{th:deduction}).
In Section \ref{sec:Beyond Godel} we illustrate a proof for \eqref{thm:trivialcom} and in Section \ref{sec:generalcompleteness},  one for Theorem \ref{thm:completeness}. As expected, the proof for \eqref{thm:Godel} is more laborious and it is  divided in two parts, illustrated in Sections \ref{sec:proofPart1} and \ref{sec:proofPart2}.

\medskip

Before delving into the proof of \eqref{thm:Godel}, it is convenient to have an overview.
First, we must say that our proof is a  faithful adaptation  of the proof of Henkin's~\cite{henkin_1949} to $\NPR$. Henkin's proof starts with the following two notions. %
\begin{defi}\label{def:henkin wit}
Let $\T{T}=(\sign, \mathbb{I})$ be a theory.
\begin{itemize}
\item $\T{T}$ is \emph{syntactically complete} if for all $c \colon 0 \to 0$ either $\id[+][0] \;\precongR{\T{T}}\; c$ or $\id[+][0] \;\precongR{\T{T}}\; \overline{c}$.
 \item $\T{T}$ has \emph{Henkin witnesses} if for all $c \colon 1 \to 0$ there is a map $k \colon 0 \to 1$ such that
    \begin{equation}\label{eq:hen-wit}\tag{H-wit}
        {
    \InputIfFileExists{henkinWit1.tikz}{}{\input{tikz/henkinWit1.tikz}}
} \syninclusionT{\T{T}} {
    \InputIfFileExists{henkinWit2.tikz}{}{\input{tikz/henkinWit2.tikz}}
}.
    \end{equation}
    \end{itemize}
\end{defi}
\noindent 
These properties do not hold for the theories we have considered so far. In terms of $\FOL$, syntactic completeness means that closed formulas either hold in all models of the theory or in none. A Henkin witness is a term $k$ such that $c(k)$ holds: a theory has Henkin witnesses if for every true formula $\exists x.c(x)$, there exists such a $k$.
We shall see in Theorem~\ref{thm:nontrivialALL} that non-trivial theories can be expanded to have Henkin witnesses, be non-contradictory and syntactically complete. The key idea of Henkin's proof, Theorem~\ref{thm:Hekinismodel}, is that these three properties yield a model, known as  Henkin's model.%

\subsection{G\"odel completeness: Part I}\label{sec:proofPart1}
Before introducing Henkin's interpretation, %
it is convenient to recall that, by Lemma \ref{prop:nary maps}, in $\LCB$, any map $k\colon 0 \to n$ can be decomposed as $k_1 \tensor[+] \dots \tensor[+] k_n$ where each $k_i\colon 0 \to 1$ is a map. %
 We thus write such $k$ as $\nterm{k}$, depicted as $\constCirc{\nterm{k}}[n]\!\!$, to make explicit the fact that it is a vector of $n$ constants.
This  allows for a generalisation to $n$-ary maps of the Henkin witness property.
 \begin{lem}\label{lemma:henkinn}
    Let $\T{T}$ be a theory with Henkin witnesses. For all $c \colon n \to 0$, there is a map $\nterm{k} \colon 0 \to n$ such that
    \[
    \InputIfFileExists{henkinWit1.tikz}{}{\input{tikz/henkinWit1.tikz}}
 \syninclusionT{\T{T}} 
    \InputIfFileExists{henkinWitn.tikz}{}{\input{tikz/henkinWitn.tikz}}
\text{.}\]
\end{lem}
\begin{proof}
The proof goes by induction on $n$. For $n=0$, take $\nterm{k}=\id[+][0]$. For $n+1$, we have that:

\input{tikz/proofs/henkinn.tex}

\end{proof}
\noindent 
 Now we have all the necessary equipment for being able to define a peculiar interpretation of monoidal signatures in $\Rel$.

\begin{defi}\label{def:henkin struct}
Let $\T{T}=(\Sigma,\mathbb{I})$ be a theory.
The \emph{Henkin interpretation} $\Hen$ of $\sign$, consists of a set $X \defeq \texttt{Map}(\LCB[\T{T}])[0,1]$ and a function $\rho$, defined for all $R  \colon n \to m \in \sign$ as:
    \[
        \rho(R) \defeq \{ (\nterm{k},\nterm{l}) \in X^n \times X^m \mid \scalebox{0.8}{\emptyCirc[+]} \syninclusionT{\T{T}} \scalebox{0.8}{\closedFormulaCirc{\nterm{k}}{R}{\nterm{l}}} \}.
    \]
\end{defi}
 The domain is the set of constants of the theory. Indeed $\texttt{Map}(\LCB[\T{T}])[0,1]$ is the set of all maps $k\colon 0 \to 1$ in $\LCB[\T{T}]$. Then $R\colon n \to m$ is mapped to all pairs $(\nterm{k},\nterm{l})$ of vectors that make $R$ true in $\T{T}$.
 The following characterisation of $\HenFunctor \colon \LCB \to \Rel$ is crucial.

\begin{prop}\label{prop:henkin}
    Let $\T{T}=(\Sigma,\mathbb{I})$ be a non-contradictory, syntactically complete theory with Henkin witnesses. Then, for any $c \colon n \to m$,
    $\HenFunctor(c) = \{ (\nterm{k},\nterm{l}) \in X^n \times X^m \mid \scalebox{0.8}{\emptyCirc[+]} \syninclusionT{\T{T}} \scalebox{0.8}{\closedFormulaCirc{\nterm{k}}{c}{\nterm{l}}} \}$.
\end{prop}
\begin{proof}%
    The proof goes by induction on $c$.

    Consider the case $c=\id[+][1]$. Observe that, for all $k,l \in X$ (i.e., $\texttt{Map}(\LCB[\T{T}])[0,1]$),
    \begin{equation}\label{eqinsideHenkinproof}
    \id[+][0] \syninclusionT{\T{T}} k\seq[+]\op{l} \text{ iff } k=l
    \end{equation}
    Indeed, if $k=l$, then $\id[+][0] \syninclusionT{\T{T}} k\seq[+]\op{l}$ by Proposition \ref{prop:opcartesianfunctor}. Vice versa, if $\id[+][0] \syninclusionT{\T{T}} k\seq[+]\op{l}$, then
    \begin{align*}
    k &\structuralcong  \id[+][0] \seq[+] k \tag{SMC}\\
   & \syninclusionT{\T{T}}  k\seq[+]\op{l} \seq[+] k \tag{$\id[+][0] \syninclusionT{\T{T}} k\seq[+]\op{l}$}\\
   & =  \op{(k\seq[+]\op{l})} \seq[+] k \tag{Table \ref{fig:sugar}}\\
    &=  l \seq[+] \op{k} \seq[+] k \tag{Table \ref{table:daggerproperties}}\\
    &\syninclusionT{\T{T}}  l \tag{Proposition \ref{prop:opcartesianfunctor}}
    \end{align*}
We thus have that
      \begin{align*}
      \HenFunctor(\id[+][1]) = & \id[+][X] \tag{\ref{fig:semantics}}\\
      =&\{(k,l) \mid k=l\} \tag{\ref{eq:idRel}}\\
      =&\{(k,l) \mid \id[+][0] \syninclusionT{\T{T}} k\seq[+]\op{l}\} \tag{\ref{eqinsideHenkinproof}}\\
      \structuralcong&\{(k,l) \mid \id[+][0] \syninclusionT{\T{T}} k\seq[+]\id[+][1]\seq[+] \op{l}\} \tag{SMC}
      \end{align*}
The proofs for the other constants of the white fragment follow analogous arguments. Note that none of the hypothesis about the theory is used here.%

Instead, for the case $c=\id[-][0]$ we use the hypothesis that $\T{T}$ is not contradictory. Suppose that there exist map $k,l \colon 0 \to 0$ such that $\id[+][0]\syninclusionT{\T{T}} k \seq[+] \id[-][0] \seq[+] \op{l}$. By Lemma \ref{lemma:cb maps}, $\id[+][0]$ is the only map of type $0 \to 0$ and thus, it should be the case that $\id[+][0] \syninclusionT{\T{T}} \id[+][0] \seq[+] \id[-][0] \seq[+] \id[+][0]$. Since $\id[+][0] \seq[+] \id[-][0] \seq[+] \id[+][0] \structuralcong \id[-][0]$, we have that $\id[+][0] \syninclusionT{\T{T}} \id[-][0]$, against the hypothesis that 
$\T{T}$ is non-contradictory. Thus $\{(k,l) \mid \id[+][0] \syninclusionT{\T{T}} k \seq[+] \id[-][0] \seq[+] \op{l}\} = \varnothing$ which, by \eqref{fig:semantics}, is $\HenFunctor(\id[-][0])$.

The remaining constants of the black fragment follow a recurring pattern, using the hypothesis that $\T{T}$ is syntactically complete. We show only the case $c=\scalebox{0.8}{\boxOpCirc[-]{R}[m][n]}$.
    {\allowdisplaybreaks
    \begin{align*}
        \HenFunctor(\scalebox{0.8}{\boxOpCirc[-]{R}})
            &= \tag{\ref{fig:semantics}} \{ (\nterm{l}, \nterm{k}) \in X^n \times X^m \mid (\nterm{k}, \nterm{l}) \notin \HenFunctor(\scalebox{0.8}{\boxCirc[+]{R}}) \} \\[5pt]
            &= \tag{Definition~\ref{def:henkin struct}} \{ (\nterm{l}, \nterm{k}) \in X^n \times X^m \mid \scalebox{0.8}{\emptyCirc[+]} \not\syninclusionT{\T{T}} \scalebox{0.8}{\closedFormulaCirc{\nterm{k}}{R}{\nterm{l}}} \} \\[5pt]
            &= \tag{$\T{T}$ is syntactically complete} \{ (\nterm{l}, \nterm{k})\in X^n \times X^m \mid \scalebox{0.8}{\emptyCirc[+]} \syninclusionT{\T{T}} \scalebox{0.8}{\closedFormulaCirc[-]{\nterm{k}}{R}{\nterm{l}}} \} \\[5pt]
            &= \tag{Table~\ref{table:daggerproperties}} \{ (\nterm{l}, \nterm{k}) \in X^n \times X^m \mid \op{(\scalebox{0.8}{\emptyCirc[+]})} \syninclusionT{\T{T}} \op{(\scalebox{0.8}{\closedFormulaCirc[-]{\nterm{k}}{R}{\nterm{l}}})} \} \\[5pt]
            &= \tag{Table~\ref{table:daggerproperties}} \{ (\nterm{l}, \nterm{k}) \in X^n \times X^m \mid \scalebox{0.8}{\emptyCirc[+]} \syninclusionT{\T{T}} \scalebox{0.8}{\closedOpFormulaCirc[-]{\nterm{k}}{R}{\nterm{l}}} \} \\[8pt]
            &= \tag{Proposition~\ref{prop:maps}} \{ (\nterm{l}, \nterm{k}) \in X^n \times X^m \mid \scalebox{0.8}{\emptyCirc[+]} \syninclusionT{\T{T}}
            \scalebox{0.8}{\begin{tikzpicture}
                \begin{pgfonlayer}{nodelayer}
                    \node [{boxOpStyle/-}] (107) at (-0.025, 0) {$R$};
                    \node [{funcStyle/+}] (117) at (-2.125, 0) {$\nterm{l}$};
                    \node [style=none] (122) at (2.875, 1) {};
                    \node [style=none] (123) at (2.875, -1) {};
                    \node [style=none] (124) at (-2.825, -1) {};
                    \node [style=none] (125) at (-2.825, 1) {};
                    \node [{funcOpStyle/+}] (126) at (2.125, 0) {$\nterm{k}$};
                    \node [style=none] (127) at (1.125, 0.75) {};
                    \node [style=none] (128) at (1.125, -0.75) {};
                    \node [style=none] (129) at (-1.125, -0.75) {};
                    \node [style=none] (130) at (-1.125, 0.75) {};
                    \node [style=none] (131) at (1.125, 0) {};
                    \node [style=none] (132) at (-1.125, 0) {};
                \end{pgfonlayer}
                \begin{pgfonlayer}{edgelayer}
                    \draw [{bgStyle/+}] (124.center)
                            to (123.center)
                            to (122.center)
                            to (125.center)
                            to cycle;
                    \draw [{bgStyle/-}] (129.center)
                            to (128.center)
                            to (127.center)
                            to (130.center)
                            to cycle;
                    \draw [style={wStyle/+}] (117) to (132.center);
                    \draw [style={wStyle/+}] (131.center) to (126);
                    \draw [style={wStyle/-}] (131.center) to (107);
                    \draw [style={wStyle/-}] (107) to (132.center);
                \end{pgfonlayer}
            \end{tikzpicture}} \}
    \end{align*}}

The most interesting part is the inductive case $c \seq[+] d$, where one exploits the hypothesis that $\T{T}$ has Henkin witnesses.
    Suppose $c \colon n \to o$ and $d \colon o \to m$, then observe that the following holds:
    \input{tikz/proofs/henkinProp/seqPlus.tex}
Similarly for the case $c \tensor[+] d$: the reader can check it using \eqref{fig:semantics}, the induction hypothesis, \eqref{eq:tensorREL}, \eqref{ax:comPlusLaxNat},\eqref{ax:discPlusLaxNat} and Lemma \ref{lemma:cb maps}.

    For the inductive case $c \seq[-] d$, assume $c \colon n \to o$ and $d \colon o \to m$, then observe that:
    \input{tikz/proofs/henkinProp/seqMinus.tex}

    \noindent 
    The proof above relies on Lemma~\ref{lm:hen dagger adj} and the previous inductive case of $c \seq[+] d$. The case of $c \tensor[-] d$ follows the exact same reasoning but, as expected, this time one has to exploit the proof of $c \tensor[+] d$.
\end{proof}

\begin{thm}\label{thm:Hekinismodel}
Let $\T{T}=(\Sigma,\mathbb{I})$ be a non-contradictory, syntactically complete theory with Henkin witnesses. Then $\Hen$ is a model.
\end{thm}
\begin{proof} %
    We show that $c \syninclusionT{\T{T}} d$ gives $\HenFunctor(c) \subseteq \HenFunctor(d)$.
    If $(\nterm{k}, \nterm{l}) \in \HenFunctor(c)$ then $\scalebox{0.8}{\emptyCirc[+]} \!\!\!\syninclusionT{\T{T}}\!\!\! \scalebox{0.8}{\closedFormulaCirc{\nterm{k}}{c}{\nterm{l}}}$ by Prop.~\ref{prop:henkin}. Since $c \syninclusionT{\T{T}} d$,  $\scalebox{0.8}{\emptyCirc[+]} \!\!\!\syninclusionT{\T{T}}\!\!\! \scalebox{0.8}{\closedFormulaCirc{\nterm{k}}{c}{\nterm{l}}} \!\!\!\syninclusionT{\T{T}}\!\!\! \scalebox{0.8}{\closedFormulaCirc{\nterm{k}}{d}{\nterm{l}}}$ and by Prop.~\ref{prop:henkin}, $(\nterm{k}, \nterm{l}) \in \HenFunctor(d)$.
\end{proof}

\subsection{G\"odel completeness: Part II}\label{sec:proofPart2}
Theorem \ref{thm:Hekinismodel} states that any theory with Henkin witness that is syntactically complete and non-contradictory has a model. To prove \eqref{thm:Godel}, we now need show that from a non-trivial theory $\T{T}=(\Sigma, \mathbb{I})$ one can always generate a theory $\T{T'}=(\Sigma', \mathbb{I}')$ which enjoy these three properties and such that $\Sigma\subseteq \Sigma '$ and $\mathbb{I} \subseteq \mathbb{I}'$ (formally stated in Theorem \ref{thm:nontrivialALL}).

We begin by illustrating a procedure that allows us to add Henkin witnesses.
To add a witness for $c\colon 1 \to 0$, one adds a constant $k\colon 0\to 1$ and the axiom $\mathbb{W}_k^c$ below, asserting that $k$ is a witness. %
\[\mathbb{W}_k^c \defeq \{(\; \emptyCirc[+] \, , \scalebox{0.8}{
    \InputIfFileExists{proofs/nonTrivialCompleteness/newAxiom.tikz}{}{\input{tikz/proofs/nonTrivialCompleteness/newAxiom.tikz}}
})\}  \]
Now, we focus on proving the following key result. %
\begin{lem}[Witness Addition]\label{lemma:addingHenkin}%
Let $\T{T}=(\Sigma, \mathbb{I})$ be a theory and consider an arbitrary $c\colon 1\to 0$.
Let $\T{T'} = ( \Sigma \cup \{k\colon 0 \to 1\}, \mathbb{I} \cup \TMAP{k} \cup\mathbb{W}_k^c )$. %
If \,$\T{T}$ is non-trivial then  $\T{T'}$ is non-trivial.
\end{lem}

\begin{rem}%
Before proving Lemma \ref{lemma:addingHenkin}, it is worth to observe that the distinction between trivial and contradictory theories is essential for the lemma. Indeed, under the conditions of Lemma \ref{lemma:addingHenkin}, it does \emph{not} hold that
\begin{center}if $\T{T}$ is non-contradictory, then  $\T{T'}$ is non-contradictory.\end{center}
As  counter-example, take as $\T{T}$ the theory consisting only of the trivialising axiom $(tr) \defeq (\scalebox{0.8}{\codiscardCirc[+]} , \scalebox{0.8}{\codiscardCirc[-]})$. By definition $\T{T}$ is trivial but non-contradictory. Instead $\T{T}'$ is contradictory: %
\begin{equation}\label{eq:remcounter}
    \emptyCirc[+] \stackrel{\eqref{eq:bone}}{\syninclusionT{\T{T'}}} 
    \InputIfFileExists{axioms/cb/plus/codiscDisc.tikz}{}{\input{tikz/axioms/cb/plus/codiscDisc.tikz}}
 \stackrel{(tr)}{\syninclusionT{\T{T'}}} 
    \InputIfFileExists{axioms/linadj/cobangPbang.tikz}{}{\input{tikz/axioms/linadj/cobangPbang.tikz}}
 \stackrel{\eqref{ax:gammaDiscardPlus}}{\syninclusionT{\T{T'}}} \emptyCirc[-].
\end{equation}
This shows that adding Henkin witnesses to a non-contradictory theory may end up in a contradictory theory. Therefore, the usual Henkin proof for $\FOL$ works just for our \emph{non-trivial} theories. %
\end{rem}

In order to prove Lemma \ref{lemma:addingHenkin} and then Theorem \ref{thm:nontrivialALL}, we
need to show that \emph{adding} constants to a non-trivial theory results in a non-trivial theory. To do this, it is useful to have a procedure for \emph{erasing} constants. This is defined as follows.

\begin{defi}\label{def:phi}
    Let $\sign$ be a signature and $\sign' = \sign \;\cup\; \{ k \colon 0 \to 1 \}$. The function $\phi \colon \LCB[\sign'][n,m] \to \LCB[\sign][1+n,m]$ is inductively defined as follows:
    \begin{equation}\label{eq:def-phi}\tag{def-$\phi$}
    \begin{array}{r@{\;}c@{\;}l@{\qquad\quad}r@{\;}c@{\;}l}
        \phi(k^\circ) &\defeq& \idCirc[+] & \phi(k^\bullet) &\defeq& \idCirc[-] \\[10pt]
        \phi(g^\circ) &\defeq& \phiCirc[+]{g} & \phi(g^\bullet) &\defeq& \phiCirc[-]{g^\bullet}  \\[10pt]
        \phi(c \seq[+] d) &\defeq& \phiSeqCirc[+]{c}{d} & \phi(c \seq[-] d) &\defeq& \phiSeqCirc[-]{c}{d} \\[10pt]
        \phi(c \tensor[+] d) &\defeq& \phiTensorCirc[+]{c}{d} & \phi(c \tensor[-] d) &\defeq& \phiTensorCirc[-]{c}{d}
    \end{array}
    \end{equation}
where $g^\circ \in \{\copier[+][1], \discard[+][1], R^\circ, \codiscard[+][1], \cocopier[+][1], \id[+][0], \id[+][1], \symm[+][1][1] \}$ and $g^\bullet \in \{\copier[-][1], \discard[-][1], R^\bullet, \codiscard[-][1], \cocopier[-][1], \id[-][0], \id[-][1], \symm[-][1][1] \}$.
\end{defi}

\begin{lem}\label{lm:phi characterization}
    Let $c \colon n \to m$ be a diagram of $\LCB$, then $\phi(c) = \phiCirc[+]{c}$.
\end{lem}
\begin{proof}
    \input{tikz/proofs/phiCharacterization.tex}
\end{proof}

\noindent 
The proof of the following result goes by induction on \eqref{eq:pc} and relies on Lemma \ref{lm:phi characterization}. The interested reader can find its proof in Appendix \ref{app:completeness}.

\begin{lem}[Constant Erasure]\label{lemma:phi}
    Let $\T{T} = (\sign, \T{I})$ be a theory and $\T{T'} = (\sign', \T{I'})$ be the theory where $\sign' = \sign \cup \{ k \colon 0 \to 1 \}$ and $\T{I'} = \T{I} \cup \TMAP{k}$.
    Then, for any $c,d \colon n \to m$ in $\LCB[\sign ']$ if $c \syninclusionT{\T{T'}} d$ then $\phi(c) \syninclusionT{\T{T}} \phi(d)$.
\end{lem}

\noindent 
Now we are ready to prove Lemma~\ref{lemma:addingHenkin}, namely that witness addition preserves non-triviality.

\begin{proof}[Proof of Lemma~\ref{lemma:addingHenkin}]

We prove that if $\T{T'}$ is trivial, then also $\T{T}$ is trivial. Let $\T{T''} = \{ \Sigma \cup k, \mathbb{I} \cup \TMAP{k}\}$ and assume $\T{T'}$ to be trivial, i.e.\ $\scalebox{0.8}{\codiscardCirc[+]} \!\!\syninclusionT{\T{T'}}\!\! \scalebox{0.8}{\codiscardCirc[-]}$, then:
\begin{enumerate}
    \item by the Deduction Theorem (\ref{th:deduction}) we have $\scalebox{0.8}{
    \InputIfFileExists{proofs/nonTrivialCompleteness/newAxiom.tikz}{}{\input{tikz/proofs/nonTrivialCompleteness/newAxiom.tikz}}
} \precongR{\T{T''}} \scalebox{0.8}{
    \InputIfFileExists{axioms/cb/minus/codiscDisc.tikz}{}{\input{tikz/axioms/cb/minus/codiscDisc.tikz}}
}$;
    \item thus, by Lemma \ref{lemma:phi}, $\phi( \scalebox{0.8}{
    \InputIfFileExists{proofs/nonTrivialCompleteness/newAxiom.tikz}{}{\input{tikz/proofs/nonTrivialCompleteness/newAxiom.tikz}}
} \!) \,\syninclusionT{\T{T}}\, \phi( \scalebox{0.8}{
    \InputIfFileExists{axioms/cb/minus/codiscDisc.tikz}{}{\input{tikz/axioms/cb/minus/codiscDisc.tikz}}
} \!)$;
    \item and, by Definition~\ref{def:phi} and Lemma \ref{lm:phi characterization},
$\scalebox{0.8}{
    \InputIfFileExists{proofs/nonTrivialCompleteness/phiNewAxiom.tikz}{}{\input{tikz/proofs/nonTrivialCompleteness/phiNewAxiom.tikz}}
} \!\syninclusionT{\T{T}}\! \scalebox{0.8}{
    \InputIfFileExists{proofs/nonTrivialCompleteness/phiBone.tikz}{}{\input{tikz/proofs/nonTrivialCompleteness/phiBone.tikz}}
}$.
\end{enumerate}
To conclude, apply Lemma~\ref{lm:implications} and observe that
\input{tikz/proofs/nonTrivialCompleteness/mainProof.tex}
which, by Lemma~\ref{lm:implications} again, is exactly that $\scalebox{0.8}{\codiscardCirc[+]} \precongR{\T{T}} \scalebox{0.8}{\codiscardCirc[-]}$. Namely $\T{T}$ is trivial.
\end{proof}

By iteratively using Lemma \ref{lemma:addingHenkin}, one can transform a non-trivial theory into a non-trivial theory with Henkin witnesses. This was our main technical effort in this part of the proof. Now, the procedure to obtain a syntactically complete theory closely follows the standard  well-known arguments (reported e.g. in \cite{lascar2001mathematical}) and, for this reason, we defer the remaining proofs to  Appendix~\ref{app:completeness}.

\begin{prop}\label{prop:non-contraddiction}\label{prop:non-trivialchain}
    Let $I$ be a linearly ordered set and for all $i \in I$ let $\mathbb{T}_i=(\Sigma_i, \mathbb{I}_i)$ be first-order theories such that if $i\leq j$, then $\Sigma_i \subseteq \Sigma_j$ and $\mathbb{I}_i \subseteq \mathbb{I}_j$. Let $\mathbb{T}$ be the theory \[(\bigcup_{i\in I}\Sigma_i, \bigcup_{i\in I}\mathbb{I}_i)\text{.}\]
    \begin{enumerate}
    \item If all $\mathbb{T}_i$ are non-contradictory, then $\mathbb{T}$ is non-contradictory.
    \item If all $\mathbb{T}_i$ are non-trivial, then $\mathbb{T}$ is non-trivial.
    \end{enumerate}
\end{prop}
\noindent 
By means of the above result and Zorn's Lemma~\cite{zorn1935remark}, one can obtain the desired syntactically complete theory.
\begin{prop}\label{prop:syntacticallycomplete}
    Let $\mathbb{T}=(\Sigma, \mathbb{I})$ be a non-contradictory theory. There exists a theory $\mathbb{T'}=(\Sigma, \mathbb{I}')$ that is syntactically complete, non-contradictory and $\mathbb{I}\subseteq \mathbb{I}'$.
\end{prop}
\noindent 
Now, we obtain the desired result by following the standard Henkin argument which iteratively applies Lemma \ref{lemma:addingHenkin}, wisely combined with Propositions \ref{prop:non-contraddiction} and \ref{prop:syntacticallycomplete}.

\begin{thm}\label{thm:nontrivialALL}
Let $\T{T}=(\Sigma, \mathbb{I})$ be a non-trivial theory. There exists a theory $\T{T'}=(\Sigma', \mathbb{I'})$ such that $\Sigma \subseteq \Sigma'$ and $\mathbb{I}\subseteq \mathbb{I'}$; $\T{T'}$ has Henkin witnesses; $\T{T'}$ is syntactically complete; $\T{T'}$ is non-contradictory.
\end{thm}

Theorems \ref{thm:nontrivialALL} and \ref{thm:Hekinismodel} give us a proof for \eqref{thm:Godel}. %
\begin{proof}[Proof of \eqref{thm:Godel}]
Let $\T{T'}=(\Sigma', \mathbb{I'})$ be obtained via Theorem \ref{thm:nontrivialALL}.
Since $\T{T'}$ has Henkin witnesses, is syntactically complete and non-contradictory, Theorem \ref{thm:Hekinismodel} ensures that $\Hen$ is a model for $\T{T'}$. Since $\Sigma \subseteq \Sigma'$ and $\mathbb{I}\subseteq \mathbb{I'}$, then $\Hen$ is also a model for $\T{T}$. \qedhere
\end{proof}

\subsection{Propositional completeness}\label{sec:Beyond Godel}
Now, we would like to conclude Theorem \ref{thm:completeness} by means of \eqref{thm:Godel}, but this is not possible since, for the former one needs a model for all non-contradictory theories, while \eqref{thm:Godel} provides it only for non-trivial ones. %
Thankfully, the Henkin interpretation $\Hen$ (Definition \ref{def:henkin struct}) gives us, once more, a model (Proposition~\ref{prop:booleanisamodel}) that allows us to prove
\begin{equation}%
\text{\!\!if $\T{T}$ is a trivial and non-contradictory theory,  then $\T{T}$ has a model.} \tag{Prop}
\end{equation}

We commence by illustrating $\Hen$ for a trivial and non-contradictory theory.
\begin{lem}\label{lemma:Henkinboolean}
    Let $\T{T}=(\Sigma,\mathbb{I})$ be a theory that is trivial and non-contradictory and let $\booleaninterpretation$ be the Henkin interpretation of $\Sigma$.
    Then, the domain $X$ of $\booleaninterpretation$ is $\varnothing$ and
        \[\rho(R)= \begin{cases}  \{(\star,\star)\} & \text{ if $\id[+][0] \precongR{\T{T}} R^\circ$ } \\ \varnothing & \text{otherwise.}\end{cases}\]
    \end{lem}
    \begin{proof}%
    Recall by Definition \ref{def:henkin struct}, that the domain $X$ of $\booleaninterpretation$ is defined as the set $\texttt{Map}(\LCB[\T{T}])[0,1]$. This set should be necessarily empty since, if there exists some map $k\colon 0 \to 1$, then by  \eqref{eq:remcounter} $\T{T}$ would be contradictory, against the hypothesis. Thus $\texttt{Map}(\LCB[\T{T}])[0,1]= \varnothing$.

    By Lemma\ref{lemma:cb maps}.(2), one has also that $\texttt{Map}(\LCB[\T{T}])[0,n+1]= \varnothing$. We thus may only have maps in $\texttt{Map}(\LCB[\T{T}])[0,0]$.
    By Lemma \ref{lemma:cb maps}, there is only one map in $\texttt{Map}(\LCB[\T{T}])[0,0]$, which is exactly $\id[+][0] \colon 0 \to 0$.
    Recall that by Definition \ref{def:henkin struct},  \[\rho(R) = \{ (\nterm{k},\nterm{l}) \in X^n \times X^m \mid  \id[+][0] \syninclusionT{\T{T}} \nterm{k} \seq[+] R^\circ \seq[+] \op{(\nterm{k})}\} \]  for all $R\in \sign$.  Since our only map is  $\id[+][0] \colon 0 \to 0$, we have that
    \[\rho(R) = \{ (\star,\star) \in \singleton \times \singleton \mid \id[+][0] \syninclusionT{\T{T}} R^\circ \}\text{.} \qedhere \]
    \end{proof}
    \begin{lem}\label{lemma:interpretation}
    Let $\T{T}=(\Sigma,\mathbb{I})$ be a theory that is trivial and non-contradictory and let $c\colon n \to m+1$ and $d\colon n+1 \to m$ be arrows of $\LCB[\T{T}]$. Thus $\booleaninterpretationFunctor (c) = \varnothing$ and $\booleaninterpretationFunctor (d) = \varnothing$.
    \end{lem}
    \begin{proof}
    Recall that for any interpretation $\interpretation$ with domain $X$, $\interpretationFunctor(c)   \subseteq X^{n} \times X^{m+1} = X^n \times X^m \times X$.
    By Lemma \ref{lemma:Henkinboolean}, the domain of $\booleaninterpretation$ is  $\varnothing$ and thus $\booleaninterpretationFunctor (c) \subseteq   \varnothing^n \times \varnothing^m\times \varnothing$, i.e., $\booleaninterpretationFunctor (c) =  \varnothing $. The proof for $\booleaninterpretationFunctor (d)$ is identical.
    \end{proof}

    \begin{lem}\label{lemma:biglemmaboolean}
  Let $\T{T}$ be a trivial, syntactically complete and non-contradictory theory.  Let $c \colon 0 \to 0$ be an arrow of $\LCB[\T{T}]$. If $\booleaninterpretationFunctor (c) = \{(\star, \star)\}$ then $c =_{\T{T}} \id[+][0]$.
    \end{lem}
    \begin{proof}
    By induction on $c$. For the base cases, there are only four constants $c\colon 0 \to 0$.
    \begin{itemize}
    \item $c=\id[+][0]$. Then, trivially, $c =_{\T{T}} \id[+][0]$.
    \item $c=\id[-][0]$. Then, by \eqref{fig:semantics}, $\booleaninterpretationFunctor (c) = \varnothing$ against the hypothesis.
    \item  $c=R^\circ$. Then, by \eqref{fig:semantics}, $\rho(R)=\booleaninterpretationFunctor (R^\circ) =  \{(\star, \star)\}$. By Lemma \ref{lemma:Henkinboolean}, $\id[+][0] =_{\T{T}} R^\circ$.
    \item  $c=R^\bullet$. Then, by \eqref{fig:semantics}, $\nega{\rho(R)} = \booleaninterpretationFunctor (R^\bullet) = \{(\star, \star)\}$. Thus, by Lemma \ref{lemma:Henkinboolean}, $\id[+][0] \not \hspace{-0.2cm} \precongR{\T{T}} R^\circ$. Since $\T{T}$ is syntactically complete $\id[+][0] \precongR{\T{T}} R^\bullet$.
    \end{itemize}
    \noindent 
    We now consider the usual four inductive cases.
    \begin{itemize}
    \item $c=c_1 \tensor[+] c_2$. Since $c\colon 0 \to 0$, then also $c_1$ and $c_2$ have type $0 \to 0$. By definition, $\booleaninterpretationFunctor(c) = \booleaninterpretationFunctor(c_1) \tensor[+] \booleaninterpretationFunctor(c_2)$. By definition of $\tensor[+]$ in $\Rel$ both $\booleaninterpretationFunctor(c_1)$ and $\booleaninterpretationFunctor(c_2)$ must be $\{(\star, \star)\}$. We can thus apply the inductive hypothesis to deduce that $c_1=_{\T{T}}\id[+][0]$ and  $c_2=_{\T{T}}\id[+][0]$. Therefore $c=c_1 \tensor[+] c_2=_{\T{T}} \id[+][0] \tensor[+] \id[+][0] =_{\T{T}} \id[+][0]$.

    \item $c=c_1 \seq[+]c_2$. There are two possible cases: either $c_1\colon 0 \to n+1$ and $c_2\colon n+1 \to 0$, or  $c_1\colon 0 \to 0$ and $c_2\colon 0 \to 0$. In the former case, we have by Lemma \ref{lemma:interpretation}, that $\booleaninterpretationFunctor (c)= \booleaninterpretationFunctor (c_1) \seq[+] \booleaninterpretationFunctor (c_2)=\varnothing \seq[+] \varnothing = \varnothing$. Against the hypothesis. Thus the second case should hold: $c_1\colon 0 \to 0$ and $c_2\colon 0 \to 0$. In this case we just observe that $c_1 \seq[+] c_2$ is, by the laws of symmetric monoidal categories, equal to $c_1 \tensor[+] c_2$. We can thus reuse the point above.

    \item $c=c_1 \tensor[-] c_2$. Since $c\colon 0 \to 0$, then also $c_1$ and $c_2$ have type $0 \to 0$.  Consider the case where $\booleaninterpretationFunctor (c_1)= \varnothing  = \booleaninterpretationFunctor (c_2)$. Thus  $\booleaninterpretationFunctor (c)= \varnothing$, against the hypothesis. Therefore either $\booleaninterpretationFunctor (c_1)= \{ (\star,\star)\}$ or  $\booleaninterpretationFunctor (c_2)= \{ (\star,\star)\}$.  If $\booleaninterpretationFunctor (c_1)= \{ (\star,\star)\}$, then by induction hypothesis $c_1=_{\T{T}} \id[+][0]$. Therefore $c=c_1 \tensor[-]c_2 = c_1 \sqcup c_2 =_{\T{T}} \id[+][0] \sqcup c_2 =_{\T{T}} \top \sqcup c_2 =_{\T{T}} \top =_{\T{T}} \id[+][0]$. The case for $\booleaninterpretationFunctor (c_2)= \{ (\star,\star)\}$ is symmetric.

     \item $c=c_1 \seq[-] c_2$. The proof is analogous to the case $c=c_1 \seq[+]c_2$. \qedhere
    \end{itemize}
    \end{proof}

\noindent 
From the above result, one easily obtains its dual.

    \begin{lem}\label{lemma:biglemmaboolean2}
  Let $\T{T}$ be a trivial, syntactically complete and non-contradictory theory.  Let $c \colon 0 \to 0$ be an arrow of $\LCB[\T{T}]$. If $\booleaninterpretationFunctor (c) = \varnothing$ then $c =_{\T{T}} \id[-][0]$.
    \end{lem}
    \begin{proof}
    If $\booleaninterpretationFunctor (c) = \varnothing$, then by Lemma \ref{lm:hen dagger adj}, $\booleaninterpretationFunctor (\nega{c}) = \nega{\varnothing} =\{(\star, \star)\}$. Thus by Lemma \ref{lemma:biglemmaboolean}, $\nega{c}=_{\T{T}} \id[+][0]$ and thus $c=_{\T{T}} \id[-][0]$.
    \end{proof}

The following result is the key to prove \eqref{thm:trivialcom}.

    \begin{prop}\label{prop:booleanisamodel}
    Let $\T{T}$ be a trivial, syntactically complete and non-contradictory theory. Then $\booleaninterpretation$ is a model of $\T{T}$. %
    \end{prop}
    \begin{proof}
    Recall that Proposition \ref{prop:soundnessoftheories}, $\booleaninterpretation$ is a model iff for all $c,d \colon n \to m$ in $\LCB$, if $c\precongR{\T{T}} d$, then $\booleaninterpretationFunctor (c) \subseteq \booleaninterpretationFunctor (d)$.
    We prove that if $\booleaninterpretationFunctor (c) \not \subseteq \booleaninterpretationFunctor (d)$, then $c\not \hspace{-0.2cm} \precongR{\T{T}} d$.

    If $c\colon n \to m+1$ or $c \colon n+1 \to m$, then by Lemma \ref{lemma:interpretation}, $\booleaninterpretationFunctor (c)= \varnothing$ and thus it is not the case that $\booleaninterpretationFunctor (c) \not \subseteq \booleaninterpretationFunctor (d)$. Thus we need to consider only the case where $c,d\colon 0 \to 0$.

    For $c,d \colon 0 \to 0$ if $\booleaninterpretationFunctor (c) \not \subseteq \booleaninterpretationFunctor (d)$, then $\booleaninterpretationFunctor (c)=\{(\star,\star)\}$ and $\booleaninterpretationFunctor (d) = \varnothing$. By Lemmas~\ref{lemma:biglemmaboolean} and \ref{lemma:biglemmaboolean2}, we thus have $c=_{\T{T}} \id[+][0]$ and $d=_{\T{T}} \id[-][0]$. Since $\T{T}$ is non-contradictory, then $c\not \hspace{-0.2cm} \precongR{\T{T}} d$.
    \end{proof}

Analogously to the proof \eqref{thm:Godel}, we can  exploit Proposition \ref{prop:syntacticallycomplete}, but now combined with the proposition above to prove \eqref{thm:trivialcom}.
\begin{proof}[Proof of \eqref{thm:trivialcom}]
    Since $\T{T}=(\Sigma, \mathbb{I})$ is non-contradictory, by Proposition \ref{prop:syntacticallycomplete} there exists a syntactically complete non-contradictory theory $\T{T'}=(\Sigma, \mathbb{I'})$ such that $\mathbb{I}\subseteq \mathbb{I'}$. Since $\codiscard[+][1] \precongR{\T{T}} \codiscard[-][1]$, then $\codiscard[+][1] \precongR{\T{T'}} \codiscard[-][1]$, namely $\T{T'}$ is also trivial. We can thus use Proposition \ref{prop:booleanisamodel}, to deduce that $\booleaninterpretation$ is a model for $\T{T'}$. Since  $\mathbb{I}\subseteq \mathbb{I'}$, then $\booleaninterpretation$ is also a model of  $\T{T}$.
\end{proof}

\subsection{General Completeness}\label{sec:generalcompleteness}
From \eqref{thm:trivialcom} and \eqref{thm:Godel} we can prove general completeness
\begin{equation}%
\text{if $\T{T}$ is  non-contradictory theory, then  $\T{T}$ has a model.}\tag{General}
\end{equation}
\begin{proof}[Proof of \eqref{cor:gencompleteness}] To prove \eqref{cor:gencompleteness} take $\T{T}$ to be a non-contradictory theory. If $\T{T}$ is trivial, then it has a model by  \eqref{thm:trivialcom}. Otherwise, it has a model by \eqref{thm:Godel}. %
\end{proof}

Before proving Theorem~\ref{thm:completeness}, we illustrate the following result which simply rephrases standard arguments of completeness for first-order logic.

\begin{lem}\label{lemma:beforened}
Let $\T{T}=(\sign, \T{I})$ be a theory and $c\colon 0 \to 0$ be a diagram in $\LCB[\T{T}]$.
If, for all models $\interpretation$ of $\T{T}$, $\{(\star,\star)\} \subseteq \interpretationFunctor(c)$, then $\id[+][0] \syninclusionT{\T{T}} c$.
\end{lem}
\begin{proof}%
Suppose that $\id[+][0] \not  \syninclusionT{\T{T}} c$. Then, by Corollary \ref{cor:deduction},  $\T{T}'=(\sign, \T{I}\cup \{(\id[+][0], \nega{c} ) \} )$ is non-contradictory. Thus, by  \eqref{cor:gencompleteness}, $\T{T}'$ has a model $\mathcal{I}$. Since $\T{I}\subseteq \T{I}\cup \{(\id[+][0], \nega{c} )\}$, $\mathcal{I}$ is also a model of $\T{T}$.
Now note that
\begin{align*}
\{(\star,\star)\} = &  \interpretationFunctor(\id[+][0]) \tag{\ref{fig:semantics}}\\
\subseteq & \interpretationFunctor(\nega{c}) \tag{$\interpretation$ is a model of $\T{T}'$}\\
= &  \nega{\interpretationFunctor(c)}\text{.} \tag{Lemma \ref{lm:hen dagger adj}}
\end{align*}
Thus $\interpretationFunctor(c)\subseteq \nega{\{(\star,\star)\}} = \varnothing$, against the hypothesis.
\end{proof}

Using Lemma \ref{lm:implications}, we can extends the above result to arbitrary morphisms $c,d\colon n \to m$.

\begin{prop}\label{prop:moregeneral completeness}
Let $\T{T}=(\sign, \T{I})$ be a theory and $c,d\colon n \to m$ be diagrams in $\LCB[\T{T}]$.
If, for all models $\interpretation$ of $\T{T}$, $\interpretationFunctor(c) \subseteq \interpretationFunctor(d)$, then $c \syninclusionT{\T{T}} d$.
\end{prop}
\begin{proof}
Since $\Rel$ is a fo-bicategory, one can safely exploit  Lemma \ref{lm:implications} to show that
\begin{align*}
 \{(\star,\star)\}  \subseteq & \circleCirc{\interpretationFunctor c }{\interpretationFunctor d }[b] \tag{Lemma \ref{lm:implications}}\\
= & \interpretationFunctor (\circleCirc{c}{d}[b]) \text{.} \tag{Lemma \ref{lm:hen dagger adj}}
\end{align*}
Thus by  Lemma \ref{lemma:beforened}, $\emptyCirc[+] \syninclusionT{\T{T}} \circleCirc{c}{d}[b]$. Again, by Lemma \ref{lm:implications}, $c \syninclusionT{\T{T}} d$.
\end{proof}

\begin{proof}[Proof of Theorem~\ref{thm:completeness}]
From Proposition \ref{prop:moregeneral completeness} when taking $\T{I}=\varnothing$.
\end{proof}

\newcommand{\varcontext}{\vdash}
\newcommand{\varlist}{\mathbf{x}_n}
\newcommand{\varlistred}{\mathbf{x}_{n-1}}
\newcommand{\divider}{\ \rightsquigarrow \ }
\newcommand{\from}{\mathrel{:}}

\section{First-Order Logic with Equality}\label{sec:fol}

As we already mentioned in the introduction the white fragment of $\NPR$ is as expressive as the existential-conjunctive fragment of first-order logic with equality $(\FOL)$. The semantic preserving encodings between the two fragments are illustrated in \cite{GCQ}. From the fact that the full $\NPR$ can express negation, we immediately get semantic preserving encodings between $\NPR$ and the full $\FOL$. In this section we illustrate a translation $\enc{\cdot} \colon \FOL \to \NPR$ to emphasise the subtle differences between the two. %

To ease the presentation, we consider $\FOL$ formulas $\varphi$ to be typed in the context of a list of variables that are allowed (but not required) to appear in $\varphi$. Fixing $\varlist \defeq \{x_1,\dots,x_{n}\}$ we write $\varlist \varcontext \varphi$ if all free variables of $\varphi$ are contained in $\varlist$.
It is standard to present the syntax of $\FOL$ in two steps: first terms and then formulas. For every function symbol $f$ of arity $m$ in $\FOL$, we have a symbol $f\colon m \to 1$ in the signature $\sign$ together with the equations $\TMAP{f}$ forcing $f$ to be interpreted as a function. The translation of $\varlist \varcontext t$ to a $\LCB$ diagram $n\to 1$ is inductively given as follows.
\[
  \enc{\varlist \varcontext x_i} \defeq 
    \InputIfFileExists{encodingFOL2/xi.tikz}{}{\input{tikz/encodingFOL2/xi.tikz}}
 \qquad\qquad \enc{\varlist \varcontext f(t_1,\dots,t_m)} \defeq \scalebox{0.8}{
    \InputIfFileExists{encodingFOL2/t.tikz}{}{\input{tikz/encodingFOL2/t.tikz}}
}
\]

Formulas $\varlist \varcontext \varphi$ translate to $\LCB$ diagrams $n\to 0$. For every $n$-ary predicate symbol $R$ in $\FOL$ there is a symbol $R\colon n \to 0\in\sign$. In order not to over-complicate the presentation with bureaucratic details, we assume that it is always the last variable that is quantified over. Additional variable manipulations can be introduced easily (see, for example,~\cite{GCQ}).

\[
\begin{array}{c}
  \begin{array}{rclrcl}
    \enc{\varlist \varcontext \top} &\defeq& \!\!\!\discardCirc[+][n]  \quad &\quad \enc{(\varlist \varcontext \bot)} &\defeq& \!\!\!\discardCirc[-][n]  \\[5pt]
    \enc{\varlist \varcontext t_1 = t_2} &\defeq& \!\!\!\scalebox{0.8}{
    \InputIfFileExists{encodingFOL2/eq.tikz}{}{\input{tikz/encodingFOL2/eq.tikz}}
} \quad &\quad \enc{\varlist \varcontext R(t_1,\dots,t_m)}& \defeq & \!\!\!\scalebox{0.8}{
    \InputIfFileExists{encodingFOL2/R.tikz}{}{\input{tikz/encodingFOL2/R.tikz}}
}  \\[10pt]
    \enc{\varlist \varcontext \varphi_1 \wedge \varphi_2} &\defeq& \!\!
    \InputIfFileExists{encodingFOL2/and.tikz}{}{\input{tikz/encodingFOL2/and.tikz}}
  \quad &\quad \enc{\varlist \varcontext \varphi_1 \vee \varphi_2} &\defeq& \!\!
    \InputIfFileExists{encodingFOL2/or.tikz}{}{\input{tikz/encodingFOL2/or.tikz}}
  \\[3pt]
    \enc{\varlistred \varcontext \exists x_n.\, \varphi} &\defeq& \!\!\!
    \InputIfFileExists{encodingFOL2/exists.tikz}{}{\input{tikz/encodingFOL2/exists.tikz}}
  \quad &\quad \enc{\varlistred \varcontext \forall x_n.\, \varphi} &\defeq& \!\!
    \InputIfFileExists{encodingFOL2/forall.tikz}{}{\input{tikz/encodingFOL2/forall.tikz}}

  \end{array}
  \\
  \enc{\varlist \varcontext \neg \varphi} \defeq 
    \InputIfFileExists{encodingFOL2/neg.tikz}{}{\input{tikz/encodingFOL2/neg.tikz}}

  \end{array}
\]

The above encoding should give the reader the spirit of the correspondence between $\NPR$ and traditional syntax. There is one aspect of the above translation that merits additional attention.
\begin{rem}\label{rmk:ambiguity}
By the definition of $\discard[][n]$ in Table \ref{fig:sugar}, we have that:
\[
  \folEnc{0}{\top} \defeq \emptyCirc[+]  \qquad \qquad \folEnc{0}{\bot} \defeq \emptyCirc[-]
\]
Thus $\top$ and $\bot$ translate to, respectively $\id[+][0]$, $\id[-][0]$ in the absence of free variables or to $\discard[+][n]$, $\discard[-][n]$, respectively, when $n>0$. This can be seen as an ambiguity in the traditional $\FOL$ syntax, which obscures the distinction between inconsistent and trivial theories in traditional accounts, and as a side effect requires the assumption on non-empty models in formal statements of G\"odel completeness. Instead, the syntax of $\NPR$ ensures that this pitfall is side-stepped.
\end{rem}

To conclude our analysis of the relationship between $\NPR$ and $\FOL$ we show how to translate from diagrams of $\NPR$ to formulas of $\FOL$.

Note that in general terms of $\NPR$ feature ``dangling'' wires both on the left and on the right of a term. While this is inconsequential from the point of view of expressivity, since terms can always be ``rewired'' using the self-dual compact closed structure of cartesian bicategories, this separation is convenient for composing terms in a flexible manner. Therefore, in the translation in Figure~\ref{fig:fol reverse encoding}, we keep two separate lists of free variables in the context, denoted as $n;m$, where $n$ and $m$ are the lengths of the two lists.
\begin{figure*}[htb!]
  \centering
  \resizebox{\textwidth}{!}{
  $
  \begin{array}{ccc}
    \begin{prooftree}
    \justifies
    \emptyCirc[+] \divider \scriptstyle \folFormula{0; 0}{\top}%
    \using (\id[+])
    \end{prooftree}
    &
    \begin{prooftree}
      \justifies
      \idCirc[+] \divider \scriptstyle \folFormula{1; 1}{x_1 = y_1}
      \using (\id[+][1])
      \end{prooftree}
    &
    \begin{prooftree}
    \justifies
    \symmCirc[+] \divider \scriptstyle \folFormula{2; 2}{x_1=y_2 \wedge x_2=y_1} %
    \using (\symm[+])
    \end{prooftree}
    \\[20pt]
      \begin{prooftree}
      \justifies
      \emptyCirc[-] \divider \scriptstyle \folFormula{0; 0}{\bot}%
      \using (\id[-])
      \end{prooftree}
      &
      \begin{prooftree}
      \justifies
      \idCirc[-] \divider \scriptstyle \folFormula{1; 1}{x_1 \neq y_1} %
      \using (\id[-][1])
      \end{prooftree}
        &
        \begin{prooftree}
        \justifies
        \symmCirc[-] \divider \scriptstyle \folFormula{2; 2}{x_1 \neq y_2 \vee x_2 \neq y_1} %
        \using (\symm[-])
        \end{prooftree}
        \\[20pt]
      \begin{prooftree}
        \justifies
        \copierCirc[+] \divider \scriptstyle \folFormula{1; 2}{x_1=y_1 \wedge x_1=y_2} %
        \using (\copier[+])
        \end{prooftree}
        &
        \begin{prooftree}
        \justifies
        \discardCirc[+] \divider \scriptstyle \folFormula{1; 0}{\top} %
        \using (\discard[+])
        \end{prooftree}
        &
        \begin{prooftree}
        \justifies
        \boxCirc[+]{R}[n][m] \divider \scriptstyle \folFormula{n; m}{R(x_n, y_m)} %
        \using (R^{\switchLabelS{+}})
        \end{prooftree}
        \\[20pt]
      \begin{prooftree}
        \justifies
        \copierCirc[-] \divider \scriptstyle \folFormula{1; 2}{x_1\neq y_1 \vee x_1\neq y_2} %
        \using (\copier[-])
        \end{prooftree}
        &
        \begin{prooftree}
        \justifies
        \discardCirc[-] \divider \scriptstyle \folFormula{1; 0}{\bot} %
        \using (\discard[-])
        \end{prooftree}
        &
        \begin{prooftree}
        \justifies
        \boxOpCirc[-]{R}[n][m] \divider \scriptstyle \folFormula{m; n}{\neg R(x_n, y_m)} %
        \using (R^{\switchLabelS{-}})
        \end{prooftree}
        \\[20pt]
        \multicolumn{3}{c}{
          \begin{array}{cc}
            \inferrule*[right=({\!\!\! \seqOld[+] \!\!\!})]{\boxCirc[+]{p}[n][k] \!\!\!\divider \scriptstyle \folFormula{n;k}{\varphi}
    \;
    \boxCirc[+]{q}[k][m] \!\!\!\divider \scriptstyle \folFormula{k;m}{\psi}}{\seqCirc[+]{p}{q}[n][m] \divider \scriptstyle \folFormula{n;m}{\exists \mathbf{z}_k.\, \varphi \wedge \psi}}
    &
    \inferrule*[right=({\!\!\! \seqOld[-] \!\!\!})]{\boxCirc[+]{p}[n][k] \!\!\!\divider \scriptstyle \folFormula{n;k}{\varphi}
    \;
    \boxCirc[+]{q}[k][m] \!\!\!\divider \scriptstyle \folFormula{k;m}{\psi}}{
        \seqCirc[-]{p}{q}[n][m] \divider \scriptstyle \folFormula{n;m}{\forall \mathbf{z}_k.\, \varphi \vee \psi}
    }
    \\[20pt]
    \begin{prooftree}
      \boxCirc[+]{p}[n][m] \!\!\!\divider \scriptstyle \folFormula{n;m}{\varphi}
      \;
      \boxCirc[+]{q}[l][k] \!\!\!\divider \scriptstyle \folFormula{l;k}{\psi} \vspace*{0.4em}
      \justifies
      \tensorCirc[+]{p}{q}[n][m][l][k] \divider \scriptstyle \folFormula{n+l;m+k}{\varphi \wedge \psi}
      \using (\tensor[+])
      \end{prooftree}
      &
      \begin{prooftree}
      \boxCirc[+]{p}[n][m] \!\!\!\divider \scriptstyle \folFormula{n;m}{\varphi}
      \;
      \boxCirc[+]{q}[l][k] \!\!\!\divider \scriptstyle \folFormula{l;k}{\psi} \vspace*{0.4em}
      \justifies
      \tensorCirc[-]{p}{q}[n][m][l][k] \divider \scriptstyle \folFormula{n+l;m+k}{\varphi \vee \psi}
      \using (\tensor[-])
      \end{prooftree}   
          \end{array}
        }
  \end{array}
$
}
\caption{Encoding of $\NPR$ diagrams as $\FOL$ formulas.}
\label{fig:fol reverse encoding}
\end{figure*}

\section{Back to Peirce, Tarski and Quine}\label{sec:encodings}

In this section we illustrate how to encode  the calculus of binary relations (Section \ref{ssec:CRrevisited}),  Quine's predicate functor logic (Section \ref{sec:quine}) and Peirce's existential graphs (Section \ref{sec:egraphs}) within the calculus of neo-Peircean relations. 

\subsection{A positive answer to Tarski's question}\label{ssec:CRrevisited}
In Section \ref{sec:calculusrelations}, we have recalled Tarski's question about axiomatizability of $\minorExpression$.
Our unconventional answer is the following:
\begin{center}
\emph{
By leaving the traditional (cartesian) syntax of the calculus of binary relations, \\ for the diagrammatic (monoidal) syntax of the calculus of neo-Peircean relations, \\ one has a complete axiomatisation.}
\end{center}

\noindent 
To make this formal, in Table~\ref{table:cr encoding} we inductively define an encoding $\enc{\cdot} \colon \CRS \to \NPR $ assigning to each expressions $E$ of $\CRS$ a term $\enc{E}\colon 1 \to 1$ of $\NPR$. 
\begin{table*}[!htb]
  \centering
  \footnotesize{
  $
      \begin{array}{llll}
          \toprule
            \enc{\id[+]} \defeq \id[+][1]
          & \enc{E_1 \seq[+] E_2} \defeq \enc{E_1} \seq[+] \enc{E_2} 
          & \enc{\top} \defeq \discard[+][1] \seq[+] \codiscard[+][1]
          & \enc{E_1 \cap E_2} \defeq \copier[+][1] \seq[+] (\enc{E_1} \tensor[+] \enc{E_2}) \seq[+] \cocopier[+][1] 
          \\
            \enc{\id[-]} \defeq \id[-][1]
          & \enc{E_1 \seq[-] E_2} \defeq \enc{E_1} \seq[-] \enc{E_2}
          & \enc{\bot} \defeq \discard[-][1] \seq[-] \codiscard[-][1]
          & \enc{E_1 \cup E_2} \defeq \copier[-][1] \seq[-] (\enc{E_1} \tensor[-] \enc{E_2}) \seq[-] \cocopier[-][1] 
          \\
          \multicolumn{4}{c}{
            \begin{array}{l@{\qquad}l@{\qquad}l}
                \enc{R} \defeq R^\circ
                &
                \enc{\nega{E}} \defeq \nega{\enc{E}} 
                &
                \enc{\op{E}} \defeq \op{\enc{E}}
            \end{array}
          }
          \\
          \bottomrule
      \end{array}
  $
  }
  \caption{The encoding $\enc{\cdot} \colon \CRS \to \NPR $}\label{table:cr encoding}
\end{table*}

Note that the occurrences of $\op{(\cdot)}$ and $\nega{(\cdot)}$ on the right-hand side of the equations are those defined in \eqref{eqdagger} and \eqref{def:nega}.
As expected, $\enc{\cdot}$ preserves the semantics.
\begin{prop}\label{prop:cr interpretation}
  For all expressions $E$ of $\CRS$ and interpretations $\interpretation$, $\dsemRel{E} = \interpretationFunctor{(\enc{E})}$.
\end{prop}
We report the straightforward inductive proof of the above result in Appendix \ref{app:encoding}.

\begin{cor}\label{cor:cr completeness}
    For all $E_1, E_2$, $E_1 \minorExpression E_2$ iff $\enc{E_1} \syninclusion \enc{E_2}$.
\end{cor}
\begin{proof}%
  \begin{align*}
      E_1 \minorExpression E_2 &\iff \forall \interpretation . \; \dsemRel{E_1} \subseteq \dsemRel{E_2} \tag{Definition of $\minorExpression$} \\
                               &\iff \forall \interpretation . \; \interpretationFunctor{(\enc{E_1})} \subseteq \interpretationFunctor{(\enc{E_2})} \tag{Proposition~\ref{prop:cr interpretation}} \\
                               &\iff \enc{E_1} \seminclusion \enc{E_2} \tag{Definition of $\seminclusion$} \\
                               &\iff \enc{E_1} \syninclusion \enc{E_2} \tag*{(Theorem \ref{thm:completeness})\qedhere}
  \end{align*}
\end{proof}
In other words, one can check inclusions of expressions of $\CRS$ by translating them into $\NPR$ via $\enc{\cdot}$ and then using the axioms in Figures \ref{fig:cb axioms}, \ref{fig:cocb axioms}, \ref{fig:closed lin axioms} and \ref{fig:fo bicat axioms}.

\subsection{Quine's Predicate Functor Logic}\label{sec:quine}
Inspired by combinatory logic, Quine~\cite{QUINE1971309} introduced \emph{predicate functor logic}, $\PFL$ for short, as a quantifier-free treatment of first-order logic with equality. Several flavours of the logic have been proposed by Quine and others, here we focus on the treatment by Kuhn~\cite{kuhn1983}. Using the terminology of that thread of research, for
each $n\geq 0$ there is a collection of atomic $n$-ary predicates, corresponding to traditional $\FOL$ predicate symbols together with an additional binary predicate $I$ (identity). The term (predicate) constructors are called \emph{functors} -- here the terminology is unrelated to the notion of functor in category theory. These are divided into unary operations $\mathbf{p},\mathbf{P},[,]$  called \emph{combinatory functors} that, in the absence of explicit variables, capture the combinatorial aspects of handling variable lists as well as (existential) quantification. To get full expressivity of FOL, there are two additional \emph{alethic functors}: a binary conjunction and unary negation.

\begin{table}[!htb]
  \begin{center}
  \resizebox{\textwidth}{!}{%
  \tiny
    \begin{tabular}{c}
      \toprule
   $
   \begin{array}{rcl}
  P & ::=&\; R  \; \mid \; I \; \mid  \;  \mathbf{p}P \mid \; \mathbf{P}P \; \mid  \;  [P \; \mid \; ]P \; \mid \;  P \cap P  \; \mid \;  \neg P, \;\;\; \text{ where }\; R \in \sign
  \end{array}
  $ \\
   \midrule
  $
  \inferrule{-}{I\colon 2} 
  \qquad
  \inferrule{\ari(R)=n}{R \colon n} 
  \qquad
  \inferrule{P\colon n \; n\geq 2}{  \mathbf{p}P \colon n } 
  \qquad
  \inferrule{P\colon 1}{  \mathbf{p}P \colon 2 } 
  \qquad
  \inferrule{P\colon 0}{  \mathbf{p}P \colon 2 } 
  \qquad
  \inferrule{P \colon n}{ \mathbf{P}P \colon n}
  $\\
  $\inferrule{P_1 \colon n \;\; P_2\colon m \;\; n\geq m}{P_1 \cap P_2 \colon n} 
  \quad
  \inferrule{P_1 \colon n \;\; P_2\colon m\;\; n<m}{P_1 \cap P_2 \colon m} 
  \quad
  \inferrule{P \colon n}{ \neg P \colon n}  
  \quad
  \inferrule{P \colon n}{ [ P \colon n+1}  
  \quad 
  \inferrule{P \colon n+1}{ ] P \colon n} 
  \quad 
  \inferrule{P \colon 0}{ ] P \colon 0}$
  \\
  \midrule
  $\begin{array}{@{}l@{}l}
    \begin{array}{@{}l@{}l}
      \dsemRel{R} \defeq \{\tau \mid (\tau_1, \dots, \tau_{n}) \in \rho(R)\} &\;\;\quad \dsemRel{I} \defeq \{\tau \mid \tau_1=\tau_2\} \\
      \dsemRel{ P_1 \cap P_2} \defeq \dsemRel{ P_1} \cap \dsemRel{ P_2} &\;\;\quad \dsemRel{ \neg P} \defeq \{\tau \mid \tau \notin \dsemRel{P}\}
    \end{array}
    & \;\;
    \begin{array}{@{}l}
      \dsemRel{]P} \defeq \{\tau \mid \tau_2 \cdot \tau_3 \cdots \in \dsemRel{P})\}\\
      \dsemRel{[P} \defeq \{x_0 \cdot \tau_1 \cdot \tau_2 \cdots \mid x_0\in X, \tau_1 \cdot \tau_2 \dots \in \dsemRel{P}\}
    \end{array} \\
    \dsemRel{\mathbf{P}P} \defeq \{ \tau \mid \tau_n \cdot \tau_2\cdots \tau_{n-1} \cdot \tau_1 \cdot \tau_{n+1} \cdots \in \dsemRel{P} \} & \;\;\dsemRel{ \mathbf{p}P} \defeq \{\tau  \mid \tau_2 \cdot \tau_1\cdots \in \dsemRel{P}\}
  \end{array}
  $
  \\
  \bottomrule
  \end{tabular}%
  }
  \end{center}
  \caption{$\PFL$: (top) syntax; (mid) typing rules; (bottom) semantics w.r.t. an interpretation $\interpretation =(X,\rho)$.}\label{fig:typingrulesQuine}
  \end{table}

The syntax is reported on the top of Table \ref{fig:typingrulesQuine} where $R$ belongs to $\sign$, a set of symbols with an associated arity. Similarly to $\NPR$, only the predicates that can be typed according to the rules in Table \ref{fig:typingrulesQuine} are considered. The semantics, on the bottom, is defined w.r.t. an interpretation $\interpretation$ consisting of a \emph{non-empty} set $X$ and a set $\rho(R)\subseteq X^n$ for all $R\in \sign$ of arity $n$. For all predicates $P$, $\dsemRel{P}$ is a subset of $X^\omega \defeq \{\tau_1 \cdot \tau_2 \cdots \mid \tau_i \in X \text{ for all }i\in\nat^+\}$. 

\begin{table*}[!htb]
  $\tiny{
  \begin{array}{c@{\qquad\;}c@{\qquad\;}c@{\qquad\;}c@{\qquad\;}c}
    \toprule
    \inferrule{-}{\enc{I} \defeq \capCirc[+]} &
    \inferrule{\ari(R)=n}{\enc{R} \defeq 
    \InputIfFileExists{quine/R.tikz}{}{\input{tikz/quine/R.tikz}}
} &
    \inferrule{P\colon n \; n\geq 2}{  \enc{\mathbf{p}P} \defeq 
    \InputIfFileExists{quine/symm2.tikz}{}{\input{tikz/quine/symm2.tikz}}
 } &
    \inferrule{P\colon 1}{  \enc{\mathbf{p}P} \defeq 
    \InputIfFileExists{quine/symm1.tikz}{}{\input{tikz/quine/symm1.tikz}}
 } &
    \inferrule{P\colon 0}{  \enc{\mathbf{p}P} \defeq 
    \InputIfFileExists{quine/symm0.tikz}{}{\input{tikz/quine/symm0.tikz}}
 } 
    \\[25pt]
    \multicolumn{5}{c}{
      \begin{array}{c@{\qquad\qquad}c@{\qquad\qquad}c}
        \inferrule{P \colon n}{ \enc{\mathbf{P}P} \defeq 
    \InputIfFileExists{quine/symmn.tikz}{}{\input{tikz/quine/symmn.tikz}}
}  &
        \inferrule{P_1 \colon n \;\; P_2\colon m \;\; n\geq m}{\enc{P_1 \cap P_2} \defeq 
    \InputIfFileExists{quine/cap1.tikz}{}{\input{tikz/quine/cap1.tikz}}
} &
        \inferrule{P_1 \colon n \;\; P_2\colon m\;\; n<m}{\enc{P_1 \cap P_2} \defeq 
    \InputIfFileExists{quine/cap2.tikz}{}{\input{tikz/quine/cap2.tikz}}
}
      \end{array}
    } \\[21pt]
    \multicolumn{5}{c}{
      \begin{array}{c@{\qquad\qquad}c@{\qquad\qquad}c@{\qquad\qquad}c}
        \inferrule{P \colon n}{ \enc{\neg P} \defeq 
    \InputIfFileExists{quine/nega.tikz}{}{\input{tikz/quine/nega.tikz}}
}  &
        \inferrule{P \colon n}{ \enc{[ P} \defeq 
    \InputIfFileExists{quine/inj.tikz}{}{\input{tikz/quine/inj.tikz}}
}  &
        \inferrule{P \colon n+1}{ \enc{] P} \defeq 
    \InputIfFileExists{quine/proj.tikz}{}{\input{tikz/quine/proj.tikz}}
} &
        \inferrule{P \colon 0}{ \enc{] P} \defeq 
    \InputIfFileExists{quine/proj0.tikz}{}{\input{tikz/quine/proj0.tikz}}
}
      \end{array}
    }
    \\
    \bottomrule
  \end{array}
  }$
  \caption{The encoding $\enc{\cdot} \colon \PFL \to \NPR $.}\label{table:quine encoding}
\end{table*}

From $\interpretation =(X,\rho)$, one can define an interpretation of $\NPR$ $\mathcal{I}_p\defeq(X,\rho_p)$  where $\rho_p(R)\defeq \{(x,\star) \mid x\in \rho(R)\} \subseteq X^n\times \singleton$ for all $R \in \Sigma$ of arity $n$. The encoding of $\PFL$ into $\NPR$ is given in Table~\ref{table:quine encoding} where $
    \InputIfFileExists{quineSymm.tikz}{}{\input{tikz/quineSymm.tikz}}
$ is a suggestive representation for the permutation formally defined as $\symm[+][1][n-1]\seq[+] (\symm[+][n-2][1]\tensor[+] \id[+][1])$ for $n\geq 2$, $\id[+][n]$ for $n<2$. The following result (proved in Appendix \ref{app:encoding}) ensures that the encoding preserve the semantics.

\begin{prop}\label{prop:Quine}
Let $P\colon n$ be a predicate of $\PFL$. Then
\[\dsemRel{P} =\{\tau \mid ((\tau_1, \dots, \tau_n),\star)\in\interpretationFunctorP(\enc{P}) \}\text{.}\]
\end{prop}

\subsection{Peirce's Existential Graphs}\label{sec:egraphs}
The diagrammatic notation of $\LCB$ is closely related to system $\beta$ of Peirce's $\EG$~\cite{peirce2020_the-logic-of-the-future:-history-and-applications, roberts1973_the-existential-graphs-of-charles-s.-peirce}. Consider the two diagrams below corresponding to the closed $\FOL$ formula $\exists x.\, p(x) \wedge \forall y.\, p(y) \to q(y)$.

\[

    \InputIfFileExists{eg.tikz}{}{\input{tikz/eg.tikz}}

\qquad \leftrightsquigarrow \qquad

    \InputIfFileExists{eg2.tikz}{}{\input{tikz/eg2.tikz}}

\]

In existential graph notation the circle enclosure (dubbed `cut' by Peirce) signifies negation. To move from $\EG$ to diagrams of $\LCB$ it suffices to treat lines and predicate symbols in the obvious way and each cut as a color switch.

\begin{figure}[H]
    \centering
       \begin{subfigure}{0.5\textwidth}
          \centering
           $
    \InputIfFileExists{peirce/itLHS.tikz}{}{\input{tikz/peirce/itLHS.tikz}}
 \synequivalence 
    \InputIfFileExists{peirce/itRHS.tikz}{}{\input{tikz/peirce/itRHS.tikz}}
$
       \end{subfigure}
       \begin{subfigure}{0.4\textwidth}
           \centering
           \raisebox{-0.45\height}{\includegraphics*[scale=0.3]{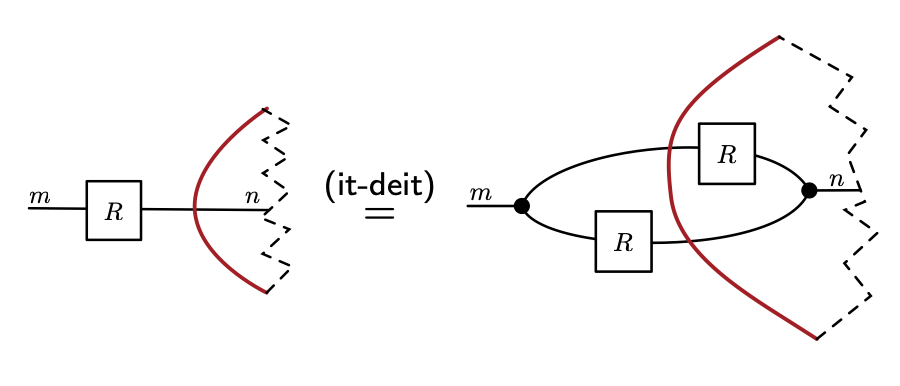}}
       \end{subfigure}
    \caption{Peirce's (de)iteration rule in $\LCB$ (left) and in~\cite{Haydon2020} (right).}
    \label{fig:itdeit}
\end{figure}

A string diagrammatic approach to existential graphs appeared in~\cite{Haydon2020}. This calculus exploits the white fragment of $\LCB$ with a primitive negation operator rendered as Peirce's cut, namely a circle around diagrams. However, this inhibits a fully compositional treatment since, for instance, negation is not functorial. %
As an example consider Peirce's (de)iteration rule in Figure~\ref{fig:itdeit}: in $\LCB$ on the left, and in~\cite{Haydon2020} on the right. 
Note that the diagrams on the right require open cuts, a notational trick, allowing to express the rule for arbitrary contexts, i.e. any diagram eventually appearing inside the cut. In $\LCB$ this ad-hoc treatment of contexts is not needed as negation is not a primitive operation, but a derived one.
Moreover, observe that in both Peirce's $\EG$ and the calculus in~\cite{Haydon2020}, the (de)iteration rule is taken as an axiom, while in $\LCB$ the rule is derivable, as shown below.
\begin{proof}[Proof of Peirce's (de)iteration rule in $\LCB$]
  \input{tikz/peirce/itproof.tex}
\end{proof}

\section{A tribute to Charles S. Peirce}\label{app:peirce}

We have chosen the name ``Neo-Peircean Relations'' to emphasize several connections with the work of Charles S. Peirce. 

First of all, $\NPR$ and  the calculus of relations in `Note B' \cite{peirce1883_studies-in-logic.-by-members-of-the-johns-hopkins-university} share the same underlying philosophy: they both propose a relational analogue to Boole's algebra of classes.

Second, Peirce's presentation in `Note B' contains already several key ingredients of $\NPR$. As we have stressed, it singles out the two forms of composition ($\seq[+]$ and $\seq[-]$), presents linear distributivity (\eqref{ax:leftLinDistr} and \eqref{ax:rightLinDistr}) and linear adjunctions (\eqref{ax:tauRPlus}, \eqref{ax:tauRMinus}, \eqref{ax:gammaRPlus}, and \eqref{ax:gammaRMinus}), and even the cyclic conditions of Lemma \ref{lm:implications}.(2)-(3). With respect to the rules for linear distributivity and linear adjunction, Peirce states that the latter are ``highly important" and that the former are ``so constantly used that hardly anything can be done without them" (p. 192 \& 190). %

At around the same time as `Note B' Peirce gave a systematic study of residuation \cite[see ``On the Logic of Relatives"]{peirce1989_writings-of-charles-s.-peirce:-a-chronological-edition-volume-4:-18791884} and listed a set of equivalent expressions that includes Lemma \ref{lm:residuation}: \[c\seq[+] a \leq b \text{ iff } c\leq b\seq[-]\rla{a}\text{.}\]
In Peirce's words:
\begin{quote}
``Hence the rule is that having a formula of the form $[c \seq[+]a \leq b]$, the three letters may be cyclically advanced one place in the order of writing, those which are carried from one side of the copula to the other being both negatived and converted.'' \cite[p. 341]{peirce1989_writings-of-charles-s.-peirce:-a-chronological-edition-volume-4:-18791884}
\end{quote}

\noindent 
Peirce took the principal defect of the presentation in ‘Note B’ to be its focus on binary relations \cite[8:831]{peirce1931_collected-papers-of-charles-sanders}. He went on to emphasize the \emph{teri-} or \emph{tri-}identity relation (in $\NPR$, $\copierCirc[+]$), arising from adding a `branch’ to the identity relation ($\idCirc[+]$), as the key to moving from binary to arbitrary relations. Having the advantage now of ``treating triadic and higher relations as easily as dyadic relations... it's superiority to the human mind as an instrument of logic", he writes, ``is overwhelming" \cite[p. 173]{peirce2022_the-logic-of-the-future:-pragmaticism}. %

By moving from binary to arbitrary relations, Peirce felt the importance of a graphical syntax and developed the existential graphs.
\begin{quote}
``One of my earliest works was an enlargement of Boole’s idea so as to take into account ideas of relation, --- or at least of all ideas of existential relation… \emph{I was finally led to prefer what I call a diagrammatic syntax}. It is a way of setting down on paper any assertion, however intricate'' [MS515, emphasis in original, 1911]. %
\end{quote}

\noindent 
We refer the reader to \cite{Haydon2020} for a detailed explanation of Peirce's topological intuitions behind the Frobenius equations and the correspondence of some inference rules for 
$\EG$s with those of (co)cartesian bicategories.
Moreover, we now know that Peirce continued to study and draw graphs of residuation \cite{haydon2021_residuation-in-existential-graphs} and --- as affirmed in Fig.~\ref{fig:correspondencepropositionalcalculus} --- we know the rules for propositional $\EG$s comprise a deep inference system \cite{ma2017_proof-analysis-of-peirces-alpha-system-of-graphs}.

In short, Peirce's development of $\EG$s shares many of the features that $\NPR$ has over other approaches, such as Tarski's presentation of relation algebra. We like to think that if Peirce had known category theory then he would have presented $\NPR$. %

\section{Concluding Remarks}

We introduced $\NPR$, a calculus of relations with the expressivity of first-order logic. We showed that it enjoys a sound and complete axiomatisation that arises through the interaction of two well-known categorical structures: cartesian bicategories and linear bicategories. We characterised these and dubbed the result fo-bicategories. Below we list some further related and future work.

In \S~\ref{sec:encodings} we already compared $\NPR$ to~\cite{Haydon2020}.
Other diagrammatic calculi, reminiscent of Peirce's $\EG$s, appear in~\cite{mellies2016bifibrational} and~\cite{brady2000string}. The categorical treatment goes, respectively, through the notions of chiralities and doctrines. The former consider a pair of categories $(\Rel_{\bullet}, \Rel_{\circ})$ that are significantly different from our $\Relp$ and $\Relm$. %

Instead, as recently shown in~\cite{DBLP:conf/mfcs/Bonchi0T24}, fo-bicategories are equivalent to Boolean hyperdoctrines that satisfy two additional conditions --extensive diagonals and the Rule of Unique Choice~\cite{maietti2013quotient}-- which are commonly employed in hyperdoctrines to capture the extensionality of equality. Intuitively, since in hyperdoctrines formulas and terms are distinct entities, the hyperdoctrine in which the formula $t_1 = t_2$ holds is different from the one in which $t_1$ and $t_2$ are identified as terms. Beyond representing extensional equality, fo-bicategories possess the crucial feature of being equationally presented, with morphisms that are simply structure-preserving maps.

The connection with allegories~\cite{freyd1990categories} is also worth exploring: since cartesian bicategories are equivalent to unitary pretabular allegories, Proposition~\ref{prop:bool homsets} suggests that fo-bicategories are closely related to Peirce allegories~\cite{olivier1997peirce}.

\subsection{Future Work}
\label{sec:future}

To conclude, it is worth outlining possible directions for future research.
\paragraph{Combinatorial characterisation, rewriting and deep inference}Throughout this work, diagrams have not been formally defined as combinatorial structures, as it is instead common in most diagrammatic languages~\cite{joyal1991geometry,Gadducci1998,DBLP:journals/jacm/BonchiGKSZ22,ghica2024stringdiagramslambdacalculifunctional,di2025rewriting}. In particular, \cite{DBLP:journals/jacm/BonchiGKSZ22} characterises diagrams for monoidal categories equipped with Frobenius bimonoids as cospans of hypergraphs. Furthermore, 2-cells between such cospans capture the ordering in cartesian bicategories (see \cite{GCQ} for the categorical formulation and~\cite{chandra1977optimal} for an analogous result in terms of conjunctive query inclusion). Building on these results, we aim to develop a combinatorial characterisation of the diagrams introduced in this paper. Relevant insights in this direction may also be drawn from~\cite{DBLP:conf/birthday/Rensink025}.

Such a combinatorial characterisation could serve two main purposes. First, it may provide a new categorical proof of completeness --recalling that the completeness of cartesian bicategories for regular logic was established via the characterisation in~\cite{GCQ}. Second, these combinatorial structures could form the basis for rewriting techniques and, potentially, the development of graphical proof assistants.
Finally, we also envisage defining a deep inference system whose rules correspond to the inequalities in our axiomatisation, and studying its proof-theoretic properties in comparison with existing systems such as~\cite{brunnler2006cut,ralph2019modular,hughes2021combinatorial}.

\paragraph{Bicategories beyond the poset-enrichment}
In this paper, we have focused on fo-bicategories whose hom-sets are posets.
For all \(c,d : X \to Y\), the ordering \(c \leq d\) indicates that there exists a proof of \(d\) from \(c\), but distinct proofs cannot be differentiated.
To distinguish between proofs and thereby develop a form of proof theory, one must move to \emph{proper bicategories}, that is, bicategories whose hom-sets are genuine categories rather than posets.
Cartesian bicategories were first introduced in the poset-enriched setting in \cite{carboni1987cartesian} and later extended to the proper bicategorical case in \cite{carboni2007cartesianbicategoriesii}.
Conversely, \emph{linear bicategories}, introduced in \cite{cockett2000introduction}, are defined as proper bicategories and can be viewed as a horizontal categorification of linearly distributive categories \cite{de1991dialectica,cockett1997weakly}.
Closed linear bicategories, in turn, correspond to non-symmetric $*$-autonomous categories \cite{barr2006autonomous}.
Interestingly, the Frobenius conditions that appear on 1-cells in cartesian bicategories also arise at the level of 2-cells in the definition of linear adjoints \cite[Definition~3.1]{cockett2000introduction}.
Understanding how the proper bicategorical structures of \cite{carboni2007cartesianbicategoriesii} and \cite{cockett2000introduction} interact---particularly the role played by Frobenius pseudomonoids at different levels---and relaxing the definition of fo-bicategories to avoid classical behaviour could lead to a notion of \emph{proof-relevant fo-bicategories}.
Further insights may be drawn from \cite{day2003quantumcategoriesstarautonomy,Dunn_2019}.

\section*{Acknowledgment}
\noindent This research was partly funded by the Advanced Research + Invention Agency (ARIA) Safeguarded AI Programme. Bonchi is supported by the Ministero dell'Università e della Ricerca of Italy grant PRIN 2022 PNRR No. P2022HXNSC - RAP (Resource Awareness in Programming) and  by the National Centre on HPC, Big Data and Quantum Computing - SPOKE 10 (Quantum Computing) and received funding from the European Union Next-GenerationEU - National Recovery and Resilience Plan (NRRP) – MISSION 4 COMPONENT 2, INVESTMENT N. 1.4 – CUP N. I53C22000690001. Di Giorgio and Soboci\'nski are supported by the European Union under Grant No. 101087529. Soboci\'nski is additionally supported by the Estonian Research Council via grants PRG3215 and TEM-TA5, and the Estonian Center of Excellence in Artificial Intelligence (EXAI).

\bibliographystyle{alphaurl}
\bibliography{main.bib}

\newcommand{\etalchar}[1]{$^{#1}$}
\begin{thebibliography}{CKWW07}

\bibitem[Bar06]{barr2006autonomous}
Michael Barr.
\newblock {\em *-Autonomous categories}, volume 752.
\newblock Springer, 2006.

\bibitem[BDM96]{bird1996algebra}
Richard Bird and Oege De~Moor.
\newblock The algebra of programming.
\newblock {\em NATO ASI DPD}, 152:167--203, 1996.

\bibitem[BE15]{BaezErbele-CategoriesInControl}
John Baez and Jason Erbele.
\newblock Categories in control.
\newblock {\em Theory and Applications of Categories}, 30:836--881, 2015.
\newblock URL: \url{http://www.tac.mta.ca/tac/volumes/30/24/30-24abs.html}.

\bibitem[BFS20]{doumane2020non}
Benedikt Bollig, Alain Finkel, and Amrita Suresh.
\newblock Bounded {{Reachability Problems Are Decidable}} in {{FIFO Machines}}.
\newblock In Igor Konnov and Laura Kov{\'a}cs, editors, {\em 31st
  {{International Conference}} on {{Concurrency Theory}} ({{CONCUR}} 2020)},
  volume 171 of {\em Leibniz {{International Proceedings}} in {{Informatics}}
  ({{LIPIcs}})}, pages 49:1--49:17, {Dagstuhl, Germany}, 2020. {Schloss
  Dagstuhl\textendash Leibniz-Zentrum f\"ur Informatik}.
\newblock \href {https://doi.org/10.4230/LIPIcs.CONCUR.2020.49}
  {\path{doi:10.4230/LIPIcs.CONCUR.2020.49}}.

\bibitem[BGHS24]{DBLP:conf/lics/Bonchi0H024}
Filippo Bonchi, Alessandro~Di Giorgio, Nathan Haydon, and Pawel Sobocinski.
\newblock Diagrammatic algebra of first order logic.
\newblock In Pawel Sobocinski, Ugo~Dal Lago, and Javier Esparza, editors, {\em
  Proceedings of the 39th Annual {ACM/IEEE} Symposium on Logic in Computer
  Science, {LICS} 2024, Tallinn, Estonia, July 8-11, 2024}, pages 16:1--16:15.
  {ACM}, 2024.
\newblock \href {https://doi.org/10.1145/3661814.3662078}
  {\path{doi:10.1145/3661814.3662078}}.

\bibitem[BGK{\etalchar{+}}22]{DBLP:journals/jacm/BonchiGKSZ22}
Filippo Bonchi, Fabio Gadducci, Aleks Kissinger, Pawel Sobocinski, and Fabio
  Zanasi.
\newblock String diagram rewrite theory {I:} rewriting with frobenius
  structure.
\newblock {\em J. {ACM}}, 69(2):14:1--14:58, 2022.
\newblock \href {https://doi.org/10.1145/3502719} {\path{doi:10.1145/3502719}}.

\bibitem[BGT24]{DBLP:conf/mfcs/Bonchi0T24}
Filippo Bonchi, Alessandro~Di Giorgio, and Davide Trotta.
\newblock {When Lawvere Meets Peirce: An Equational Presentation of Boolean
  Hyperdoctrines}.
\newblock In Rastislav Kr{\'{a}}lovic and Anton{\'{\i}}n Kucera, editors, {\em
  49th International Symposium on Mathematical Foundations of Computer Science,
  {MFCS} 2024, August 26-30, 2024, Bratislava, Slovakia}, volume 306 of {\em
  LIPIcs}, pages 30:1--30:19. Schloss Dagstuhl - Leibniz-Zentrum f{\"{u}}r
  Informatik, 2024.
\newblock \href {https://doi.org/10.4230/LIPICS.MFCS.2024.30}
  {\path{doi:10.4230/LIPICS.MFCS.2024.30}}.

\bibitem[BHP{\etalchar{+}}19]{DBLP:journals/pacmpl/BonchiHPSZ19}
Filippo Bonchi, Joshua Holland, Robin Piedeleu, Pawe{\l} Soboci{\'n}ski, and
  Fabio Zanasi.
\newblock Diagrammatic algebra: From linear to concurrent systems.
\newblock {\em Proceedings of the ACM on Programming Languages},
  3(POPL):25:1--25:28, January 2019.
\newblock \href {https://doi.org/10.1145/3290338} {\path{doi:10.1145/3290338}}.

\bibitem[BJ79]{bohm1966flow}
Corrado B{\"o}hm and Giuseppe Jacopini.
\newblock Flow diagrams, {{Turing}} machines and languages with only two
  formation rules.
\newblock In {\em Classics in Software Engineering}, pages 11--25. {Yourdon
  Press}, {USA}, January 1979.
\newblock URL: \url{https://dl.acm.org/doi/abs/10.5555/1241515.1241517}.

\bibitem[Boo47]{boole1847mathematical}
George Boole.
\newblock {\em The mathematical analysis of logic}.
\newblock Philosophical Library, 1847.

\bibitem[BPS25]{DBLP:journals/corr/abs-1711-08699}
Filippo Bonchi, Dusko Pavlovic, and Paweł Sobociński.
\newblock Functorial semantics for relational theories.
\newblock {\em Polynesian Journal of Mathematics}, 2(6):1--56, 10 2025.
\newblock \href {https://doi.org/10.69763/polyjmath.2.6}
  {\path{doi:10.69763/polyjmath.2.6}}.

\bibitem[Br{\"{u}}03]{DBLP:phd/de/Brunnler2003}
Kai Br{\"{u}}nnler.
\newblock {\em Deep inference and symmetry in classical proofs}.
\newblock PhD thesis, Dresden University of Technology, Germany, 2003.

\bibitem[Br{\"{u}}06]{brunnler2006cut}
Kai Br{\"{u}}nnler.
\newblock Cut elimination inside a deep inference system for classical
  predicate logic.
\newblock {\em Studia Logica}, 82(1):51--71, 2006.
\newblock \href {https://doi.org/10.1007/s11225-006-6605-4}
  {\path{doi:10.1007/s11225-006-6605-4}}.

\bibitem[BSS18]{GCQ}
Filippo Bonchi, Jens Seeber, and Pawel Sobocinski.
\newblock {Graphical Conjunctive Queries}.
\newblock In Dan Ghica and Achim Jung, editors, {\em 27th EACSL Annual
  Conference on Computer Science Logic (CSL 2018)}, volume 119 of {\em Leibniz
  International Proceedings in Informatics (LIPIcs)}, pages 13:1--13:23,
  Dagstuhl, Germany, 2018. Schloss Dagstuhl--Leibniz-Zentrum fuer Informatik.
\newblock \href {https://doi.org/10.4230/LIPIcs.CSL.2018.13}
  {\path{doi:10.4230/LIPIcs.CSL.2018.13}}.

\bibitem[BSZ15]{Bonchi2015}
Filippo Bonchi, Pawel Sobocinski, and Fabio Zanasi.
\newblock Full {{Abstraction}} for {{Signal Flow Graphs}}.
\newblock In {\em Proceedings of the 42nd {{Annual ACM SIGPLAN-SIGACT
  Symposium}} on {{Principles}} of {{Programming Languages}}}, {{POPL}} '15,
  pages 515--526, {New York, NY, USA}, January 2015. {Association for Computing
  Machinery}.
\newblock \href {https://doi.org/10.1145/2676726.2676993}
  {\path{doi:10.1145/2676726.2676993}}.

\bibitem[BSZ17]{interactinghopf}
Filippo Bonchi, Pawe{\l} Soboci\'nski, and Fabio Zanasi.
\newblock Interacting {H}opf algebras.
\newblock {\em Journal of Pure and Applied Algebra}, 221(1):144--184, 2017.

\bibitem[BT98]{brady2000string}
Geraldine Brady and Todd~H. Trimble.
\newblock {A String Diagram Calculus for Predicate Logic and C. S. Peirce's
  System Beta}.
\newblock \url{https://ncatlab.org/nlab/files/BradyTrimbleString.pdf}, 1998.

\bibitem[CD11]{CoeckeDuncanZX2011}
Bob Coecke and Ross Duncan.
\newblock Interacting quantum observables: Categorical algebra and
  diagrammatics.
\newblock {\em New Journal of Physics}, 13(4):043016, April 2011.
\newblock \href {https://doi.org/10.1088/1367-2630/13/4/043016}
  {\path{doi:10.1088/1367-2630/13/4/043016}}.

\bibitem[CFC{\etalchar{+}}58]{curry1958combinatory}
Haskell~Brooks Curry, Robert Feys, William Craig, J~Roger Hindley, and
  Jonathan~P Seldin.
\newblock {\em Combinatory logic}, volume~1.
\newblock North-Holland Amsterdam, 1958.

\bibitem[CKS00]{cockett2000introduction}
J.~Robin~B. Cockett, J{\"u}rgen Koslowski, and Robert~AG Seely.
\newblock Introduction to linear bicategories.
\newblock {\em Mathematical Structures in Computer Science}, 10(2):165--203,
  2000.

\bibitem[CKWW07]{carboni2007cartesianbicategoriesii}
A.~Carboni, G.~M. Kelly, R.~F.~C Walters, and R.~J. Wood.
\newblock Cartesian bicategories ii, 2007.
\newblock URL: \url{https://arxiv.org/abs/0708.1921}, \href
  {https://arxiv.org/abs/0708.1921} {\path{arXiv:0708.1921}}.

\bibitem[CM77]{chandra1977optimal}
Ashok~K. Chandra and Philip~M. Merlin.
\newblock Optimal implementation of conjunctive queries in relational data
  bases.
\newblock In {\em Proceedings of the Ninth Annual {{ACM}} Symposium on
  {{Theory}} of Computing}, {{STOC}} '77, pages 77--90, {New York, NY, USA},
  May 1977. {Association for Computing Machinery}.
\newblock \href {https://doi.org/10.1145/800105.803397}
  {\path{doi:10.1145/800105.803397}}.

\bibitem[Cod83]{codd1983relational}
Edgar~Frank Codd.
\newblock A relational model of data for large shared data banks.
\newblock {\em Communications of the ACM}, 26(1):64--69, 1983.

\bibitem[CS97a]{cockett1997proof}
J~Robin~B Cockett and Robert~AG Seely.
\newblock Proof theory for full intuitionistic linear logic, bilinear logic,
  and mix categories.
\newblock {\em Theory and Applications of categories}, 3(5):85--131, 1997.

\bibitem[CS97b]{cockett1997weakly}
J~Robin~B Cockett and Robert~AG Seely.
\newblock Weakly distributive categories.
\newblock {\em Journal of Pure and Applied Algebra}, 114(2):133--173, 1997.

\bibitem[CW87]{carboni1987cartesian}
A.~Carboni and R.~F.~C. Walters.
\newblock Cartesian bicategories {{I}}.
\newblock {\em Journal of Pure and Applied Algebra}, 49(1):11--32, November
  1987.
\newblock \href {https://doi.org/10.1016/0022-4049(87)90121-6}
  {\path{doi:10.1016/0022-4049(87)90121-6}}.

\bibitem[DGGZ25]{di2025rewriting}
Alessandro Di~Giorgio, Dan~R Ghica, and Fabio Zanasi.
\newblock Rewriting for traced monoidal closed categories.
\newblock In {\em International Conference on Graph Transformation}, pages
  24--43. Springer, 2025.

\bibitem[dP91]{de1991dialectica}
Valeria de~Paiva.
\newblock The dialectica categories. university of cambridge.
\newblock {\em Computer Lab Technical Report, PhD thesis}, 2:3, 1991.

\bibitem[DS03]{day2003quantumcategoriesstarautonomy}
Brian Day and Ross Street.
\newblock Quantum categories, star autonomy, and quantum groupoids, 2003.
\newblock URL: \url{https://arxiv.org/abs/math/0301209}, \href
  {https://arxiv.org/abs/math/0301209} {\path{arXiv:math/0301209}}.

\bibitem[DV19]{Dunn_2019}
Lawrence Dunn and Jamie Vicary.
\newblock Coherence for frobenius pseudomonoids and the geometry of linear
  proofs.
\newblock {\em Logical Methods in Computer Science}, Volume 15, Issue 3, July
  2019.
\newblock \href {https://doi.org/10.23638/lmcs-15(3:5)2019}
  {\path{doi:10.23638/lmcs-15(3:5)2019}}.

\bibitem[Fox76]{fox1976coalgebras}
T.~Fox.
\newblock Coalgebras and cartesian categories.
\newblock {\em Communications in Algebra}, 4(7):665--667, 1976.
\newblock \href {https://doi.org/10.1080/00927877608822127}
  {\path{doi:10.1080/00927877608822127}}.

\bibitem[FPT99]{fiore1999abstract}
Marcelo Fiore, Gordon Plotkin, and Daniele Turi.
\newblock Abstract syntax and variable binding.
\newblock In {\em Proceedings. 14th Symposium on Logic in Computer Science
  (Cat. No. PR00158)}, pages 193--202. IEEE, 1999.

\bibitem[FS90]{freyd1990categories}
Peter Freyd and Andre Scedrov.
\newblock {\em Categories, {{Allegories}}}, volume~39 of {\em North-{{Holland
  Mathematical Library}}}.
\newblock {Elsevier B.V}, 1990.

\bibitem[FS20]{DBLP:journals/corr/abs-2009-06836}
Brendan Fong and David Spivak.
\newblock String diagrams for regular logic (extended abstract).
\newblock In John Baez and Bob Coecke, editors, {\em Applied Category Theory
  2019}, volume 323 of {\em Electronic Proceedings in Theoretical Computer
  Science}, pages 196--229. Open Publishing Association, Sep 2020.
\newblock \href {https://doi.org/10.4204/eptcs.323.14}
  {\path{doi:10.4204/eptcs.323.14}}.

\bibitem[FSR16]{Fong2015}
Brendan Fong, Pawe{\l} Soboci{\'n}ski, and Paolo Rapisarda.
\newblock A categorical approach to open and interconnected dynamical systems.
\newblock In {\em Proceedings of the 31st {{Annual ACM}}/{{IEEE Symposium}} on
  {{Logic}} in {{Computer Science}}}, {{LICS}} '16, pages 495--504, {New York,
  NY, USA}, July 2016. {Association for Computing Machinery}.
\newblock \href {https://doi.org/10.1145/2933575.2934556}
  {\path{doi:10.1145/2933575.2934556}}.

\bibitem[GH97]{Gadducci1998}
Fabio Gadducci and Reiko Heckel.
\newblock An inductive view of graph transformation.
\newblock In {\em WADT 1997}, pages 223--237, 1997.

\bibitem[GJ16]{Ghica2016}
Dan~R. Ghica and Achim Jung.
\newblock Categorical semantics of digital circuits.
\newblock In {\em 2016 Formal Methods in Computer-Aided Design (FMCAD)}, pages
  41--48, 2016.
\newblock \href {https://doi.org/10.1109/FMCAD.2016.7886659}
  {\path{doi:10.1109/FMCAD.2016.7886659}}.

\bibitem[GKO{\etalchar{+}}16]{gaboardi2016combining}
Marco Gaboardi, Shin-ya Katsumata, Dominic Orchard, Flavien Breuvart, and Tarmo
  Uustalu.
\newblock Combining effects and coeffects via grading.
\newblock {\em ACM SIGPLAN Notices}, 51(9):476--489, 2016.

\bibitem[GMS{\etalchar{+}}23]{goncharov2023towards}
Sergey Goncharov, Stefan Milius, Lutz Schr{\"o}der, Stelios Tsampas, and
  Henning Urbat.
\newblock Towards a higher-order mathematical operational semantics.
\newblock {\em Proceedings of the ACM on Programming Languages},
  7(POPL):632--658, 2023.

\bibitem[GP02]{DBLP:journals/fac/GabbayP02}
Murdoch Gabbay and Andrew~M. Pitts.
\newblock A new approach to abstract syntax with variable binding.
\newblock {\em Formal Aspects Comput.}, 13(3-5):341--363, 2002.
\newblock \href {https://doi.org/10.1007/s001650200016}
  {\path{doi:10.1007/s001650200016}}.

\bibitem[Gug07]{guglielmi2007system}
Alessio Guglielmi.
\newblock A system of interaction and structure.
\newblock {\em ACM Transactions on Computational Logic (TOCL)}, 8(1):1--es,
  2007.

\bibitem[GZ24]{ghica2024stringdiagramslambdacalculifunctional}
Dan Ghica and Fabio Zanasi.
\newblock String diagrams for $\lambda$-calculi and functional computation,
  2024.
\newblock URL: \url{https://arxiv.org/abs/2305.18945}, \href
  {https://arxiv.org/abs/2305.18945} {\path{arXiv:2305.18945}}.

\bibitem[Hen49]{henkin_1949}
Leon Henkin.
\newblock The completeness of the first-order functional calculus.
\newblock {\em The Journal of Symbolic Logic}, 14(3):159--166, 1949.
\newblock \href {https://doi.org/10.2307/2267044} {\path{doi:10.2307/2267044}}.

\bibitem[HJ86]{hoare1986weakest}
CAR Hoare and He~Jifeng.
\newblock The weakest prespecification, part {I}.
\newblock {\em Fundamenta Informaticae}, 9(1):51--84, 1986.

\bibitem[HM00]{hodkinson2000axiomatizability}
Ian Hodkinson and Szabolcs Mikul{\'a}s.
\newblock Axiomatizability of reducts of algebras of relations.
\newblock {\em Algebra Universalis}, 43(2):127--156, August 2000.
\newblock \href {https://doi.org/10.1007/s000120050150}
  {\path{doi:10.1007/s000120050150}}.

\bibitem[Hof99]{hofmann1999semantical}
Martin Hofmann.
\newblock Semantical analysis of higher-order abstract syntax.
\newblock In {\em Proceedings. 14th Symposium on Logic in Computer Science
  (Cat. No. PR00158)}, pages 204--213. IEEE, 1999.

\bibitem[HP21]{haydon2021_residuation-in-existential-graphs}
Nathan Haydon and Ahti-Veikko Pietarinen.
\newblock Residuation in existential graphs.
\newblock In Amrita Basu, Gem Stapleton, Sven Linker, Catherine Legg, Emmanuel
  Manalo, and Petrucio Viana, editors, {\em Diagrammatic Representation and
  Inference}, pages 229--237, Cham, 2021. Springer International Publishing.

\bibitem[HS20]{Haydon2020}
Nathan Haydon and Pawe{\l} Soboci\'{n}ski.
\newblock Compositional diagrammatic first-order logic.
\newblock In {\em 11th International Conference on the Theory and Application
  of Diagrams (DIAGRAMS 2020)}, 2020.

\bibitem[HSW21]{hughes2021combinatorial}
Dominic~JD Hughes, Lutz Stra{\ss}burger, and Jui-Hsuan Wu.
\newblock Combinatorial proofs and decomposition theorems for first-order
  logic.
\newblock In {\em 2021 36th Annual ACM/IEEE Symposium on Logic in Computer
  Science (LICS)}, pages 1--13. IEEE, 2021.

\bibitem[JS91]{joyal1991geometry}
Andr{\'e} Joyal and Ross Street.
\newblock The geometry of tensor calculus, {{I}}.
\newblock {\em Advances in Mathematics}, 88(1):55--112, July 1991.
\newblock \href {https://doi.org/10.1016/0001-8708(91)90003-P}
  {\path{doi:10.1016/0001-8708(91)90003-P}}.

\bibitem[JY24]{johnson2024bimonoidal}
Niles Johnson and Donald Yau.
\newblock {\em Bimonoidal Categories, $ E\_n $-Monoidal Categories, and
  Algebraic $ K $-Theory: Volume III: From Categories to Structured Ring
  Spectra}, volume 285.
\newblock American Mathematical Society, 2024.

\bibitem[KN12]{krauss2012regular}
Alexander Krauss and Tobias Nipkow.
\newblock Proof pearl: Regular expression equivalence and relation algebra.
\newblock {\em Journal of Automated Reasoning}, 49(1):95--106, 2012.
\newblock \href {https://doi.org/10.1007/s10817-011-9223-4}
  {\path{doi:10.1007/s10817-011-9223-4}}.

\bibitem[Kuh83]{kuhn1983}
Steven~T. Kuhn.
\newblock {An axiomatization of predicate functor logic.}
\newblock {\em Notre Dame Journal of Formal Logic}, 24(2):233 -- 241, 1983.
\newblock \href {https://doi.org/10.1305/ndjfl/1093870313}
  {\path{doi:10.1305/ndjfl/1093870313}}.

\bibitem[Lac04]{Lack2004a}
Stephen Lack.
\newblock Composing {PROPs}.
\newblock {\em Theory and Application of Categories}, 13(9):147--163, 2004.
\newblock URL: \url{http://www.tac.mta.ca/tac/volumes/13/9/13-09abs.html}.

\bibitem[Las98]{lassen1998relational}
S{\o}ren~B Lassen.
\newblock Relational reasoning about contexts.
\newblock {\em Higher order operational techniques in semantics}, 91, 1998.

\bibitem[Law63]{LawvereOriginalPaper}
F.~W. Lawvere.
\newblock {\em Functorial {{Semantics}} of {{Algebraic Theories}}}.
\newblock PhD thesis, Columbia University, {New York, NY, USA}, 1963.

\bibitem[LG22]{DBLP:journals/pacmpl/LagoG22a}
Ugo~Dal Lago and Francesco Gavazzo.
\newblock A relational theory of effects and coeffects.
\newblock {\em Proc. {ACM} Program. Lang.}, 6({POPL}):1--28, 2022.
\newblock \href {https://doi.org/10.1145/3498692} {\path{doi:10.1145/3498692}}.

\bibitem[L{\"o}w15]{lowenheim1915moglichkeiten}
Leopold L{\"o}wenheim.
\newblock {\"U}ber m{\"o}glichkeiten im relativkalk{\"u}l.
\newblock {\em Mathematische Annalen}, 76(4):447--470, 1915.

\bibitem[LP01]{lascar2001mathematical}
Daniel Lascar and Donald~H Pelletier.
\newblock {\em Mathematical Logic: A Course with Exercises Part I:
  Propositional Calculus, Boolean Algebras, Predicate Calculus, Completeness
  Theorems}.
\newblock Oxford University Press, 2001.

\bibitem[MCG18]{DBLP:conf/lics/MuroyaCG18}
Koko Muroya, Steven W.~T. Cheung, and Dan~R. Ghica.
\newblock The geometry of computation-graph abstraction.
\newblock In Anuj Dawar and Erich Gr{\"{a}}del, editors, {\em Proceedings of
  the 33rd Annual {ACM/IEEE} Symposium on Logic in Computer Science, {LICS}
  2018, Oxford, UK, July 09-12, 2018}, pages 749--758. {ACM}, 2018.
\newblock \href {https://doi.org/10.1145/3209108.3209127}
  {\path{doi:10.1145/3209108.3209127}}.

\bibitem[Mil09]{milner2009space}
Robin Milner.
\newblock {\em The space and motion of communicating agents}.
\newblock Cambridge University Press, 2009.

\bibitem[ML78]{mac_lane_categories_1978}
S.~Mac~Lane.
\newblock {\em Categories for the {{Working Mathematician}}}, volume~5 of {\em
  Graduate {{Texts}} in {{Mathematics}}}.
\newblock {Springer-Verlag}, {New York}, second edition, 1978.
\newblock URL: \url{https://www.springer.com/gb/book/9780387984032}.

\bibitem[Mon64]{monk}
Donald Monk.
\newblock {On representable relation algebras}.
\newblock {\em Michigan Mathematical Journal}, 11(3):207 -- 210, 1964.
\newblock \href {https://doi.org/10.1307/mmj/1028999131}
  {\path{doi:10.1307/mmj/1028999131}}.

\bibitem[Mor60]{de1860syllogism}
Augustus~De Morgan.
\newblock {\em On the Syllogism, No. {IV}. And on the Logic of Relations}.
\newblock Printed by C.J. Clay at the University Press, 1860.

\bibitem[MP17]{ma2017_proof-analysis-of-peirces-alpha-system-of-graphs}
Minghui Ma and Ahti-Veikko Pietarinen.
\newblock {Proof Analysis of Peirce's Alpha System of Graphs}.
\newblock {\em Studia Logica}, 105(3):625--647, 2017.
\newblock \href {https://doi.org/10.1007/s11225-016-9703-y}
  {\path{doi:10.1007/s11225-016-9703-y}}.

\bibitem[MR13]{maietti2013quotient}
Maria~Emilia Maietti and Giuseppe Rosolini.
\newblock Quotient completion for the foundation of constructive mathematics.
\newblock {\em Logica Universalis}, 7:371--402, 2013.

\bibitem[MZ16]{mellies2016bifibrational}
Paul-Andr{\'e} Melli{\`e}s and Noam Zeilberger.
\newblock {A bifibrational reconstruction of Lawvere's presheaf hyperdoctrine}.
\newblock In {\em Proceedings of the 31st Annual ACM/IEEE Symposium on Logic in
  Computer Science}, pages 555--564, 2016.

\bibitem[Nae25]{naeimabadiconstructing}
Shayesteh Naeimabadi.
\newblock {\em Cartesian Linear Bicategories}.
\newblock PhD thesis, University of Ottawa, 2025.
\newblock \href {https://doi.org/10.20381/ruor-30846}
  {\path{doi:10.20381/ruor-30846}}.

\bibitem[OLEI19]{orchard2019quantitative}
Dominic Orchard, Vilem-Benjamin Liepelt, and Harley Eades~III.
\newblock Quantitative program reasoning with graded modal types.
\newblock {\em Proceedings of the ACM on Programming Languages}, 3(ICFP):1--30,
  2019.

\bibitem[OS97]{olivier1997peirce}
Jean-Pierre Olivier and Dany Serrato.
\newblock Peirce allegories. identities involving transitive elements and
  symmetrical ones.
\newblock {\em Journal of Pure and Applied Algebra}, 116(1-3):249--271, 1997.

\bibitem[PE88]{pfenning1988higher}
Frank Pfenning and Conal Elliott.
\newblock Higher-order abstract syntax.
\newblock {\em ACM sigplan notices}, 23(7):199--208, 1988.

\bibitem[Pei83]{peirce1883_studies-in-logic.-by-members-of-the-johns-hopkins-university}
Charles~S. Peirce.
\newblock {\em Studies in logic. By members of the Johns Hopkins university}.
\newblock Little, Brown, and Company, 1883.

\bibitem[Pei97]{peirce1897_the-logic-of-relatives}
Charles~S. Peirce.
\newblock The logic of relatives.
\newblock {\em The Monist}, 7(2):161--217, 1897.
\newblock URL: \url{http://www.jstor.org/stable/27897407}.

\bibitem[Pei66]{peirce1931_collected-papers-of-charles-sanders}
Charles~S. Peirce.
\newblock {\em Collected Papers of Charles Sanders Peirce}.
\newblock Cambridge, MA: Harvard University Press, 1958-1966.

\bibitem[Pei89]{peirce1989_writings-of-charles-s.-peirce:-a-chronological-edition-volume-4:-18791884}
Charles~S. Peirce.
\newblock {\em Writings of Charles S. Peirce: A Chronological Edition, Volume
  4: 1879--1884}.
\newblock Indiana University Press, 1989.
\newblock URL: \url{http://www.jstor.org/stable/j.ctt16gz8j1}.

\bibitem[Pei20]{peirce2020_the-logic-of-the-future:-history-and-applications}
Charles~S. Peirce.
\newblock {\em The Logic of the Future: History and Applications}, volume~1.
\newblock De Gruyter, Berlin, Boston, 2020.
\newblock \href {https://doi.org/doi:10.1515/9783110651409}
  {\path{doi:doi:10.1515/9783110651409}}.

\bibitem[Pei22]{peirce2022_the-logic-of-the-future:-pragmaticism}
Charles~S. Peirce.
\newblock {\em The Logic of the Future: Pragmaticism}, volume 3/1.
\newblock De Gruyter, 2022.

\bibitem[Pit13]{pitts2013nominal}
Andrew~M Pitts.
\newblock {\em Nominal sets: Names and symmetry in computer science}.
\newblock Cambridge University Press, 2013.

\bibitem[Pou13]{pous2013kleene}
Damien Pous.
\newblock Kleene algebra with tests and coq tools for while programs.
\newblock In {\em Interactive Theorem Proving: 4th International Conference,
  ITP 2013, Rennes, France, July 22-26, 2013. Proceedings 4}, pages 180--196.
  Springer, 2013.

\bibitem[Pou16]{pous2016automata}
Damien Pous.
\newblock {\em Automata for relation algebra and formal proofs}.
\newblock PhD thesis, ENS Lyon, 2016.

\bibitem[Pou18]{DBLP:conf/stacs/Pous18}
Damien Pous.
\newblock On the positive calculus of relations with transitive closure.
\newblock In Rolf Niedermeier and Brigitte Vall{\'{e}}e, editors, {\em 35th
  Symposium on Theoretical Aspects of Computer Science, {STACS} 2018, February
  28 to March 3, 2018, Caen, France}, volume~96 of {\em LIPIcs}, pages
  3:1--3:16. Schloss Dagstuhl - Leibniz-Zentrum f{\"{u}}r Informatik, 2018.
\newblock \href {https://doi.org/10.4230/LIPIcs.STACS.2018.3}
  {\path{doi:10.4230/LIPIcs.STACS.2018.3}}.

\bibitem[Pra76]{pratt1976semantical}
Vaughan~R Pratt.
\newblock {Semantical considerations on Floyd-Hoare logic}.
\newblock In {\em 17th Annual Symposium on Foundations of Computer Science
  (sfcs 1976)}, pages 109--121. IEEE, 1976.

\bibitem[PZ21]{Piedeleu2021}
Robin Piedeleu and Fabio Zanasi.
\newblock A {{String Diagrammatic Axiomatisation}} of {{Finite-State
  Automata}}.
\newblock In Stefan Kiefer and Christine Tasson, editors, {\em Foundations of
  {{Software Science}} and {{Computation Structures}}}, Lecture {{Notes}} in
  {{Computer Science}}, pages 469--489, {Cham}, 2021. {Springer International
  Publishing}.
\newblock \href {https://doi.org/10.1007/978-3-030-71995-1_24}
  {\path{doi:10.1007/978-3-030-71995-1_24}}.

\bibitem[Qui71]{QUINE1971309}
W.V. Quine.
\newblock Predicate-functor logics.
\newblock In J.E. Fenstad, editor, {\em Proceedings of the Second Scandinavian
  Logic Symposium}, volume~63 of {\em Studies in Logic and the Foundations of
  Mathematics}, pages 309--315. Elsevier, 1971.
\newblock \href {https://doi.org/10.1016/S0049-237X(08)70850-4}
  {\path{doi:10.1016/S0049-237X(08)70850-4}}.

\bibitem[Ral19]{ralph2019modular}
Benjamin Ralph.
\newblock {\em Modular Normalisation of Classical Proofs}.
\newblock PhD thesis, University of Bath, 2019.

\bibitem[RC24]{DBLP:conf/birthday/Rensink025}
Arend Rensink and Andrea Corradini.
\newblock On categories of nested conditions.
\newblock In Nils Jansen, Sebastian Junges, Benjamin~Lucien Kaminski, Christoph
  Matheja, Thomas Noll, Tim Quatmann, Mari{\"{e}}lle Stoelinga, and Matthias
  Volk, editors, {\em Principles of Verification: Cycling the Probabilistic
  Landscape - Essays Dedicated to Joost-Pieter Katoen on the Occasion of His
  60th Birthday, Part {I}}, volume 15260 of {\em Lecture Notes in Computer
  Science}, pages 393--418. Springer, 2024.
\newblock \href {https://doi.org/10.1007/978-3-031-75783-9\_16}
  {\path{doi:10.1007/978-3-031-75783-9\_16}}.

\bibitem[Red64]{redko1964defining}
Valentin~N Redko.
\newblock On defining relations for the algebra of regular events.
\newblock {\em Ukrainskii Matematicheskii Zhurnal}, 16:120--126, 1964.

\bibitem[Rob73]{roberts1973_the-existential-graphs-of-charles-s.-peirce}
Don~D. Roberts.
\newblock {\em The Existential Graphs of Charles S. Peirce}.
\newblock De Gruyter Mouton, 1973.

\bibitem[See20]{seeber2020logical}
Jens Seeber.
\newblock {\em Logical completeness for string diagrams}.
\newblock PhD thesis, 2020.

\bibitem[Sel10]{Selinger2009}
P.~Selinger.
\newblock A {{Survey}} of {{Graphical Languages}} for {{Monoidal Categories}}.
\newblock In B.~Coecke, editor, {\em New {{Structures}} for {{Physics}}},
  volume 813 of {\em Lecture {{Notes}} in {{Physics}}}, pages 289--355.
  {Springer, Berlin, Heidelberg}, 2010.
\newblock \href {https://doi.org/10.1007/978-3-642-12821-9_4}
  {\path{doi:10.1007/978-3-642-12821-9_4}}.

\bibitem[Tar41]{tarski1941calculus}
Alfred Tarski.
\newblock On the calculus of relations.
\newblock {\em The Journal of Symbolic Logic}, 6(3):73--89, September 1941.
\newblock \href {https://doi.org/10.2307/2268577} {\path{doi:10.2307/2268577}}.

\bibitem[TG88]{tarski1988formalization}
Alfred Tarski and Steven~R Givant.
\newblock {\em A formalization of set theory without variables}, volume~41.
\newblock American Mathematical Soc., 1988.

\bibitem[Zor35]{zorn1935remark}
Max Zorn.
\newblock A remark on method in transfinite algebra.
\newblock {\em Bulletin of the American Mathematical Society}, 41(10):667--670,
  1935.

\end{thebibliography}

\newpage

\appendix
\section{Complete axiomatisation of $\NPR$}\label{app:additional}
\begin{figure}[H]
  \centering
  {\tiny
  $
      \begin{array}{@{}r@{\,\,}c@{\,\,}l @{\qquad} r@{\,\,}c@{\,\,}l@{}}
        \toprule
          \copier[+][1] \seq[+] (\id[+][1] \tensor[+] \copier[+][1]) &\stackrel{(\copier[+]\text{-as})}{=}& \copier[+][1] \seq[+] (\copier[+][1] \tensor[+] \id[+][1]) & (\id[+][1] \tensor[+] \cocopier[+][1]) \seq[+] \cocopier[+][1] &\stackrel{(\cocopier[+]\text{-as})}{=}& (\cocopier[+][1] \tensor[+] \id[+][1]) \seq[+] \cocopier[+][1] \\
          \copier[+][1] \seq[+] (\id[+][1] \tensor[+] \discard[+][1]) &\stackrel{(\copier[+]\text{-un})}{=}& \id[+][1]  & (\id[+][1] \tensor[+] \codiscard[+][1]) \seq[+] \cocopier[+][1] &\stackrel{(\cocopier[+]\text{-un})}{=}& \id[+][1] \\
          \copier[+][1] \seq[+] \symm[+][1][1] &\stackrel{(\copier[+]\text{-co})}{=}& \copier[+][1] & \symm[+][1][1] \seq[+] \cocopier[+][1] &\stackrel{(\cocopier[+]\text{-co})}{=}& \cocopier[+][1] \\
          \addlinespace[0.6em]
          \midrule
          \addlinespace[0.6em]
          (\copier[+][1] \tensor[+] \id[+][1]) \seq[+] (\id[+][1] \tensor[+] \cocopier[+][1]) &\stackrel{(\text{F}^{\switchLabelS{+}})}{=}& (\id[+][1] \tensor[+] \copier[+][1]) \seq[+] (\cocopier[+][1] \tensor[+] \id[+][1]) & \copier[+][1] \seq[+] \cocopier[+][1] &\stackrel{(\text{S}^{\switchLabelS{+}})}{=}& \id[+][1] \\
          \addlinespace[0.6em]
          \midrule
          \addlinespace[0.6em]
          \multicolumn{6}{c}{
              \begin{array}{c@{\qquad}c}
                  \begin{array}{r@{\,\,}c@{\,\,}l @{\qquad} r@{\,\,}c@{\,\,}l}
                      \codiscard[+][1] \seq[+] \discard[+][1]
                      &\stackrel{\eqref*{ax:plusCodiscDisc}}{\leq}&
                      \id[+][0]
                      &
                      \cocopier[+][1] \seq[+] \copier[+][1]
                      &\stackrel{\eqref*{ax:plusCocopyCopy}}{\leq}&
                      (\id[+][1] \tensor[+] \id[+][1])
                      \\
                      \id[+][1]
                      &\stackrel{\eqref*{ax:plusDiscCodisc}}{\leq}&
                      \discard[+][1] \seq[+] \codiscard[+][1]
                      &
                      \id[+][1]
                      &\stackrel{\eqref*{ax:plusCopyCocopy}}{\leq}&
                      \copier[+][1] \seq[+] \cocopier[+][1] \\
                  \end{array}
                  &
                  \begin{array}{r@{\,\,}c@{\,\,}l r@{\,\,}c@{\,\,}l}
                  c \seq[+] \copier[+][m] &\stackrel{\eqref*{ax:comPlusLaxNat}}{\leq}& \copier[+][n] \seq[+] (c \tensor[+] c)
                  \\
                  c \seq[+] \discard[+][m] &\stackrel{\eqref*{ax:discPlusLaxNat}}{\leq}& \discard[+][n]
                  \end{array}
              \end{array}
          }
          \\
          \addlinespace[0.6em]
          \toprule
          \addlinespace[0.6em]
          \copier[-][1] \seq[-] (\id[-][1] \tensor[-] \copier[-][1]) &\stackrel{(\copier[-]\text{-as})}{=}& \copier[-][1] \seq[-] (\copier[-][1] \tensor[-] \id[-][1]) & (\id[-][1] \tensor[-] \cocopier[-][1]) \seq[-] \cocopier[-][1] &\stackrel{(\cocopier[-]\text{-as})}{=}& (\cocopier[-][1] \tensor[-] \id[-][1]) \seq[-] \cocopier[-][1] \\
          \copier[-][1] \seq[-] (\id[-][1] \tensor[-] \discard[-][1]) &\stackrel{(\copier[-]\text{-un})}{=}& \id[-][1]  & (\id[-][1] \tensor[-] \codiscard[-][1]) \seq[-] \cocopier[-][1] &\stackrel{(\cocopier[-]\text{-un})}{=}& \id[-][1] \\
          \copier[-][1] \seq[-] \symm[-][1][1] &\stackrel{(\copier[-]\text{-co})}{=}& \copier[-][1] & \symm[-][1][1] \seq[-] \cocopier[-][1] &\stackrel{(\cocopier[-]\text{-co})}{=}& \cocopier[-][1] \\
          \addlinespace[0.6em]
          \midrule
          \addlinespace[0.6em]
          (\copier[-][1] \tensor[-] \id[-][1]) \seq[-] (\id[-][1] \tensor[-] \cocopier[-][1]) &\stackrel{(\text{F}^{\switchLabelS{-}})}{=}& (\id[-][1] \tensor[-] \copier[-][1]) \seq[-] (\cocopier[-][1] \tensor[-] \id[-][1]) & \copier[-][1] \seq[-] \cocopier[-][1] &\stackrel{(\text{S}^{\switchLabelS{-}})}{=}& \id[-][1] \\
          \addlinespace[0.6em]
          \midrule
          \addlinespace[0.6em]
           \multicolumn{6}{c}{
            \begin{array}{c@{\qquad}c}
              \begin{array}{r@{\,\,}c@{\,\,}l @{\qquad} r@{\,\,}c@{\,\,}l}
                \discard[-][1] \seq[-] \codiscard[-][1]
                &\stackrel{\eqref*{ax:minusDiscCodisc}}{\leq}&
                \id[-][1]
                &
                \copier[-][1] \seq[-] \cocopier[-][1]
                &\stackrel{\eqref*{ax:minusCopyCocopy}}{\leq}&
                \id[-][1]
                \\
                \id[-][0]
                &\stackrel{\eqref*{ax:minusCodiscDisc}}{\leq}&
                \codiscard[-][1] \seq[-] \discard[-][1]
                &
                \cocopier[-][1] \seq[-] \copier[-][1]
                &\stackrel{\eqref*{ax:minusCocopyCopy}}{\leq}&
                (\id[-][1] \tensor[-] \id[-][1])
            \end{array}
            &
              \begin{array}{r@{\,\,}c@{\,\,}l r@{\,\,}c@{\,\,}l}
                \copier[-][n] \seq[-] (c \tensor[-] c) &\stackrel{\eqref*{ax:comMinusLaxNat}}{\leq}& c \seq[-] \copier[-][m]
                \\
                \discard[-][n] &\stackrel{\eqref*{ax:discMinusLaxNat}}{\leq}& c \seq[-] \discard[-][m]
              \end{array}
            \end{array}
           }\\
          \addlinespace[0.6em]
          \toprule
          \addlinespace[0.6em]
          a \seq[+] (b \seq[-] c) &\stackrel{\eqref*{ax:leftLinDistr}}{\leq}& (a \seq[+] b) \seq[-] c & (a \seq[-] b) \seq[+] c &\stackrel{\eqref*{ax:rightLinDistr}}{\leq}& a \seq[-] (b \seq[+] c)
          \\
          \addlinespace[0.6em]
          \midrule
          \addlinespace[0.6em]
          \multicolumn{6}{c}{
              \begin{array}{cc}
                  \begin{array}{r@{\,\,}c@{\,\,}l r@{\,\,}c@{\,\,}l}
                      \id[+][n+m] &\stackrel{\eqref*{ax:tauSymmPlus}}{\leq}& \symm[+][n][m] \seq[-] \symm[-][m][n]
                      &
                      \symm[-][n][m] \seq[+] \symm[+][m][n] &\stackrel{\eqref*{ax:gammaSymmPlus}}{\leq}& \id[-][n+m]
                      \\
                      \id[+][n+m] &\stackrel{\eqref*{ax:tauSymmMinus}}{\leq}& \symm[-][n][m] \seq[-] \symm[+][m][n]
                      &
                      \symm[+][n][m] \seq[+] \symm[-][m][n] &\stackrel{\eqref*{ax:gammaSymmMinus}}{\leq}& \id[-][n+m]
                  \end{array}
                  &
                  \begin{array}{r@{\,\,}c@{\,\,}l r@{\,\,}c@{\,\,}l}
                      \id[+][n] &\stackrel{\eqref*{ax:tauRPlus}}{\leq}& R^\circ \seq[-] R^\bullet
                      &
                      R^\bullet \seq[+] R^\circ &\stackrel{\eqref*{ax:gammaRPlus}}{\leq}& \id[-][m]
                      \\
                      \id[+][m] &\stackrel{\eqref*{ax:tauRMinus}}{\leq}& R^\bullet \seq[-] R^\circ
                      &
                      R^\circ \seq[+] R^\bullet &\stackrel{\eqref*{ax:gammaRMinus}}{\leq}& \id[-][n]
                  \end{array}
              \end{array}
          }
          \\
          \addlinespace[0.6em]
          \midrule
          \addlinespace[0.6em]
          \id[+][n+m] &\stackrel{\eqref*{ax:tensorMinusIdPlus}}{\leq}& \id[+][n] \tensor[-] \id[+][m]
          &
          \id[-][n] \tensor[+] \id[-][m] &\stackrel{\eqref*{ax:tensorPlusIdMinus}}{\leq}& \id[-][n+m]
          \\
          (a \seq[-] b) \tensor[+] (c \seq[-] d) &\stackrel{\eqref*{ax:linStrn1}}{\leq}& (a \tensor[+] c) \seq[-] (b \tensor[-] d)
          &
          (a \tensor[-] c) \seq[+] (b \tensor[+] d) &\stackrel{\eqref*{ax:linStrn3}}{\leq}& (a \seq[+] b) \tensor[-] (c \seq[+] d)
          \\
          (a \seq[-] b) \tensor[+] (c \seq[-] d) &\stackrel{\eqref*{ax:linStrn2}}{\leq}& (a \tensor[-] c) \seq[-] (b \tensor[+] d)
          &
          (a \tensor[+] c) \seq[+] (b \tensor[-] d) &\stackrel{\eqref*{ax:linStrn4}}{\leq}& (a \seq[+] b) \tensor[-] (c \seq[+] d)
          \\
          \addlinespace[0.6em]
          \toprule
          \addlinespace[0.6em]
          \multicolumn{6}{c}{

          }
          \\
          \bottomrule
      \end{array}
  $}
  \caption{Axioms for $\NPR$. Here $a,b,c,d$ are properly typed terms.}
  \label{fig:textual axioms}
\end{figure}

\newpage

\begin{figure}[H]
  \centering
  {\tiny
  \resizebox{\columnwidth}{!}{$
  %
$}
  }
  \caption{Diagrammatic axioms for $\NPR$. Here $a,b,c,d$ are properly typed terms.}
  \label{fig:diagrammatic axioms}
\end{figure}

\newpage
\section{Dictionary}\label{app:dictionary}
\begin{table}[H]
\centering
    \begin{minipage}{\linewidth}
    \centering
    \def\arraystretch{1.5}
    \resizebox{\linewidth}{!}{
    %
    }
    \end{minipage}
    \caption{Correspondence between term and graphical notation; and diagrammatic conventions.}
\end{table}
\clearpage
\begin{table}[H]
    \centering
    \begin{subtable}{\textwidth}
        %
        \caption{Inductive definitions of the white structure.}
    \end{subtable}
    \centering
    \begin{subtable}{\textwidth}
        %
        \end{tabular}
        \caption{Inductive definitions of the black structure.}
    \end{subtable}
    \caption{}
    \label{tab:dictionary}
\end{table}

\newpage
 
\section{Appendix to Section \ref{sec:linbic}}\label{app:linbic}

\begin{proof}[Proof of Proposition \ref{prop:rlamorphism}]
    First, we prove that for all $a,b\colon X \to Y$ it holds that
    \begin{enumerate}\setcounter{enumi}{-1}
    \item if $a\leq b$ then $\rla{a} \geq \rla{b}$
    \end{enumerate}

        \noindent The proof is illustrated below.

    \begin{enumerate}\setcounter{enumi}{-1}
    \item
    \begin{align*}
            \rla{b} & =   \rla{b}\seq[+]\id[+][Y]  \\
             & \leq \rla{b}\seq[+] (a \seq[-] \rla{a} )  \tag{$\rla{a}\Vdash a$} \\
             & \leq ( \rla{b}\seq[+] a ) \seq[-] \rla{a}  \tag{\ref{ax:leftLinDistr}}  \\
             & \leq ( \rla{b}\seq[+] b ) \seq[-] \rla{a}  \tag{$a \leq b$} \\
             &  \leq  \id[-][Y] \seq[-] \rla{a} \tag{$\rla{b}\Vdash b$} \\
             &  =  \rla{a}  \\
    \end{align*}
    \end{enumerate}

    \noindent 
    We next illustrate that for all $a\colon X \to Y$ and $b\colon Y \to Z$
    \begin{enumerate}
    \item $\rla{(\id[+][X])}=\id[-][X]$ 
    \item $\rla{(\id[-][X])}=\id[+][X]$
    \item $\rla{(a \seq[+] b)} = \rla{b} \seq[-] \rla{a}$ 
    \item $\rla{(a \seq[-] b)} = \rla{b} \seq[+] \rla{a}$
    \end{enumerate}
    The proofs are displayed below.

    \begin{enumerate}
    
    \item Observe that $\id[+][X] = \id[+][X] \seq[-] \id[-][X]$ and $\id[-][X] \seq[+] \id[+][X] = \id[-][X]$. Thus, by Lemma \ref{lemma:uniquenessla}, $\rla{(\id[+][X])}=\id[-][X]$.
    
    \item Similarly, $\id[+][X] = \id[-][X] \seq[-] \id[+][X]$ and $\id[+][X] \seq[+] \id[-][X] = \id[-][X]$. Again, by Lemma \ref{lemma:uniquenessla}, $\rla{(\id[-][X])}=\id[+][X]$.

    \item The following two derivations
    
    \noindent\begin{minipage}{0.44\linewidth}
    \begin{align*}
            \id[+][X] \leq \;  &a \seq[-] \rla{a} \tag{$\rla{a}\Vdash a$} \\
             =\;  &(a \seq[+] \id[+][Y]) \seq[-] \rla{a} \\
             \leq\;  &(a \seq[+] (b \seq[-]\rla{b}) ) \seq[-] \rla{a}  \tag{$\rla{b}\Vdash b$} \\
             \leq\; &( (a \seq[+] b) \seq[-]\rla{b} ) \seq[-] \rla{a} \tag{\ref{ax:leftLinDistr}} \\
             =\;  &(a \seq[+] b) \seq[-] (\rla{b}  \seq[-] \rla{a})
    \end{align*}
    \end{minipage}\;\;\vline
    \begin{minipage}{0.54\linewidth}
    \begin{align*}
             (\rla{b}  \seq[-] \rla{a}) \seq[+] (a \seq[+] b)  =\; &( (\rla{b}  \seq[-] \rla{a}) \seq[+] a) \seq[+] b  \\
             \leq\;  &( \rla{b}  \seq[-] ( \rla{a} \seq[+] a)) \seq[+] b \tag{\ref{ax:rightLinDistr}}\\
             \leq\;  &( \rla{b}  \seq[-] \id[-][Y]) \seq[+] b \tag{$\rla{a}\Vdash a$} \\
             =\;  &\rla{b}  \seq[+] b \\
             \leq\; &\id[-][Z] \tag{$\rla{b}\Vdash b$}
     \end{align*}
    \end{minipage}\\
    show that $(\rla{b}  \seq[-] \rla{a}) \Vdash (a \seq[+] b)$. Thus, by Lemma \ref{lemma:uniquenessla}, $\rla{(a \seq[+] b)} = \rla{b} \seq[-] \rla{a}$.
    
    \item The following two derivations
    
    \noindent\begin{minipage}{0.44\linewidth}
    \begin{align*}
            \id[+][X] \leq\;&   a \seq[-] \rla{a} \tag{$\rla{a}\Vdash a$} \\
             =\;  &a \seq[-] (\id[+][Y] \seq[+] \rla{a}) \\
             \leq\; &  a \seq[-] ((b \seq[-]\rla{b})  \seq[+] \rla{a})  \tag{$\rla{b}\Vdash b$} \\
             \leq\; &  a \seq[-] (b \seq[-] (\rla{b}  \seq[+] \rla{a})) \tag{\ref{ax:rightLinDistr}} \\
             =\; &  (a \seq[-] b) \seq[-] (\rla{b}  \seq[+] \rla{a})
    \end{align*}
    \end{minipage}\;\;\vline
    \begin{minipage}{0.54\linewidth}
    \begin{align*}
            (\rla{b}  \seq[+] \rla{a}) \seq[+] (a \seq[-] b) =\;  &\rla{b}  \seq[+] (\rla{a} \seq[+] (a \seq[-] b))  \\
            =\;  &\rla{b}  \seq[+] (( \rla{a} \seq[+] a) \seq[-] b ) \tag{\ref{ax:leftLinDistr}} \\
            \leq\;  &\rla{b}  \seq[+] ( \id[-][Y] \seq[-] b) \tag{$\rla{a}\Vdash a$} \\
            =\; &   \rla{b}  \seq[+] b \\
            \leq\; & \id[-][Z] \tag{$\rla{b}\Vdash b$}
     \end{align*}
    \end{minipage}\\
    show that $(\rla{b}  \seq[+] \rla{a}) \Vdash (a \seq[-] b)$. Thus, by Lemma \ref{lemma:uniquenessla}, $\rla{(a \seq[-] b)} = \rla{b} \seq[+] \rla{a}$.
     \end{enumerate}    
\noindent 
The remaining cases are illustrated in the main text.
\end{proof}

\section{Appendix to Section \ref{sec:fobic}}

\subsection{Proof of Proposition \ref{prop:rlafunctor}}\label{app:functors}

In this appendix, we illustrate a proof of Proposition \ref{prop:rlafunctor}.
 In order to prove that $\op{(\cdot)} \colon \Cat{C} \to \opposite{\Cat{C}}$ is a morphism of fo-bicategories, it is convenient to observe that in a cocartesian bicategory there is an analogue operation $\opp{(\cdot)} \colon \Cat{C} \to \opposite{\Cat{C}}$ that is an identity-on-objects isomorphism, mapping every arrow $c \colon X \to Y$ to $\opp{c} \colon Y \to X$ defined as below on the left.

\begin{table}[H]
    \begin{minipage}{0.38\textwidth}
    \begin{center}
        $\opp{c} \defeq \daggerCirc[-]{c}[Y][X]$
    \end{center}
\end{minipage}
\hfill
\begin{minipage}{0.60\textwidth}
    \begin{center}
        \scalebox{0.85}{
        $
        \renewcommand{\arraystretch}{1.2} 
        \begin{array}{@{}cccc@{}}
            \toprule
            \multicolumn{4}{c}{
                \text{Properties of $\opp{(\cdot)} \colon (\Cat{C},\copier[+]\!, \cocopier[+]\!,  \copier[-]\!, \cocopier[-]) \to (\opposite{\Cat{C}},\cocopier[+]\!, \copier[+]\!,  \cocopier[-]\!, \copier[-])$}
            }
            \\
            \midrule
            \multicolumn{2}{c}{
                \text{if }c\leq d\text{ then }\opp{c} \leq \opp{d}
            }
            &
            \multicolumn{2}{c}{
                \opp{(\opp{c})}= c
            }
            \\
        \opp{(c \seq[+] d)} = \opp{d} \seq[+] \opp{c}
        &\opp{(\id[+][X])}=\id[+][X] 
        &\opp{(\cocopier[+][X])}= \copier[+][X]
        & \opp{(\codiscard[+][X])}= \discard[+][X]
        \\
        \opp{(c \tensor[+] d)} = \opp{c} \tensor[+] \opp{d} 
        & \opp{(\symm[+][X][Y])} = \symm[+][Y][X] 
        & \opp{(\copier[+][X])}= \cocopier[+][X]
        & \opp{(\discard[+][X])}= \codiscard[+][X]
        \\
        \midrule
        \opp{(c \seq[-] d)} = \opp{d} \seq[-] \opp{c}
        & \opp{(\id[-][X])} = \id[-][X]
        &\opp{(\cocopier[-][X])}= \copier[-][X]
        & \opp{(\codiscard[-][X])}= \discard[-][X]
        \\
        \opp{(c \tensor[-] d)} = \opp{c} \tensor[-] \opp{d} 
        & \opp{(\symm[-][X][Y])} = \symm[-][Y][X] 
        & \opp{(\copier[-][X])}= \cocopier[-][X]
        & \opp{(\discard[-][X])}= \codiscard[-][X] 
        \\
        \bottomrule
        \end{array}
        $}
    \end{center}
    \caption{}
    \label{table:oppproperties}
\end{minipage}
\end{table}

\begin{prop}\label{prop:oppcartesianfunctor}
$\opp{(\cdot)}\colon \Cat{C} \to \opposite{\Cat{C}}$ is an isomorphism of cocartesian bicategories, that is the rules in the first three rows of Table \ref{table:oppproperties} hold.
\end{prop}
\begin{proof}
	See Theorem 2.4 in~\cite{carboni1987cartesian}.
\end{proof}

\noindent 
In a fo-bicategory, it turns out that $\op{(\cdot)}$ and $\opp{(\cdot)}$ coincide. For example, in $\Rel$, both operations correspond to taking the converse of a relation. To establish this result, it is helpful to first examine their behaviour on identities (Lemma~\ref{lm:id dagger}) and their interaction with $\rla{(\cdot)}$ (Lemma~\ref{lm:daggeradj}).

\begin{lem}\label{lm:id dagger}
For all $X$ it holds that $\id[-][X] = \op{(\id[-][X])}$ and $\id[+][X] = (\id[+][X])^\ddagger$.
\end{lem}
\begin{proof}
    First we prove the following equalities:

    \begin{center}
        \begin{enumerate*}[itemjoin={\qquad}]
            \item $\copierCirc[-] = 
    \InputIfFileExists{mixedSpiders.tikz}{}{\input{tikz/mixedSpiders.tikz}}
$
            \item $
    \InputIfFileExists{axioms/linadj/comPmon.tikz}{}{\input{tikz/axioms/linadj/comPmon.tikz}}
 = \idCirc[-]$
            \item $
    \InputIfFileExists{axiomsNEW/linadj2/comPmon2.tikz}{}{\input{tikz/axiomsNEW/linadj2/comPmon2.tikz}}
 = \idCirc[-]$.
        \end{enumerate*}
    \end{center}
    The proofs are shown below.

    \begin{enumerate}
        \item is proved as follows:  $\copierCirc[-] \stackrel{\eqref{ax:comPlusUnit}}{=} 
    \InputIfFileExists{mixedSpiders2.tikz}{}{\input{tikz/mixedSpiders2.tikz}}
 \stackrel{\eqref{ax:bwFrob2}}{=} 
    \InputIfFileExists{mixedSpiders.tikz}{}{\input{tikz/mixedSpiders.tikz}}
$.
        \item is \eqref{ax:gammaCocopierPlus} for the left-to-right inclusion and the other inclusion is proved below:
        \input{tikz/proofs/mixedSpecFrob.tex}
        \item is proved analogously, but exploiting \eqref{ax:leftLinDistr} and \eqref{ax:gammaCopierPlus}.
    \end{enumerate}

    \medskip

	\noindent Now we are ready to show that $\id[-][X] = \op{(\id[-][X])}$.
	\input{tikz/proofs/idDagger}
    The proof for $\id[+][X] = \opp{(\id[+][X])}$ is analogous.
\end{proof}

\begin{lem}\label{lm:daggeradj}
  For all $a \colon X \to Y$ it holds that $\rla{(a^\dagger)} = (\rla{a})^\ddagger$.
\end{lem}
\begin{proof}
  The proof follows from the fact that $(\copier[+], \discard[+])$ is right linear adjoint to $(\cocopier[-], \codiscard[-])$, Proposition~\ref{prop:rlamorphism} and the definition of $(\cdot)^\dagger$ and $(\cdot)^\ddagger$.
\end{proof}

\begin{lem}\label{lm:dagger}
  For all $a \colon X \to Y$ it holds that $a^\dagger = a^\ddagger$.
\end{lem}

\begin{proof}
	We prove the inclusion $a^\dagger \leq a^\ddagger$ (left) by means of Lemma~\ref{lm:residuation} and the other inclusion (right) directly: \\
	\begin{minipage}{0.48\linewidth}
		\begin{align*}
		(a^\ddagger \seq[-] \rla{(a^\dagger)}) =\;  &a^\ddagger \seq[-] (\rla{a})^\ddagger \tag{Lemma~\ref{lm:daggeradj}}\\
			 =\;    &(\rla{a} \seq[-] a)^\ddagger  \tag{Table~\ref{table:oppproperties}}   \\
			 \geq\; &(\id[+][Y])^\ddagger   \tag{$\rla{a}\Vdash a$}\\
			=\;     &\id[+][Y] \tag{Lemma \ref{lm:id dagger}}
		\end{align*}
	\end{minipage}
	\;\;\vline
	\begin{minipage}{0.50\linewidth}
		\begin{align*}
			\opp{a} =\; &\opp{(\op{(\op{a})})} \tag{$\op{(\cdot)}$ is an iso} \\
			\leq\; &\opp{(\opp{(\op{a})})} \tag{$\op{a} \leq \opp{a}$} \\
			=\; &\op{a} \tag{$\opp{(\cdot)}$ is an iso}
		\end{align*}\qedhere
	\end{minipage}
	\end{proof}

\begin{lem}\label{lemma:opfunctor}
$\op{(\cdot)} \colon \Cat{C} \to \opposite{\Cat{C}}$ is an isomorphism of fo-bicategories, namely all the laws in Table \ref{table:daggerproperties}.(a) hold.
\end{lem}
\begin{proof}
Follows from Lemma~\ref{lm:dagger} and the fact that $\op{(\cdot)}$ preserves the positive structure (Proposition \ref{prop:opcartesianfunctor})  and  $(\cdot)^\ddagger$ preserve the negative structure (Proposition \ref{prop:oppcartesianfunctor}). For instance,
to prove that  $\op{(a \seq[-] b)} = \op{b} \seq[-] \op{a}$, it is enough to observe that  $\op{(a \seq[-] b)} = (a \seq[-]b)^\ddagger$ and that $ (a \seq[-]b)^\ddagger = b^\ddagger \seq[-] a^\ddagger$.
\end{proof}

\begin{cor}\label{corrlaop}
$\rla{(\op{c})} = \op{(\rla{c})}$.
\end{cor}
\begin{proof}
By Lemma \ref{lemma:opfunctor} and Lemma \ref{lm:adjfunctor}.
\end{proof}

\begin{lem}\label{lemma:doublelinearadjoint}
For all $a\colon X \to Y$ it holds that $\rla{(\rla{a})}=a$.
\end{lem}
\begin{proof}
The following two derivations

\noindent\begin{minipage}{0.44\linewidth}
\begin{align*}
		\id[+][Y]
		 = &\op{(\id[+][Y])} \tag{Proposition \ref{prop:opcartesianfunctor}} \\
		 \leq &\op{( \op{a} \seq[-]\rla{(\op{a})})} \tag{$\rla{(\op{a})} \Vdash \op{a}$}\\
		 = &\op{( \op{a} \seq[-]\op{(\rla{a})})} \tag{Corollary \ref{corrlaop}} \\
		 =  &\op{(\op{( \rla{a} \seq[-]a)})}  \tag{Lemma \ref{lemma:opfunctor}} \\
		=  &\rla{a} \seq[-] a  \tag{Proposition \ref{prop:opcartesianfunctor}}
\end{align*}
\end{minipage}\;\;\vline
\begin{minipage}{0.54\linewidth}
\begin{align*}
    \id[-][X]
	= &\op{(\id[-][X])} \tag{Lemma \ref{lemma:opfunctor}} \\
	\geq &\op{( \rla{(\op{a})} \seq[+]\op{a})} \tag{$\rla{(\op{a})} \Vdash \op{a}$}\\
	= &\op{( \op{(\rla{a})}  \seq[+] \op{a}  )} \tag{Corollary \ref{corrlaop}} \\
	=  &\op{(\op{( a \seq[+] \rla{a} )})}  \tag{Proposition \ref{prop:opcartesianfunctor}} \\
	=  &a \seq[+] \rla{a}  \tag{Proposition \ref{prop:opcartesianfunctor}}
\end{align*}
\end{minipage}\\
prove that the right linear adjoint of $\rla{a}$ is $a$. Thus, by Lemma \ref{lemma:uniquenessla}, $\rla{(\rla{a})}=a$. \qedhere

\end{proof}

\begin{lem}\label{lemma:rlafunctor}
$\rla{(\cdot)}\colon \Cat{C} \to \opposite{(\co{\Cat{C}})}$ is an isomorphisms of fo-bicategories, namely all the laws in Table \ref{table:rlaproperties}.(b) hold.
\end{lem}

\begin{proof}
By  Proposition \ref{prop:rlamorphism}, $\rla{(\cdot)}\colon \Cat{C} \to \opposite{(\co{\Cat{C}})}$ is a morphism of linear bicategories.
Observe that $\opposite{(\co{\Cat{C}})}$ carries the structure of a cartesian bicategory where the positive comonoid is $(\cocopier[-], \codiscard[-])$ and the positive monoid is $(\copier[-], \discard[-])$. By Definition \ref{def:linear bicategory}.4, one has that $\rla{(\copier[+])} = \cocopier[-]$, $\rla{(\discard[+])}=\codiscard[-]$ and $\rla{(\cocopier[+])} = \copier[-]$, $\rla{(\codiscard[+])}=\discard[-]$. Thus $\rla{(\cdot)}\colon \Cat{C} \to \opposite{(\co{\Cat{C}})}$ is a morphism of cartesian bicategories.

By Lemma \ref{lemma:doublelinearadjoint}, we also immediately know that $\rla{(\copier[-])} = \cocopier[+]$, $\rla{(\discard[-])}=\codiscard[+]$ and $\rla{(\cocopier[-])} = \copier[+]$, $\rla{(\codiscard[-])}=\discard[+]$. Thus, $\rla{(\cdot)}\colon \Cat{C} \to \opposite{(\co{\Cat{C}})}$ is a morphism of cocartesian bicategories. Thus, it is a morphism of fo-bicategories.

The fact that it is an isomorphism is immediate by Lemma \ref{lemma:doublelinearadjoint}.
\end{proof}

\begin{proof}[Proof of Proposition \ref{prop:rlafunctor}]
By Lemmas \ref{lemma:opfunctor} and \ref{lemma:rlafunctor}.
\end{proof}

\subsection{Appendix to Section \ref{sec:freely}}\label{sec:appfreely}
In order to prove Proposition \ref{prop:LCBfo} it is convenient to use the following function on diagrams and then prove that it maps every diagram in its  right (Lemma \ref{lemma:alphar}) and left (Lemma \ref{lemma:alphal}) linear adjoint.

\begin{defi}\label{defalpha}
The function $\alpha\colon \NPR \to \NPR$ is inductively defined as follows.
\[
	\begin{array}{@{}ccccc@{}}
        \alpha{(c \seq[+] d)} \defeq \alpha{(d)} \seq[-] \alpha{(c)}   %
        &\alpha{(\id[+][1])}\defeq\id[-][1]  %
        &\alpha{(\cocopier[+][1])}\defeq \copier[-][1] %
        & \alpha{(\codiscard[+][1])}\defeq \discard[-][1] %
        & \alpha{(R^\circ)} \defeq R^\bullet
        \\
        \alpha{(c \tensor[+] d)} \defeq \alpha{(c)} \tensor[-] \alpha{(d)}  %
        & \alpha{(\symm[+][1][1])} \defeq \symm[-][1][1]  %
        & \alpha{(\copier[+][1])}\defeq \cocopier[-][1] %
        & \alpha{(\discard[+][1])}\defeq \codiscard[-][1] %
        \\[1.2em]
        \alpha{(c \seq[-] d)} \defeq \alpha{(d)} \seq[+] \alpha{(c)} %
        & \alpha{(\id[-][1])} \defeq \id[+][1] %
        &\alpha{(\cocopier[-][1])}\defeq \copier[+][1]%
        & \alpha{(\codiscard[-][1])}\defeq \discard[-][1] %
        & \alpha(R^\bullet) \defeq R^\circ
        \\
        \alpha{(c \tensor[-] d)} \defeq \alpha{(c)} \tensor[+] \alpha{(d)}  %
        & \alpha{(\symm[-][1][1])} \defeq \symm[+][1][1]  %
        & \alpha{(\copier[-][1])}\defeq \cocopier[+][1] %
        & \alpha{(\discard[-][1])}\defeq \codiscard[+][1] %
    \end{array}
\]
\end{defi}

\begin{lem}\label{lemma:alphar}
For all terms $c\colon n \to m$ in $\NPR$, $\id[+][n] \syninclusion c \seq[-]\alpha(c)$ and $\alpha(c) \seq[+] c \syninclusion \id[-][m]$.
\end{lem}
\begin{proof}
The proof goes by induction on $c$.
For the base cases of black and white (co)monoid, it is immediate by the axioms in the first block of Figure \ref{fig:fo bicat axioms}.%
For $R^\circ$, $R^\bullet$, $\symm[+]$ and $\symm[-]$, it is immediate by the axioms in the bottom Figure \ref{fig:closed lin axioms}.
For $\id[+]$ and $\id[-]$ is trivial. For the inductive cases of $\seq[+]$, $\seq[-]$, $\tensor[+]$ and $\tensor[-]$ one can reuse exactly the proof of Proposition \ref{prop:rlamorphism}.
\end{proof}

\begin{lem}\label{lemma:alphainv}
For all terms $c\colon n \to m$ in $\NPR$, $\alpha(\alpha(c))=c$.
\end{lem}
\begin{proof}
The proof goes by induction on $c$. For the base cases, it is immediate by Definition \ref{defalpha}. For the inductive cases, one has just to use the definition and the inductive hypothesis. For instance $\alpha(\alpha(a\seq[+] b))$ is, by Definition \ref{defalpha}, $\alpha (\alpha(a) \seq[-] \alpha(b))$ which, by Definition \ref{defalpha}, is $\alpha(\alpha(a)) \seq[+] \alpha(\alpha(b))$ that, by induction hypothesis, is $a \seq[+] b$.
\end{proof}

\begin{lem}\label{lemma:alphaco}
For all terms $c,d\colon n \to m$ in $\NPR$, if $c \syninclusion d$, then $\alpha(d) \syninclusion \alpha (c)$.
\end{lem}
\begin{proof}
Observe that the axioms in Figures \ref{fig:cb axioms}, \ref{fig:cocb axioms}, \ref{fig:closed lin axioms} and \ref{fig:fo bicat axioms} are closed under $\alpha$, namely if $c\leq d$ is an axiom also $\alpha(d) \leq \alpha(c)$ is an axiom.
\end{proof}

\begin{lem}\label{lemma:alphal}
For all terms $c\colon n \to m$ in $\NPR$, $\id[+][m] \syninclusion \alpha(c) \seq[-] c$ and $c  \seq[+] \alpha(c) \syninclusion \id[-][n]$.
\end{lem}
\begin{proof}
By Lemma \ref{lemma:alphar}, it holds that \[ \id[+][n] \syninclusion c \seq[-]\alpha(c) \text{ and } \alpha(c) \seq[+] c \syninclusion \id[-][m]\text{.}\]
By Lemma \ref{lemma:alphaco}, one can apply $\alpha$ to all the sides of the two inequalities to get
 \[   \alpha( c \seq[-]\alpha(c) ) \syninclusion \alpha( \id[+][n])  \text{ and }   \alpha( \id[-][m] ) \syninclusion \alpha( \alpha(c) \seq[+] c)  \text{.}\]
That, by Definition \ref{defalpha} gives exactly
 \[   \alpha(\alpha(c) )  \seq[+] \alpha( c)  \syninclusion  \id[-][n]  \text{ and }   \id[+][m] \syninclusion  \alpha (c) \seq[-]  \alpha( \alpha(c))  \text{.}\]
By Lemma \ref{lemma:alphainv}, one can conclude that
 \[   c  \seq[+] \alpha( c)  \syninclusion  \id[-][n]  \text{ and }   \id[+][m] \syninclusion  \alpha (c) \seq[-] c  \text{.}\qedhere\]
\end{proof}

\begin{proof}[Proof of Proposition \ref{prop:LCBfo}]
By Lemmas \ref{lemma:alphar} and \ref{lemma:alphal}, the diagram $\alpha(c)$ is both the right and the left linear adjoint of any diagram $c$. Thus $\LCB$ is a closed linear bicategory.

Next, we show that $(\LCB^\circ, \copier[+], \cocopier[+])$ is a cartesian bicategory: for all objects $n\in \nat$, $\copier[+][n]$, $\discard[+][n]$, $\cocopier[+][n]$ and $\codiscard[+][n]$ are inductively defined as in Table \ref{fig:sugar}. Observe that such definitions guarantees that the coherence conditions in Definition \ref{def:cartesian bicategory}.(5) are satisfied. The conditions in Definition \ref{def:cartesian bicategory}.(1).(2).(3).(4) are the axioms in Figure \ref{fig:cb axioms} (and appear in the term version in Figure \ref{fig:textual axioms}) that we have used to generate $\syninclusion$. %

Similarly, $(\LCB^\bullet, \copier[-], \cocopier[-])$ is a cocartesian bicategory: for all objects $n\in \nat$, $\copier[-][n]$, $\discard[-][n]$, $\cocopier[-][n]$ and $\codiscard[-][n]$ are inductively defined as in Table \ref{fig:sugar}. Again,  the coherence conditions are satisfied by construction. The other conditions are  the axioms in Figure \ref{fig:cocb axioms} (and appear in the term version in Figure \ref{fig:textual axioms}) that, by construction, are in $\syninclusion$.
To conclude that $\LCB$ is a first-order bicategory we have to check that the conditions in Definition \ref{def:fobicategory}.(4),(5). But these are exactly the axioms in Figure \ref{fig:fo bicat axioms} (and appear in the term version in Figure \ref{fig:textual axioms}).
\end{proof}

\section{Additional results on the trivial theories of propositional calculus}\label{app:trivialprop}

\begin{lem}\label{lemma:trivalallequal}
Let $\T{T}$ be a trivial theory and $c\colon n \to m+1, d\colon m+1 \to n$ be arrows of $\LCB[\T{T}]$.
Then it holds that:
\[ {
    \InputIfFileExists{tikz/trivialCollapse/nm1Top.tikz}{}{\input{tikz/tikz/trivialCollapse/nm1Top.tikz}}
} \!\syninclusionT{\T{T}}\! {
    \InputIfFileExists{tikz/trivialCollapse/c.tikz}{}{\input{tikz/tikz/trivialCollapse/c.tikz}}
} \!\syninclusionT{\T{T}}\! {
    \InputIfFileExists{tikz/trivialCollapse/nm1Bot.tikz}{}{\input{tikz/tikz/trivialCollapse/nm1Bot.tikz}}
} \;\; \text{ and } \;\; {
    \InputIfFileExists{tikz/trivialCollapse/m1nTop.tikz}{}{\input{tikz/tikz/trivialCollapse/m1nTop.tikz}}
} \syninclusionT{\T{T}} {
    \InputIfFileExists{tikz/trivialCollapse/d.tikz}{}{\input{tikz/tikz/trivialCollapse/d.tikz}}
} \syninclusionT{\T{T}} {
    \InputIfFileExists{tikz/trivialCollapse/m1nBot.tikz}{}{\input{tikz/tikz/trivialCollapse/m1nBot.tikz}}
}. \]
\end{lem}
\begin{proof}
    \input{tikz/proofs/generalTrivial.tex}
    The proof for $d$ follows a similar reasoning.
\end{proof}

\begin{prop}\label{prop:propcalc} %
For every diagram $a\colon 0 \to 0$ in $\LCB[\T{P}]$ there exists a $\synequivalenceT{\T{P}}$-equivalent diagram generated by the following grammar where $R\in \sign$.%
\[ \Circ{c} ::= \propVar[+]{} \; \mid \; \propVar[-]{} \; \mid \; \propVar[+]{R} \; \mid \; \propOpVar[-]{R} \; \mid \; \propSeqCirc[+]{c}{c} \; \mid \; \propSeqCirc[-]{c}{c} \] 
\end{prop}
\begin{proof}
By induction on $a \colon 0 \to 0$.
Observe that there are only four base cases:  $\id[+][0]$, $\id[-][0]$, $R^\circ$ and $R^\bullet$. These already appear in the grammar above.
We have the usual four inductive cases:
\begin{enumerate}
\item $a= c\seq[+] d$. There are two sub-cases: either $c,d\colon 0 \to 0$ or $c\colon 0 \to n+1$ and $d\colon n+1 \to 0$. In the former we can use the inductive hypothesis to get $c'$ and $d'$ generated by the above grammar such that $c'\synequivalenceT{\T{P}} c$ and $d' \synequivalenceT{\T{P}}d$. Thus $a$ is $\synequivalenceT{\T{P}}$-equivalent to $c' \seq[+] d'$ that is generated by the above grammar. 

Consider now the case where $c\colon 0 \to n+1$ and $d\colon n+1 \to 0$. By Lemma \ref{lemma:trivalallequal}, $c\synequivalenceT{\T{P}} \codiscard[+][n+1]$ and $d\synequivalenceT{\T{P}} \discard[-][n+1]$. By axiom \eqref{ax:gammaDiscardMinus}, $\codiscard[+][n+1] \seq[+] \discard[-][n+1] \synequivalenceT{\T{P}} \id[-][0]$. Thus $a \synequivalenceT{\T{P}} \id[-][0]$.
\item $a=c \tensor[+] d$. Note that, in this case both $c$ and $d$ must have type $0 \to 0$. Thus we can use the inductive hypothesis to get $c'$ and $d'$ generated by the above grammar such that $c'\synequivalenceT{\T{P}} c$ and $d' \synequivalenceT{\T{P}}d$.  Thus $a \synequivalenceT{\T{P}} c' \tensor[+]d' \structuralcong c' \seq[+] d'$. Note that $c' \seq[+] d'$ is generated by the above grammar.
\item $a=c \seq[-] d$. The proof follows symmetrical arguments to the case $c\seq[+] d$.
\item $a=c \tensor[-] d$. The proof follows symmetrical arguments to the case $c\tensor[+] d$. \qedhere
\end{enumerate}
\end{proof}

\section{Appendix to Section \ref{sec:completeness}}\label{app:completeness}

\begin{proof}[Proof of Lemma \ref{lemma:phi}]
    The proof goes by induction on the rules in~\eqref{eq:pc}.

For the rule $(id)$ we have three cases: either $(c,d) \in \T{I}$ or $(c,d) \in \syninclusionT{\sign'}$ or $(c,d) \in \TMAP{k}$. 

If $(c,d) \in \T{I}$ then, by Lemma~\ref{lm:phi characterization}, $\phi(c) = \phiCirc[+]{c} \syninclusionT{\T{T}} \phiCirc[+]{d} = \phi(d)$.

If $(c,d) \in \syninclusionT{\sign'}$ then $(c,d)$ has been obtained by instantiating the axioms in Figures~\ref{fig:cb axioms},\ref{fig:cocb axioms} and \ref{fig:closed lin axioms} with diagrams containing $k$. Therefore, we need to show that $\phi$ preserves these axioms. In the following we show only a few of them. The remaining ones follow similar reasonings.

For $\eqref{ax:comPlusLaxNat}$ the following holds:
\input{tikz/proofs/phiLemma/copier.tex}

For $\eqref{ax:discPlusLaxNat}$ the following holds:
\input{tikz/proofs/phiLemma/discard.tex}

For $\eqref{ax:tauRPlus}$ and $\eqref{ax:gammaRPlus}$ the following holds:
\input{tikz/proofs/phiLemma/linAdjRel.tex}

For $\eqref{ax:leftLinDistr}$ the following holds:
\input{tikz/proofs/phiLemma/linDistr.tex}

For $\eqref{ax:linStrn2}$ the following holds:
\input{tikz/proofs/phiLemma/linStr.tex}

Similar to the previous argument, if $(c,d) \in \TMAP{k}$ then it is enough to show that $\phi$ preserves the axioms in $\TMAP{k}$, namely that
\input{tikz/proofs/phiLemma/copierK.tex}
and
\input{tikz/proofs/phiLemma/discardK.tex}

The base case $(r)$ is trivial, while the proof for the remaining rules follows a straightforward inductive argument.
\end{proof}
    
\begin{proof}[Proof of Proposition \ref{prop:non-contraddiction}]
    By using the well-known fact that $\pcong{\cdot}$ preserves   chains, one can easily see that
    \begin{equation}\label{eq:bla}
    \syninclusionT{\mathbb{T}} = \bigcup_{i \in I} \syninclusionT{\mathbb{T}_i}
    \end{equation}
    The interested reader can find all the details in Appendix \ref{app:onprec}, Lemma \ref{lemma:noncontalltogether}.
    \begin{enumerate}
    \item Suppose that $\mathbb{T}$ is contradictory. By definition $\id[+][0] \syninclusionT{\mathbb{T}} \id[-][0]$ and then, by \eqref{eq:bla}, $(\id[+][0], \id[-][0]) \in \bigcup_{i \in I} \syninclusionT{\mathbb{T}_i}$. Thus there exists an $i\in I$ such that $\id[+][0] \syninclusionT{\mathbb{T}_i} \id[-][0]$. Against the hypothesis.
    \item Suppose that $\mathbb{T}$ is trivial. By definition $\codiscard[+][1] \syninclusionT{\mathbb{T}} \codiscard[-][1]$ and then, by \eqref{eq:bla}, $(\codiscard[+][1], \codiscard[-][1]) \in \bigcup_{i \in I} \syninclusionT{\mathbb{T}_i}$. Thus there exists an $i\in I$ such that $\codiscard[+][1] \syninclusionT{\mathbb{T}_i} \codiscard[-][1]$. Against the hypothesis.\qedhere
    \end{enumerate}
\end{proof}    

\begin{proof}[Proof of Proposition \ref{prop:syntacticallycomplete}]
    The proof of this proposition relies on Zorn's Lemma~\cite{zorn1935remark} which states that if, in a non empty poset $L$ every chain has a least upper bound, then $L$ has at least one maximal element.
    
    We consider the set $\Gamma$ of all non-contradictory theories on $\Sigma$ that include $\mathbb{I}$, namely 
    \[\Gamma \defeq \{\mathbb{T}=(\Sigma, \mathbb{J})  \mid \mathbb{T} \text{ is non-contradictory and }\mathbb{I}\subseteq \mathbb{J}\}\text{.}\]
    Observe that the set $\Gamma$ is non empty since there is at least $\mathbb{T}$ which belongs to $\Gamma$.
    
    Let $\Lambda \subseteq \Gamma$ be a chain, namely $\Lambda = \{\mathbb{T}_i=(\Sigma, \mathbb{J}_i) \in \Gamma \mid i \in I\}$ for some linearly ordered set $I$ and if $i\leq j$, then $\mathbb{J}_i \subseteq \mathbb{J}_j$. By Proposition \ref{prop:non-contraddiction}, the theory $(\Sigma, \bigcup_{i\in I}\mathbb{J}_i)$ is non-contradictory and thus it belongs to $\Gamma$.
    
    We can thus use Zorn's Lemma: the set $\Gamma$ has a maximal element $\mathbb{T'}=(\Sigma, \mathbb{I}')$. By definition of $\Gamma$, $\mathbb{I}\subseteq\mathbb{I'}$ and, moreover,  $\mathbb{T'}$ is non-contradictory.
    
    We only need to prove that $\mathbb{T'}$ is syntactically complete, i.e., for all $c \colon 0 \to 0$, either $\id[+][0] \precongR{\T{T'}} c$ or $\id[+][0] \precongR{\T{T'}} \overline{c}$.
    Assume that $\id[+][0] \not \hspace{-0.2cm}\precongR{\T{T}} c$. Thus $\mathbb{I}'$ is \emph{strictly} included into $\mathbb{I}' \cup \{(\id[+][0], c) \}$. By maximality of $\mathbb{T'}$ in $\Gamma$, we have that the theory $\mathbb{T''}=(\Sigma, \mathbb{I}' \cup \{(\id[+][0], c) \})$ is contradictory, i.e., $\id[+][0]\precongR{\T{T}''} \id[-][0]$. By the deduction theorem (Theorem \ref{th:deduction}), $c \precongR{\T{T'}} \id[-][0]$. Therefore $\id[+][0] \precongR{\T{T'}} \nega{c}$.
\end{proof}

\begin{proof}[Proof of Theorem \ref{thm:nontrivialALL}]
    This proof reuses the well-known arguments reported e.g. in \cite{lascar2001mathematical}.
    
    We first illustrate a procedure to add Henkin witnesses without losing the property of being non-trivial.
    
    Take an enumeration of diagrams in $\LCB[\Sigma] [1,0]$ and write $c_i$ for the $i$-th diagram.
    
    For all natural numbers $n \in \nat$, we define 
    \[
        \begin{array}{c@{\qquad}c@{\qquad}c}
            \Sigma^n \defeq \Sigma \cup \{k_i\colon 0 \to 1 \mid i \leq n\}  
            &  
            \mathbb{I}^n \defeq \mathbb{I} \cup \TMAP{k_i} \cup \bigcup_{i \leq n} \mathbb{W}_{k_i}^{c_i}
            &
            \T{T}^n \defeq (\Sigma^n, \mathbb{I}^n).
        \end{array}
    \]
    By applying Lemma \ref{lemma:addingHenkin} $n$-times, one has that $\T{T}^n$ is non-trivial. %
    Define now
    \[ \Sigma_0 \defeq \bigcup_{i\in \nat}\Sigma^i  \qquad  \mathbb{I}_0 \defeq \bigcup_{i \in \nat} \mathbb{I}^j  \qquad \T{T}_0 \defeq (\Sigma_0, \mathbb{I}_0).\]
    
    Since $\T{T}^0 \subseteq \T{T}^1 \subseteq \dots \subseteq \T{T}^n \subseteq \dots$ are all non-trivial, then by Proposition \ref{prop:non-trivialchain}.2, we have that $\T{T}_0$ is non-trivial. One must not jump to the conclusion that $\T{T}_0$ has Henkin witnesses: all the diagrams in $\LCB[\Sigma] [1,0]$ have Henkin witnesses, but in $\T{T}_0$ we have more diagrams, since we have added the constants $k_i$ to $\Sigma_0$.
    
    We thus repeat the above construction, but now for diagrams in  $\LCB[\Sigma_0] [1,0]$. We define
    \[
        \begin{array}{c@{\qquad}c@{\qquad}c}
            \Sigma_1 \defeq \Sigma_0 \cup \{k_c \mid c \in \LCB[\Sigma_0] [1,0]\} &  \mathbb{I}_1 \defeq \mathbb{I}_0 \cup \TMAP{{k_c}} \cup \mathbb{W}_{k_c}^{c} & \T{T}_1 \defeq (\Sigma_1, \mathbb{I}_1).
        \end{array}
    \]
    The theory $\T{T}_1$ is non-trivial but has Henkin witnesses only for the diagrams in $\LCB[\Sigma_0]$.

    Thus, for all natural numbers $n \in \nat$, we define 
    \[
        \begin{array}{c@{\qquad}c@{\qquad}c}
            \Sigma_{n+1} \defeq \Sigma_{n} \cup \{k_c \mid c \in \LCB[\Sigma_n] [1,0]\} &  \mathbb{I}_{n+1} \defeq \mathbb{I}_{n} \cup \TMAP{{k_c}} \cup \mathbb{W}_{k_c}^{c} & \T{T}_{n+1} \defeq (\Sigma_{n+1}, \mathbb{I}_{n+1})
        \end{array}
    \]
    and
    \[ \Sigma' \defeq \bigcup_{i\in \nat}\Sigma_i  \qquad  \mathbb{I}' \defeq \bigcup_{i \in \nat} \mathbb{I}_i  \qquad \T{T}' \defeq (\Sigma ', \mathbb{I}'). \]
    Since $\T{T}_0 \subseteq \T{T}_1 \subseteq \dots \subseteq \T{T}_n \subseteq \dots$ are all non-trivial, then by Proposition \ref{prop:non-trivialchain}.2, we have that $\T{T}'$ is also non-trivial. Now $\T{T}'$ has Henkin witnesses: if $c\in \LCB[\Sigma'] [0,1]$, then there exists $n \in \nat$ such that $c\in \LCB[\Sigma_n] [0,1]$. By definition of $\mathbb{I}_n$, it holds that $\mathbb{W}_{k_c}^{c} \subseteq \mathbb{I}_{n+1}$ and thus $\mathbb{W}_{k_c}^{c} \subseteq \mathbb{I}'$. 
    
    Summarising, we manage to build a theory $\T{T}'=(\Sigma',\mathbb{I}')$ that has Henkin witnesses and it is non-trivial. By Lemma \ref{lemma:contraddictoryimpliestrivial}, $\T{T}'$ is non-contradictory. We can thus use Proposition \ref{prop:syntacticallycomplete}, to obtain a theory $\T{T}''=(\Sigma',\mathbb{I}'')$ that is syntactically complete and non-contradictory. Observe that $\T{T}''$ has Henkin witnesses, since the signature $\Sigma'$ is the same as in $\T{T}'$ and $\mathbb{I}'\subseteq \mathbb{I}''$.
\end{proof}

\section{Appendix to Section \ref{sec:encodings}}\label{app:encoding}
\begin{proof}[Proof of Proposition \ref{prop:cr interpretation}]
  The proof is by induction on $E$. The base cases are trivial. The inductive cases are shown below.
  {\allowdisplaybreaks
  \begin{align*}
      \interpretationFunctor{(\enc{E_1 \seq[+] E_2})} &= \interpretationFunctor{(\enc{E_1} \seq[+] \enc{E_2})} \tag{Table~\ref{table:cr encoding}}  \\
                                                      &= \interpretationFunctor{(\enc{E_1})} \seq[+] \interpretationFunctor{(\enc{E_2})} \tag{\ref{fig:semantics}} \\
                                                      &\tag{\text{Ind. hyp.}}{=} \dsemRel{E_1} \seq[+] \dsemRel{E_2}  \\
                                                      &\tag{\ref{eq:sematicsExpr}}{=} \dsemRel{E_1 \seq[+] E_2} 
                                                      \\[8pt]
      \interpretationFunctor{(\enc{E_1 \seq[-] E_2})} & \tag{\text{Table }\ref{table:cr encoding}}{=} \interpretationFunctor{(\enc{E_1} \seq[-] \enc{E_2})}  \\
                                                      &\tag{\ref{fig:semantics}}{=} \interpretationFunctor{(\enc{E_1})} \seq[-] \interpretationFunctor{(\enc{E_2})} \\
                                                      &\tag{\text{Ind. hyp.}}{=} \dsemRel{E_1} \seq[-] \dsemRel{E_2}  \\
                                                      &\tag{\ref{eq:sematicsExpr}}{=} \dsemRel{E_1 \seq[-] E_2} 
                                                      \\[8pt]
      \interpretationFunctor{(\enc{E_1 \cap E_2})} & \tag{\text{Table }\ref{table:cr encoding}}{=} \interpretationFunctor{(\copier[+][1] \! \seq[+] (\enc{E_1} \tensor[+] \enc{E_2}) \seq[+] \! \cocopier[+][1])} \\
                                                   &\tag{\ref{fig:semantics}}{=} \interpretationFunctor{(\copier[+][1])} \seq[+] (\interpretationFunctor{(\enc{E_1})} \tensor[+] \interpretationFunctor{(\enc{E_2})}) \seq[+] \interpretationFunctor{(\cocopier[+][1])} \\
                                                   & \tag{\ref{fig:semantics}}{=} \copier[+][X] \seq[+] (\interpretationFunctor{(\enc{E_1})} \tensor[+] \interpretationFunctor{(\enc{E_2})}) \seq[+] \cocopier[+][X] \\
                                                   &\tag{\text{Ind. hyp.}}{=} \copier[+][X] \seq[+] (\dsemRel{E_1} \tensor[+] \dsemRel{E_2}) \seq[+] \cocopier[+][X] \\
                                                   & \tag{\ref{eq:def:cap}}{=} \dsemRel{E_1} \cap \dsemRel{E_2} \\
                                                   &\tag{\ref{eq:sematicsExpr}}{=}  \dsemRel{E_1 \cap E_2} 
                                                   \\[8pt]
      \interpretationFunctor{(\enc{E_1 \cup E_2})} & \tag{\text{Table }\ref{table:cr encoding}}{=} \interpretationFunctor{(\copier[-][1] \! \seq[-] (\enc{E_1} \tensor[-] \enc{E_2}) \seq[-] \! \cocopier[-][1])} \\
                                                   &\tag{\ref{fig:semantics}}{=} \interpretationFunctor{(\copier[-][1])} \seq[-] (\interpretationFunctor{(\enc{E_1})} \tensor[-] \interpretationFunctor{(\enc{E_2})}) \seq[-] \interpretationFunctor{(\cocopier[-][1])} \\
                                                   & \tag{\ref{fig:semantics}}{=} \copier[-][X] \seq[-] (\interpretationFunctor{(\enc{E_1})} \tensor[-] \interpretationFunctor{(\enc{E_2})}) \seq[-] \cocopier[-][X] \\
                                                   &\tag{\text{Ind. hyp.}}{=} \copier[-][X] \seq[-] (\dsemRel{E_1} \tensor[-] \dsemRel{E_2}) \seq[-] \cocopier[-][X] \\
                                                   & \tag{\ref{eq:def:cup}}{=} \dsemRel{E_1} \cup \dsemRel{E_2} \\
                                                   &\tag{\ref{eq:sematicsExpr}}{=}  \dsemRel{E_1 \cup E_2} 
                                                   \\[8pt]                                                     
      \interpretationFunctor{(\enc{\op{E}})}       & \tag{\text{Table }\ref{table:cr encoding}}{=} \interpretationFunctor{(\op{(\enc{E})})} \\
                                                   &\tag{\text{Lemma }\ref{lm:opfunctor}}{=} \op{(\interpretationFunctor{(\enc{E})})} \\
                                                   &\tag{\text{Ind. hyp.}}{=} \op{\dsemRel{E}} \\
                                                   &\tag{\ref{eq:sematicsExpr}}{=} \dsemRel{\op{E}} 
                                                   \\[8pt]
      \interpretationFunctor{(\enc{\nega{E}})}     & \tag{\text{Table }\ref{table:cr encoding}}{=} \interpretationFunctor{(\nega{(\enc{E})})} \\
                                                   &{=}\interpretationFunctor{(\op{(\rla{(\enc{E})})})} \tag{Definition of $\nega{(\cdot)}$}  \\
                                                   &\tag{\text{Lemmas }\ref{lm:opfunctor}, \ref{lm:adjfunctor}}{=} \op{(\rla{\interpretationFunctor{(\enc{E})}})} \\
                                                   &\tag{\text{Ind. hyp.}}{=} \op{(\rla{(\dsemRel{E})})} \\
                                                   &= \nega{\dsemRel{E}} \tag{Definition of $\nega{(\cdot)}$} \\
                                                   &\tag*{(\ref{eq:sematicsExpr})\qedhere}{=} \dsemRel{\nega{E}}
  \end{align*}}
\end{proof}

\begin{proof}[Proof of Proposition \ref{prop:Quine}]
  The proof goes by induction on the typing rules. For the base cases we have the following:
  \begin{itemize}
  \item $I\colon 2$.  By definition $\dsemRel{I}=\{\tau \mid \tau_1=\tau_2\}$ and $\interpretationFunctorP (\enc{I})=\{((x_1,x_2),\star)\mid x_1=x_2\}$. Thus  $\dsemRel{I} =\{ \tau \mid ((\tau_1,\tau_2),\star)\in \interpretationFunctorP(\enc{I})\}$.
  \item $R\colon n$. Assume $\ari(R)=n$. By definition $\dsemRel{R} = \{\tau \mid (\tau_1, \dots, \tau_{n}) \in \rho(R)\}$ and $\interpretationFunctorP (\enc{R})=\{((x_1,\dots, x_n),\star)\mid (x_1, \dots x_n) \in \rho(R)\}$. Thus $\dsemRel{R} = \{ \tau \mid ((\tau_1,\dots, \tau_n),\star)\in \interpretationFunctorP(\enc{R})\}$.
  \end{itemize}
  The inductive cases follow always the same argument. We report below only the most interesting ones.
  \begin{itemize}
  \item $P_1 \cap P_2$. Assume $P_1\colon n$, $P_2 \colon m$ and $n\geq m$. %
  \begin{align*}
    \dsemRel{P_1 \cap P_2} 
  =\;\; &\dsemRel{P_1} \cap \dsemRel{P_2} \tag{Definition of $\dsemRel{\cdot}$}\\
  =\;\; &\begin{aligned}
      &\{ \tau \mid ((\tau_1, \dots, \tau_n),\star)\in\interpretationFunctorP(\enc{P_1}) \} \\[-5px]
      & \qquad \cap  \{\tau \mid ((\tau_1, \dots, \tau_m),\star)\in\interpretationFunctorP(\enc{P_2})\}
    \end{aligned} \tag{Ind. hyp.}\\
  =\;\; &\begin{aligned}
    &\{ \tau \mid ((\tau_1, \dots, \tau_n),\star)\in\interpretationFunctorP(\enc{P_1}) \\[-5px]
    &\qquad\wedge ((\tau_1, \dots, \tau_m),\star)\in\interpretationFunctorP(\enc{P_2})\}
  \end{aligned} \\
  =\;\; &\{ \tau \mid ((\tau_1, \dots, \tau_n),\star)\in \interpretationFunctorP(\enc{P_1 \cap P_2} \} \tag{Definition of $\enc{\cdot}$ and $\interpretationFunctorP(\cdot)$}
  \end{align*}
  \item $\mathbf{p}P\colon 2$. Assume $P\colon 1$.
  \begin{align*}
  \dsemRel{\mathbf{p}P} &=  \{\tau  \mid \tau_2,\tau_1,\tau_3, \tau_4 \dots \in \dsemRel{P}\} \tag{Definition of $\dsemRel{\cdot}$}\\
  &=  \{\tau  \mid \tau_2,\tau_1,\dots \in \{ \tau \mid (\tau_1,\star)\in \interpretationFunctorP(\enc{P})\} \; \; \}  \tag{Ind. hyp.}\\
  &=  \{\tau  \mid  (\tau_2,\star) \in \interpretationFunctorP(\enc{P})\}   \\
  & =  \{\tau  \mid ((\tau_1,\tau_2),\star) \in \interpretationFunctorP(\enc{\mathbf{p}P}) \} \tag{Definition of $\enc{\cdot}$ and $\interpretationFunctorP(\cdot)$}
  \end{align*}
  \item $]P\colon 0$. Assume $P\colon 0$. 
  \begin{align*}
  \dsemRel{]P} &=  \{\tau  \mid \tau_2,\tau_3,\dots \in \dsemRel{P}\} \tag{Definition of $\dsemRel{\cdot}$}\\
  &=  \{\tau  \mid \tau_2,\tau_3,\dots \in \{ \tau \mid (\star,\star)\in \interpretationFunctorP(\enc{P})\} \; \; \}  \tag{Ind. hyp.}\\
  &=  \{\tau  \mid  (\star,\star) \in \interpretationFunctorP(\enc{P})\}   \\
  & =  \{\tau  \mid (\star,\star) \in \interpretationFunctorP(\enc{]P}) \} \tag*{(Definition of $\enc{\cdot}$ and $\interpretationFunctorP(\cdot)$)\qedhere}
  \end{align*}
  \end{itemize}
  
  \end{proof}

\section{Some well known facts about chains in a lattice}\label{app:continuous} 
A \emph{chain} on a complete lattice $(L, \sqsubseteq)$ is a family $\{x_i\}_{i\in I}$ of elements of $L$ indexed by a linearly ordered set $I$ such that $x_i \sqsubseteq x_j$ whenever $i \leq j$.
A monotone map $f\colon L \to L$ is said to \emph{preserve chains} if
\[f(\bigsqcup_{i\in I}x_i) = \bigsqcup_{i \in I}f(x_i)\]
We write $id \colon L \to L$ for the identity function and $f \sqcup g\colon L \to L$ for the pointwise join of $f\colon L \to L$ and $g\colon L \to L$, namely $f\sqcup g (x) \defeq f(x) \sqcup g(x)$ for all $x\in L$.  For all natural numbers $n \in \nat$, we define $f^n\colon L \to L$ inductively as $f^0 =id$ and $f^{n+1} = f^n;f$. We fix $f^\omega \defeq \bigsqcup_{n\in \nat} f^n$.

\begin{lem}\label{lemma:dchaindecomposition}
Let $f,g \colon L \to L$ be monotone maps preserving  chains. Then
\begin{enumerate}
\item $id\colon L \to L$ preserves   chains;
\item $f \sqcup g\colon L \to L$ preserves   chains;
\item $f^\omega \colon L \to L$ preserves   chains.
 \end{enumerate}
\end{lem}
\begin{proof}
\begin{enumerate}
\item Trivial.
\item By hypothesis we have that
$f(\bigsqcup_{i\in I}x_i) = \bigsqcup_{i \in I}f(x_i)$ and $g(\bigsqcup_{i\in I}x_i) = \bigsqcup_{i \in I}g(x_i)$.
Thus
\begin{align*}
			 f \sqcup g (\bigsqcup_{i\in I}x_i) &= f(\bigsqcup_{i\in I}x_i) \sqcup g(\bigsqcup_{i\in I}x_i)   \\
			&= \bigsqcup_{i \in I}f(x_i) \sqcup \bigsqcup_{i \in I}g(x_i)  \\
			&= \bigsqcup_{i \in I}( f(x_i) \sqcup g(x_i) )  \\
			&=  \bigsqcup_{i \in I}(f\sqcup g)(x_i)
\end{align*}
\item We prove $f^n(\bigsqcup_{i\in I}x_i) = \bigsqcup_{i\in I}f^n(x_i)$ for all $n \in \nat$. We proceed by induction on $n$.

For $n=0$, $f^0(\bigsqcup_{i\in I}x_i) = \bigsqcup_{i\in I}x_i = \bigsqcup_{i\in I}f^0(x_i)$.

For $n+1$, we use the hypothesis that $f$ preserves   chain and thus
\begin{align*}
			f^{n+1}((\bigsqcup_{i\in I}x_i) &= f (\; f^{n+1}((\bigsqcup_{i\in I}x_i)\; )   \\
			&= f (\,\bigsqcup_{i \in I}f^n(x_i) \,) \tag{induction hypothesis}  \\
			&=  \bigsqcup_{i \in I}f (\, f^n(x_i) \,)\\
			&=  \bigsqcup_{i \in I}f^{n+1}(x_i) \tag*{\qedhere}
\end{align*}
\end{enumerate}

\end{proof}

\begin{lem}\label{lemma:fomega2}
Let $f,g\colon L \to L$ be monotone maps preserving   chains such that $g\sqsubseteq f$. Then $f^\omega ; g \sqsubseteq f^\omega$
\end{lem}
\begin{proof}
For all $x\in L$, $f^\omega ; g (x) = g (\bigsqcup_{n\in \nat} f^n(x)) = \bigsqcup_{n \in \nat}g( f^n(x)) \sqsubseteq \bigsqcup_{n \in \nat} f^{n+1}(x) \sqsubseteq \bigsqcup_{n \in \nat} f^n(x) =f^\omega (x)$.
\end{proof}

\begin{lem}\label{lemma:omegaomega}
Let $f\colon L \to L$ be a monotone map preserving   chains. Thus $f^\omega = f^\omega ; f^\omega$
\end{lem}
\begin{proof}
$f^\omega = f^\omega ; id \sqsubseteq f^\omega ; f^\omega$.
For the other direction we prove that $f^\omega ; f^n \sqsubseteq f^\omega$ for all $n \in \nat$.  We proceed by induction on $n$. For $n=0$ is trivial. For $n+1$, $f^\omega ; f^{n+1} = f^\omega ; f^ n ; f \sqsubseteq f^\omega ; f \sqsubseteq f^\omega$. For the last inequality we use Lemma \ref{lemma:fomega2}.
\end{proof}

\begin{lem}\label{lemma:fg}
Let $f,g\colon L \to L$ be monotone maps preserving   chains. Then $(f\sqcup g)^\omega = (f^\omega \sqcup g)^\omega$
\end{lem}
\begin{proof}
Since $f=f^1 \sqsubseteq f^\omega$ and since $(\cdot)^\omega$ is monotone, it holds that $(f\sqcup g)^\omega \sqsubseteq (f^\omega \sqcup g)^\omega$.

For the other inclusion, we prove that $(f^\omega \sqcup g)^n \sqsubseteq (f\sqcup g)^\omega$ for all $n \in \nat$. We proceed by induction on $n\in \nat$.
For $n=0$, $(f^\omega \sqcup g)^0 = id \sqsubseteq (f\sqcup g)^\omega$.

For $n+1$, observe that $f^\omega \sqsubseteq (f\sqcup g)^\omega $  and than $g\sqsubseteq (f\sqcup g)^\omega$. Thus
\begin{equation}\label{eq:prooflattice}
(f^\omega \sqcup g) \sqsubseteq (f\sqcup g)^\omega
\end{equation}
We conclude with the following derivation.
\begin{align*}
(f^\omega \sqcup g)^{n+1} &= (f^\omega \sqcup g)^{n} ; (f^\omega \sqcup g) \\
& \sqsubseteq  (f\sqcup g)^\omega ; (f^\omega \sqcup g) \tag{Induction Hypothesis}\\
& \sqsubseteq  (f\sqcup g)^\omega ; (f\sqcup g)^\omega   \tag{\eqref{eq:prooflattice}} \\
& =  (f\sqcup g)^\omega \tag*{(Lemma \ref{lemma:omegaomega})\qedhere}
\end{align*}
\end{proof}

\subsection{Some well known facts about precongruence closure}\label{app:onprec}

Let $X = \{X[n,m]\}_{n,m\in \nat}$ be a family of sets indexed by pairs of natural numbers $(n,m)\in \nat \times \nat$. A well-typed relation $\mathbb{R}$ is a family of relation $\{R_{n,m}\}_{n,m\in \nat}$ such that each $R_{n,m} \subseteq X[n,m] \times X[n,m]$. We consider the set $\WTRel_X$ of well typed relations over $X$. It is easy to see that $\WTRel_X$ forms a complete lattice with join given by union $\cup$. Hereafter we fix an arbitrary well-typed relation $\mathbb{I}$ and the well-typed identity relation $\Delta$.

We define the following monotone maps for all $\mathbb{R} \in \WTRel_X$:
\begin{itemize}
\item $(id)\colon \WTRel_X \to \WTRel_X$ defined as the identity function;
\item $(\mathbb{I}) \colon \WTRel_X \to \WTRel_X$ defined as the constant function $\mathbb{R}\mapsto\mathbb{I}$;
\item $(r) \colon \WTRel_X \to \WTRel_X$ defined as the constant function $\mathbb{R}\mapsto\Delta$;
\item $(t) \colon \WTRel_X \to \WTRel_X$ defined as $\mathbb{R}\mapsto\{(x,z) \mid \exists y. (x,y)\in \mathbb{R} \wedge (y,z)\in \mathbb{R}\}$;
\item $(s) \colon \WTRel_X \to \WTRel_X$ defined as $\mathbb{R}\mapsto\{(x,y) \mid (y,x)\in \mathbb{R}\}$;
\item $(\seq[]) \colon \WTRel_X \to \WTRel_X$ defined as $\mathbb{R}\mapsto\{(x_1\seq[]y_1, x_2 \seq[] y_2) \mid (x_1,x_2)\in \mathbb{R} \wedge (y_1,y_2)\in \mathbb{R}\}$;
\item $(\tensor[]) \colon \WTRel_X \to \WTRel_X$ defined as $\mathbb{R}\mapsto\{(x_1\tensor[]y_1, x_2 \tensor[] y_2) \mid (x_1,x_2)\in \mathbb{R} \wedge (y_1,y_2)\in \mathbb{R}\}$;
\end{itemize}
\noindent 
Observe that the function $(id)$, $(r)$, $(t)$, $(\seq[])$ and $(\tensor[])$ are exactly the inference rules used in the definition of $\pcong{\cdot}$ given in \eqref{eq:pc}. Indeed the function
$\pcong{\cdot}\colon \WTRel_X \to \WTRel_X$ can be decomposed as
\[\pcong{\cdot} = (\, (id) \cup (r) \cup (t) \cup (\seq[]) \cup (\tensor) \,)^\omega\]
where $f^\omega$ stands the $\omega$-iteration of a map $f$ defined in the standard way (see Appendix \ref{app:continuous} for a definition).

Similarly the congruence closure $\congr{\cdot}\colon \WTRel_X \to \WTRel_X$ can be decomposed as \[\congr{\cdot} = (\, (id) \cup (r) \cup (t) \cup (s) \cup (\seq[]) \cup (\tensor) \,)^\omega\]
These decompositions allow us to prove several facts in a modular way. For instance, to prove that $\pcong{\cdot}$ preserves   chains is enough to prove the following.

\begin{lem}\label{lemma:componentspreserves}
The monotone maps $(id)$, $(\mathbb{I})$, $(r)$, $(s)$, $(t)$, $(\seq[])$ and $(\tensor[])$ defined above preserve   chains.
\end{lem}
\begin{proof}
All the proofs are straightforward, we illustrate as an example the one for $(\tensor)$.

Let $I$ be a linearly ordered set and $\{\mathbb{R}_i\}_{i\in I}$ be a family of well-typed relations such that if $i\leq j$, then $R_i\subseteq R_j$.
We need to prove that $(\tensor) (\bigcup_{i\in I}\mathbb{R}_i) = \bigcup_{i \in I}(\tensor)(\mathbb{R}_i)$.

The inclusion $(\tensor) (\bigcup_{i\in I}\mathbb{R}_i) \supseteq \bigcup_{i \in I}(\tensor)(\mathbb{R}_i)$ trivially follows from monotonicity of $(\tensor)$ and the universal property of union. For the inclusion $(\tensor) (\bigcup_{i\in I}\mathbb{R}_i) \subseteq \bigcup_{i \in I}(\tensor)(\mathbb{R}_i)$, we take an arbitrary $(a,b)\in (\tensor) (\bigcup_{i\in I}\mathbb{R}_i)$. By definition of $(\tensor)$, there exist $x_1,x_2,y_1,y_2$ such that
\[a=x_1 \tensor[] y_1 \quad b = x_2 \tensor[] y_2 \quad (x_1,x_2)\in \bigcup_{i\in I}\mathbb{R}_i \quad (y_1,y_2)\in \bigcup_{i\in I}\mathbb{R}_i\]
By definition of union, there exist $i,j\in I$ such that $(x_1,y_1)\in R_i$ and $(x_2,y_2)\in R_j$.
Since $I$ is linearly ordered, there are two cases: either $i\leq j$ or $i \geq j$.

If $i\leq j$, then $R_i \subseteq R_j$ and thus $(x_1,y_1)\in R_j$. By definition of $(\tensor[])$, we have $(x_1 \tensor x_2, y_1 \tensor y_2)\in R_j$ and thus $(a,b)\in R_j$. Since $R_j \subseteq \bigcup_{i\in I}\mathbb{R}_i$, then $(a,b) \in \bigcup_{i\in I}\mathbb{R}_i$. The case for $j\leq i$ is symmetric.
\end{proof}

\begin{prop}
The monotone maps $\pcong{\cdot},\congr{\cdot}\colon \WTRel_X \to \WTRel_X$ preserve   chains.
\end{prop}
\begin{proof}
Follows immediately from Lemma \ref{lemma:componentspreserves} and Lemma \ref{lemma:dchaindecomposition} in Appendix \ref{app:continuous}.
\end{proof}

\begin{lem}\label{lemma:pcJcont}
For all well-typed relations $\mathbb{J}$, the  map $\pcong{\mathbb{J} \cup \cdot}\colon \WTRel_X \to \WTRel_X$ preserves   chains.
\end{lem}
\begin{proof}
Follows immediately from Lemma \ref{lemma:componentspreserves} and Lemma \ref{lemma:dchaindecomposition} in Appendix \ref{app:continuous}.
\end{proof}

\begin{lem}\label{lemma:pcongdis}
For all well-typed relations $\mathbb{I}$ and $\mathbb{J}$,
$\pcong{\mathbb{I} \cup \mathbb{J}} =  \pcong{\pcong{\mathbb{I}} \cup \mathbb{J}}$
\end{lem}
\begin{proof}
Let $(\mathbb{J}) \colon \WTRel_X \to \WTRel_X $ be the constant function to $\mathbb{J}$ and define
$f,g\colon \WTRel_X$\\
$ \to \WTRel_X$ as
\[f \defeq   (id) \cup (r) \cup (t) \cup (\seq[]) \cup (\tensor)  \qquad g \defeq (\mathbb{J})\]
From Lemma \ref{lemma:componentspreserves} and Lemma \ref{lemma:dchaindecomposition}, both $f$ and $g$ preserve   chains.
Observe that $f^\omega(\mathbb{I}) = \pcong{\mathbb{I}}$, that $(f\cup g)^\omega = \pcong{\mathbb{I} \cup\mathbb{J}}$ and that $(f^\omega \cup g)^\omega(\mathbb{I})=\pcong{\pcong{\mathbb{I}} \cup\mathbb{J}}$.  Conclude with Lemma \ref{lemma:fg} in Appendix \ref{app:continuous}.
\end{proof}

\begin{lem} Let $\mathbb{T}=(\Sigma, \mathbb{I})$ be a first-order theory.
Then $\syninclusionT{\mathbb{T}} = \pcong{\mathbb{FOB} \cup \mathbb{I}}$
\end{lem}
\begin{proof}
By definition  $\syninclusionT{\mathbb{T}} = \pcong{ \syninclusion \cup \mathbb{I}}$. Recall that $\syninclusion = \pcong{\mathbb{FOB}}$. Thus $\syninclusionT{\mathbb{T}} = \pcong{ \pcong{\mathbb{FOB}} \cup \mathbb{I}}$. By Lemma \ref{lemma:pcongdis}, $\syninclusionT{\mathbb{T}} = \pcong{ \mathbb{FOB} \cup \mathbb{I}}$.

\end{proof}

\begin{lem}\label{lemma:noncont}
Let $I$ be a linearly ordered set and, for all $i \in I$, let $\mathbb{T}_i=(\Sigma, \mathbb{I}_i)$ be first-order theories such that if $i\leq j$, then $\mathbb{I}_i \subseteq \mathbb{I}_j$. Let $\mathbb{T}$ be the theory $(\Sigma, \bigcup_{i\in I}\mathbb{I}_i)$.
Then $\syninclusionT{\mathbb{T}} = \bigcup_{i \in I} \syninclusionT{\mathbb{T}_i}$.
\end{lem}
\begin{proof}
By definition $\syninclusionT{\mathbb{T}} = \pcong{\syninclusion \cup \bigcup_{i\in I}\mathbb{I}_i}$. Since $\mathbb{I}_i $ form a   chain, by Lemma \ref{lemma:pcJcont},\newline\indent $\pcong{\syninclusion \cup \bigcup_{i\in I}\mathbb{I}_i} = \bigcup_{i \in I} \pcong{\syninclusion \cup \mathbb{I}_i} $. The latter is, by definition, $\bigcup_{i \in I} \syninclusionT{\mathbb{I}_i}$.
\end{proof}

\begin{lem}\label{lemma:noncontsyntax}
Let $I$ be a linearly ordered set and, for all $i \in I$, let $\mathbb{T}_i=(\Sigma_i, \mathbb{I})$ be first-order theories such that if $i\leq j$, then $\Sigma_i \subseteq \Sigma_j$. Let $\mathbb{T}$ be the theory $(\bigcup_{i\in I} \Sigma_i, \mathbb{I})$.
Then $\syninclusionT{\mathbb{T}} = \bigcup_{i \in I} \syninclusionT{\mathbb{T}_i}$.
\end{lem}
\begin{proof}
By Lemma \ref{lemma:componentspreserves}, the monotone map $\pcongr{\cdot} \defeq (\, (id) \cup (\mathbb{I}) \cup (t) \cup (\seq[]) \cup (\tensor) \,)^\omega$ preserves chains.
Let $\Delta_i$ be the well-typed identity relation on $\LCB[\Sigma_i]$.
Observe that $\syninclusionT{\mathbb{T}_i} = \pcongr{\Delta_i}$ and that $\syninclusionT{\mathbb{T}}= \pcongr{\bigcup_{i\in I} \Delta_i}$.  To summarise:
\begin{align*}
\syninclusionT{\mathbb{T}} & = \pcongr{\bigcup_{i\in I} \Delta_i}\\
&=\bigcup_{i\in I}  \pcongr{\Delta_i} \tag{preserve chains}\\
& = \bigcup_{i\in I}  \syninclusionT{\mathbb{T}_i} \tag*{\qedhere}
\end{align*}
\end{proof}

\begin{lem}\label{lemma:noncontalltogether}
Let $I$ be a linearly ordered set and, for all $i \in I$, let $\mathbb{T}_i=(\Sigma_i, \mathbb{I}_i)$ be first-order theories such that if $i\leq j$, then $\Sigma_i \subseteq \Sigma_j$ and $\mathbb{I}_i\subseteq\mathbb{I}_j$. Let $\mathbb{T}$ be the theory $(\bigcup_{i\in I} \Sigma_i, \bigcup_{i\in I} \mathbb{I}_i)$.
Then $\syninclusionT{\mathbb{T}} = \bigcup_{i \in I} \syninclusionT{\mathbb{T}_i}$.
\end{lem}
\begin{proof}
Immediate by Lemma \ref{lemma:noncontsyntax} and Lemma \ref{lemma:noncont}.
\end{proof}

\end{document}